\documentclass[prb,twocolumn,10,aps,superscriptaddress,longbibliography]{revtex4-1}
\usepackage[T1]{fontenc}
\usepackage[latin9]{inputenc}
\usepackage{color}
\usepackage{float}
\usepackage{amsmath}
\usepackage{amssymb}
\usepackage{graphicx}
\usepackage{esint}
\usepackage[colorlinks,bookmarks=false,citecolor=blue,linkcolor=blue,urlcolor=blue,hyperfootnotes=true]{hyperref}

\makeatletter
\pdfoutput = 1
\usepackage{amsfonts}
\usepackage[normalem]{ulem}

\newcommand{\bd}{\bm}
\newcommand{\sgn}{\text{sgn}}

\newcommand{\pf}{2{\bf {p}}_{\text{F}}}
\newcommand{\qv}{{\bf {q}}}
\newcommand{\kv}{{\bf {k}}}
\newcommand{\pv}{{\bf {P}}}
\newcommand{\kf}{{\bf {k}}_{\text{F}}}

\newcommand{\vk}{{\bf{v}}_{{\bf {k}}}}
\newcommand{\vf}{{\bf{v}}_{\text{F}}}

\newcommand{\nb}{{n}_{\text{B}}}

\newcommand{\xa}{\text{\tiny $\parallel$}}
\newcommand{\xe}{\text{\tiny $\perp$}}

\makeatother

\begin{document}

\title{Effective SU(2) theory for the pseudogap state}

\author{X.\ Montiel}
\affiliation{IPhT, L'Orme des Merisiers, CEA-Saclay, 91191 Gif-sur-Yvette, France }
\affiliation{Department of Physics, Royal Holloway, University of London, Egham,
Surrey TW20 0EX, United Kingdom}

\author{T. Kloss}
\affiliation{IPhT, L'Orme des Merisiers, CEA-Saclay, 91191 Gif-sur-Yvette, France }
\affiliation{INAC-PHELIQS, Universit\'{e} Grenoble Alpes and CEA, 38000 Grenoble, France}

\author{C.P\'{e}pin}
\affiliation{IPhT, L'Orme des Merisiers, CEA-Saclay, 91191 Gif-sur-Yvette, France }

\begin{abstract}
This paper exposes in a detailed manner the recent findings about
the SU(2) scenario for the underdoped phase of the Cuprate superconductors.
The SU(2) symmetry is formulated as a rotation between the $d$-wave
SC phase and a $d$-wave charge order. We define
the operators responsible for the SU(2) rotations and we derive the non-linear $\sigma$-model associated
with it. In this framework, we demonstrate that SU(2)
fluctuations are massless in finite portions of the Brillouin Zone
corresponding to the anti-nodal regions ($0,$$\pi$), ($\pi,0$).
We argue that the presence of SU(2) fluctuations in the anti-nodal
region leads to the opening of Fermi arcs around the Fermi surface
and to the formation of the pseudo-gap. Moreover,
we show that SU(2) fluctuations lead, in turn, to the emergence of
a finite momentum SC order -or Pair Density Wave (PDW)- and more importantly
to a new kind of excitonic particle-hole
pairs liquid, the Resonant Excitonic State (RES), which is made of
patches of preformed particle-hole pairs with multiple momenta. When
the RES liquid becomes critical, we demonstrate that
electronic scattering through the critical modes leads to anomalous
transport properties. This new finding can account
for the Strange Metal (SM) phase at finite temperature, on the right
hand side of the SC dome, shedding light on another notoriously mysterious
part of the phase diagram of the Cuprates. 
\end{abstract}
\maketitle

\tableofcontents

\makeatletter
\let\toc@pre\relax
\let\toc@post\relax
\makeatother 

\section{Introduction }

When doping a Mott insulator, the systems becomes a superconductor
at high temperature. This phenomenon remains one of the most enduring
mysteries of material science . The origin of the pseudo-gap (PG)
phase \cite{Alloul89,Warren89}, which shows a loss of electronic
density of states at finite temperatures above the SC state, in the
underdoped regime, has generated some intense debate in the past thirty
years, and still remains an open issue \cite{Norman03,Carlson:2004hn,Lee06,LeHur:2009iw,Rice12,Norman14,Keimer:2015vy,Carbotte:2011ip,Eschrig:2006ky,Fradkin:2015ch}.
The mystery of the PG phase is maybe better seen within the Angle
Resolved Photo Emission (ARPES) measurements, in which we observe
a continuous evolution from small hole pockets at low oxygen doping
$x<0.05$, to Fermi arcs at intermediate doping ( or underdoped region)
$0.08<x<0.19$, to finally the opening of a larger Fermi surface in
the over-doped region $0.20<x$. The notion of Fermi ``arcs'' instead
of closed Fermi surface of electrons has a ground-breaking character
because it breaks the Luttinger theorem relating the counting of the
conduction electrons with the ``volume'' of the Fermi surface \cite{Campuzano98,Campuzano:1999ed,Vishik12,Vishik14,Shen:2005ir,Kaminski:2002wl}.
The theories of the PG can be divided into two major lines of thoughts.
In the first line of thoughts the emphasis is given to the proximity
to the Mott insulator at zero doping ($x=0$) , and argue that the
considerable strength of the Coulomb interactions for these systems
produce strong correlations between the electrons, from the scale
of $1eV$ down to the lowest energy scales \cite{Lee06}. Exotic states
are created, the most notoriously famous of them being the Resonating
Valence Bond (RVB) state proposed in the early days, just after the
discovery of the YBCO \cite{Anderson87,Kotliar88a,Kotliar88b,Lee98}.
This approach has also lead to many numerical advances including the
celebrated Dynamical Mean Field Theory (DMFT) \cite{Georges96,Gull:2013hh,Sordi2012,Sordi:2012jc,Gull:2014bl},
designed to capture the proximity to the Mott transition, as well
as field theory treatment including gauge field\cite{Nagaosa:1990ut,Nagaosa:1992vj,Ferraz:2015voa,Franz:2001hk,Tesanovic:2002il,Castellani94,Tsvelik07,Chubukov:2007gk,Yang:2006eq},
with U(1), or SU(2) symmetries \cite{Lee06,Kotliar88a,Kotliar88b}.
The second type of theories assumes the existence of a singularity
in the phase diagram, for example with the presence of a quantum critical
point (QCP) -also called a zero temperature phase transition, where
the quantum fluctuations dominate the thermal ones \cite{Barzykin:1995fr,abanov03,Onufrieva:2000tk,Onufrieva:2002uf,Pines2002,Sachdev98,Sur:2014kd,Sur:2015fi,Schattner:2016dw,Gerlach:2016ux,Lederer14,Wang:2016tr,Dicastro96a}. While the correlations
between electrons are not very strong at the UV scale, they drastically
grow when the temperature is reduced, leading to a strong coupling
in the vicinity of the QCP.

The importance of phase fluctuations for small hole concentration
when approaching the Mott transition, was outlined in a seminal study
of the under-doped regime of Cuprates \cite{EmeryVJ:1995dr}. The
main argument is simply that when the electron density gets locked
at the brink of localization, the phase fluctuates within the phase-density
duality relation. Three types of fluctuations were identified : the
quantum phase fluctuations arising the Heisenberg uncertainty principle,
the classical- thermal- phase fluctuations, and the fluctuations of
the amplitude of the order parameter promoted by some extra degree
of freedom. This line of approach was explored in details in the ``preformed
pairs'' scenario, where Cooper pairs are forming at a temperature
$T>T_{c}$, with the phase coherence setting precisely at $T_{c}$
\cite{NormanKanigel07,Banerjee:2011bz,Banerjee:2011cu,Norman:1995dd,Franz:1998et,Kampf:2003kx},
as well as in scenarios involving phase separation in real space with,
for example, the formation of stripes \cite{Zaanen89,Zaanen:1998cl,Machida1989192,Machida1990,Emery99,Kivelson:1998ir,Kivelson:2002ug,Kivelson03,Chakraborty:2014iq}.
It has to be noticed that a scenario has already attributed the opening
of the PG to fluctuating charge order\cite{Dicastro96a}. In this
scenario, the Cooper pairing enable the opening of the PG in the anti-nodal
region, allowing the formation of the Fermi arcs in the nodal part
of the Fermi surface.

Despite very intense and focused experimental search, preformed pairs
were not observed at the PG energy scale $T^{*}$, and phase fluctuations
were found only in a window of $15K$ above $T_{c}$ \cite{Corson:1999bo,Alloul:2010ko,Benfatto:2007df,Benfatto00}.
A question then naturally arises : where has gone the enormous amount
of classical phase fluctuations that should be present in the under-doped
regime ?

In this paper, we argue that a new type of pairing
fluctuations has to be considered in the under-doped region, governed
by an emergent SU(2) symmetry which rotates the superconducting state
towards the charge sector \cite{Metlitski10b,Hayward14,Efetov13,Kloss15,Kloss:2016hu,Chowdhury:2014cp,Chowdhury:2015gx}.
Within the SU(2) paradigm, pairing fluctuations do not only involve
the phase of the U(1) superconducting order parameter, but also ''pairing''
fluctuations towards the charge sector as well as charge phase fluctuations.
For example these operators can rotate a Pair Density Wave (PDW) -
or finite momentum superconducting order, into a charge density wave
(CDW) state with the same wave vector, as was recently reported \cite{Hamidian16},
but it can also rotate a standard $d$-wave superconducting state
into a new kind of excitonic state. Support for the concept of an
underlying SU(2) symmetry in the background of the under-doped region
comes from the recent findings of CDW in the phase diagram of the
cuprates, and subsequent theoretical investigations over this findings
\cite{Allais14c,Atkinson15,Atkinson:2015kk,Meier13,Meier14,Einenkel14,Agterberg:2014wf,Wang14,Wang15a,Wang15b,Wang15c,Pepin14,Freire:2015kg,Caprara:2016gs,Caprara:2016vh}.
This started around a decade ago with a first observation of modulations
inside vortices in Bi2212 \cite{Hoffman02,Fujita14}. Subsequent studies
with Fermi surface reconstruction showed that this feature was generic
\cite{Vershinin:2004gk,daSilvaNeto:2014vy}, also verified in Bi-2201
\cite{He14,Wise08}, and that the charge patterns corresponded to
two axial wave vectors $\left(0,Q_{y}\right)$ and $\left(Q_{x},0\right)$,
incommensurate with the lattice periodicity, and which magnitude of
the wave vectors growing with oxygen-doping. Quantum Oscillations
in YBCO \cite{Doiron-Leyraud07,LeBoeuf07}, NMR \cite{Wu11,Wu13a,Wu14}
and \cite{Chang16} X-rays studies, hard \cite{Chang12,Blackburn13a}
and soft \cite{Ghiringhelli12,LeTacon11,Blanco-Canosa13,Blanco-Canosa14},
provided a new understanding in the nature of the charge ordering,
as a reasonably long ranged excitation ($\sim20\,a_{0}$
where $a_{0}$ is the elementary cell parameter of
the square lattice) stabilized to a true long range order upon a
magnetic field larger than $17$T\cite{Gerber:2015gx,Chang16}.

Maybe the strongest suggestion that d-wave charge order and SC are
mysteriously related by a symmetry, comes from the phase diagram showing
the response of charge ordering as a function of temperature and magnetic
field, in the underdoped region \cite{Doiron-Leyraud07,Sebastian10,Wu11,LeBoeuf13,Gerber:2015gx,Chang16}.
Similar energy scales are observed for both orders, with a sharp (
and flat) transition at $H_{0}=17$T, very suggestive of a ``spin-flop''-type
transition between the two states.

Bulk probe spectroscopies also hint towards the presence of a collective
mode in the underdoped phase of the cuprates. It has been argued that
the $A_{1g}$-mode in Raman scattering \cite{Gallais04,Blanc:2009vo,Blanc:2010tm,Sacuto13,Sacuto2015TS,SacutoSidis02}
can be associated with the presence of SU(2) symmetry \cite{Montiel15a,Chubukov06}.
Likewise a theory \cite{Eschrig:2006ky,Eschrig:2000bf,Chubukov:479596,Norman07,NormanChub01}
for the PG state shall address the long standing observation by
Inelastic Neutron Scattering (INS) of a finite energy resonance around
the AF wave vector $\left(\pi,\pi\right)$ in both the SC and PG states
of those compounds \cite{Rossat,Sidis2001,Hinkov07,Hinkov04,Greven2016,Chan:2016vk,Bourges2005}.

The formation of the PG state is accompanied by $\mathbf{Q}=\mathbf{0}$
orders as observed by INS techniques \cite{Bourges11,Fauque06} and
transport measurements \cite{Cyr-Choiniere15}. These orders have been
interpreted as loop-currents \cite{Bourges11,Fauque06} or nematicity
\cite{Cyr-Choiniere15}, which have led to recent theoretical developments
\cite{Varma06,Aji:2013eo,AjiVarma07,Nie13,Nie:2015bm,Lee:2016jc,Carvalho15,deCarvalho:2014tj}.

Typically the constraint in the non-linear $\sigma$-model associated
with SU(2) fluctuations, creates a strong coupling between the two
channels, which in turn generates phase separation \cite{Schmalian:2000fn,Terletska:2011dd}.
We succinctly describes this situation in the second part of this
paper, with the creation of patches- or droplets, of excitonic particle-hole
pairs. The statistics of such objects is analogous
to the phase separation of polarons in an electronic medium \cite{Micnas:1990ee},
and is also related to the emergence of skyrmions in the pseudo-spin
space, which come out of the non-linear $\sigma$-model. The detailed
link between these approaches is deferred to a future work.

In this paper, the SU(2) symmetry emerges from
short range AF correlations, which is a more realistic starting point
for the phase diagram of the cuprates than our previous study \cite{Efetov13}
where the proximity to an AF QCP was assumed.

Although a few of the essential ideas developed
in this paper have already been introduced elsewhere \cite{Kloss15a}-
like the idea of particle-hole ''droplet'', or excitonic patches,
the detailed calculations behind these ideas have never been presented
so far. The description of the non-linear $\sigma$-model is given
for the first time, directly starting from a realistic short range
AF correlation and a realistic electronic dispersion rather than from
a more idealistic eight-hot-spot model close to an AF QCP. It is shown
that the coupling between the non-linear $\sigma$-model and the underlying
fermions restricts the SU(2) fluctuations to the anti-nodal region
of the BZ, which is a crucial new feature of the theory. The rotation
of the charge ordering wave vectors from the diagonal to the axes
is explained for the first time. The symmetries of the emerging orders,
CDW, PDW are clarified. Moreover the study of the strange metal, and
the implications of our proposal for the PG to anomalous transport
properties in this region of the phase diagram are given here for
the first time.

The paper is organized as follows. In section \ref{sec:The-SU(2)-symmetry},
we introduce the pseudo-spin operators relevant to our study, and
the triplet representation on which they apply, which rotates the
d-wave SC state to a d-wave CDW. In section \ref{sec:The-SU(2)-dome},
we give a mean-field decoupling of a Hamiltonian pertaining to the
solution of cuprate superconductors, which retains mainly short range
AF interactions. The decoupling in the charge and SC channels gives
a degeneracy (at the hot spots) between the two channels, for a wide
range of doping. It defines the temperature scale below which on
can get SU(2) fluctuations. In section \ref{sec:SU(2)-fluctuations-coupled},
we start our study of the fluctuations between the two states, introduce
the effective Lagrangian with its symmetric part and symmetry breaking
part. In section \ref{sec:Non-linear--model} we use the SU(2) symmetric
part of the Lagrangian to perform the integration over the fermionic
degrees of freedom, leading to the standard expression for the non
linear $\sigma$-model. In section \ref{sec:The-SU(2)-line}, we focus
on the symmetry breaking term, and show that massless SU(2) fluctuations
occur only on specific loci of the Brillouin Zone, that we call SU(2)
lines. Everywhere else in the Brillouin zone the fluctuations are
heavily massive. In section \ref{sec:Rotation-of-the}, we start to
study the effect of the SU(2) fluctuations on the charge and SC channels.
We show that SU(2) pairing fluctuations induce a nematic response
and, importantly, tilt the charge ordering modulation wave vector
from the diagonal $\left(Q_{0,}Q_{0}\right)$ to the axes $\left(Q_{0},0\right)$and
$\left(0,Q_{0}\right)$. In section \ref{sec:Effect-of-the}, we discuss
the possibility that SU(2) fluctuations lead to the emergence of preformed
excitonic (particle-hole) pairs owing many $\pf$ wave vectors, whereas
similar study in the SC channels leads to the emergence of a small
Pair Density wave contribution with the same wave vectors $\left(Q_{0},0\right)$and
$\left(0,Q_{0}\right)$ as in the CDW channel. Finally, in section
\ref{sec:Global-phase-diagram}, we depict a global phase diagram
for the physics of the underdoped region of the cuprates using heuristic
arguments from the SU(2) theory. We also study the strange metal regime
at optimal doping and show that our pictures provides very anomalous
transport exponents, with in particular a resistivity going like $\rho\sim T/\log T$
in three spatial dimensions.

\section{The SU(2) symmetry\label{sec:The-SU(2)-symmetry} }

The paradigm of emerging symmetry is not new \cite{Shen:1996hn} and
possibly one of its most famous proponents is the SO(5)-theory for
cuprate superconductors \cite{Demler95,Zhang:1997ew,Zhang:1999tq,Demler04,Demler:1998iw,denHertog:1999dk}
where it was proposed that the d-wave SC state can be rotated into
the AF sector. Thermal fluctuations between the two states, described
by the non-linear $\sigma$-model were shown to become massively dominant
in the under-doped region of the phase diagram and it was suggested
that they were responsible for the formation of the pseudo-gap.

The present study is based on the assumption that an underlying SU(2)
symmetry governs the phase diagram in the underdoped region of the
cuprates. In contrast to SO(5) symmetry described above, the SU(2) symmetry we talk about here  connects the SC  and CDW sectors. This concept of pseudo-spin symmetry is not new and can
be traced back to the Yang and Zhang for Hubbard-model at half filling
\cite{Yang89,Yang:1990cf}. A set of pseudo-spin operators were introduced,
which rotate the d-wave SC state into a d-wave modulated charge order.
The pseudo-spin idea was later used in the context of the d-Density
Wave (DDW) \cite{Nayak00} and nematic states \cite{Kee:2008gw},
using as well the SU(2) pseudo-spin operators in order to rotate the
d-wave SC state towards one of those two. Recently, the ubiquitous
presence of charge excitations in the underdoped region, and the stabilization
of long range CDW in high magnetic fields ($B>17$T ) lead to the
revival of the idea of emerging SU(2) symmetry, and the pseudo-spin
operators in this case rotate the d- SC state towards the charge sector. 

In this section, we give the mathematical definitions
of the pseudo-spin operators of the SU(2) symmetry and describe explicitly
the $l=1$ minimal representation. We rapidly review previous work
on the eight hot-spots model, generalization to the more realistic
model including short range AF correlations are given in the section
\ref{sec:The-SU(2)-dome}.

\subsection{The ``eight hot-spots'' model\label{sub:The-eight-hot-spots}}

\begin{figure}[h]
\includegraphics[width=75mm]{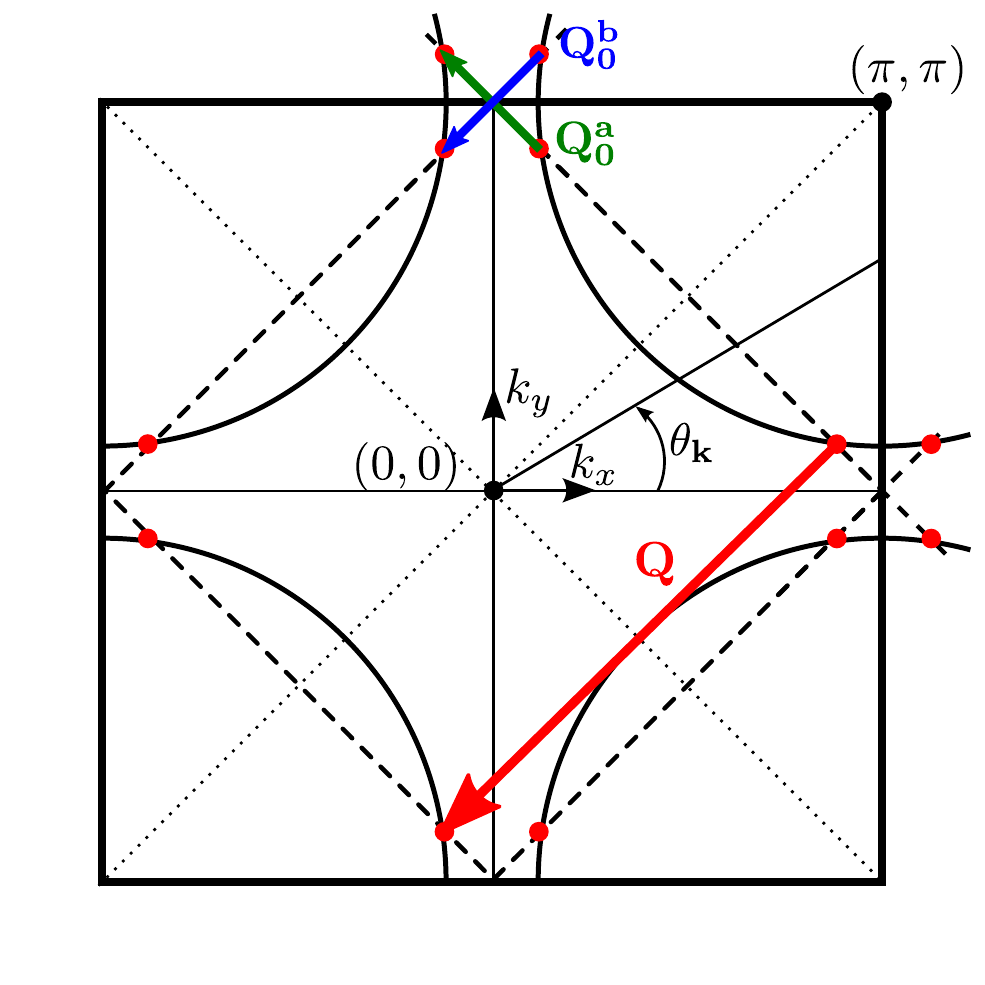} \caption{\label{fig:hs} (Color online)\textcolor{black}{{} Schematic representation
of a hole-doped cuprate Fermi surface in the first BZ. The \textquotedbl{}Hot-Spots\textquotedbl{}
(red points) are the point of the FS close to the critical AFM modes
and connected by the vector $\mathbf{Q}=(\pi,\pi)$. Two different
ordering vectors $\mathbf{Q}_{0}^{a}$ and $\mathbf{Q}_{0}^{b}$ (green
and blue), coupling hot spots between two opposed FS in the anti-nodal
region are shown. The angle $\theta_{{\bf {k}}}$ localizes the points
in the first BZ.}}
\end{figure}

The SU(2) symmetry rotating the d-wave superconductor to the charge
channel was first derived in the context of the eight hot-spots model
\cite{Metlitski10b,Efetov13}, where the Fermi surface is reduced
to eight points related two by two by the wave vector $\mathbf{Q}=\left(\pi,\pi\right)$
as depicted in Fig. \ref{fig:hs}. In this model, electrons interact
through critical bosonic modes following the Lagrangian $L=L_{\psi}+L_{\phi}$

\begin{align}
L_{\psi} & =\psi^{\dagger}\left(\partial_{\tau}+\varepsilon_{\mathbf{k}}+\lambda\phi\cdot\sigma\right)\psi,\label{eq:1-2}\\
L_{\phi} & =\phi\frac{D^{-1}}{2}\phi+\frac{g}{2}\phi^{4},
\end{align}

where $\psi$ is the electron-field with dispersion $\varepsilon_{\mathbf{k}}$around
each hot-spot, coupled to the spin fluctuation field $\phi$ evolving
through the spin-wave propagator of typical Ornstein- Zernike form
\begin{align}
D^{-1} & =\frac{\omega^{2}}{v_{s}^{2}}+\left(\mathbf{q}-\mathbf{Q}\right)^{2}+m_{a}.
\end{align}
$m_{a}$ is the mass which characterizes the distance to the quantum
critical point (QCP). $\sigma$ is the Pauli spin in Eqn.(\ref{eq:1-2}).
When the Fermi dispersion $\xi_{\mathbf{k}}$ is linearized around
each hot-spot, one obtains a composite order as a precursor of anti-ferromagnetism.
The composite order-parameter can be viewed as a non-abelian superconductor
\begin{align}
\hat{b} & =b\hat{u}, & \mbox{ with } & u=\left(\begin{array}{cc}
\chi & \Delta\\
-\Delta^{*} & \chi^{*}
\end{array}\right),\label{constrain}\\
 &  & \mbox{ and } & \left|\chi\right|^{2}+\left|\Delta\right|^{2}=1.\nonumber 
\end{align}

Instead of having a U(1)-phase as it is the case superconductors,
the operator $\hat{b}$ has now an SU(2) -phase rotating between the
d-wave SC channel $\Delta=\frac{1}{\sqrt{2}}\sum_{\mathbf{k}}d_{\mathbf{k}}\psi_{\mathbf{k}\downarrow}\psi_{-\mathbf{k}\uparrow},$
with $d_{\mathbf{k}}=2\cos\left(2\theta_{\mathbf{k}}\right)$ and
the d-wave Peierls channel $\chi=\frac{1}{2}\sum_{\mathbf{k},\sigma}d_{\mathbf{k}}\psi_{\mathbf{k}+\mathbf{Q}_{0}\sigma}^{\dagger}\psi_{\mathbf{k},\sigma}$
also called quadrupolar order \cite{Efetov13}. Within this simplified
model,\textbf{ $\mathbf{k}$} is defined in a small region around
each hot-spot and the definition of the charge wave vector $\mathbf{Q}_{0}=\left(\pm\mathbf{Q}_{a},\mathbf{\pm\mathbf{Q}_{b}}\right)$
depends on the of the hot-spot in $\mathbf{k}$-space (see Fig. \ref{fig:hs}).
$\mathbf{Q}_{0}$ is the a $\mathbf{k}$-dependent, diagonal wave
vector which relates, using an Umklapp wave vector, the two hot-spots
opposite to each other across the Fermi surface. Note that the choice
of $\mathbf{Q}_{a}$ or $\mathbf{Q}_{b}$ is tight to the precise
each hot-spot. The precursing order $\hat{b}$ thus possesses an exact
SU(2) symmetry which relates the SC channel to the charge channel,
and importantly, it is driven by AF fluctuations which dominate in
the vicinity of the QCP.

\subsection{Operators}

\begin{figure*}[p]
\begin{minipage}[c]{5.75cm}%
a)\includegraphics[width=5.5cm]{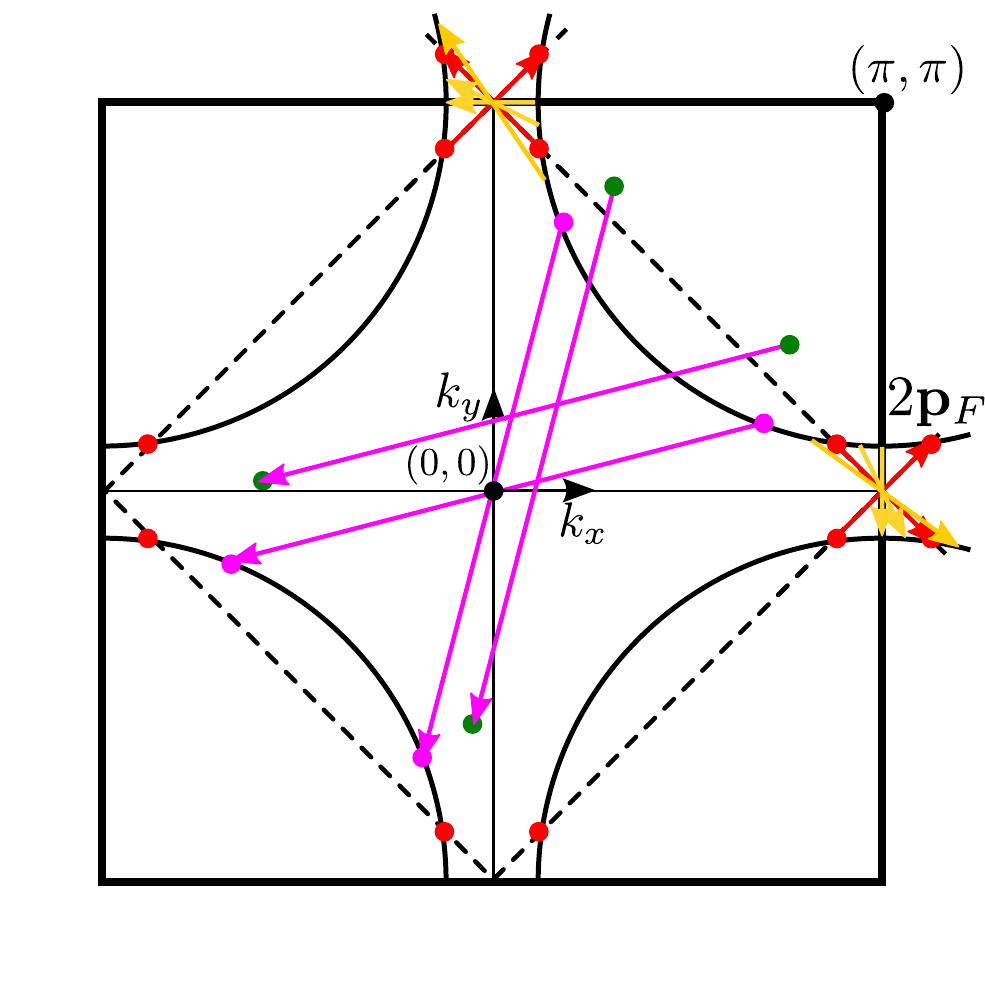} \vspace{1ex}
 a')\includegraphics[width=5.5cm]{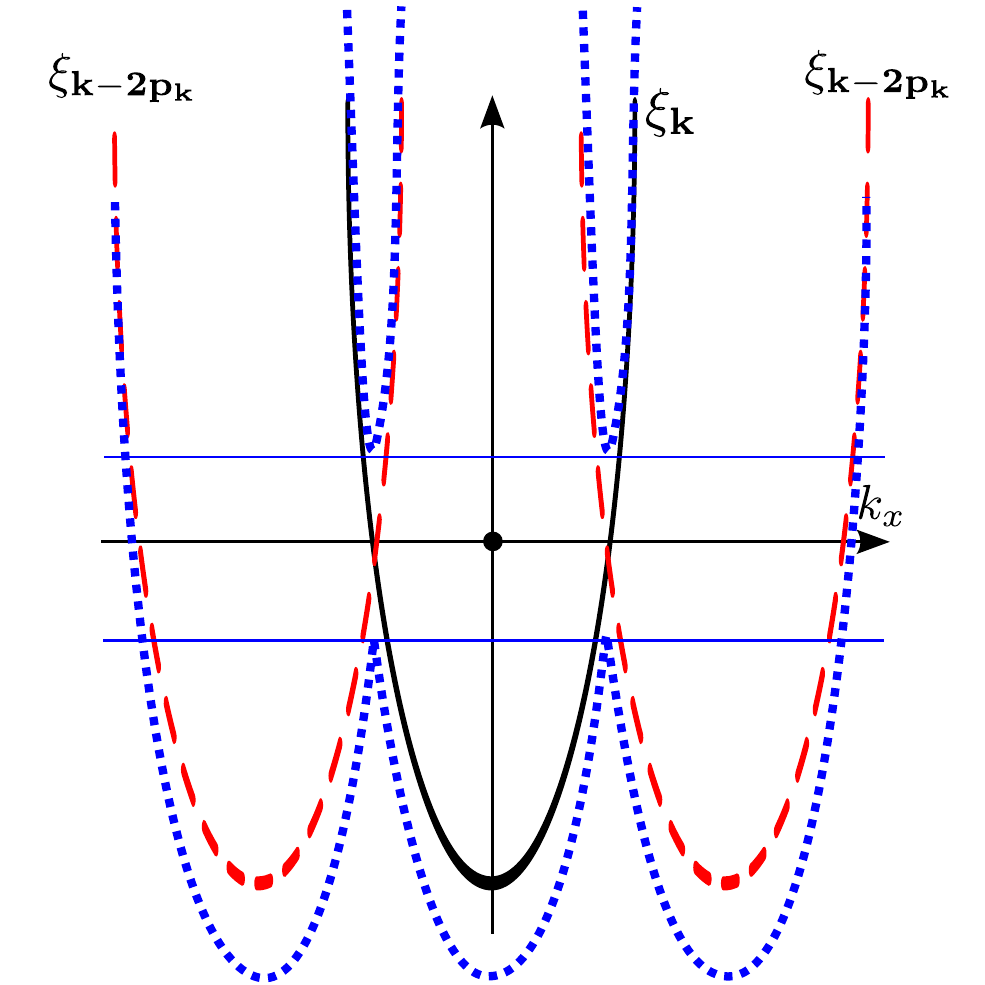} \vspace{1ex}
\end{minipage}%
\begin{minipage}[c]{5.75cm}%
b)\includegraphics[width=5.5cm]{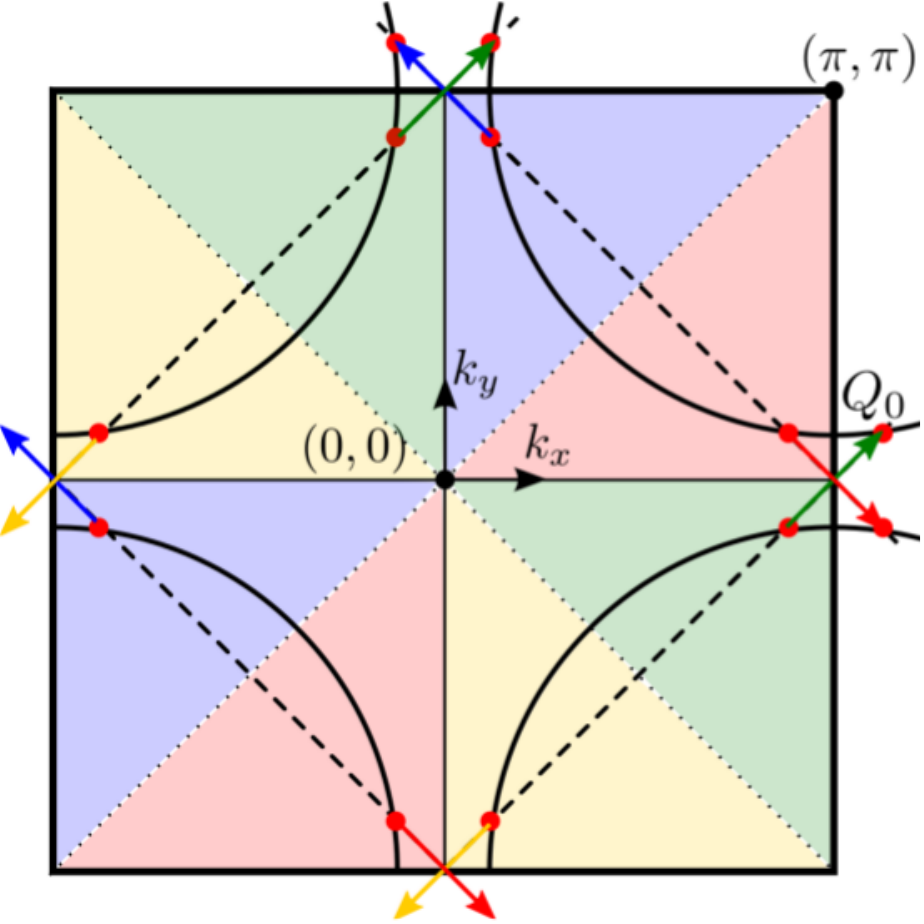} \vspace{1ex}
 b')\includegraphics[width=5.5cm]{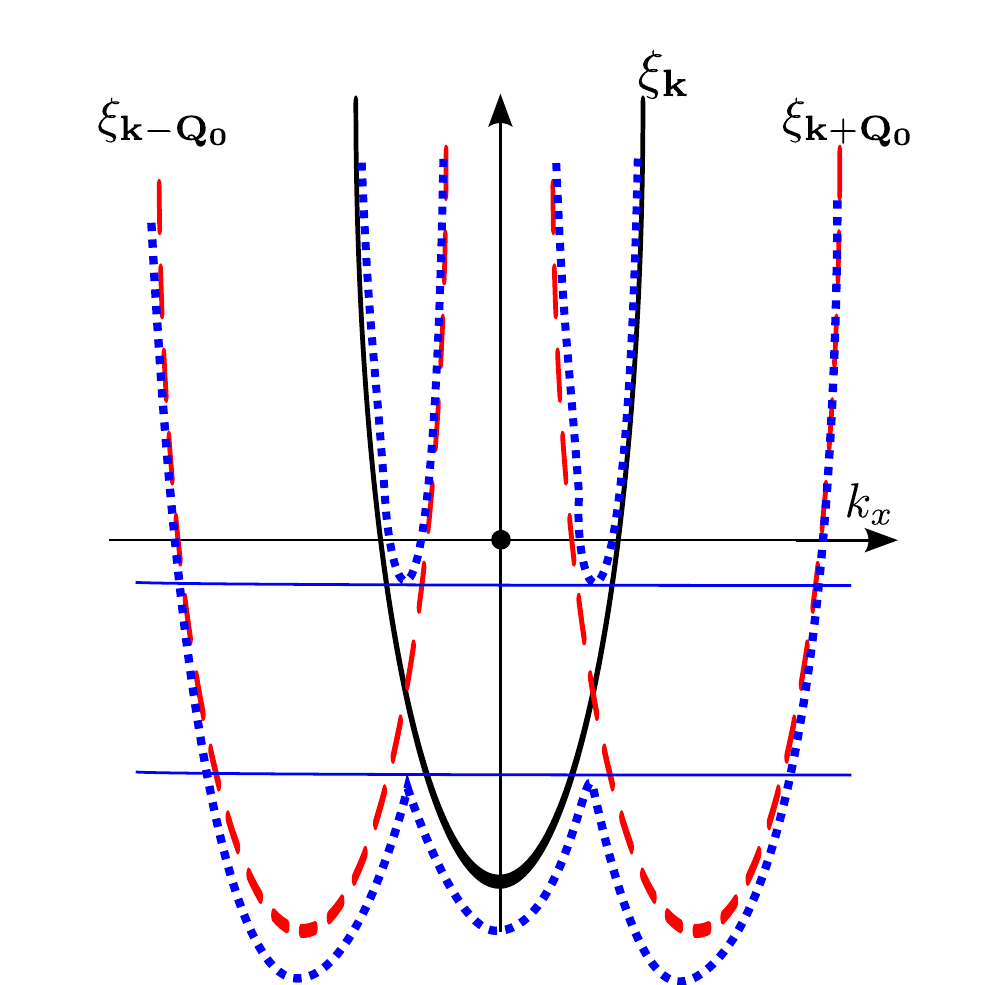} \vspace{1ex}
\end{minipage}%
\begin{minipage}[c]{5.75cm}%
c)\includegraphics[width=5.5cm]{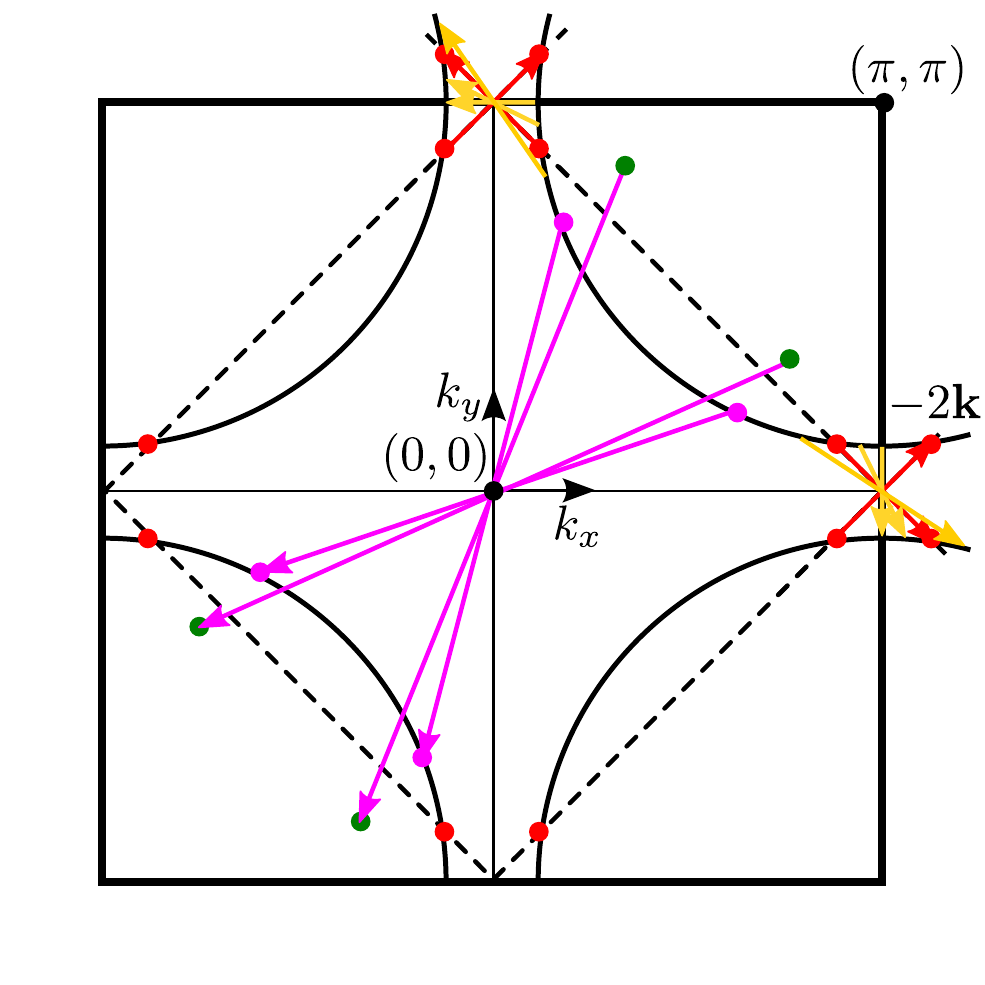} \vspace{1ex}
 c')\includegraphics[width=5.5cm]{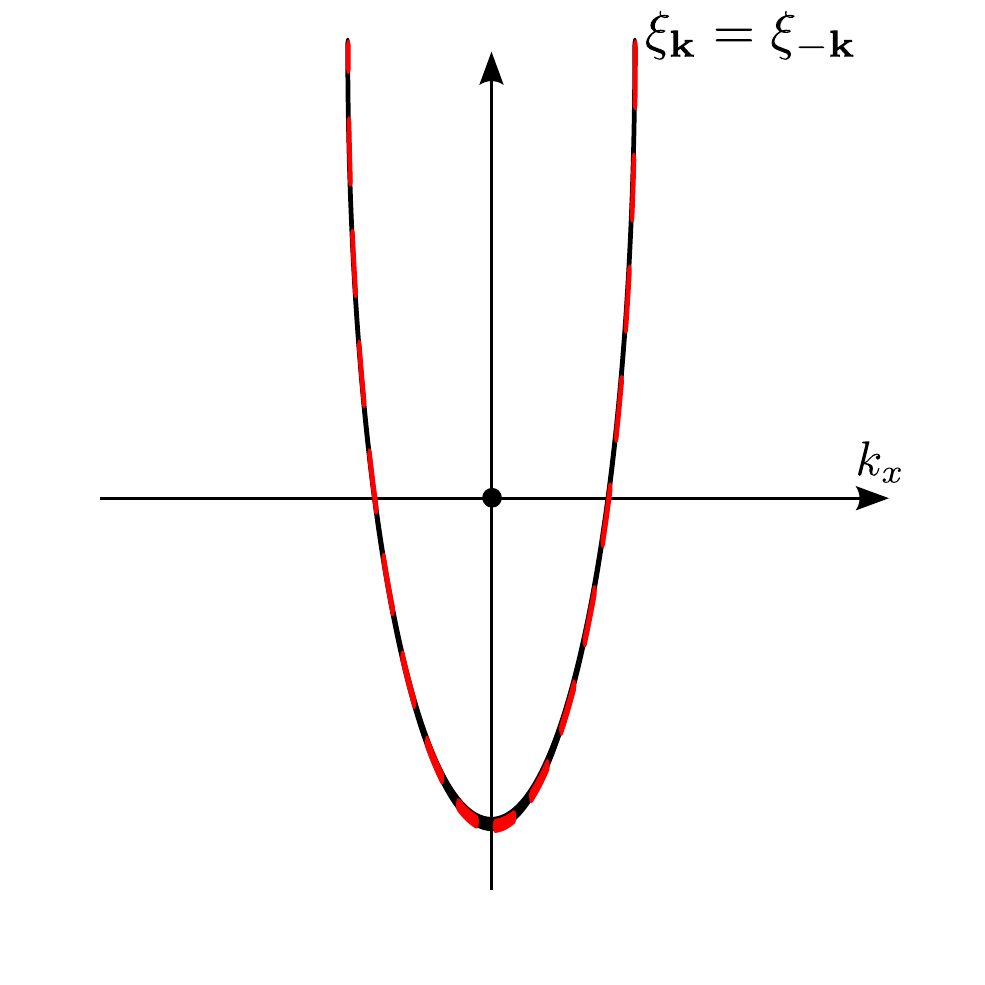} \vspace{1ex}
\end{minipage}\caption{\label{fig:spec} (Color online) \textcolor{black}{We represent in
first BZ of the different involution scenario : a)$\overline{\mathbf{k}}=-\mathbf{k}+2\mathbf{k}_{F}$
b) $\overline{\mathbf{k}}=-\mathbf{k}+\mathbf{Q_{0}}$, with $\mathbf{Q}_{0}=\left(\pm\mathbf{Q}_{a},\mathbf{\pm\mathbf{Q}_{b}}\right)$
c)$\overline{\mathbf{k}}=\mathbf{k}$. We represent the bare (solid
line), shifted (dashed line) and hybridized (dotted line) electronic
band dispersion along the $(\pi,0)$ to $(\pi,\pi)$ direction in
a')$\overline{\mathbf{k}}=-\mathbf{k}+2\mathbf{k}_{F}$ scenario b')$\mathbf{Q}_{0}=\left(\pm\mathbf{Q}_{a},\mathbf{\pm\mathbf{Q}_{b}}\right)$
scenario and c')$\overline{\mathbf{k}}=\mathbf{k}$ scenario. As drawn
in the figures a'), b'), the opening of the gap opens at the crossing
of the original and shifted spectra. In a') the opening of the gap
occurs at the Fermi surface while it is the case at only one point
for an incommensurate ordering vector. For example, in the scenario
b'), this opening occurs below the Fermi surface. In absence of symmetry
breaking, it is not able to open a gap between two identical electronic
band as presented in c').}}
\end{figure*}

In this paper, we study a generalization of the SU(2) symmetry of
the eight hot-spots model in the case of a real compound, with a generic
dispersion not reduced to the eight hot-spots, including the curvature.
The first step in this direction is to introduce the notion of involution,
implicitly present in the $\mathbf{k}$-dependence of the $\mathbf{Q}_{0}$
modulation vector os subsection \ref{sub:The-eight-hot-spots}. An
involution is a mapping which sends $\mathbf{k}\rightarrow\overline{\mathbf{k}}$
, such that for each $\mathbf{k}$-vector we have 
\begin{align}
\overline{\overline{\mathbf{k}}} & =\mathbf{k}\mbox{ and }\overline{\left(-\mathbf{k}\right)}=-\left(\overline{\mathbf{k}}\right).\label{eq:inv}
\end{align}
Such a mapping was already present in the definition of the $\mathbf{k}$-
dependent wave vector in subsection \ref{sub:The-eight-hot-spots}.
It is important for the generalization to the entire BZ, because it
ensures that the SU(2) algebra defined below is self-constrained,
and doesn't produce harmonics with each product of two operators.
Concrete examples of the involution that we use in this study, are
given in the next paragraph and are depicted in Figs \ref{fig:spec}
a), b) and c).

We now move to the definition of the pseudo-spin operators associated
with the SU(2) symmetry. The pseudo-spin operators $\eta^{+}$, $\eta^{-}=\left(\eta^{+}\right)^{\dagger}$
and $\eta^{z}$ are defined as \begin{subequations} \label{eq:1}
\begin{align}
\eta^{+} & =\sum_{\mathbf{k}}\psi_{\mathbf{k}\uparrow}^{\dagger}\psi_{\overline{\mathbf{k}}\downarrow}^{\dagger}\label{eq:1-1}\\
\eta_{z} & =\frac{1}{2}\sum_{\mathbf{k}}\left(\psi_{\mathbf{k}\uparrow}^{\dagger}\psi_{\mathbf{k}\uparrow}+\psi_{\overline{\mathbf{k}}\downarrow}^{\dagger}\psi_{\overline{\mathbf{k}}\downarrow}-1\right),\label{eq:6}
\end{align}
\end{subequations} The operators in Eqn.(\ref{eq:1}) form and SU(2)
algebra and are thus called pseudo-spin operators. They can act on
various representations, but in the present scenario for the underdoped
region, the representation chosen is a $l=1$ triplet involving two
conjugated SC operators ($\Delta_{-1}$ and $\Delta_{1}$) and a d-wave
charge sector operator $\Delta_{0}$ which are defined as:\begin{subequations}
\label{eq:1-3} 
\begin{align}
\Delta_{-1} & =\frac{1}{\sqrt{2}}\sum_{\mathbf{k}}\overline{d}_{\mathbf{k}}\psi_{\mathbf{k}\downarrow}\psi_{-\mathbf{k}\uparrow},\\
\Delta_{0} & =\frac{1}{2}\sum_{\mathbf{k},\sigma}\overline{d}_{\mathbf{k}}\psi_{\overline{\mathbf{k}}\sigma}^{\dagger}\psi_{-\mathbf{k},\sigma},\label{eq:2}\\
\Delta_{1} & =-\frac{1}{\sqrt{2}}\sum_{\mathbf{k}}\overline{d}_{\mathbf{k}}\psi_{\mathbf{k}\uparrow}^{\dagger}\psi_{-\mathbf{k}\downarrow}^{\dagger}.
\end{align}

\end{subequations} The form factor is given by $\overline{d}_{\mathbf{k}}=\left(d_{\mathbf{k}}+d_{\overline{\mathbf{k}}}\right)/2$,
with $d_{\mathbf{k}}=2\cos\left(2\theta_{\mathbf{k}}\right)$ , and
$\theta_{\mathbf{k}}$ the angle spanning the BZ. The standard SU(2)
relations 
\begin{align}
\left[\eta^{\pm},\Delta_{m}\right] & =\sqrt{l\left(l+1\right)-m\left(m\pm1\right)}\Delta_{m\pm1},\label{eq:alg1}\\
\mbox{and } & \left[\eta_{z},\Delta_{m}\right]=m\Delta_{m},
\end{align}
are valid here.

\subsection{The involutions }

\begin{figure*}
\includegraphics[width=5cm]{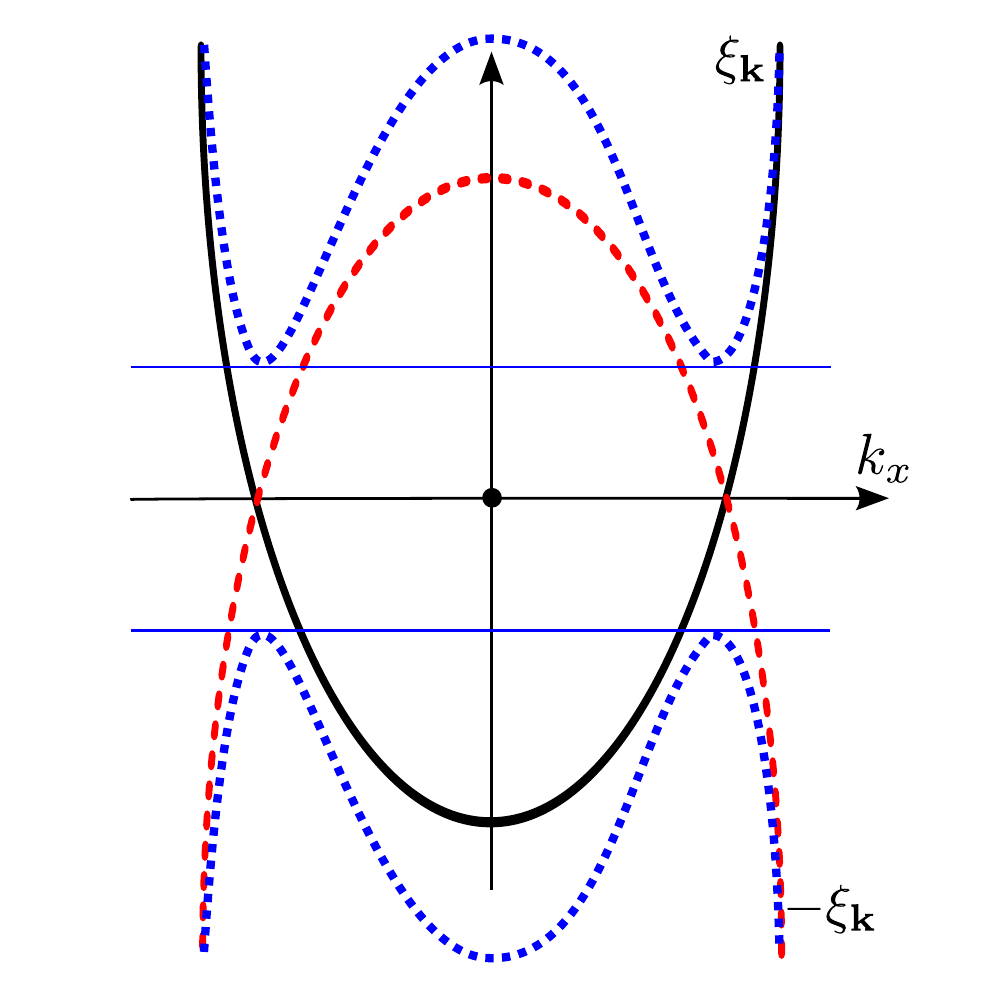} \vspace{1ex}
 \caption{\label{fig:sup-1} (Color online) \textcolor{black}{Representation
of the electronic spectrum for a superconducting scenario. The superconducting
state provide an hybridization between an electronic and a hole spectrum.
The gap opens at the Fermi surface and does not depend on the curvature.}}
\end{figure*}

\subsubsection{Definitions}

For the physics of underdoped cuprates, we consider and compare three
types of involutions depicted below: 
\begin{align}
A) &  & \overline{\mathbf{k}}=-\mathbf{k}+2\mathbf{k}_{F},\label{eq:8}
\end{align}
where $\mathbf{k}_{F}$ is the Fermi wave vector parallel to $\mathbf{k}$.
This form connects each wave vector in the BZ with a ``$2\mathbf{k}_{F}$''-partner
close to the opposite side of the Fermi surface $\psi_{\mathbf{k}}\rightarrow\psi_{\mathbf{k}-2\mathbf{k}_{F}}^{\dagger}$
(see Fig. \ref{fig:spec} a)). The pseudo-spin SU(2) symmetry is exactly
realized in the eight hot spots spin-fermion model, where the electronic
density is linearized around the hot spots \cite{Metlitski10b,Efetov13}.
In this case there are only four ''$2\mathbf{k}_{F}$'' wave vectors
denoted by 
\begin{align}
B) &  & \overline{\mathbf{k}}=-\mathbf{k}+\mathbf{Q_{0}},\label{eq:5}\\
 &  & \mbox{with }\mathbf{Q}_{0}=\left(\pm\mathbf{Q}_{a},\mathbf{\pm\mathbf{Q}_{b}}\right),\nonumber 
\end{align}
which are aligned with the diagonal of the BZ. For a generic Fermi
surface, multiple $2\mathbf{k}_{F}$ wave vectors can be chosen, as
depicted in Fig. \ref{fig:spec} a) , or alternatively we can keep
the four wave vectors defined for eight hot spots model and generalize
their action on the whole BZ as shown in Fig \ref{fig:spec} b).

An important point to stress out is that the two forms of possible
involutions Eqn.(\ref{eq:8}) and Eqn.(\ref{eq:5}) are degenerate
in the eight hot-spots model, since at the hot spots, the ``$2\mathbf{k}_{F}$''
wave vectors reduce to the four wave vectors of Eqn.(\ref{eq:5}).
In the case of a full Fermi surface, the two generalizations give
very different physics, that we will describe in the following paragraph.
Before, let's introduce a third kind of involution which corresponds
to a the particle-hole transformation $\psi_{\mathbf{k}}\rightarrow\psi_{\mathbf{k}}^{\dagger}$
and for which we have 
\begin{align}
C) &  & \overline{\mathbf{k}}=\mathbf{k}.\label{eq:9-1}
\end{align}
\textcolor{black}{The case C is presented in Fig. \ref{fig:spec}
c). This case corresponds to an ordering vector of $-2{\bf {k}}$.}

\subsubsection{Physical interpretation}

The three kinds of involutions rotate a superconducting doublet $\Delta_{-1},\Delta_{1}$
Eqns.(\ref{eq:1-3}) into an alternative channel in the charge sector
$\Delta_{0}$ Eqn.(\ref{eq:2}). The forms of $\Delta_{0}$ vary explicitly,
however in the three cases\begin{subequations} \label{eq:1-4} 
\begin{align}
A) &  & \Delta_{0} & =\frac{1}{2}\sum_{\mathbf{k},\sigma}\overline{d}_{\mathbf{k}}\psi_{\mathbf{k}-2\mathbf{k}_{F}\sigma}^{\dagger}\psi_{\mathbf{k},\sigma},\\
B) &  & \Delta_{0} & =\frac{1}{2}\sum_{\mathbf{k},\sigma}\overline{d}_{\mathbf{k}}\psi_{\mathbf{k}-\mathbf{Q}_{0}\sigma}^{\dagger}\psi_{\mathbf{k},\sigma},\\
C) &  & \Delta_{0} & =\frac{1}{2}\sum_{\mathbf{k},\sigma}\overline{d}_{\mathbf{k}}\psi_{-\mathbf{k}\sigma}^{\dagger}\psi_{\mathbf{k},\sigma}.
\end{align}

\end{subequations} The charge order parameters A, B, and C couple very differently with the conduction electrons,
 represented in Fig. \ref{fig:spec} and differently than SC order parameter represented in Fig. \ref{fig:sup-1}.

In the case of the ``Peierls'' or ``$2\mathbf{k}_{F}$'' coupling,
the electronic dispersion is translated by ``$2\mathbf{k}_{F}$''
around each point of the Fermi surface, which leads to an obvious
band-crossing and opening of a gap. The same is valid for B), where
the electronic dispersion is translated by the wave vector $\pm\mathbf{Q}_{a,b}$
around the zone edge, leading to band crossing and the opening of
a gap. The situation C), however is drastically different since without
inversion symmetry point we have $\xi_{\mathbf{k}}=\xi_{\mathbf{-k}}$
and the transformation does not lead to the opening of a gap.

For comparison, let's mention the SC parts $\Delta_{-1},\Delta_{1}$,
for which the opening of the gap is ensured by the charge conjugation
leading to a reversing of the electron energy $\xi_{\mathbf{k}}\rightarrow-\xi_{-\mathbf{k}}$
with a band-crossing locked at the Fermi level.

The picture is also drastically different in real space, and the easiest
way to see it is to Fourier transform the ladder operator $\eta^{+}$
Eqn. (\ref{eq:5}) in the three cases \begin{subequations} \label{eq:1-5}

\begin{align}
A) &  & \eta^{+} & =\sum_{\mathbf{k}}\psi_{\mathbf{k}\uparrow}^{\dagger}\psi_{-\mathbf{k}+2\mathbf{k}_{F}\downarrow}^{\dagger},\\
B) &  & \eta^{+} & =\sum_{\mathbf{k}}\psi_{\mathbf{k}\uparrow}^{\dagger}\psi_{-\mathbf{k}+\mathbf{Q}_{0}\downarrow}^{\dagger},\\
C) &  & \eta^{+} & =\sum_{\mathbf{k}}\psi_{\mathbf{k}\uparrow}^{\dagger}\psi_{\mathbf{k}\downarrow}^{\dagger}.
\end{align}

\end{subequations} In the three cases Eqns. (\ref{eq:1-5}) correspond
to a finite wave vector pairing -also called Fulde-Ferrell-Larkin-Ovshinnikov
(FFLO) pairing, at wave vectors $2\mathbf{k}_{F}$ in case A), $\mathbf{Q}_{0}$
in case B) and $2\mathbf{k}$ in case C). This leads to a re-writing
of the ladder operators as\begin{subequations} \label{eq:1-6}

\begin{align}
A) &  & \eta^{+} & =\sum_{\mathbf{k}}\sum_{i,j}e^{i2\mathbf{k}_{F}\cdot\left(r_{i}+r_{j}\right)/2}e^{i\tilde{\mathbf{k}}_{a}\cdot\left(r_{i}-r_{j}\right)}\psi_{i\uparrow}^{\dagger}\psi_{j\downarrow}^{\dagger},\\
B) &  & \eta^{+} & =\sum_{\mathbf{k}}\sum_{i,j}e^{i\mathbf{Q}_{0}\cdot\left(r_{i}+r_{j}\right)/2}e^{i\tilde{\mathbf{k}}_{b}\cdot\left(r_{i}-r_{j}\right)}\psi_{i\uparrow}^{\dagger}\psi_{j\downarrow}^{\dagger},\\
C) &  & \eta^{+} & =\sum_{\mathbf{k}}\sum_{i,j}e^{i2\mathbf{k}\cdot\left(r_{i}+r_{j}\right)/2}\psi_{i\uparrow}^{\dagger}\psi_{j\downarrow}^{\dagger}.
\end{align}

\end{subequations} In the case A) $2\mathbf{k}_{F}$ depends on $\mathbf{k}$
while $\tilde{\mathbf{k}}_{a}=\mathbf{k}-\mathbf{k}_{F}$ is small
close to the Fermi energy. In this case the summation over $\mathbf{k}$
leads to a localization of the center of mass of the pair $r_{i}+r_{j}=0$.
The same holds in case C) where the summation over $\mathbf{k}$ doesn't
affect the variable $r_{i}-r_{j}$. The case B), however, is the opposite,
since $\mathbf{Q}_{0}$ is a finite wave vector independent of $\mathbf{k}$,
while $\tilde{\mathbf{k}}_{b}=\mathbf{k}-\mathbf{Q}_{0}/2$ is $\mathbf{k}$-dependent
and locates the relative position of the pair to the be small $r_{i}-r_{j}=0$.
Case B) is similar to a standard, zero momentum superconductor, for
which we would have $\eta^{+}=\sum_{\mathbf{k}}\psi_{\mathbf{k}\uparrow}^{\dagger}\psi_{\mathbf{-k}\downarrow}^{\dagger}$,
leading to $\eta^{+}=\sum_{\mathbf{k}}\sum_{i,j}e^{i\mathbf{k}\cdot\left(r_{i}-r_{j}\right)}\psi_{\mathbf{i}\uparrow}^{\dagger}\psi_{j\downarrow}^{\dagger}$,
for which the $\mathbf{k}$-summation located $r_{i}-r_{j}=0$.

\begin{figure}
\begin{minipage}[c]{4.25cm}%
\includegraphics[width=4cm]{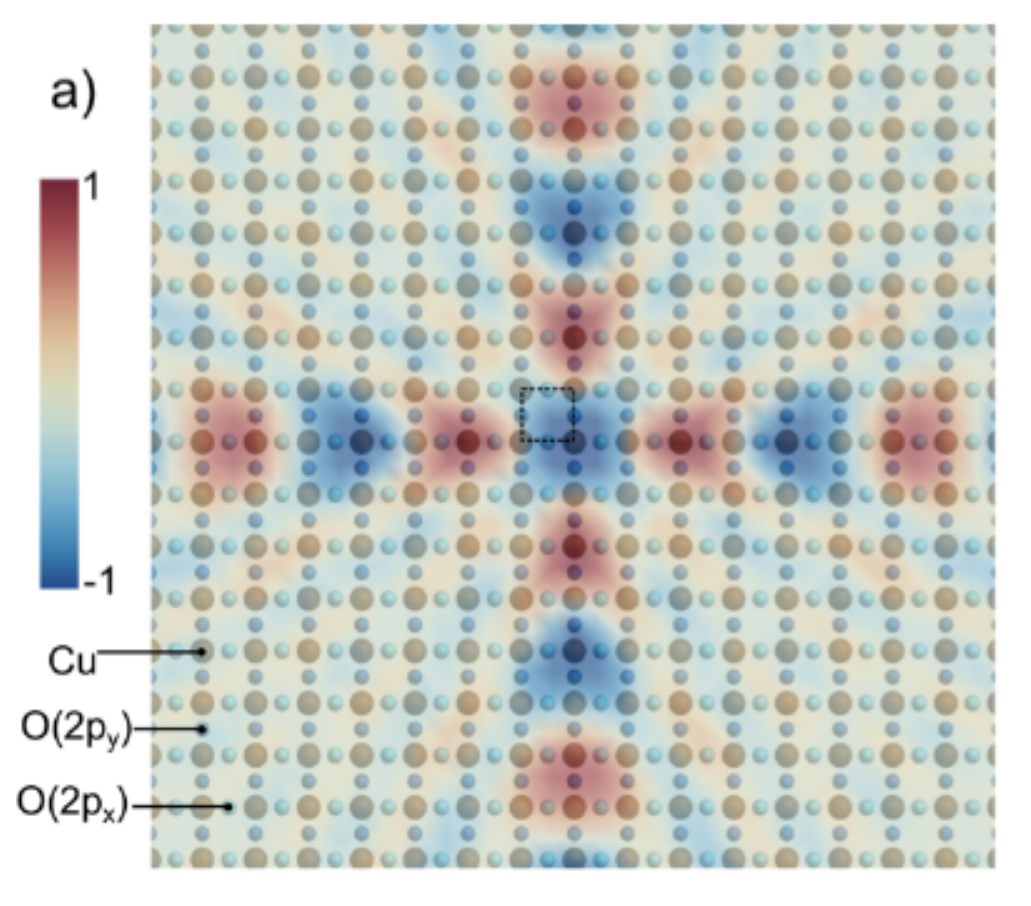} \vspace{1ex}
\end{minipage}%
\begin{minipage}[c]{4.25cm}%
\includegraphics[width=4cm]{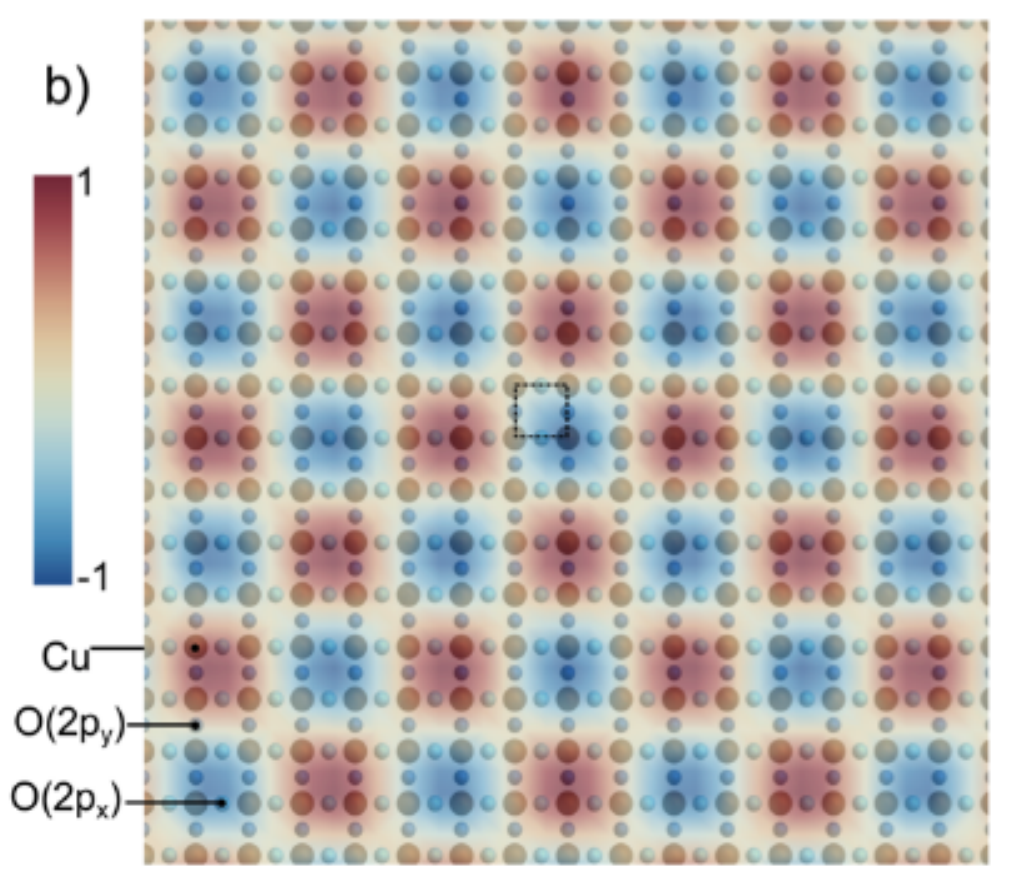} \vspace{1ex}
\end{minipage}

\begin{minipage}[c]{4.25cm}%
\includegraphics[width=4cm]{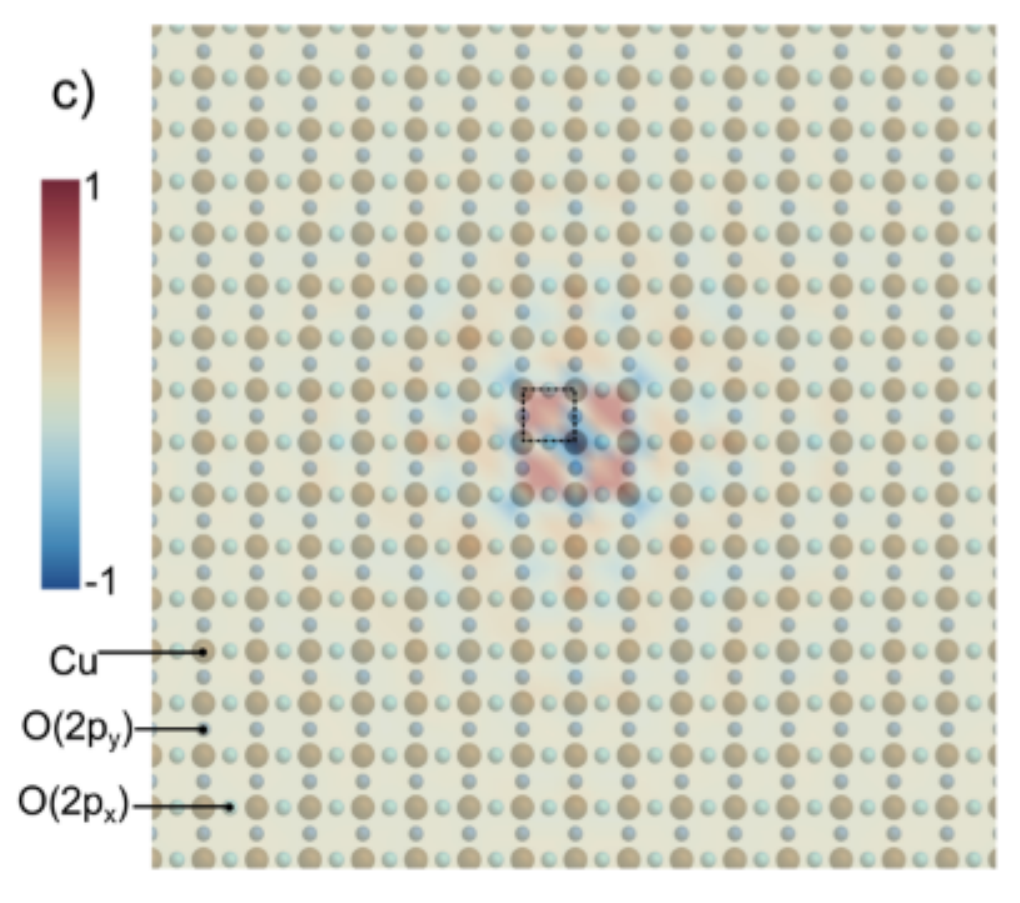} \vspace{1ex}
\end{minipage}\caption{\label{fig:dens} (Color online) \textcolor{black}{Charge density
in the real space in the different involution scenarios a) $\overline{\mathbf{k}}=-\mathbf{k}+2\mathbf{k}_{F}$
b) $\mathbf{Q}_{0}=\left(\pm\mathbf{Q}_{a},\mathbf{\pm\mathbf{Q}_{b}}\right)$
c) $\overline{\mathbf{k}}=\mathbf{k}$.}\textcolor{red}{{} }}
\end{figure}

Real space pictures illustrating the three situations are given in
Fig. \ref{fig:dens}. We note the cross-structure in a) showing the
singularity of the origin, which comes from the multiple wave vectors,
leading to a typical checkerboard structure in b), which corresponds
to and order with the superposition of the two axial wave vectors
$\left(\mbox{0,\ensuremath{Q_{\mathbf{0}}}}\right)$ and $\left(Q_{\mathbf{0}},0\right)$.
The case c) which never leads to the opening of a gap, shows a very
small typical lenghtscale.

The real space picture associated with the physics of the objects
depicted in Fig.\ref{fig:dens}a) have been described in Ref.\cite{Kloss15a},
and will be addressed again in sections \ref{sec:Effect-of-the} and
\ref{sec:Global-phase-diagram}. Noticeably, the structure depicted
in Fig.\ref{fig:dens}a) has two energy scales, one associated with
the relative distance between electron and hole in the pair, and the
other one associated to the position and extension around the center
of mass $\left(r_{i}+r_{j}\right)/2$. The summation over the multiple
$\pf$ wave vectors produces a localization of the center of mass
at the origin, which is typically associated with the formation of
a local object, with a specific modulation pattern. The study of the
physics of such objects, or patches, goes beyond the scope of this
paper, but it is interesting to see that already at the level of the
symmetries one sees a profound difference in real space between patches
of particle-hole pairs (Fig.\ref{fig:dens}a)) and uniform checkerboard
phase (Fig.\ref{fig:dens}b)).

The same game can be played with the charge states given in Eqns.
(\ref{eq:1-4})

\begin{subequations} \label{eq:1-7} 
\begin{align}
A) &  & \Delta_{0} & =\frac{1}{2}\sum_{\mathbf{k},\sigma}\sum_{i,j}e^{i2\mathbf{k}_{F}\cdot\left(r_{i}+r_{j}\right)/2}e^{i\tilde{\mathbf{k}}_{a}\cdot\left(r_{i}-r_{j}\right)}\overline{d}_{\mathbf{k}}\psi_{i\sigma}^{\dagger}\psi_{j\sigma},\\
B) &  & \Delta_{0} & =\frac{1}{2}\sum_{\mathbf{k},\sigma}\sum_{i,j}e^{i\mathbf{Q}_{0}\cdot\left(r_{i}+r_{j}\right)/2}e^{i\tilde{\mathbf{k}}_{b}\cdot\left(r_{i}-r_{j}\right)}\overline{d}_{\mathbf{k}}\psi_{i\sigma}^{\dagger}\psi_{j\sigma},\\
C) &  & \Delta_{0} & =\frac{1}{2}\sum_{\mathbf{k},\sigma}\sum_{i,j}e^{i2\mathbf{k}\cdot\left(r_{i}+r_{j}\right)/2}\overline{d}_{\mathbf{k}}\psi_{i\sigma}^{\dagger}\psi_{j\sigma}.
\end{align}

\end{subequations} Note the similarity of Eqns. (\ref{eq:1-6}) and
(\ref{eq:1-7}) which lead to the same real space interpretations.

\section{The SU(2) envelop\label{sec:The-SU(2)-dome}}

\subsection{The starting model with short range AF interactions}

There are a few models which are well-known to give rise to d-wave
superconductivity. The repulsive Hubbard-model can be mapped out onto
an effective model where the super-exchange between adjacent sites
is described via the t-J Hamiltonian where the strong Coulomb interactions
are described through a constraint of no double occupancy (see e.g.
\cite{Lee06} for a review). 
\begin{align}
H_{tJ} & =\sum_{i,j,\sigma}\psi_{i,\sigma}^{\dagger}t_{ij}\psi_{j,\sigma}+J\sum_{\left\langle i,j\right\rangle \alpha\beta}\psi_{i}^{\dagger}\vec{\sigma}_{\alpha\beta}\psi_{i\beta}\cdot\psi_{j\alpha'}^{\dagger}\vec{\sigma}_{\alpha'\beta'}\psi_{j\beta'},\label{eq:ham1}
\end{align}

where $\psi$ is the conduction electron field, $t_{ij}$ is the hopping
matrix describing the band structure of the materials, which is typically
of the order of $1$eV, $\left\langle i,j\right\rangle $ denotes
the summation over nearest neighbors typical of the AF super-exchange
term of order $0.7$eV, and $\vec{\sigma}$ is the Pauli matrix describing
the spin. The constraint of no double occupancy has to be imposed,
in order to give a good treatment to the vicinity to a Mott insulator,
but we neglect it for simplicity and consider that the main effects
treated here come from the AF short range interactions in Eqn. (\ref{eq:ham1}).
In Momentum space, the Hamiltonian reads

\begin{align}
H & =\sum_{{\bf k}\alpha}\xi_{{\bf k}}\psi_{{\bf k},\alpha}^{\dagger}\psi_{{\bf k},\alpha}-\sum\limits _{\substack{kk'\overline{q}\\
\sigma\sigma'
}
}J_{\overline{\mathbf{q}}}\psi_{\sigma,\mathbf{k}+\overline{\mathbf{q}}}^{\dagger}\psi_{\mathbf{k},\sigma'}\psi_{\mathbf{k'}-\overline{\mathbf{q}}}^{\dagger}\psi_{\mathbf{k'},\sigma},\label{eq:sint}
\end{align}

where $J_{\overline{\mathbf{q}}}=2J\cos\overline{\mathbf{q}}$, with
$\overline{\mathbf{q}}=\mbox{\textbf{Q}+\textbf{q}}$, and $\mathbf{Q}=\left(\pi,\pi\right)$
the AF wave vector. In contrast with section \ref{sub:The-eight-hot-spots},
where the AF coupling had been taken close to a QCP where it becomes
singular, we assume no such singularity here. The AF correlations are
typically found to be strong and short ranged in the cuprates, and
the Hamiltonian (\ref{eq:ham1}) is generic enough to account for
this feature. In our previous work on the eight
hot spots model\cite{Efetov13}, the proximity to AF quantum criticality
was assumed and crucial for the control of the solution. Here, although
we use mean-field like methods, the starting point is more realistic
for a general theory of the PG in cuprates.

\subsection{Charge and SC decoupling}

\begin{figure}
\begin{minipage}[c]{4.25cm}%
a)

\includegraphics[width=4cm]{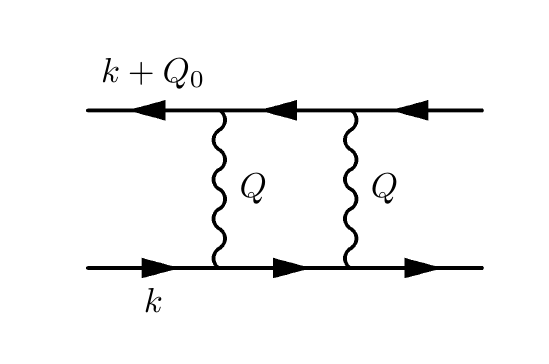} \vspace{1ex}
\end{minipage}%
\begin{minipage}[c]{4.25cm}%
b)

\includegraphics[width=4cm]{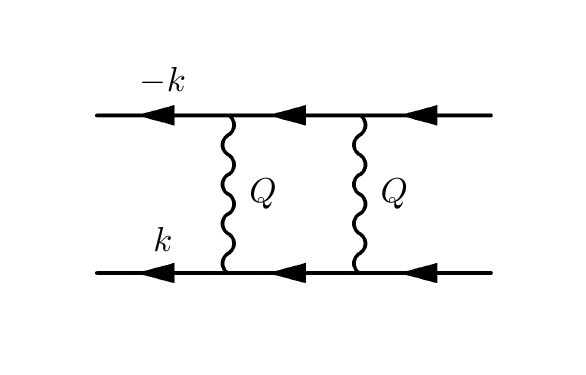} \vspace{1ex}
\end{minipage}

\textcolor{black}{\caption{\label{fig:diags1} (Color online) \textcolor{black}{Infinite ladder
series corresponding respectively to the gap equations $\left(\ref{eq:chiQ0}\right)$
for diagram a) and$\left(\ref{eq:SCgap}\right)$for diagram b).}}
} 
\end{figure}

We can now decouple the second term in Eqn. (\ref{eq:sint}) in the
charge and SC channels, which leads to two types of gap Equations 
\begin{itemize}
\item In the charge channel, the Hubbard-Stratonovich decoupling of Eqn.
(\ref{eq:sint}) leads to the effective action 
\begin{align*}
S_{\chi}^{eff} & =\int_{k,k',\overline{q}}(J_{\overline{q}}^{-1}\chi_{k,k'}\chi_{k+\overline{q},k'+\overline{q}}\\
 & +\chi_{k,k'}\sum_{\sigma}\psi_{k+\overline{q},\sigma}^{\dagger}\psi_{k'+\overline{q},\sigma}\\
 & +\chi_{k+\overline{q},k'+\overline{q}}\sum_{\sigma}\psi_{k,\sigma}^{\dagger}\psi_{k',\sigma}),
\end{align*}
where $\chi_{k,k'}=\left\langle \sum_{\sigma}\psi_{k,\sigma}^{\dagger}\psi_{k',\sigma}\right\rangle $.
Integrating the fermions out of the partition function and then differentiating
with respect to $\chi$ leads to the gap equation, in the charge sector.
Here ${\bf {k}'-}{\bf {k}}={\bf {Q}_{0}}$, where $\mathbf{Q}_{0}$
is the incommensurate charge modulation vector (see Fig. \ref{fig:diags1} (a)): 
\begin{align}
\chi_{k,k'} & =-\delta_{\mathbf{k}',\mathbf{k}+\mathbf{Q}_{0}}\ \Re T\sum_{\omega,\overline{\mathbf{q}}}J_{\overline{\mathbf{q}}}\times\label{eq:chiQ0}\\
 & \frac{\chi_{k+\overline{q},k'+\overline{q}}}{(i\epsilon+i\omega-\xi_{{\bf {k+\overline{\mathbf{q}}}}})(i\epsilon'+i\omega-\xi_{{\bf {k}'+{\bf {\overline{\mathbf{q}}}}}})-\chi_{k+\overline{q},k'+\overline{q}}^{2}}.\nonumber 
\end{align}

\item Similar action is derived in the SC channel, with 
\begin{align}
S_{\Delta}^{eff} & =\int_{k,k',\overline{q}}(J_{\overline{q}}^{-1}\Delta_{k,k'}^{\dagger}\Delta_{k+\overline{q},k'-\overline{q}}\nonumber \\
 & +\Delta_{k,k'}^{\dagger}\sum_{\sigma}\sigma\psi_{k+\overline{q},\sigma}\psi_{k'-\overline{q},-\sigma}\nonumber \\
 & +\chi_{k+\overline{q},k'-\overline{q}}\sum_{\sigma}\sigma\psi_{k,\sigma}^{\dagger}\psi_{k',-\sigma}^{\dagger}),
\end{align}
where $\Delta_{k,k'}=\left\langle \sum_{\sigma}\sigma\psi_{k,\sigma}\psi_{k',-\sigma}\right\rangle $,
and $\mathbf{k}'=-\mathbf{k}$. We get the standard SC gap equation
$\left(\Delta_{k}=\Delta_{k,-k}\right)$  (see Fig. \ref{fig:diags1} (b)): 
\end{itemize}
\begin{equation}
\Delta_{k}=-T\sum_{\omega,{\bf \overline{\mathbf{q}}}}J_{\overline{\mathbf{q}}}\frac{\Delta_{k+\overline{q}}}{\Delta_{k+\overline{q}}^{2}+\xi_{{\bf {k}+\overline{\mathbf{q}}}}^{2}+\left(\epsilon+\omega\right)^{2}},\label{eq:SCgap}
\end{equation}

\begin{figure}
\begin{minipage}[c]{4.25cm}%
a)\includegraphics[width=3.95cm]{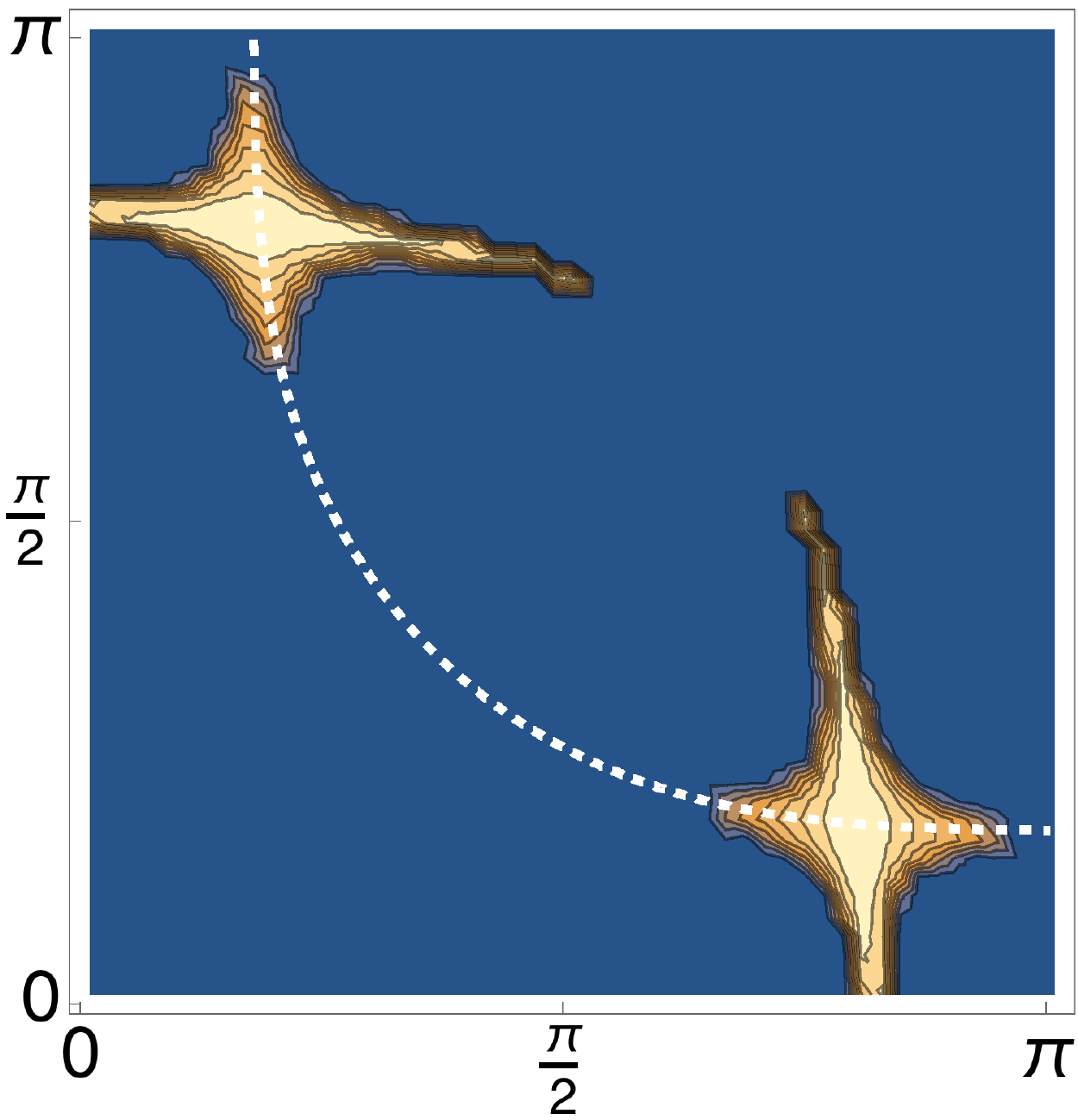} \vspace{1ex}
\end{minipage}%
\begin{minipage}[c]{4.25cm}%
b)\includegraphics[width=3.95cm]{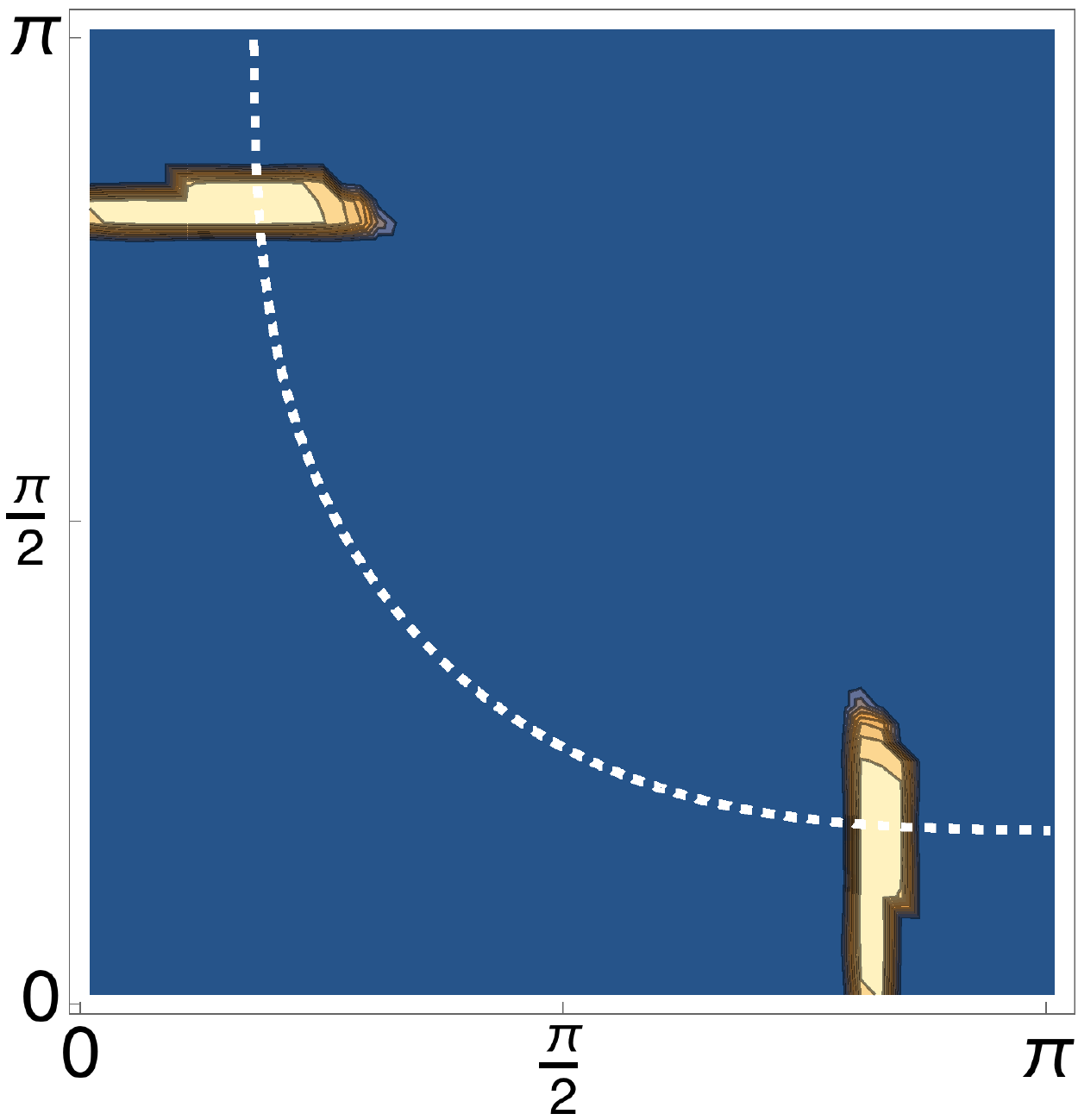} \vspace{1ex}
\end{minipage}

\begin{minipage}[c]{4.25cm}%
c)\includegraphics[width=3.95cm]{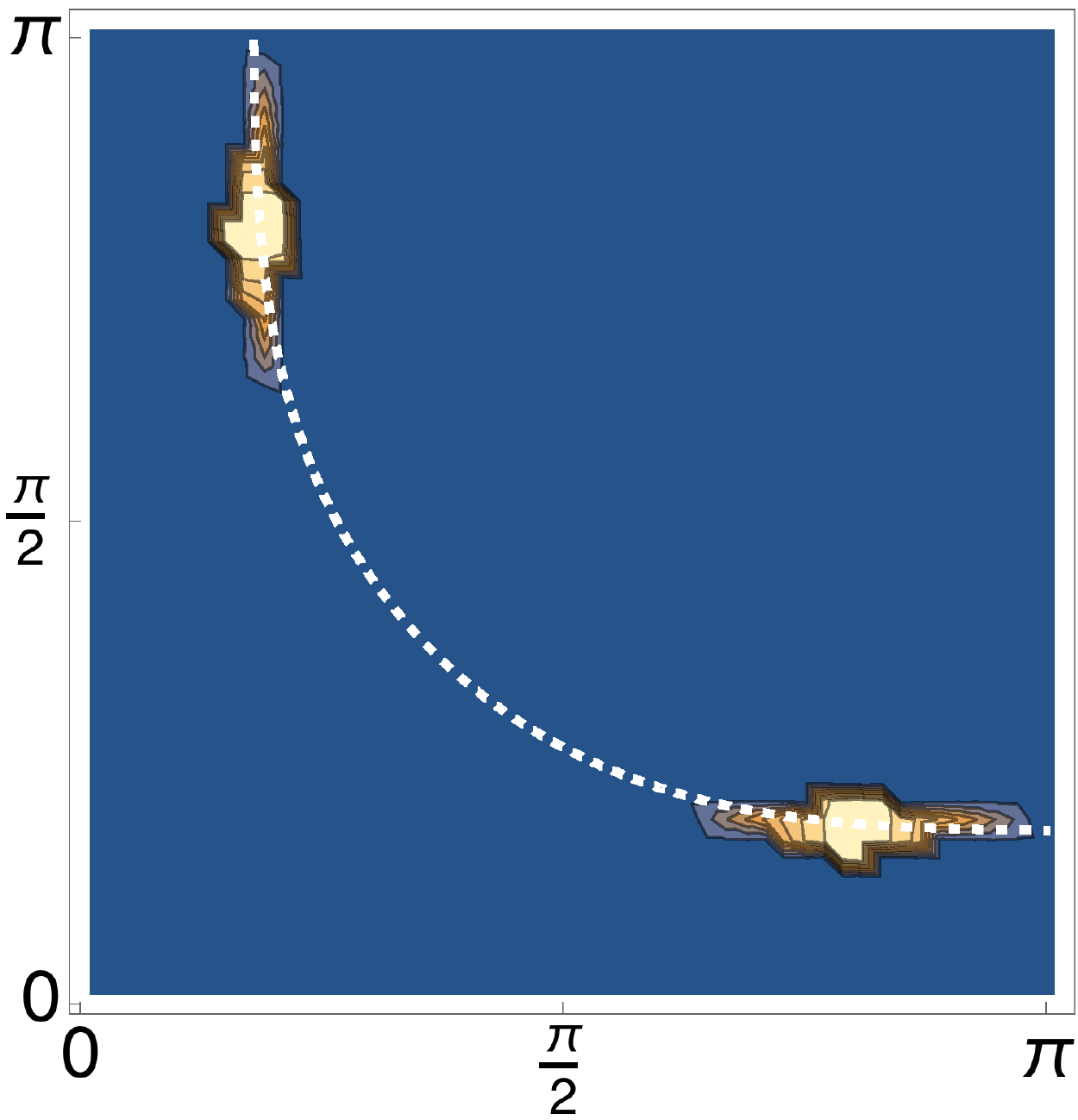} \vspace{1ex}
\end{minipage}%
\begin{minipage}[c]{4.25cm}%
d)\includegraphics[width=3.95cm]{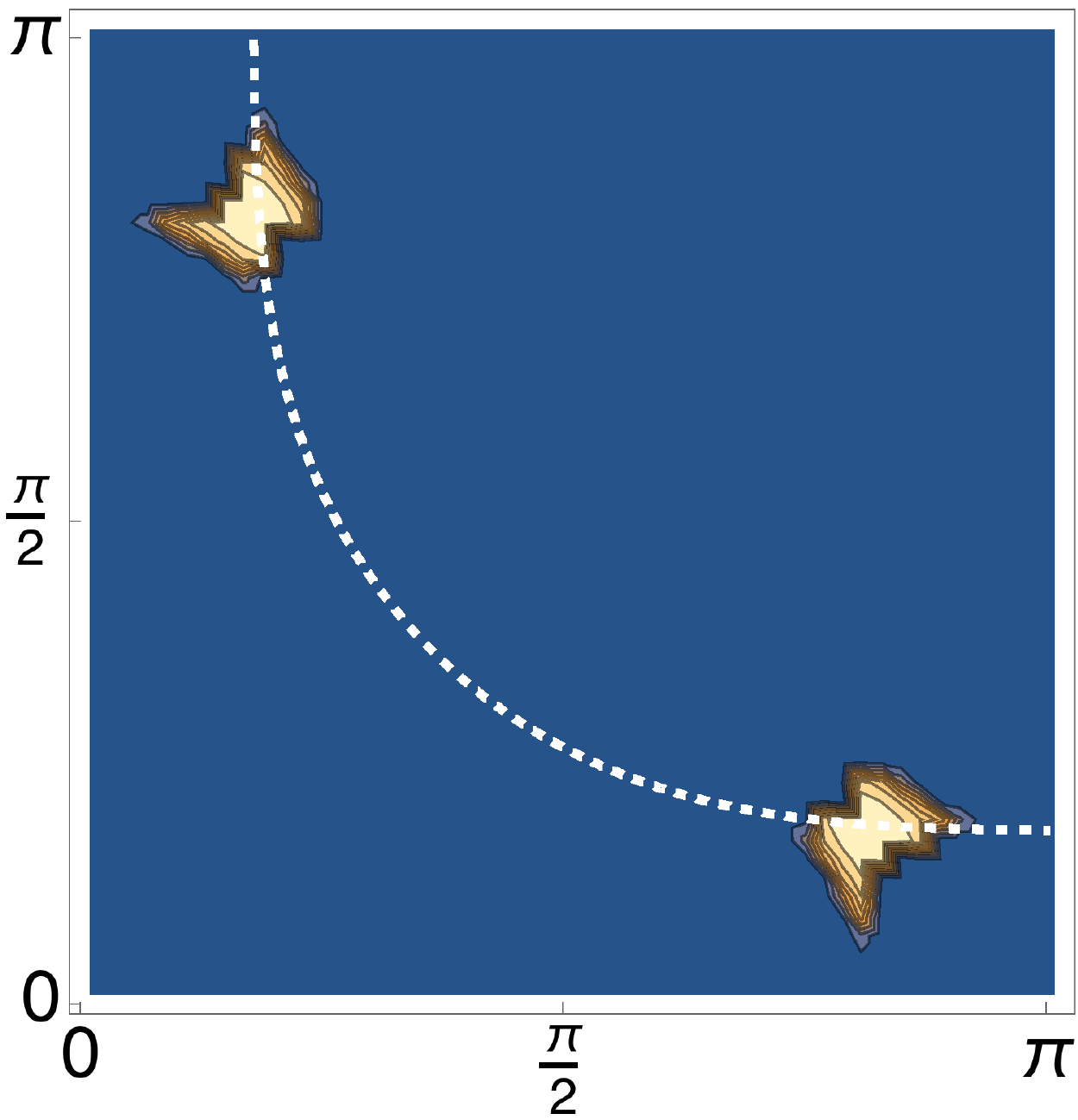} \vspace{1ex}
\end{minipage}

\begin{minipage}[c]{4.25cm}%
e)\includegraphics[width=3.95cm]{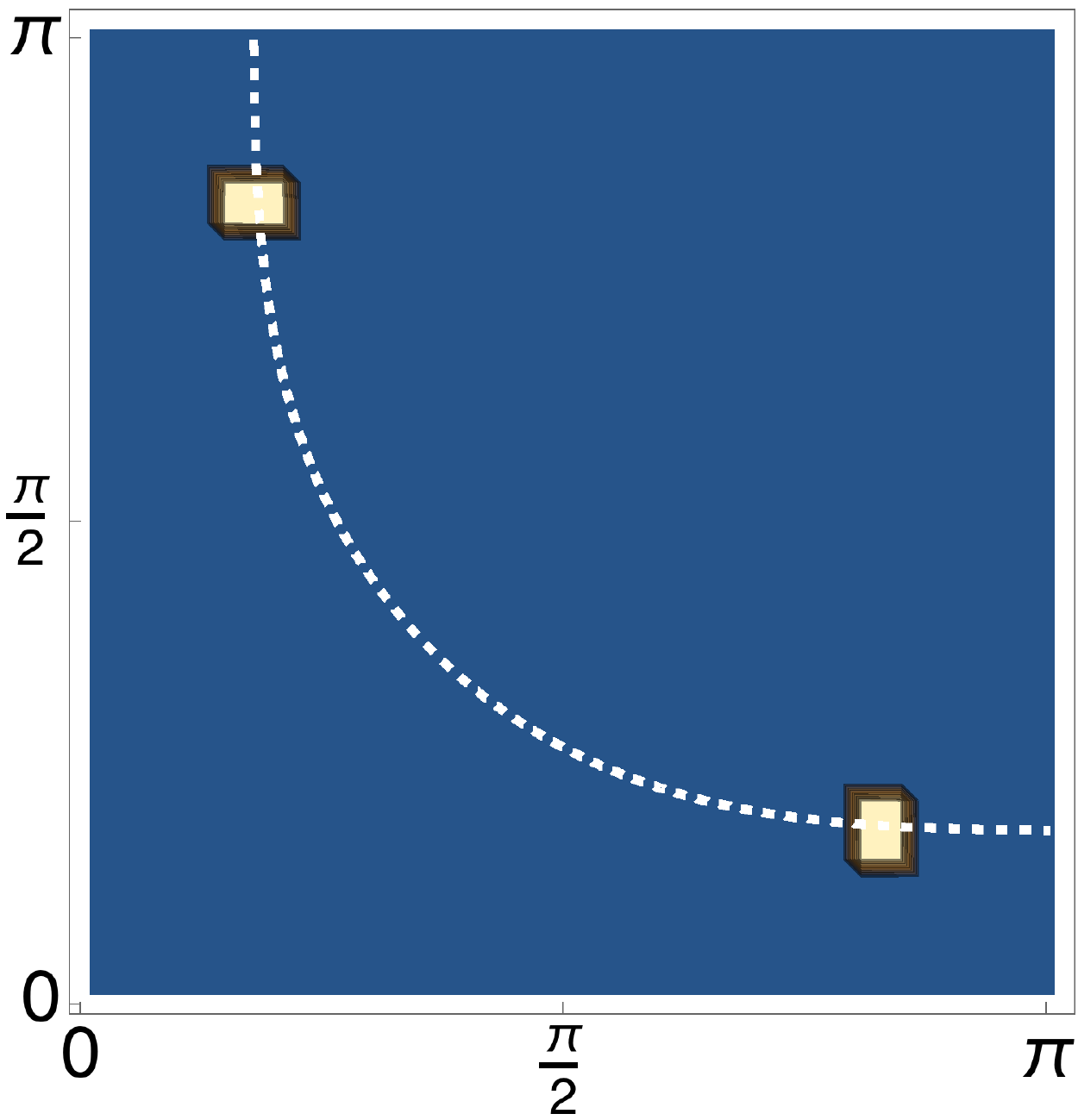} \vspace{1ex}
\end{minipage}%
\begin{minipage}[c]{4.25cm}%
f)\includegraphics[width=3.95cm]{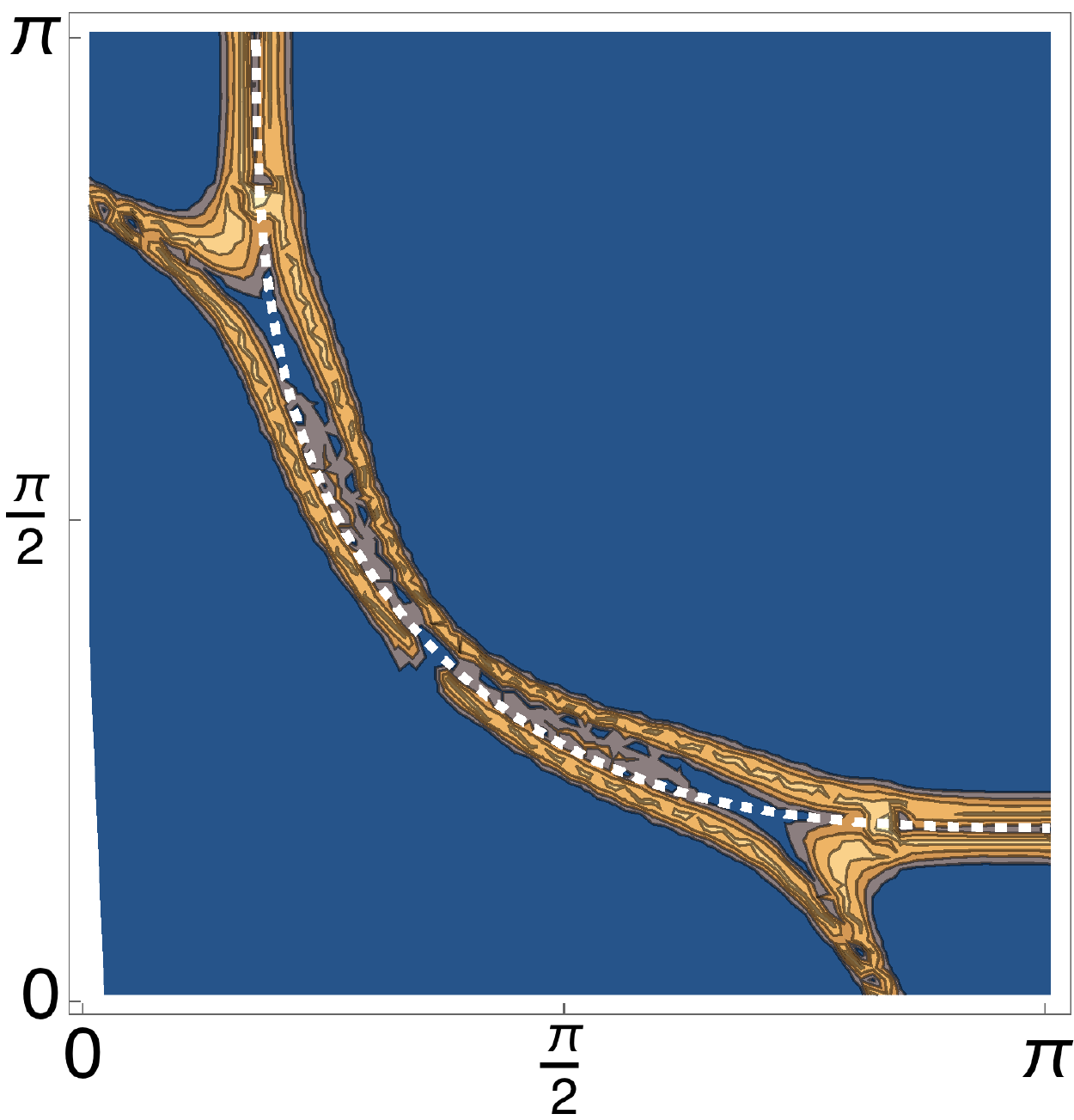} \vspace{1ex}
\end{minipage}

\begin{minipage}[c]{4.25cm}%
g)\includegraphics[width=3.95cm]{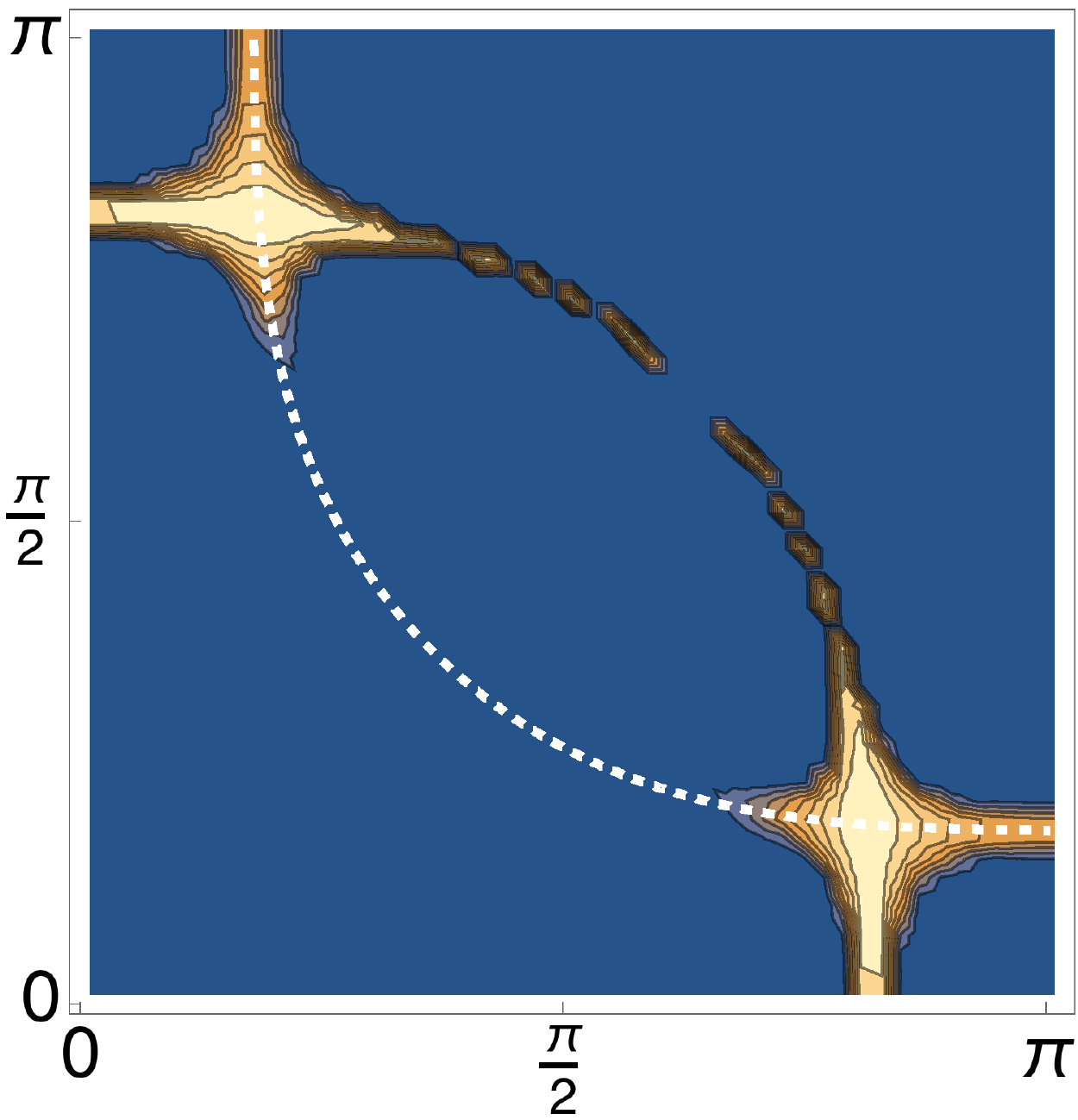} \vspace{1ex}
\end{minipage}

\caption{\label{fig:gaps1} (Color online) Solution of the
gap equations $\chi_{k,k+Q_{0}}$ from Eqn. (\ref{eq:chiQ0}) (panels
a), b), c), d) e) and f) and  the superconducting
order parameter\textcolor{black}{{} $\Delta_{k}$ from (\ref{eq:SCgap})
(pannel f)). Vanishing solutions are color-coded in blue while non
vanishing points are depicted in yellow. We took various modulation
wave vectors $\mathbf{Q}_{0}$ with a) the diagonal wave vector $\mathbf{Q}_{0}=\left(Q_{\mathbf{0}},Q_{\mathbf{0}}\right)$
linking two hot spots, b) the axial wave vector $\mathbf{Q}_{0}=\left(Q_{\mathbf{0}},0\right)$
and c) $\mathbf{Q}_{0}=\left(0,Q_{\mathbf{0}}\right)$ which are observed
experimentally, d) the AF wave vector $\mathbf{Q}_{0}=\left(\pi,\pi\right)$
and e) the null wave vector $\mathbf{Q}_{0}=\left(0,0\right)$}
and f) the $\pf$-wave vector corresponding to the involution described
in Eq.(\ref{eq:8}).\textcolor{black}{{} The solution of the SC gap
equation is given in g). The calculations are made on the band structure
of Bi2212 form Ref.\cite{Norman07} (see details in the text for the
band parameters). The calculations are made within the approximation
$J_{\overline{\mathbf{q}}}=J\delta\left(\overline{\mathbf{q}}\right)$,
with $J=0.35$, which restricts the q-integration at the vector $\left(\pi,\pi\right)$.
The energy units, if not stated otherwise, are in eV.}}
\end{figure}

Throughout the paper, if not stated otherwise, the calculations are
made for Bi2212, with a band structure taken from Ref.\textcolor{black}{\cite{Norman07}}.
Specifically we take 
\begin{align}
\xi_{\mathbf{k}} & =2t_{1}+t_{2}\left(\cos k_{x}+\cos k_{y}\right)+2t_{3}\cos k_{x}\cos k_{y}\\
 & +t_{4}\cos2k_{x}+\cos2k_{y}+t_{5}\left(\cos2k_{x}\cos k_{y}+\cos2k_{y}\cos k_{x}\right)\nonumber \\
 & +2t_{6}\cos2k_{x}\cos2ky-\mu,\nonumber 
\end{align}

with (in eV) $t_{1}=0.196$, $t_{2}=-0.6798$, $t_{3}=0.2368$, $t_{4}=-0.0794$,
$t_{5}=0.0343$ and $t_{6}=0.0011$. The solution of Eqns.(\ref{eq:chiQ0},\ref{eq:SCgap})
is given in Fig.(\ref{fig:gaps1}) for various charge modulation vectors.
The main point of this preliminary study, is that all the wave vectors
have an equivalent response at the hot-spot, which is also the same
as the SC response. In other words, all the orders
considered above are quasi-degenerate at the hot spots. The difference
between the gap solution of various wave vectors lies in its extension
in \textbf{k}-space, which is more pronounced for the SC, the $\pf$
and the diagonal $\mathbf{Q}_{0}=\left(Q_{\mathbf{0}},Q_{\mathbf{0}}\right)$
cases. The only modulation wave vectors which give a non-zero answer
are the ones relating two hot spots, or surrounding
the hot spots in the case of the SC and $\pf$ orders. An important point is that d-wave symmetry is
required to satisfied Eqns.(\ref{eq:chiQ0},\ref{eq:SCgap}). A simple
way to see this is to notice that the gap equations relate the two
antinoal zones $\mathbf{k}\rightarrow\mathbf{k}+\mathbf{Q}$, with
$\mathbf{Q=\left(\pi,\pi\right)}$ the AF wave vector. Solutions with
$\Delta_{\mathbf{k}}=-\Delta_{\mathbf{k+Q}}$ are thus stabilized.

\textcolor{black}{The case of strong coupling is treated in Appendix
\ref{sec:Mean-Field-gap-equations:}, where we see that, as the coupling
increased, the shape of the SC and CDW changes. The SC solution is
now gapping out the entire fermi surface whereas the CDW solutions
are confined within the anti-nodal regions. The development of the
SU(2) fluctuations requires the mean field decoupling to give sensibly
equal values of the order parameters in the two sectors. This is true
at the hot-spots, as seen in the next sub section \ref{sub:High-energy-dome},
but it not valid anymore far away from the hot-spot. A simple way
to quantize this effect is to define the cut-off energy scale below
which SU(2) fluctuations are present, as the mean of the gaps in the
two sectors 
\begin{align}
\Delta_{SU(2)}^{2} & =\sqrt{\chi^{2}\Delta^{2}},\label{eq:gapmean}
\end{align}
}

so that $\Delta_{SU(2)}$ naturally vanishes away from the hot-spots.

\subsection{Cut-off energy scale \label{sub:High-energy-dome}}

\begin{figure}[b]
\begin{minipage}[c]{7.25cm}%
\includegraphics[width=7cm]{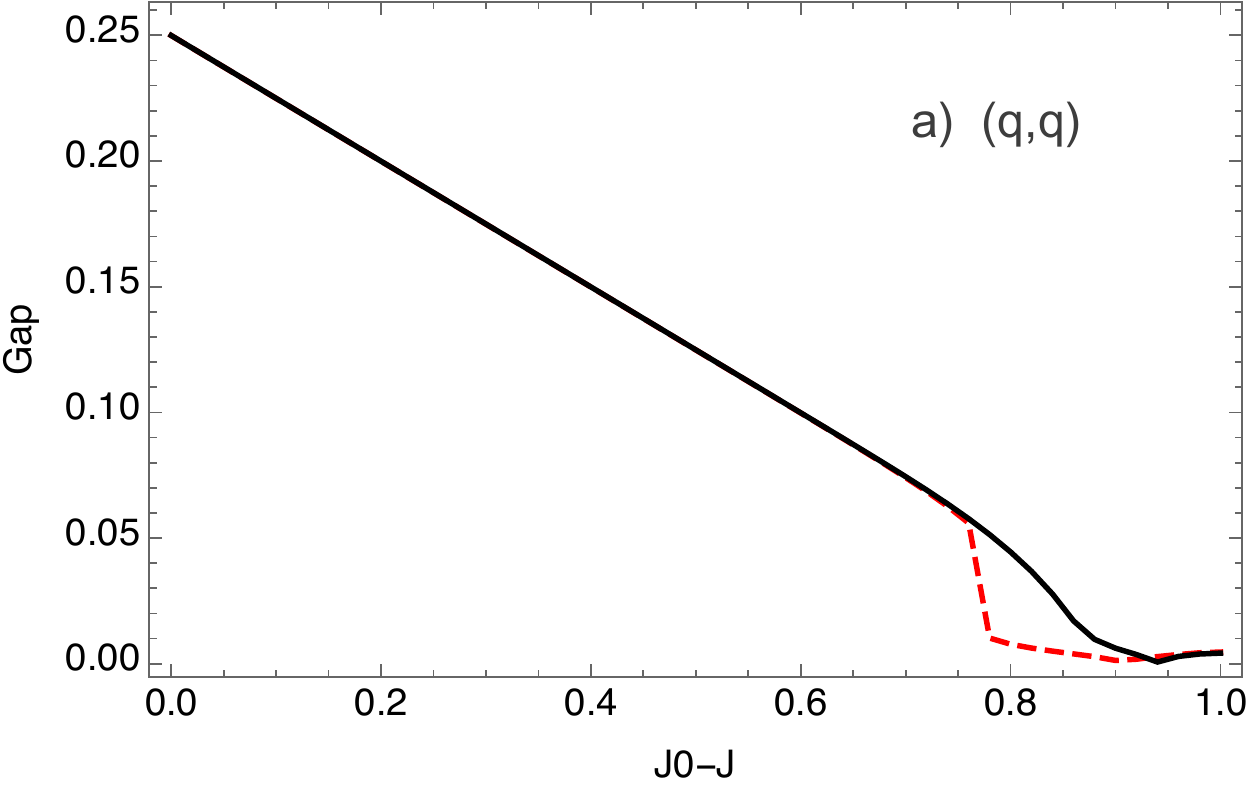} \vspace{1ex}
\end{minipage}

\begin{minipage}[c]{7.25cm}%
\includegraphics[width=7cm]{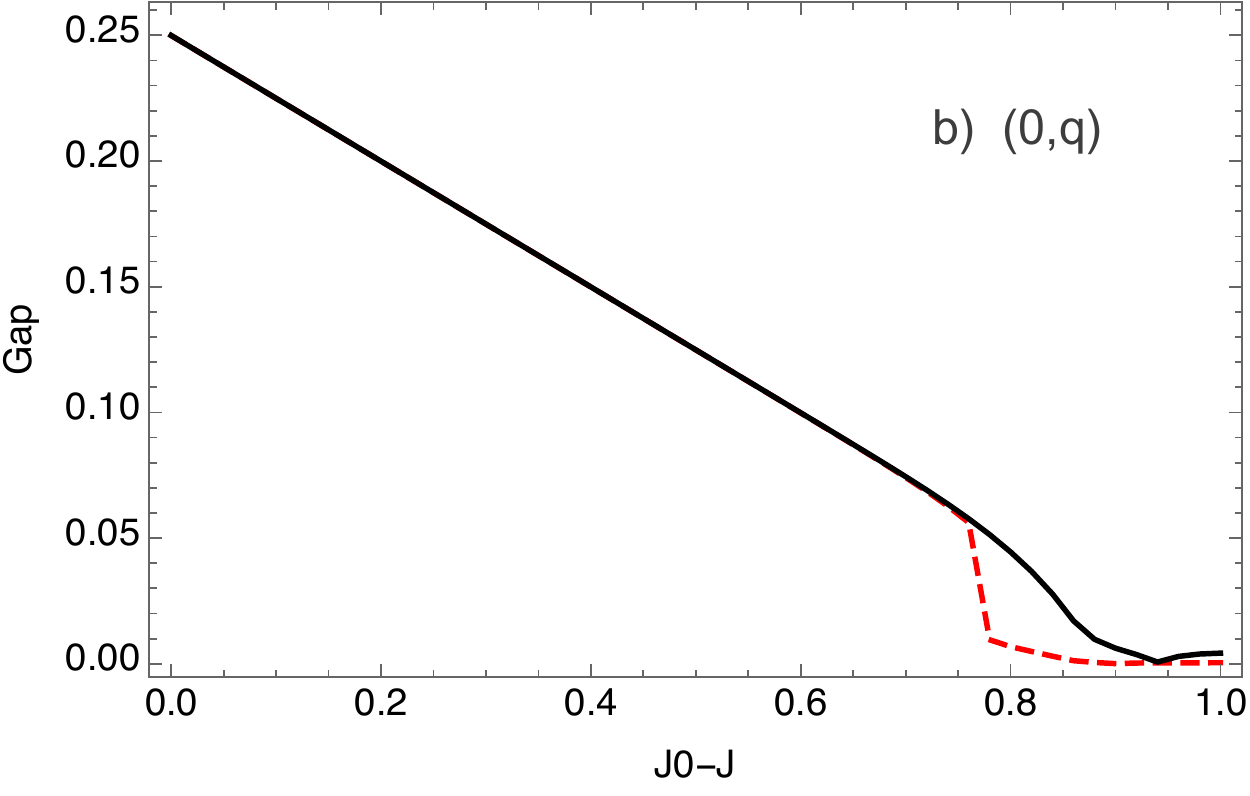} \vspace{1ex}
\end{minipage}

\begin{minipage}[c]{7.25cm}%
\includegraphics[width=7cm]{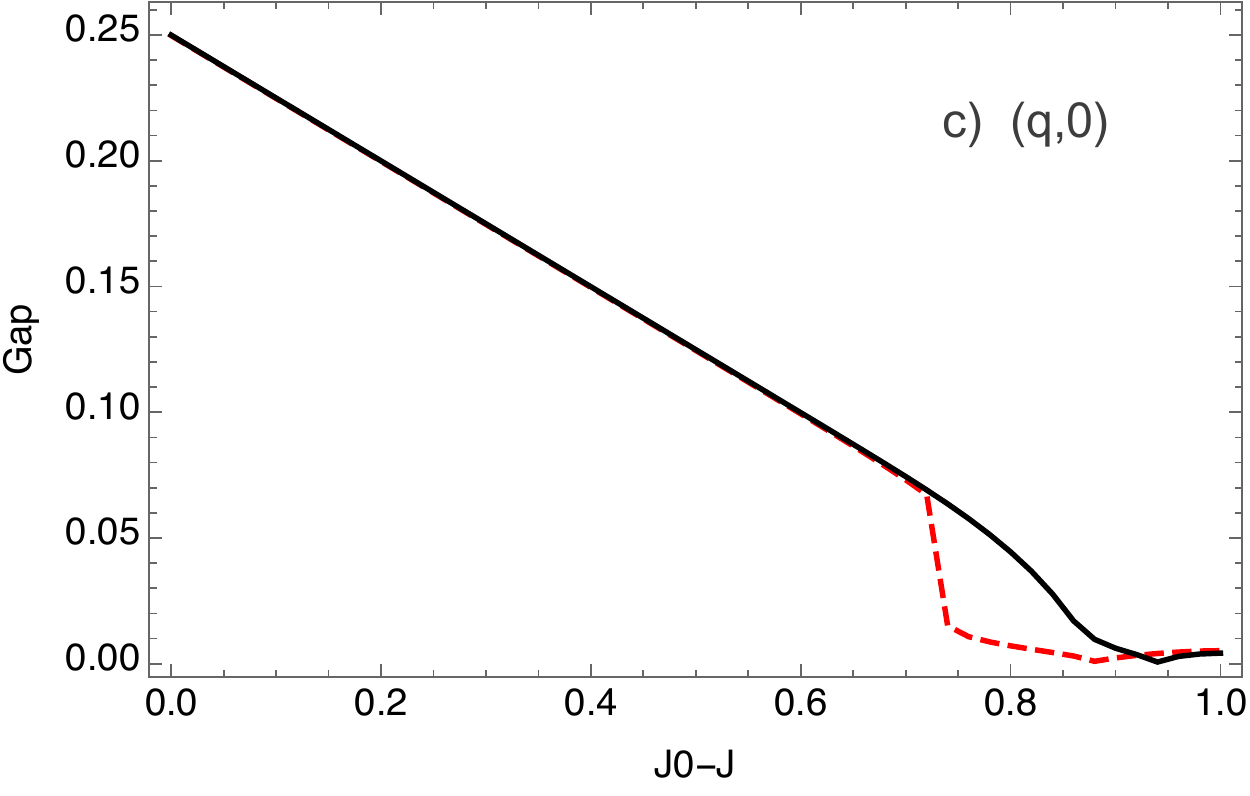} \vspace{1ex}
\end{minipage}\caption{\label{fig:dome} (Color online) \textcolor{black}{Comparison of the
d-wave charge $\chi_{k,k+Q_{0}}$ solution of Eqn.(\ref{eq:chiQ0})
(dashed red) and d-wave SC $\Delta_{k}$ solution of the Eqn.(\ref{eq:SCgap})
(black line) taken at the hot spot. We compare various modulation
wave vectors for $\chi_{k,k+Q_{0}}$ with a) the diagonal wave vector
$\mathbf{Q}_{0}=\left(Q_{\mathbf{0}},Q_{\mathbf{0}}\right)$ linking
two hot spots, b) the axial wave vector $\mathbf{Q}_{0}=\left(0,Q_{\mathbf{0}}\right)$
and c)$\mathbf{Q}_{0}=\left(Q_{\mathbf{0}},0\right)$. The evolution of the SC and CDW gaps as a function of $J-J_{0}$ has the typical form of a dome. SC
and CDW solutions at the hot spots are completely degenerate in a
with range of $J$, whereas the CDW solution is lost before the SC
one when $J\sim J_{0}$ ($J_{0}=1$).}}
\end{figure}

The starting point of our reflexion is to notice that a simple model
with short range AF correlations, which is minimal to describe the
under-doped regime of cuprate superconductors, has a few quasi-degenerate
solutions at the hot spot, including the d-wave SC and d-wave charge
orders. Our assumption, starting from now, is that this simple model
gives a good insight, and hints that an SU(2) symmetry is present
in the phase diagram of those compounds, which relates the d-wave
SC state to the d-wave charge sector. The SU(2) symmetry is broken
at low temperature, but then fluctuations will exist up to a temperature
scale which defines the SU(2)-dome. In Fig.\ref{fig:dome} the solutions
at the hot spots of the d-wave SC and d -wave CDW are given for various
wave vectors, as a functions of the decreasing AF coupling constant
$J$ present in Eqns.(\textcolor{black}{\ref{eq:chiQ0}}) and (\textcolor{black}{\ref{eq:SCgap}}).
$J$ slowly decreases from $J=J_{0}=1$ at half filling ($p=0$, where
$p$ is the hole doping), to $J\simeq0$ at larger hole doping. Assuming
a scaling relation of the type $p\sim\left(J_{0}-J\right)^{\alpha}$,
we get a form of the PG dome very close to the one experimentally
observed in cuprates. For a wide region of hole-doping, the SC solution
at the hot spot is degenerate with the CDW one. When $J\sim0$, the
CDW solution is lost whereas the SC solution survives. The phase diagrams
of Fig.\ref{fig:dome} mimic the situation in the under-doped regime
of the cuprates as a function of hole doping. The region where the
two solutions are degenerate is interpreted in our framework as the
SU(2)-envelop, below which SU(2) fluctuations are present. They will
be described in the next section.

\section{SU(2) fluctuations coupled to fermions\label{sec:SU(2)-fluctuations-coupled}}

In the previous section we have shown that short
range AF correlations give rise to a finite number of possible d-wave
order parameters which are quasi-degenerate at the hot spots. The
main idea of this paper is that, first this quasi-degeneracy is described
by an emerging SU(2) symmetry, and second, that the fluctuations associated
with this symmetry are in turn lifting the degeneracy between the
various modulation vectors. This section is devoted to the study of
the SU(2) fluctuations.

In order to proceed with the study of the SU(2) fluctuations, we must
choose one of the wave vectors associated with the charge sector.
For definiteness, we start with the diagonal wave vector\textcolor{black}{{}
$\mathbf{Q_{0}}=\left(Q_{\mathbf{0}},Q_{\mathbf{0}}\right)$}, bearing
in mind that it is not the one experimentally observed in the under-doped
region. Our starting point is to derive the SU(2) effective model
which couples to fermions. The action is comprised of three terms
: 
\begin{align}
S_{st} & =S_{\psi}^{0}+S_{int}+S_{Q}^{0}.\label{eq:B1}
\end{align}

\subsection{Bare action $S_{\psi}^{0}$}

$S_{\psi}^{0}$ is the bare action for electrons, which is defined
in SU(2) context as 
\begin{align}
S_{\psi}^{0} & =-\int_{x,x'}\overline{\Psi}_{x}{G_{0}}_{x,x'}^{-1}\Psi_{x'},\label{eq:bare}
\end{align}
where $x=\left({\bf {r},\tau,\sigma}\right)$ with $\sigma\in\{\uparrow,\downarrow\}$
the spin and $\int_{x}\equiv\int d{\bf {r}}\int_{0}^{\beta}d\tau\sum_{\sigma}$
and the free inverse propagator is 
\begin{equation}
{G_{0}}_{x,x'}^{-1}=(\partial_{\tau}-\hat{\xi}_{i\nabla_{{\bf r}}})\delta^{(d)}({\bf {r}}-{\bf {r}'})\delta(\tau-\tau')\delta_{\sigma,\sigma'}.
\end{equation}
In momentum and imaginary frequency space, the Green functions are
defined as 
\begin{equation}
G_{k,k'}\delta_{\sigma,\sigma'}=-\langle\mathcal{T}\Psi_{\sigma}(k)\bar{\Psi}_{\sigma'}(k')\rangle.\label{eq:gdef}
\end{equation}
The field $\Psi$ is written in a $4\times4$ basis in momentum space
with 
\begin{align}
\Psi_{k} & =\frac{1}{\sqrt{2}}\left(\begin{array}{cccc}
\psi_{\mathbf{k},\sigma}, & \psi_{-\mathbf{k}-\mathbf{Q_{0}},-\sigma}^{\dagger}, & \psi_{\mathbf{k}+\mathbf{Q_{0}},\sigma}, & \psi_{-\mathbf{k},-\sigma}^{\dagger}\end{array}\right),^{T}\label{eq:29}
\end{align}
where $\left(\right)^{T}$ denotes the standard transposition, and
\begin{align}
\overline{\Psi}_{k} & =\frac{1}{\sqrt{2}}\left(\begin{array}{cccc}
\psi_{\mathbf{k},\sigma}^{\dagger}, & -\psi_{-\mathbf{k}-\mathbf{Q_{0}},-\sigma}, & \psi_{\mathbf{k}+\mathbf{Q_{0}},\sigma}^{\dagger}, & -\psi_{-\mathbf{k},-\sigma}\end{array}\right),\label{eq:30}
\end{align}
where $k\equiv(i\omega_{n},{\bf k})$ and the factor $1/\sqrt{2}$
normalizes the spin summation. Note that the conjugation in the particle-hole
sector ($\tau$) is not standard, with the ''charge conjugate''
defined as $\overline{\Psi}=\Psi^{\dagger}\tau_{3}$. Throughout the
paper $\tau_{\alpha}$, $\Lambda_{\alpha}$ with $\alpha=1,3$ stand
for the Pauli matrices in each sector. In this basis, and in momentum
space, the bare electron action becomes 
\begin{equation}
S_{\psi}^{0}=-\frac{1}{\beta N}\sum_{\mathbf{k},\omega}\overline{\Psi}_{k}{G_{0}}_{k}^{-1}\Psi_{k},\label{eq:bare1}
\end{equation}
and $G_{0}^{-1}$is defined as 
\begin{align}
\hat{G}_{0,k}^{-1} & =\left(\begin{array}{cc|cc}
i\omega-\xi_{{\bf k}} & \\
 & i\omega+\xi_{-{\bf k}-{\bf \mathbf{Q_{0}}}}\\
\hline  &  & i\omega-\xi_{{\bf k}+{\bf \mathbf{Q_{0}}}}\\
 &  &  & i\omega+\xi_{-{\bf k}}
\end{array}\right)_{\Lambda},\label{eq:g0mat}
\end{align}
where $\mathbf{Q_{0}}$ is the diagonal wave vector connecting two
hot spots, as defined in Eqn.(\ref{eq:1-1}), and $\xi_{\mathbf{k}}$
is the electronic dispersion. The $4\times4$ basis can be conveniently
factorized as the direct product of two sub-spaces $\tau\otimes\Lambda$
where $\tau$ is the charge conjugation space describing the SC channel
and $\Lambda$ is the subspace corresponding to the translation by
the vector $\mathbf{Q_{0}}$. In the case where the model is reduced
to eight hot spots (see e.g. \cite{Efetov13}), the vector $\mathbf{Q_{0}}$
corresponds to the vector $2\mathbf{k}_{F}$ relating the diagonal
hot spots in the same AN region, and we have the symmetry relations
$\xi_{-{\bf k}}=\xi_{{\bf k}}$ and $\xi_{{\bf k}+\mathbf{Q_{0}}}=-\xi_{{\bf k}}$,
the latter being valid when the dispersion is linearized around the
Fermi surface and close to the hot spots. Within this approximation
we get 
\begin{align}
\hat{G}_{0,hs,k}^{-1} & =\left(\begin{array}{cc|cc}
i\omega-\xi_{{\bf k}} & \\
 & i\omega-\xi_{{\bf k}}\\
\hline  &  & i\omega+\xi_{{\bf k}}\\
 &  &  & i\omega+\xi_{{\bf k}}
\end{array}\right)_{\Lambda},\nonumber \\
 & =i\omega-\xi_{{\bf k}}\Lambda_{3}.\label{eq:disp}
\end{align}
In the form of Eqn.(\ref{eq:disp}), the SU(2) symmetry in $G_{0,hs,k}^{-1}$
is explicit. In all generality, it possible to models the term breaking
the SU(2) symmetry by noticing that the condition $\xi_{{\bf {k}+{\bf {\mathbf{Q_{0}}}}}}=-\xi_{{\bf {k}}}$
is valid only close to the hot spot and when the dispersion is linearized
around the Fermi level. If this condition is not verified, we define
\begin{align}
\overline{\xi}_{{\bf k}} & =(\xi_{{\bf {k}}}-\xi_{{\bf {k}+{\bf {\mathbf{Q_{0}}}}}})/2,\\
\Delta\xi_{{\bf k}} & =(\xi_{{\bf {k}}}+\xi_{{\bf {k}+{\bf {\mathbf{Q_{0}}}}}})/2,
\end{align}

where $\overline{\xi}_{\mathbf{k}}$ is the symmetric dispersion and
$\Delta\xi_{\mathbf{k}}$ can be understood as a curvature term. The
matrix $\hat{G}_{0}^{-1}$in Eqn.(\ref{eq:g0mat}) then takes the
form 
\begin{align}
\hat{G}_{0,k}^{-1} & =\left(\begin{array}{cc}
i\omega-\mathbf{\overline{\xi}}_{{\bf k}}-\Delta\xi_{{\bf k}}\tau_{3} & 0\\
0 & i\omega+\overline{\xi}_{{\bf k}}-\Delta\xi_{{\bf k}}\tau_{3}
\end{array}\right)_{\Lambda}\label{disp1}
\end{align}
which we can rewrite Eq.(\ref{disp1}) as 
\begin{equation}
\hat{G}_{0,k}^{-1}=i\omega-\overline{\xi}_{{\bf k}}\Lambda_{3}+\Delta\xi_{{\bf k}}\tau_{3},\label{eq:disp2}
\end{equation}
where $G_{0,s,k}^{-1}=i\omega-\overline{\xi}_{{\bf k}}\Lambda_{3}$
is SU(2)-symmetric (proportional to $\tau_{0}$) and the term $\Delta\xi_{{\bf k}}\tau_{3}$
is the SU(2) symmetry-breaking term (proportional to $\tau_{3}$).
The separation between symmetric and symmetry breaking terms in Eqns.(\ref{disp1},\ref{eq:disp2})
is the crucial step, which will be useful in describing the fluctuations
in section \ref{sec:Non-linear--model}.

\subsection{Interacting term $S_{int}$}

The interacting term can simply be taken as a two-body interaction
\begin{align}
S_{i}^{0} & =\frac{1}{4}\int_{x,x'}\gamma_{x-x'}Tr\left[\overline{\Psi}_{x}\Psi_{x}\overline{\Psi}_{x'}\Psi_{x'}\right],
\end{align}

where $\Psi$ is the two by for fields- spinor defined in Eqn.(\ref{eq:gdef}).
The form of the propagator $\gamma_{x-x'}$ is not detailed at the
moment. Using a Hubbard-Stratonovich decoupling with respect to the
field $\hat{Q}_{x,x'}$, we get $S_{i}^{0}\rightarrow S_{int}+S_{Q}^{0}$
from Eqn.(\ref{eq:B1}) with

\begin{align}
S_{int} & =\frac{1}{2}\int_{x,x'}Tr\left[\overline{\Psi}_{x}\hat{Q}_{x,x'}\Psi_{x'}\right],\label{eq:B2}\\
S_{Q}^{0} & =\frac{1}{4}\int_{x,x'}Tr\left[\overline{\hat{Q}}_{x,x'}\hat{\gamma}_{x-x'}^{-1}\hat{Q}_{x',x}\right],\label{eq:B2'}
\end{align}

where the $Tr$ runs over the matrix structure and $\hat{Q}_{x,x'}\sim\left\langle \overline{\Psi}_{x}\Psi_{x'}\right\rangle $is
a $4\times4$ matrix which can be decomposed ( within the direct product
of spaces $\tau\times\Lambda$) as 
\begin{equation}
\hat{Q}_{x,x'}=\left(\begin{array}{cc}
 & \hat{q}_{x,x'}\\
\hat{q}_{x,x'}^{\dagger}
\end{array}\right)_{\Lambda},\label{eq:9}
\end{equation}
with $\hat{q}_{x,x'}=Q_{x-x'}^{0}\hat{u}_{x,x'}$,

$\hat{u}_{x,x'}=\frac{1}{N^{2}}\sum_{\mathbf{k,}\mathbf{k'},\omega'}e^{-i\mathbf{k\cdot}r}e^{i\omega_{n}\tau}e^{i\mathbf{k'\cdot}r'}e^{-i\omega'_{n}\tau'}\hat{u}_{k,k'}$,
and 
\begin{align}
\hat{u}_{k,k'} & =\left(\begin{array}{cc}
\chi_{k,k'+Q_{0}} & -\sigma\Delta_{k,-k'}\\
\sigma\Delta_{k+Q_{0},-k'-Q_{0}}^{\dagger} & \chi_{-k-Q_{0},-k'}^{\dagger}
\end{array}\right)_{\tau}.\label{eq:10}
\end{align}

Mean field effects are obtained by taking $k'=k$. Indeed, the field
$\chi$ represents a particle-hole pair, suitable to describe the
charge modulations ($\chi_{k,k'+Q_{0}}\sim\left\langle \psi_{\mathbf{k},\sigma}^{\dagger}\psi_{\mathbf{k}+Q_{0},\sigma}\right\rangle $),
while the field $\Delta$ is the SC particle-particle pairing field
describing the formation of coherent pairs ($\Delta_{k,-k'}\sim\sigma\left\langle \psi_{\mathbf{k},-\sigma}\psi_{\mathbf{-k},\sigma}\right\rangle $).
In this limit, the matrix $\hat{u}_{k,k'}$ in Eqn. (\ref{eq:10})
writes 
\begin{align}
\hat{u}_{k} & =\left(\begin{array}{cc}
\chi_{k,k+Q_{0}} & -\sigma\Delta_{k,-k}\\
\sigma\Delta_{k+Q_{0},-k-Q_{0}}^{\dagger} & \chi_{-k-Q_{0},-k}^{\dagger}
\end{array}\right)_{\tau},
\end{align}

where 
\begin{align*}
\Delta_{k} & \sim\sigma\langle\psi_{k,-\sigma}{\psi_{-k,\sigma}}\rangle\\
\chi_{k} & \sim\langle{\psi_{k,\sigma}^{\dagger}}{\psi_{k+Q_{0},\sigma}}\rangle\\
\Delta_{k+Q_{0}}^{\dagger} & \sim\sigma\langle\psi_{k+Q_{0},-\sigma}{\psi_{-k-Q_{0},\sigma}}\rangle\\
\chi_{-k-Q_{0}}^{\dagger} & \sim\langle{\psi_{-k-Q_{0},-\sigma}^{\dagger}}{\psi_{-k,-\sigma}}\rangle
\end{align*}
The SU(2)-symmetry requires that $\Delta_{k+Q_{0},-k-Q_{0}}^{\dagger}=\Delta_{k,-k}^{\dagger}$
and $\chi_{-k-Q_{0},-k}^{\dagger}=\chi_{k,k+Q_{0}}^{\dagger}$, which
is approximately verified in the linearized regime. The SU(2) condition
then implies that $\hat{u}^{\dagger}\hat{u}=1$, which in turn requires
that $|\chi|^{2}+|\Delta|^{2}=1$.

We now expand around the mean-field values of the parameters in order
to get the small fluctuations regime. We first Fourier transform to
get $\hat{Q}_{x,x'}\rightarrow\hat{Q}_{k,k'}$ and then Wigner transform
it, which leads to 
\begin{align}
\hat{Q}_{x,x'} & \rightarrow\hat{M}_{x-x',\left(x+x'\right)/2}.
\end{align}
In Fourier space, this writes $\hat{M}_{x-x',\left(x+x'\right)/2}\rightarrow\hat{M}_{\left(k+k'\right)/2,k-k'}$.
We then decompose into fast and slow variables as $\hat{M}\sim\left\langle \overline{\Psi}_{k+q/2}\Psi_{k-q/2}\right\rangle $
with the fast momenta ${\bf {k}}\simeq\kf$ and the slow momenta ${\bf {q}}\ll\kf$.
With the change of variables as $k\rightarrow k+q/2$ and $k'\rightarrow k-q/2$
, we get $\hat{M}_{k,q}$, such as

\begin{align}
S_{int} & =\frac{1}{2}\sum_{k,q,\sigma}Tr\left[\overline{\Psi}_{k+q/2}\hat{M}_{k,q}\Psi_{k-q/2}\right],
\end{align}

where $\overline{\psi}_{k}$ and $\psi_{k}$ are given by Eqns.(\ref{eq:29},\ref{eq:30}).
We have then

\begin{align}
\hat{M}_{k,q} & =M_{k}\ \hat{U}_{k,q},\mbox{ with }\hat{U}_{k,q}=\left(\!\!\begin{array}{cc}
 & \hat{u}_{k,q}\\
\hat{u}_{k,q}^{\dagger}
\end{array}\!\!\right)_{\Lambda},\label{eq:invpropg-1}\\
 & \mbox{ and }\hat{u}_{k,q}=\left(\!\!\begin{array}{cc}
\chi_{k,q} & -\sigma\Delta_{k,q}\\
\sigma\Delta_{k+Q_{0},q}^{\dagger} & \chi_{-k-Q_{0},q}^{\dagger}
\end{array}\!\!\right)_{\tau}\nonumber 
\end{align}
where $M_{k}$ is the magnitude of the order parameter while $\hat{u}_{k,q}$
is the SU(2) non-abelian phase associated to it. We have 
\begin{align*}
\Delta_{k,q} & \sim\sigma\langle\psi_{k+q/2,-\sigma}{\psi_{-k+q/2,\sigma}}\rangle\\
\chi_{k,q} & \sim\langle{\psi_{k+q/2,\sigma}^{\dagger}}{\psi_{k+Q_{0}-q/2,\sigma}}\rangle\\
\Delta_{k+Q_{0},q}^{\dagger} & \sim\sigma\langle\psi_{k+Q_{0}+q/2,-\sigma}{\psi_{-k-Q_{0}+q/2,\sigma}}\rangle\\
\chi_{-k-Q_{0},q}^{\dagger} & \sim\langle{\psi_{-k-Q_{0}-q/2,-\sigma}^{\dagger}}{\psi_{-k+q/2,-\sigma}}\rangle
\end{align*}
As mentioned above, in the linearized regime, we have $\chi_{-k-Q_{0}}^{\dagger}=\chi_{k}$
and $\Delta_{k+Q_{_{0}}}^{\dagger}=-\Delta_{k}^{\dagger}$ which ensures
the SU(2) condition that the determinant is equal to one, such that
$|\chi|^{2}+|\Delta|^{2}=1$ and also implies that $\hat{u}_{k,q}^{\dagger}\hat{u}_{k,q}=\hat{1}$.
In this regime, we have 
\begin{align}
\hat{u}_{k,q}=\left(\!\!\begin{array}{cc}
\chi_{k,q} & -\sigma\Delta_{k,q}\\
\sigma\Delta_{k,q}^{\dagger} & \chi_{k,q}^{\dagger}
\end{array}\!\!\right)_{\tau}.
\end{align}

The decomposition of the interaction field $\hat{M}_{k,q}$ into an
amplitude $M_{k}$ and an SU(2)-``phase'' $\hat{u}_{k,q}$ in Eqn.(\ref{eq:invpropg-1})
is a second important ingredient in the study of the fluctuations
in section \ref{sec:Non-linear--model}. We will see there that the
writing of the non-linear $\sigma$-model relies on the separation
between a field depending only on fast variables $M_{k}$, whereas
the phase will depend on slow variables only $\hat{k}_{k,q}\sim\hat{u}_{q}$.

\subsection{The quadratic term $S_{Q}^{0}$}

The quadratic term in $S_{st}$ Eqn.(\ref{eq:B1}) and Eqn.(\ref{eq:B2'})
writes 
\begin{align}
S_{Q}^{0} & =\frac{1}{4}\int_{x,x'}Tr\left[\overline{\hat{Q}}_{x,x'}\hat{\gamma}_{x-x'}^{-1}\hat{Q}_{x',x}\right],\label{eq:B3}
\end{align}
where $\hat{\gamma}^{-1}$ is a bare propagator whose form is not
crucial at this stage, since it is to be renormalized by the thermal
fluctuations described in section \ref{sec:Non-linear--model}. In
the case of the eight hot spot model \cite{Efetov13}, or of the spin
Fermion model with hot regions \cite{Kloss15}, this term corresponds
to anti-ferromagnetic (AF) paramagnons mediating the formation of
the SC and CDW orders, but in the minimal version of the model which
is controlled solely by the SU(2) symmetry, it is not necessary to
mention the origin of the bare propagator.

\section{Non-linear $\sigma$-model\label{sec:Non-linear--model}}

We derive the fluctuations induced by the SU(2) structure presented
in the two preceding sections. The generic form of the O(4) non-linear
$\sigma$-model is obtained by integrating out the fermions in Eqn.(\ref{eq:B1})
and extracting the $S_{Q}$ action which renormalizes Eqn.(\ref{eq:B3}).
After formally integrating out the fermions, we get 
\begin{align}
S_{eff} & =\frac{1}{4}\int_{\mathbf{x},\mathbf{x'}}\hat{\gamma}_{x-x'}^{-1}Tr\left[\overline{\hat{Q}}_{x,x'}\hat{Q}_{x',x}-\frac{1}{2}\log\hat{G}_{x,x'}^{-1}\right],\label{eq:nlsm3}\\
\mbox{ with } & \hat{G}_{x,x'}^{-1}=\partial_{\tau}-\hat{\xi}_{{\bf x}-{\bf x}'}+\hat{Q}_{x,x'},\nonumber 
\end{align}
where the $Tr$ operates on the matrix structure except on the space
indices and $\hat{Q}$ is the SU(2) operator obtained by the Hubbard-Stratonovich
decoupling. We can now Wigner- transform Eqn.(\ref{eq:nlsm3}), which
yields

\begin{align}
S_{eff}=\frac{1}{4}\sum_{k,q,\overline{q}}\gamma_{\overline{q}}^{-1}Tr\hat{\overline{M}}_{k+\overline{q},q}\hat{M}_{k,q}-\frac{1}{2}\sum_{k,q}Tr\log\hat{G}_{k,q}^{-1},\label{eq:nlsm4}\\
\mbox{ with }\hat{G}_{k,q}^{-1}=\hat{G}_{0k}^{-1}-\frac{1}{2}\hat{M}_{k,q},\label{eq:B5}
\end{align}

with the matrix $\hat{M}$ defined in (\ref{eq:invpropg-1}).

\subsection{Separation of variables}

The fluctuations associated with the non-linear $\sigma$-model are
obtained by separating the fast momenta ${\bf {k}\sim\kf}$ and the
slow momenta ${\bf {q}\ll\kf}$, as 
\begin{equation}
\hat{M}_{k,q}=M_{0,k}\ {\hat{U}}_{q},\mbox{ and }\hat{U}_{q}=\left(\begin{array}{cc}
 & \hat{u}_{q}\\
\hat{u}_{q}^{\dagger}
\end{array}\right)_{\Lambda},\label{eq:18}
\end{equation}
with $M_{0,k}$ being the ``fast varying'' component, is a scale
comparable to the SU(2) dome $\Delta_{SU(2)}$ of subsection \ref{sub:High-energy-dome}.
The slow varying variables are taken to act only on the SU(2) matrix
$\hat{U}_{q}$. In the following we first assume, using the symmetric
part of the bare action in Eqn.(\ref{eq:disp2}), that the condition
of separation of variables Eqn.(\ref{eq:18}) is valid everywhere
and derive the effective non-linear $\sigma$-model in subsections
\ref{sub:Symmetric-part} and \ref{sub:Effective-model}. The validity
of the hypothesis of separation of variables relies on the physical
idea that the slow varying phases $\hat{U}_{q}$ can be treated as
perturbations around a larger mean-field like field $M_{0,k}$. This
idea will be tested in subsection \ref{sub:Evaluation-of-the}, and
we will discover that it is valid only in a restricted parts of the
Brillouin Zone (BZ).

The SU(2) condition is now given by 
\begin{align}
{\hat{U}}_{q}^{\dagger}\hat{U}_{q}=\hat{1},\label{eq:su2}\\
\mbox{or }\left|\Delta_{q}\right|^{2}+\left|\chi_{q}\right|^{2}=1.\nonumber 
\end{align}
Thermal fluctuations then correspond to variations of $\delta\hat{U}_{q}$,
$M_{0,k}$ being kept as a constant: 
\begin{equation}
\delta\hat{M}_{k,q}=M_{0,k}\ \delta\hat{U}_{q}.\label{eq:27}
\end{equation}
Introducing this decoupling back into Eqn.(\ref{eq:nlsm4}) we note
that the first term does not contribute because of the unitarity condition
(\ref{eq:su2}). We henceforth expand the free energy in the second
order in the Hubbard-Stratonovich fields. In real space, from Eqn.(\ref{eq:B2})
and 
\begin{equation}
Z=e^{-S_{t}},\mbox{with }S_{t}=-\frac{1}{2}\left\langle \left(S_{int}\right)^{2}\right\rangle _{\phi}.
\end{equation}
and $F=-T\ln Z$ we get 
\begin{align*}
F\left[u\right] & =\frac{T^{2}}{4}\int_{x,x',x_{1},x'_{1}}Tr\left\langle \bar{\Psi}_{x}\hat{\overline{Q}}_{x,x'}\Psi_{x'}\overline{\Psi}_{x_{1}}\hat{Q}_{x_{1},x'_{1}}\Psi_{x'_{1}}\right\rangle _{\phi},
\end{align*}

where the $Tr$ runs over the $4\times4$ matrix and the fields are
defined in Eqns.(\ref{eq:bare},\ref{eq:gdef}). Performing the Wick
pairing of the fields yields with the definition of the Green functions
in Eq.(\ref{eq:gdef}) we get 
\begin{align}
F\left[u\right]=\frac{T^{2}}{4}\int_{x,x',x_{1},x'_{1}}Tr[\hat{\overline{Q}}_{x,x'}\hat{G}_{x',x_{1}}\hat{Q}_{x_{1},x'_{1}}\hat{G}_{x'_{1},x}],\label{eq:C1}
\end{align}
which after Fourier transforming, gives 
\begin{align}
F\left[u\right] & =\frac{T^{2}}{4}\sum_{\varepsilon,\varepsilon'}\sum_{\mathbf{k},\mathbf{k}',\mathbf{k_{1}},\mathbf{k'_{1}}}Tr[\hat{\overline{Q}}_{k,k'}\hat{G}_{k',k_{1}}\hat{Q}_{k_{1},k'_{1}}\hat{G}_{k'_{1},k}],\label{eq:C2}
\end{align}

where the $Tr$ runs on the spin indices, as well as on the $4\times4$
matrices, and $\hat{G}$ is given by 
\begin{align}
\hat{G}_{k,k'}^{-1} & =\hat{G}_{0,k}^{-1}\delta_{k,k'}-\hat{Q}_{k,k'},\label{eq:C3}
\end{align}

with $\hat{G}_{0,k}$ defined in Eqn.(\ref{eq:g0mat}) and $\hat{Q}_{k,k'}$
in Eqn.(\ref{eq:9}).

\subsection{Symmetric part\label{sub:Symmetric-part}}

We start by retaining only the SU(2)-symmetric part $\hat{G}_{0,s,k}$=$\left(i\varepsilon_{n}-\overline{\xi}_{\mathbf{k}}\Lambda_{3}\right)^{-1}$,
leading to 
\begin{align}
\hat{G}_{s,k,k'}^{-1} & =\hat{G}_{0,s,k}^{-1}\delta_{k,k'}-\hat{Q}_{k,k'},\nonumber \\
 & =\left(i\varepsilon_{n}-\overline{\xi}_{\mathbf{k}}\Lambda_{3}\right)\delta_{k,k'}-\hat{Q}_{k,k'}.
\end{align}
and $\hat{G}_{0,s,k}$ given by Eqn.(\ref{eq:disp}). Equation (\ref{eq:9})
indicates that $\hat{Q}\sim\Lambda_{1}$, such that $\left\{ \hat{Q},\Lambda_{3}\right\} =0$
holds for arbitrary arguments $x,x'$ (respectively $k,k'$) of the
matrix $\hat{Q}$. Defining 
\begin{align}
\hat{\overline{G}}_{s,k,k'}^{-1} & =\left(-i\varepsilon_{n}-\overline{\xi}_{\mathbf{k}}\Lambda_{3}\right)\delta_{k,k'}+\hat{Q}_{k,k'}.
\end{align}
we find 
\begin{equation}
\hat{\overline{G}}_{s}^{-1}\hat{Q}=-\hat{Q}\hat{G}_{s}^{-1}
\end{equation}
which enables us to re-write Eqn.(\ref{eq:C2}) as 
\begin{align}
F\left[u\right] & =-\frac{T^{2}}{4}\sum_{\varepsilon,\varepsilon',\varepsilon_{1},\varepsilon_{1}'}\sum_{\mathbf{k},\mathbf{k}',\mathbf{k_{1}},\mathbf{k'_{1}}}Tr[\hat{\overline{Q}}_{k,k'}\hat{G}_{s,k',k_{1}}\hat{\overline{G}}_{s,k'_{1},k}\hat{Q}_{k_{1},k'_{1}}].\label{eq:C4}
\end{align}
Using the Wigner transformation, Eqn.(\ref{eq:C4}) can be re-cast
into the form: 
\begin{align}
F\left[u\right] & =-\frac{T^{2}}{4}\sum_{\varepsilon,\omega,\varepsilon',\omega'}\int_{\mathbf{k},\mathbf{q},\mathbf{k'},\mathbf{q'}}Tr[\hat{\overline{M}}_{k,q}\hat{G}_{s,k,q}\hat{\overline{G}}_{s,k',q'}\hat{M}_{k',q'}].\label{eq:nlsm2}
\end{align}
Using $\delta M_{k,q}$ form Eqn.(\ref{eq:27}), we get 
\begin{align}
 & F\left[\delta u\right]=-\frac{T^{2}}{4}\sum_{\varepsilon,\varepsilon'}\int_{\mathbf{k,k'}}\left|M_{0,k}\right|\left|M_{0,k'}\right|\nonumber \\
 & \quad\times\sum_{\omega}\int d\mathbf{q}Tr\left[\delta\hat{U}_{q}^{\dagger}\hat{G}_{s,k,q}\hat{G}_{s,k',q}\delta\hat{U}_{q}\right].\label{eq:nlsm2-1}
\end{align}
Expanding $\hat{G}_{0}^{-1}$ to the second order in $\omega$ and
$q$ and noting that the terms depending only on the fact variables
$\varepsilon,k$ do not contribute, we obtain 
\begin{align}
F\left[\delta u\right] & =\frac{T^{2}}{2}\sum_{\omega,\mathbf{q}}Tr\delta\hat{u}_{q}^{\dagger}\left[J_{0}\omega^{2}+J_{1}q^{2}\right]\delta\hat{u}_{q},\label{eq:24}
\end{align}
where the $tr$ runs on the SU(2) structure and with 
\begin{align}
J_{0}= & \sum_{\varepsilon,\mathbf{k}}\frac{\left|M_{0,k}\right|^{2}}{\left|G_{s}^{-1}\right|^{2}}, & J_{1}=\sum_{\varepsilon,\mathbf{k}}\frac{\left|M_{0,k}\right|^{2}v_{k}^{2}}{\left|G_{s}^{-1}\right|^{2}},\label{eq:69}
\end{align}
with $v_{k}$ the velocity at the Fermi level. The form (\ref{eq:nlsm2-1})
is generic for the non-linear $\sigma$-model.$J_{0}$ and $J_{1}$
are non-vanishing only when $\left|M_{0,k}\right|$ doesn't vanish,
which restricts the SU(2) fluctuations to be below the fluctuations
dome depicted in Fig.\ref{fig:dome}.

\subsection{Effective model\label{sub:Effective-model}}

One can put Eqns.(\ref{eq:24}) into a standard form \cite{Hayward14}
by introducing the four fields $n_{\alpha}$, $\alpha=1,4$ such that
$\Delta=n_{1}+in_{2}$ and $\chi=n_{3}+in_{4}$. The stands for action
for the non-linear $\sigma$-model then writes 
\begin{align}
S_{\sigma} & =-\frac{\rho_{s}}{T}\int d\mathbf{q}\left[\sum_{\alpha=1}^{4}\mathbf{q}^{2}n_{\alpha}^{2}-a_{0}\sum_{\alpha=1}^{2}n_{\alpha}^{2}+a_{0}\sum_{\alpha=3}^{4}n_{\alpha}^{2}\right],\label{eq:31}
\end{align}

with $\rho_{s}=T^{3}\sum_{\varepsilon,\mathbf{k}}\sum_{\omega}\left|M_{0,k}\right|^{2}v_{k}^{2}/\left|G^{-1}\right|^{2}$
, $a_{0}=a_{0}^{disp}+a_{0}^{mf}$, with $a_{0}^{disp}=\frac{T^{3}}{4\rho_{s}}\sum_{\varepsilon,\mathbf{k}}\sum_{\omega}\left|M_{0,k}\right|^{2}\left(\Delta\epsilon_{k}\right)^{2}/\left|G^{-1}\right|^{2}$,
which will be treated in details in the next section. The constant
term coming from the integration over $\omega$ in Eqn.(\ref{eq:24})
has been neglected. A small mean-field mass term $a_{0}^{mf}\ll\left|M_{0,k}\right|$
has been introduced in Eqn.(\ref{eq:31}) which can be generated,
for example, by a magnetic field ($a_{0}^{mf}<0$), which favors the
CDW state, or by the increase of the chemical potential, favoring
the SC ($a_{0}^{mf}>0$), because of the proportionality $\eta_{z}$
to the electron density Eqn.(\ref{eq:6}), but is not considered in
this work.

\section{The SU(2) line\label{sec:The-SU(2)-line}}

\subsection{Symmetry breaking term\label{sub:Evaluation-of-the}}

\begin{figure}[h]
\begin{minipage}[c]{4cm}%
a)\includegraphics[width=27mm]{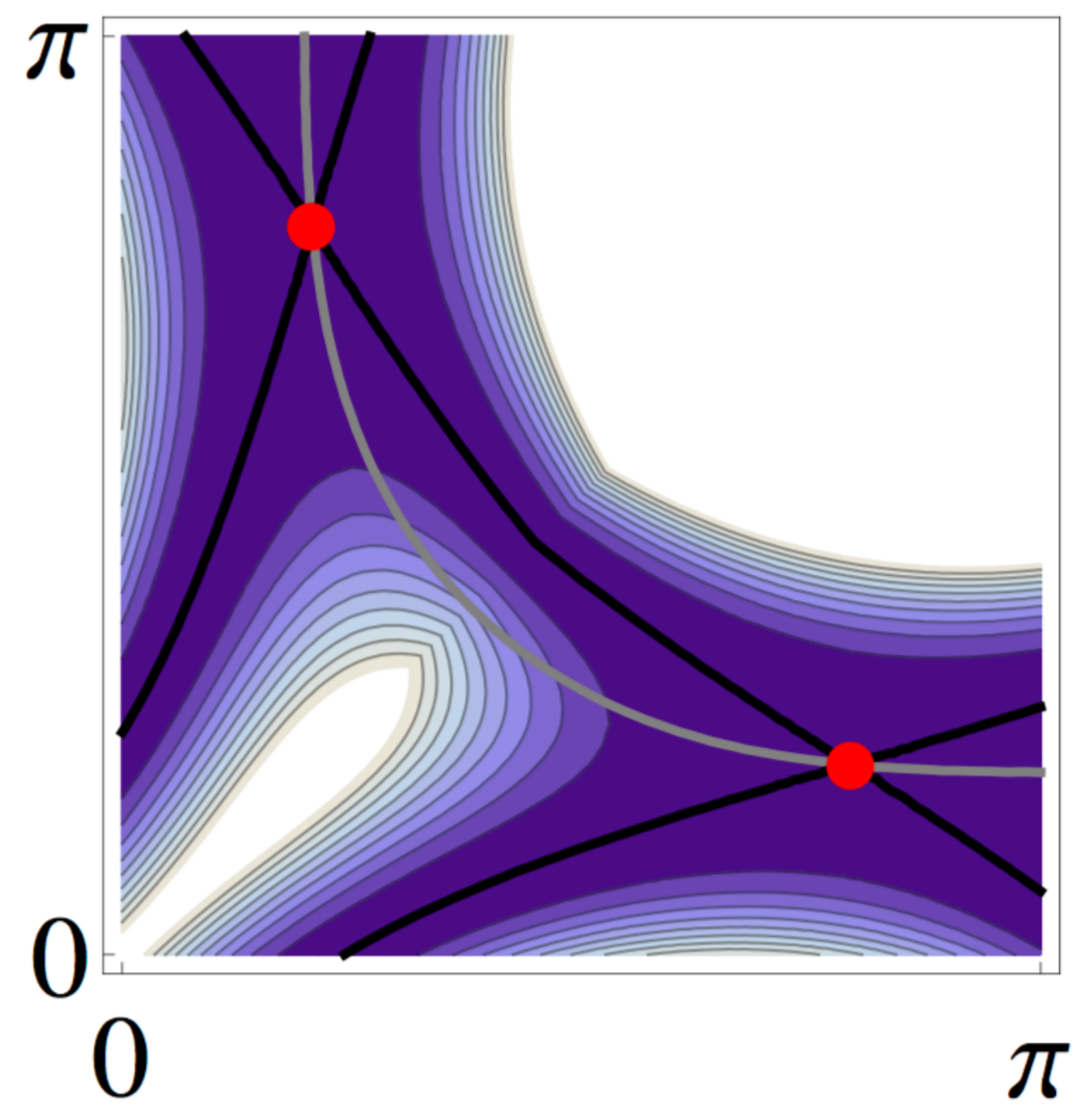} \vspace{1ex}
\end{minipage}%
\begin{minipage}[c]{4cm}%
b)\includegraphics[width=40mm]{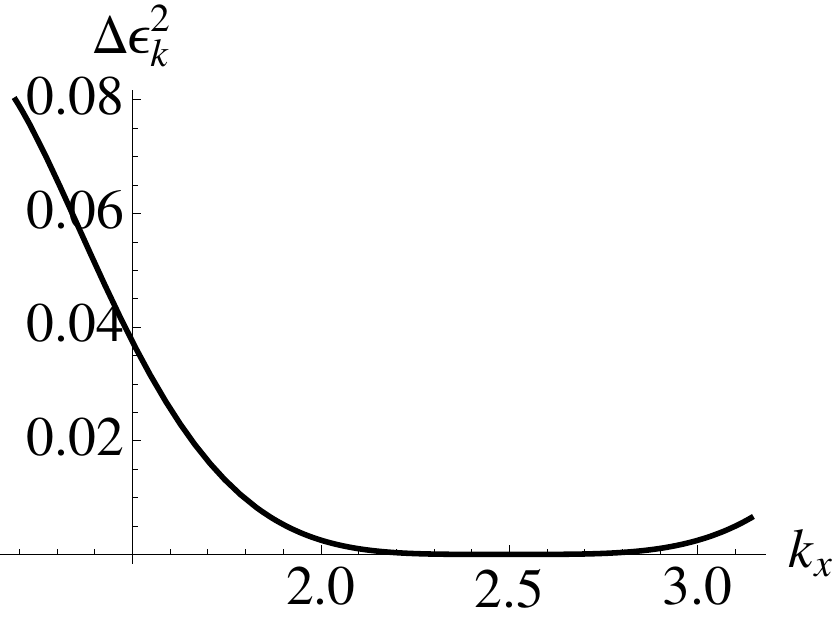} \vspace{1ex}
\end{minipage}

\begin{minipage}[c]{4cm}%
c)\includegraphics[width=27mm]{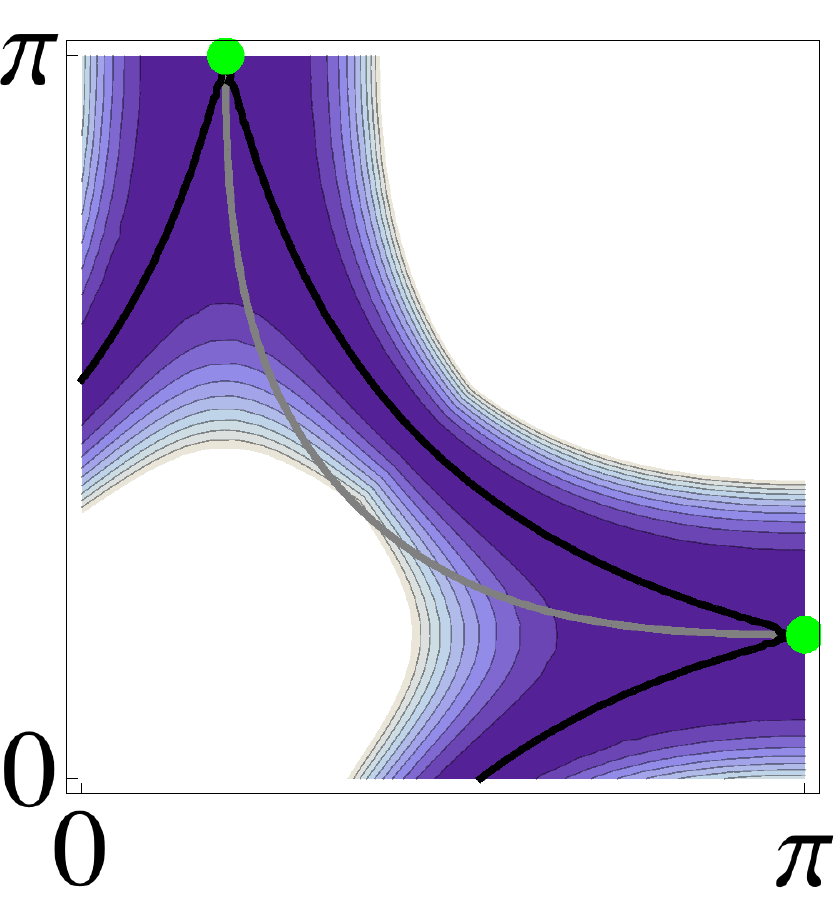} \vspace{1ex}
\end{minipage}%
\begin{minipage}[c]{4cm}%
d)\includegraphics[width=40mm]{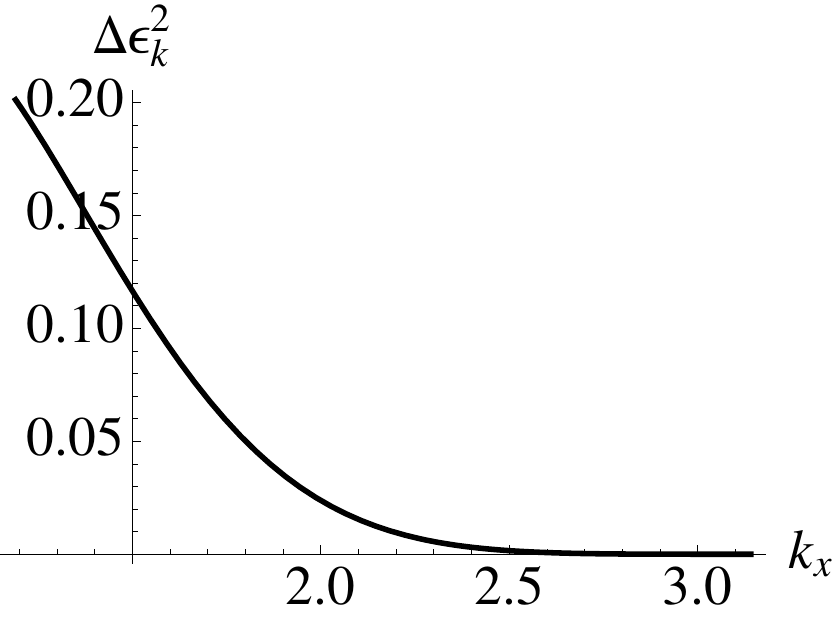} \vspace{1ex}
\end{minipage}\caption{\label{fig:massbz} (Color online) panel a) Visualization of $(\Delta\epsilon_{{\bf {k}}})^{2}$
in the positive region of the first Brillouin zone that gives rise
SU(2) symmetry breaking mass contribution. In the blue region this
contribution is small and vanishes at the two black lines as well
as on the hotspots. In the nodal region the contribution an so the
mass is large. panel b) Variation of $(\Delta\epsilon_{{\bf {k}}})^{2}$
as a function of $k_{x}$ when we follow the Fermi surface (indicated
by the grey line in panel a)). The mass $(\Delta\epsilon_{{\bf {k}}})^{2}$
vanishes at the hotspot and is small close to the zone edge but becomes
large in the nodal region. The difference between the two upper and
lower panels is the choice of the charge-ordering vectors. For panel
a) and b), the charge ordering vector connects the two hotspots, whereas
in panel c) and d) it connects the points at the zone edge. The dispersion
is modelized in tight-binding approximation for Bi2212 Ref.\ \cite{Norman07}
(parameter set tb2). }
\end{figure}

In this section we study the domain of validity of the hypothesis
of the factorization of the fields between fast and slow variables
Eqn.(\ref{eq:18}). As noticed above in Eqn. (\ref{disp1}), it is
possible to models in a simple way the symmetry breaking term. The
term proportional to $\tau_{3}$ in (\ref{disp1}) is proportional
to 
\begin{align}
\Delta\xi_{\mathbf{k}} & =\frac{1}{2}\left(\xi_{\mathbf{k}+\mathbf{Q}_{0}}+\xi_{\mathbf{k}}\right)\label{eq:71}
\end{align}
brings a mass to the free propagator leading to 
\begin{align}
F\left[\delta u\right]_{SB} & =\frac{T^{2}}{2}\sum_{\omega,\mathbf{q}}\sum_{\varepsilon,\mathbf{k}}J_{3,k}\,tr\left[\delta\hat{u}_{k,q}^{\dagger}\tau_{3}\delta\hat{u}_{k,q}\tau_{3}\right],\label{eq:nlsm2-1-1-1}\\
\mbox{where }J_{3,k}= & \frac{1}{4}\frac{\left|M_{0,k}\right|^{2}\left(\Delta\xi_{k}\right)^{2}}{\left|G_{s}^{-1}\right|^{2}}.\label{eq:64}
\end{align}
The effect of symmetry breaking is to produce a mass-term Eqn.(\ref{eq:nlsm2-1-1-1})
which becomes large in the nodal $\left(\pi,\pi\right)$-region as
depicted in Fig.\ref{fig:massbz}. We consider that the theory ceases
to be valid when the dispersion mass term $J_{3,k}$ in Eqn.(\ref{eq:64})
becomes larger than one. This corresponds physically to a situation
where the curvature effects from the Fermi surface in Eqn.(\ref{eq:71})
become stronger than the value of the dome gap $\Delta_{SU(2)}$.
In this case the hypothesis of separation of variables ceases to be
valid and the SU(2) fluctuations vanish. This can be taken into account
by replacing Eqn.(\ref{eq:27}) by 
\begin{align}
\delta\hat{M}_{k,q} & =M_{k}\ \delta\hat{U}_{q},\label{eq:27-1}\\
\mbox{with } & M_{k}=M_{0,k}\mbox{ if }\mathbf{k}\in C,\nonumber \\
\mbox{and } & M_{k}=0,\mbox{ elsewhere}.\nonumber 
\end{align}

In the Eqns.(\ref{eq:27-1}) above, $C$ is the loci of the k-points
in the BZ where the mass $J_{3,k}\ll1$, and thus where the theory
of slowly fluctuating SU(2) phases $\hat{U}_{q}$ is valid. A self-consistency
shall be introduced, and the coefficients of the non-linear $\sigma$-model
in Eqn.(\ref{eq:69}) shall be evaluated again with the fast and slow
variables decoupling of Eqn.(\ref{eq:27-1}). Alternatively, the situation
can be viewed from view point of the magnitude of the SU(2) gap $\Delta_{SU(2)}\sim M_{0,k}$
over the Fermi surface. Since the SU(2) gap requires gapping equally
in the charge (CDW) and SC d-wave channels (see Fig.\ref{fig:dome}),
we see from Fig.\ref{fig:gaps1}, that it is non-zero only in a rather
wide region surrounding the AF hot spots (see Eqn.(\ref{eq:gapmean})).
Since the evaluation of the coefficients in Eqn.(\ref{eq:69}) requires
summation over $\left|M_{0,k}\right|^{2}$, we see that the contribution
of the SU(2) fluctuations is naturally restricted to the anti-nodal
$\left(0,\pi\right)$, or $\left(\pi,0\right)$-region.

We make the claim here, that fluctuations associated with the SU(2)
fields are present in the theory, but act only in restricted areas
of the BZ. This produces SU(2)-lines of massless fluctuations along
which the electron self-energy diverges. This produces a line of zeros
in the electron Green's function, in analogy with the findings of
other theoretical approaches \cite{LeHur:2009iw,Zhang:1988tq,Zhang:1999tq}, Fig.\ref{fig:massbz}
shows two typical cases of SU(2) lines. Fig.\ref{fig:massbz} a) and
b) are concerned with a charge wave vector lying on the diagonal of
the form \textcolor{black}{$\left(Q_{\mathbf{0}},Q_{\mathbf{0}}\right)$},
while the panels \ref{fig:massbz} c) and d) show the same lines of
zeros of the Green's function, but for a wave vector located at the
ZE. We can see in Fig.\ref{fig:massbz} b) that the minimum of the
mass is located at the AF hot spots, while in Fig.\ref{fig:massbz}
d), we see that it is at located at the zone edge. A more detailed
study for various wave vectors, including the evolution with doping,
is given in the next section.

\subsection{Evolution with doping}

\begin{figure*}[t]
a)%
\begin{minipage}[c]{4.25cm}%
\includegraphics[width=4cm]{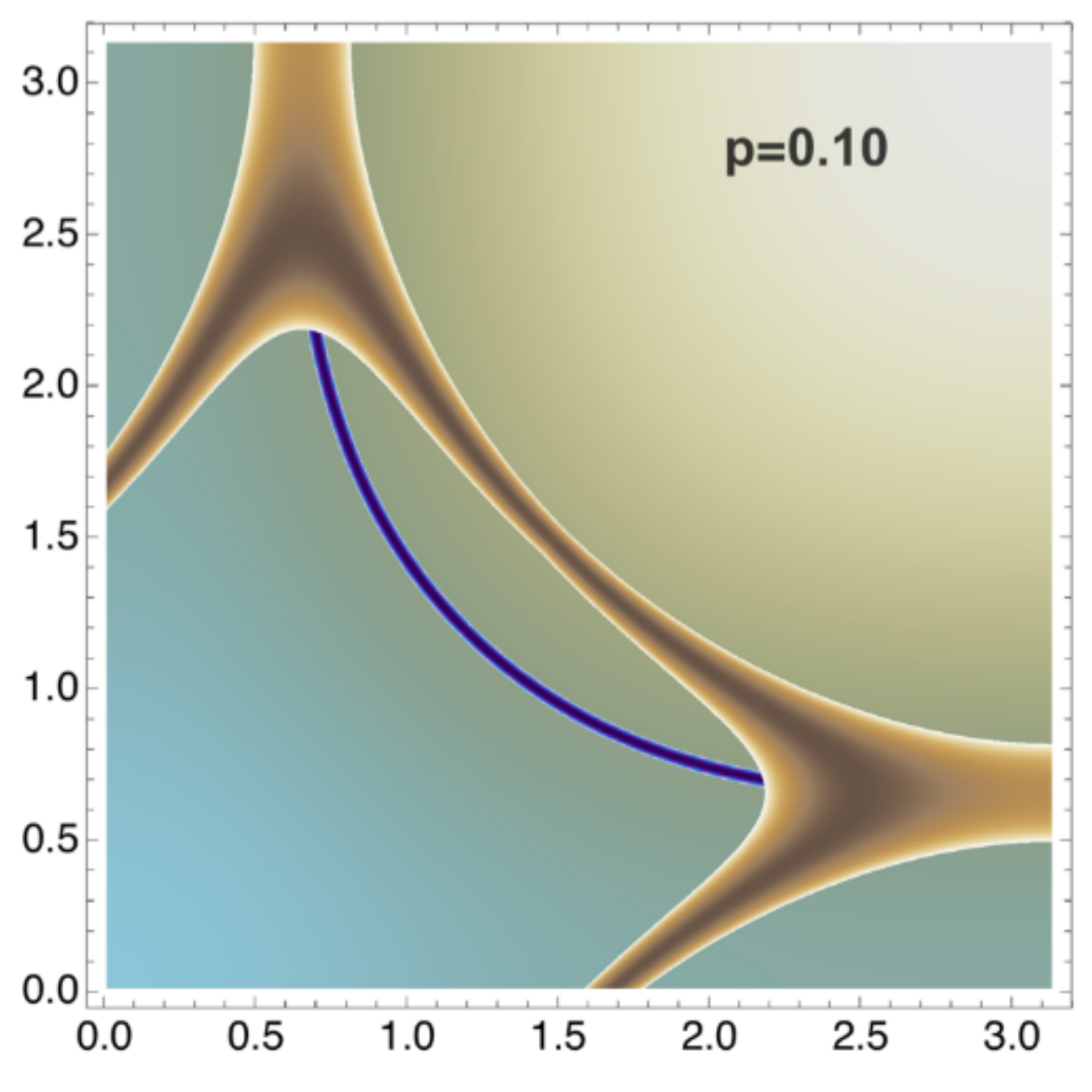} \vspace{1ex}
\end{minipage}%
\begin{minipage}[c]{4.25cm}%
\includegraphics[width=4cm]{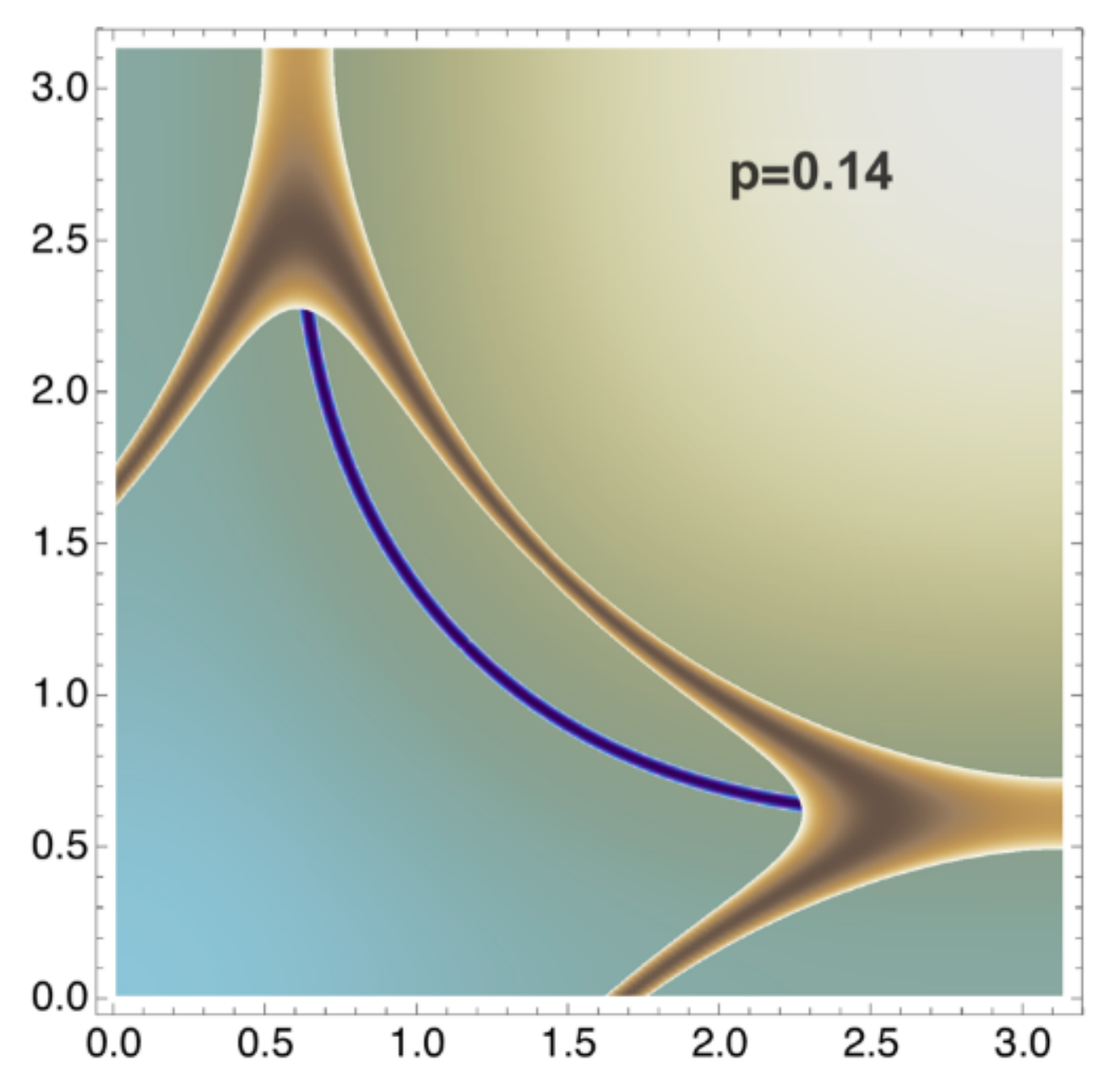} \vspace{1ex}
\end{minipage}%
\begin{minipage}[c]{4.25cm}%
\includegraphics[width=4cm]{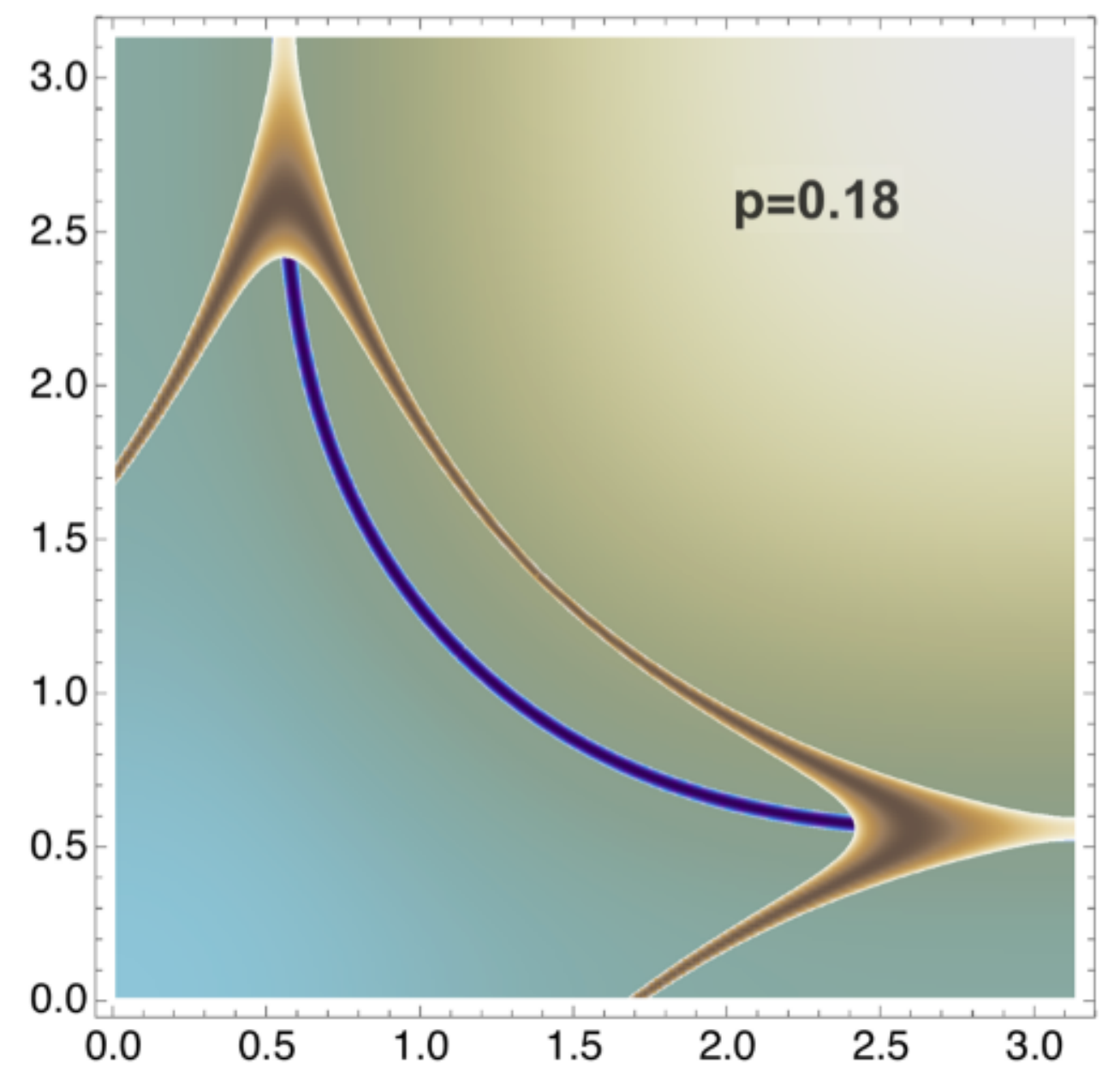} \vspace{1ex}
\end{minipage}%
\begin{minipage}[c]{4.25cm}%
\includegraphics[width=4cm]{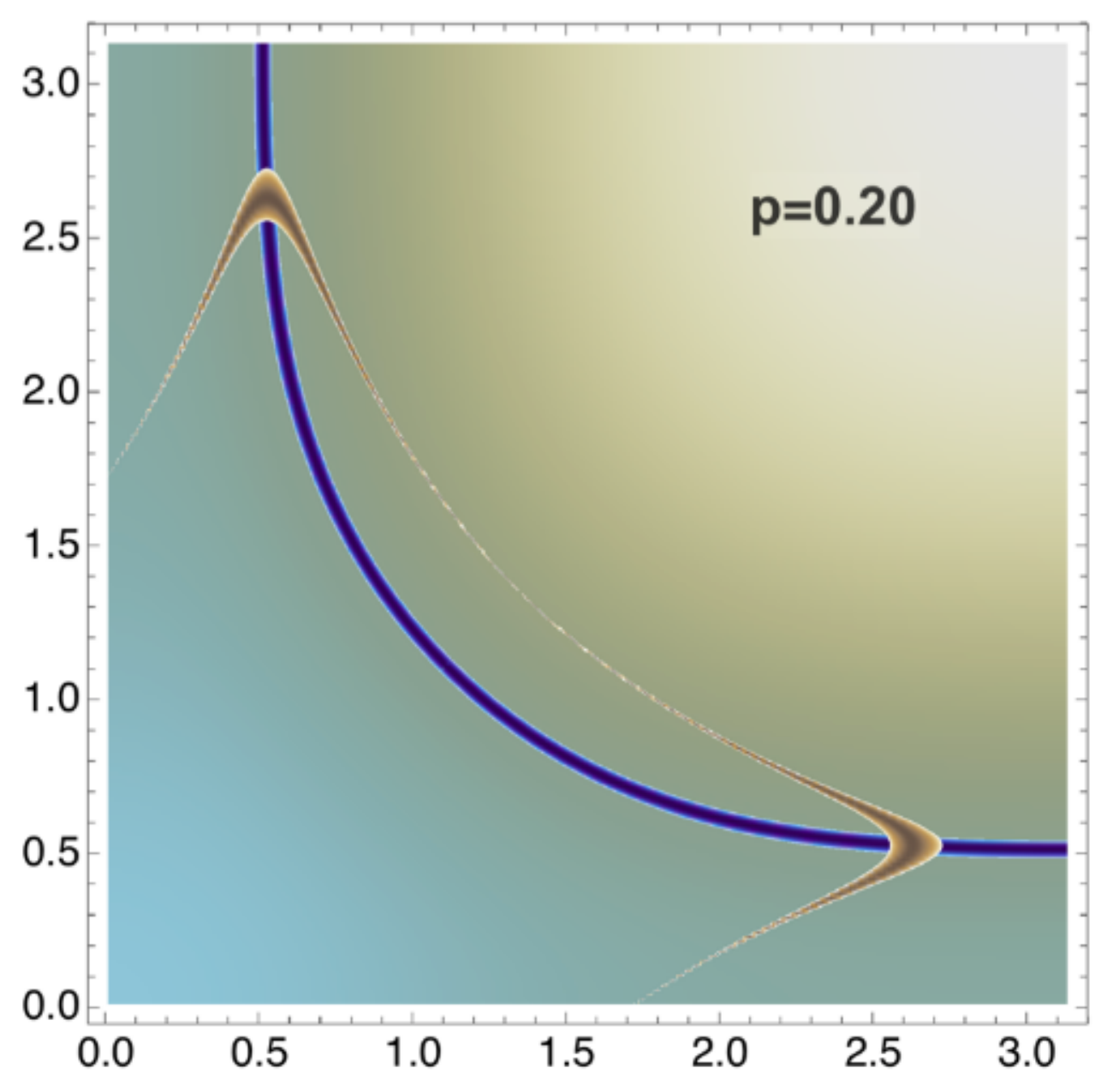} \vspace{1ex}
\end{minipage}

b)%
\begin{minipage}[c]{4.25cm}%
\includegraphics[width=4cm]{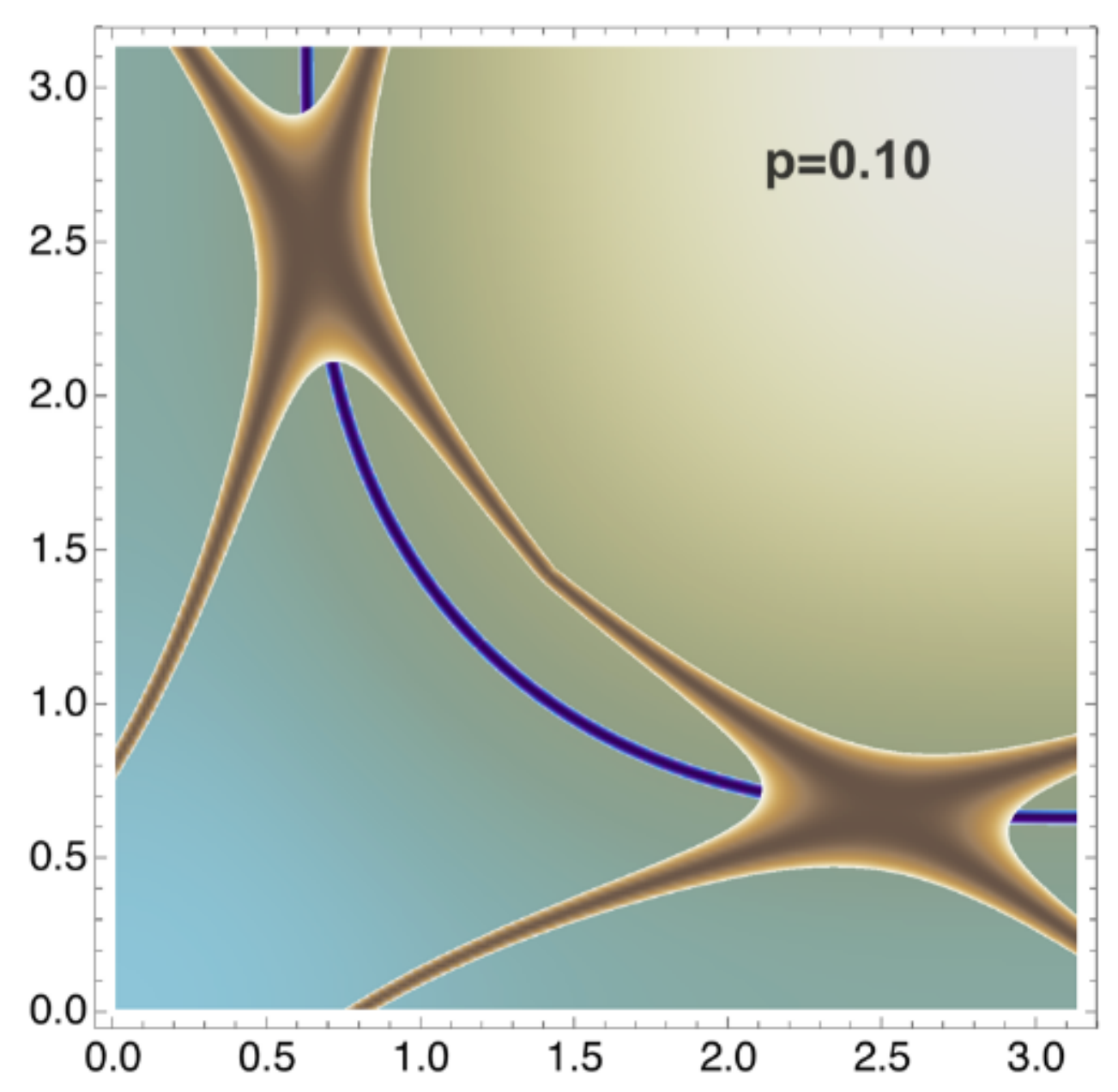} \vspace{1ex}
\end{minipage}%
\begin{minipage}[c]{4.25cm}%
\includegraphics[width=4cm]{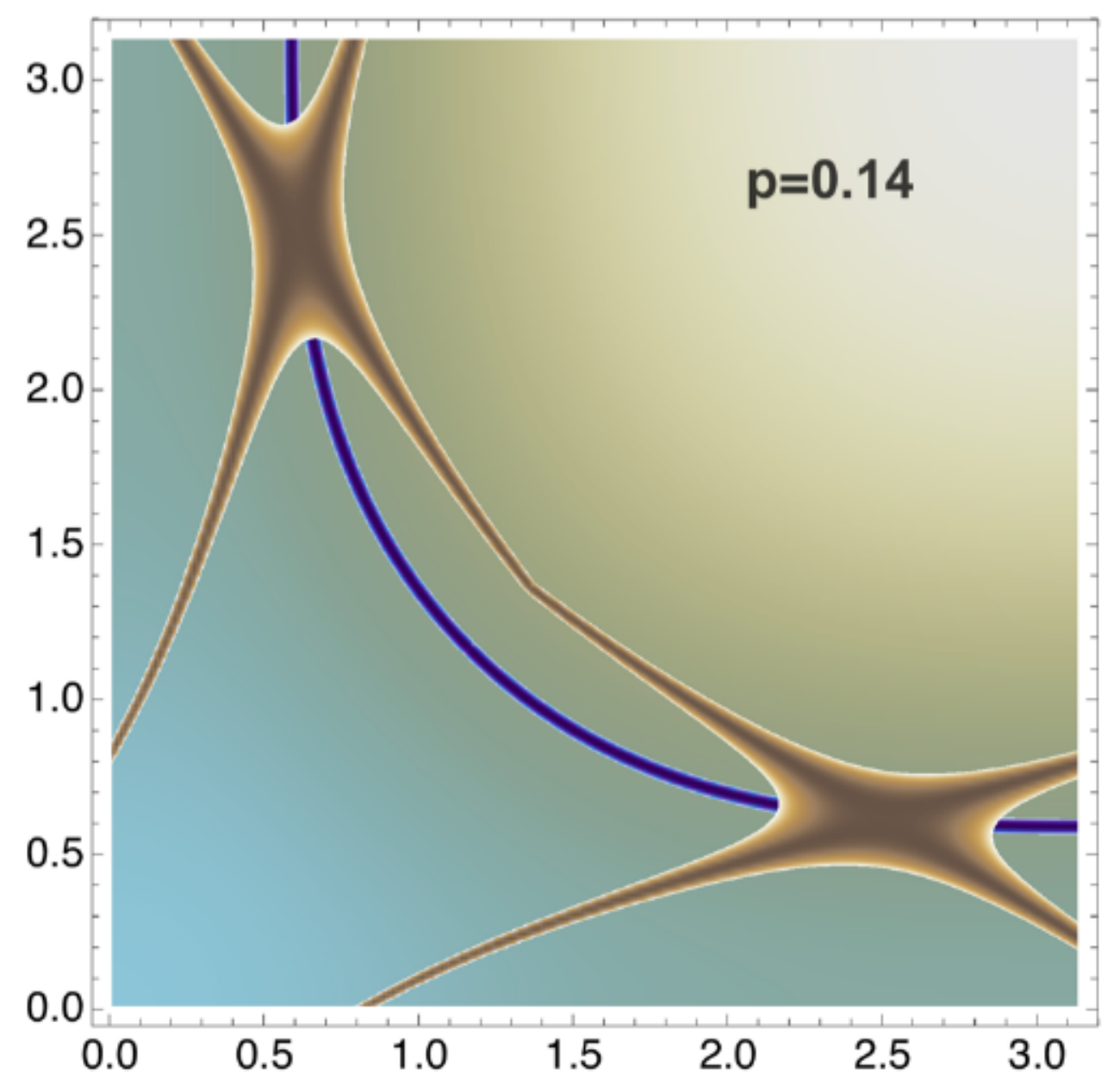} \vspace{1ex}
\end{minipage}%
\begin{minipage}[c]{4.25cm}%
\includegraphics[width=4cm]{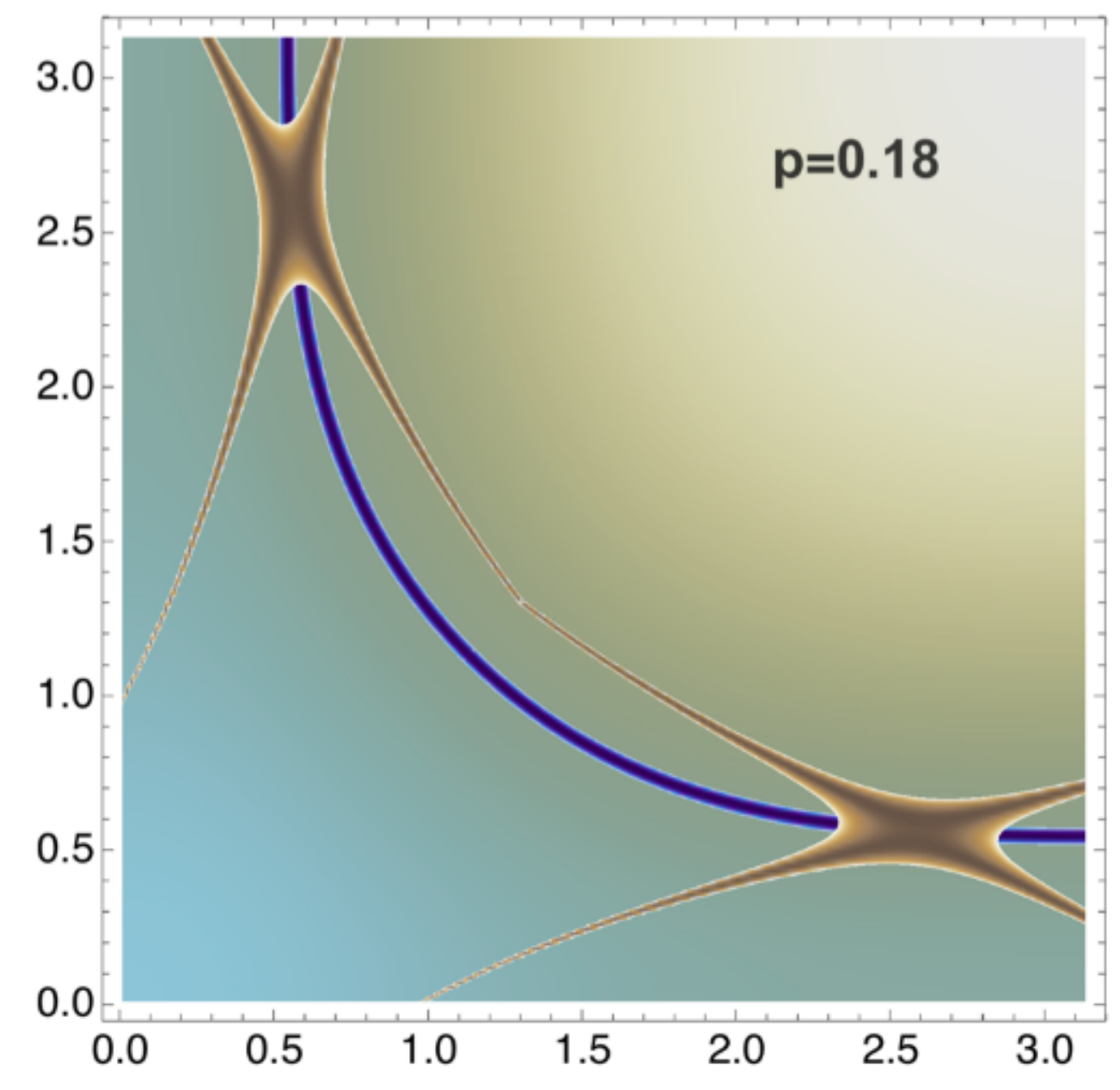} \vspace{1ex}
\end{minipage}%
\begin{minipage}[c]{4.25cm}%
\includegraphics[width=4cm]{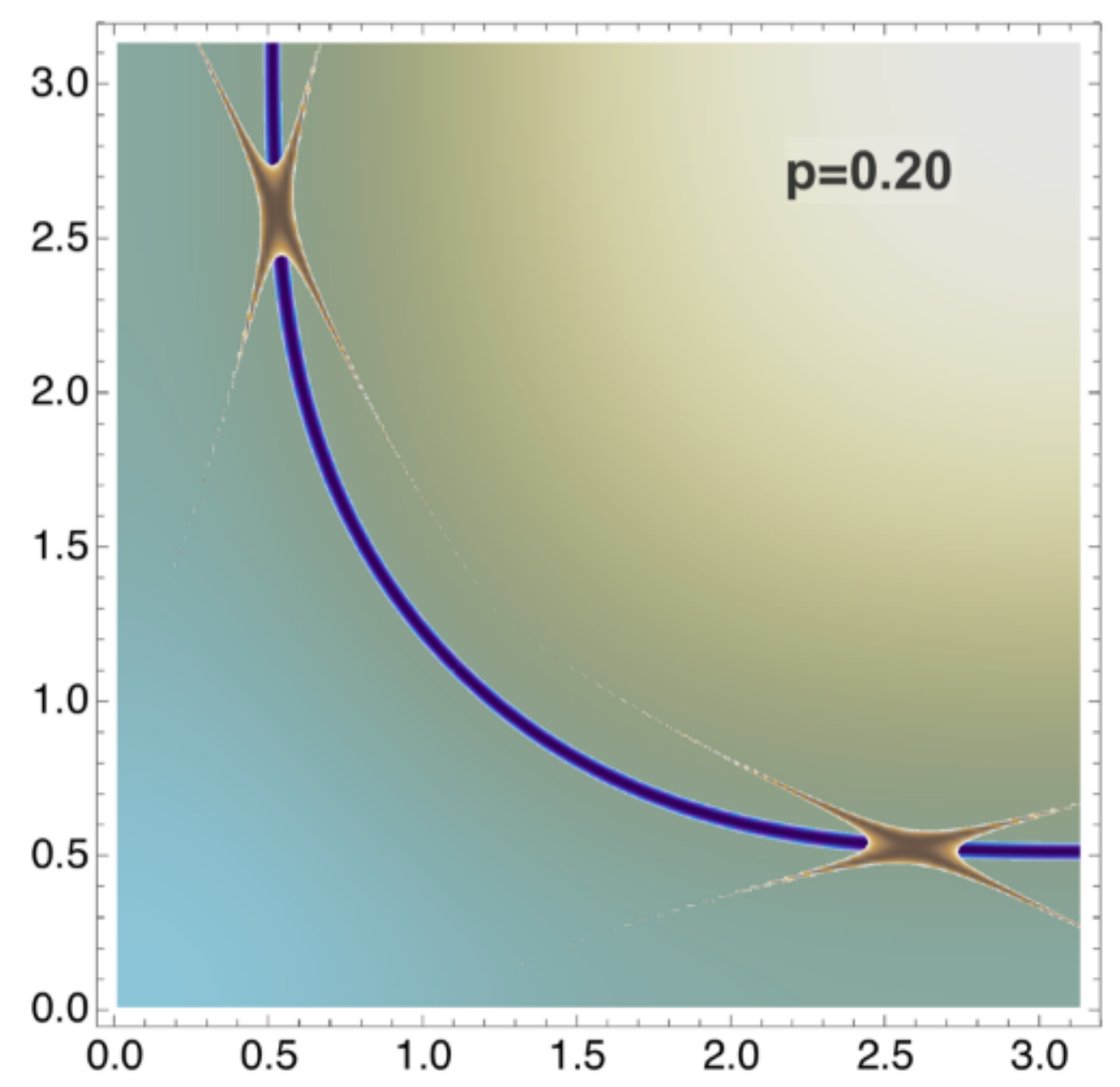} \vspace{1ex}
\end{minipage}

\caption{\label{fig:doping} (Color online) Evolution of the SU(2) fluctuations
as the function of the hole-doping. In brown is represented the $\mathbf{k}$-points
or which the SU(2) mass $J_{3,k}<1$ in Eqn.(\ref{eq:64}). This criterion
defines the ``hot-regions'' for which the SU(2) fluctuations are
important. Note that the ``hot-regions'' are centered around the
hot-spots and extend in the anti-nodal part of the BZ. We present
two set of curves: a) for the axial wave vector\textcolor{black}{{}
$\mathbf{Q}_{0}=\left(Q_{\mathbf{0}},0\right)$}, and b) for the diagonal
wave vector\textcolor{black}{{} $\mathbf{Q}_{0}=\left(Q_{\mathbf{0}},Q_{\mathbf{0}}\right)$}.
In Eqn.(\ref{eq:64}), we approximate the factor $\left|M_{0,k}/G_{s}^{-1}\right|^{2}\sim1/\Delta_{SU(2)}^{2}$,
with a phenomenological form for $\Delta_{SU(2)}=T^{*}\left(\frac{p_{c}-p}{p_{c}-p_{0}}\right)$,
with $T^{*}=3.10^{-2}$, $p_{c}=0.22$, $p_{0}=0.12$, and $p$ is
the hole doping. In brown are the regions where the dispersion mass
$J_{3,k}\ll1$ in Eqn.(\ref{eq:64}). Note that in both cases, the
dispersion mass has a minimum at the hot spots. The electron dispersion
is modelized in tight-binding approximation for Bi2212 Ref.\ \cite{Norman07}
(parameter set tb2). }
\end{figure*}

In Fig.\ref{fig:doping}, we present the evolution with doping of
the hot regions for two configurations of the charge modulation wave
vector. Fig.\ref{fig:doping}a) relates to the diagonal wave vector
$\mathbf{Q}_{0}=\left(Q_{0},Q_{0}\right)$, while Fig.\ref{fig:doping}b)
describes the axial wave vector $\mathbf{Q}_{0}=\left(Q_{0,}0\right)$
observed experimentally. Note that both sets of SU(2) fluctuations
are massless at the hot spot. According to the simple calculation
presented in Fig.\ref{fig:dome}, and assuming the relation $p\simeq J_{0}-J$,
we have modelized the PG by the following phenomenological variation
with doping 
\begin{align}
\Delta_{SU(2)} & =T^{*}\left(\frac{p_{c}-p}{p_{c}-p_{0}}\right),\label{eq:75}
\end{align}

with $p$ the hole doping, $p_{0}=0.12$ and $p_{c}=0.22$.

The brown region corresponds to the loci of the points where the SU(2)
mass $J_{3,k}$ in Eqn.(\ref{eq:64}) is less than one, and hence
where the SU(2) fluctuations are coupled to the electrons. A massless
line-or SU(2) line, is visible and crosses the Fermi surface at the
AF hot spots. Although the shape of the SU(2) line doesn't change
much with doping, its width decreases until it gets located at the
hot-spots and then disappears. This study makes very clear the fractionalization
of the Fermi surface between hot regions and cold regions. The case
for a wave vector at the Zone Edge (ZE) is presented in Appendix \ref{sec:Zone-Edge-hot}.
We see that in that case the mass minimum is located at the Zone Edge.

\section{Rotation of the charge ordering wave vector and nematicity\label{sec:Rotation-of-the}}

In section \ref{sec:SU(2)-fluctuations-coupled}, we have decided
to rotate the d-wave SC state towards a d-wave charge channel with
diagonal wave vector $\mathbf{Q}_{0}=\left(Q_{\mathbf{0}},Q_{\mathbf{0}}\right)$.
This choice was rather arbitrary, considering
that all the wave vectors considered in section \ref{sec:The-SU(2)-dome}
are quasi degenerate. We chose the diagonal wave vector simply for
historical reasons, that the eight hot-spots spin Fermion model considered
in our previous work possesses an exact SU(2) symmetry involving charge
order with diagonal wave vectors \cite{Metlitski10b,Efetov13}. In
this section, we explore the effects of the SU(2) pairing fluctuations
on the modulations wave vector in the charge sector. We
show that the main surprising effect of these fluctuations is to lift
the degeneracy between the various modulation vector, with the uniform
$\mathbf{Q}_{0}=\mathbf{0}$ and axial wave vectors $\mathbf{Q}_{0}=\left(Q_{0},0\right)$
and $\mathbf{Q}_{0}=\left(0,Q_{0}\right)$ becoming the leading ones.
This leads to the emergence of d-wave axial charge modualtions associated
with a d-wave Pomeranchuk instability, or nematic order.

\subsection{Bare polarization}

\begin{figure}
a)%
\begin{minipage}[c]{6.25cm}%
\includegraphics[width=7cm]{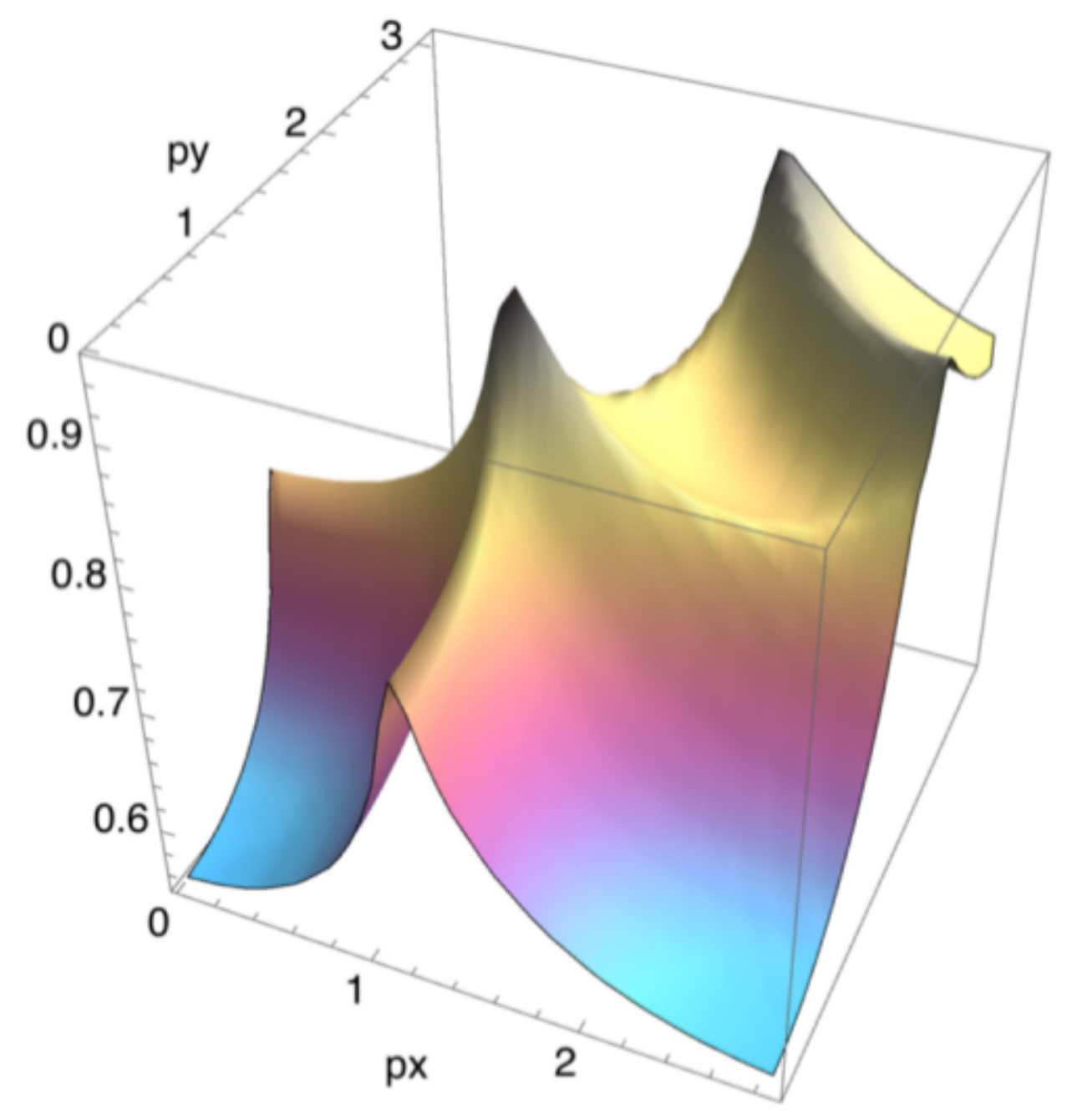} \vspace{1ex}
\end{minipage}

b)%
\begin{minipage}[c]{6.25cm}%
\includegraphics[width=7cm]{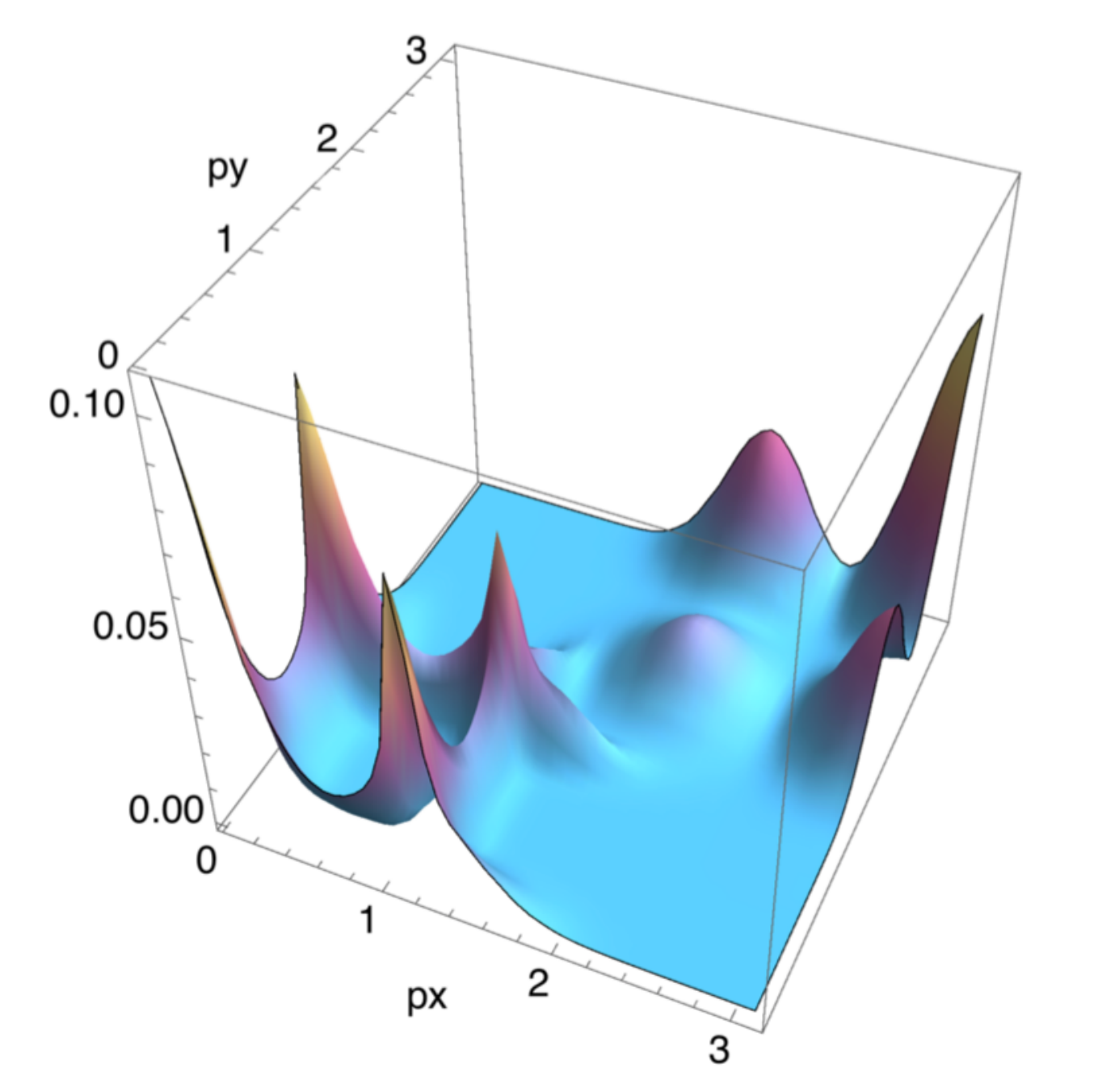} \vspace{1ex}
\end{minipage}

c) %
\begin{minipage}[c]{3.25cm}%
\includegraphics[width=2.5cm]{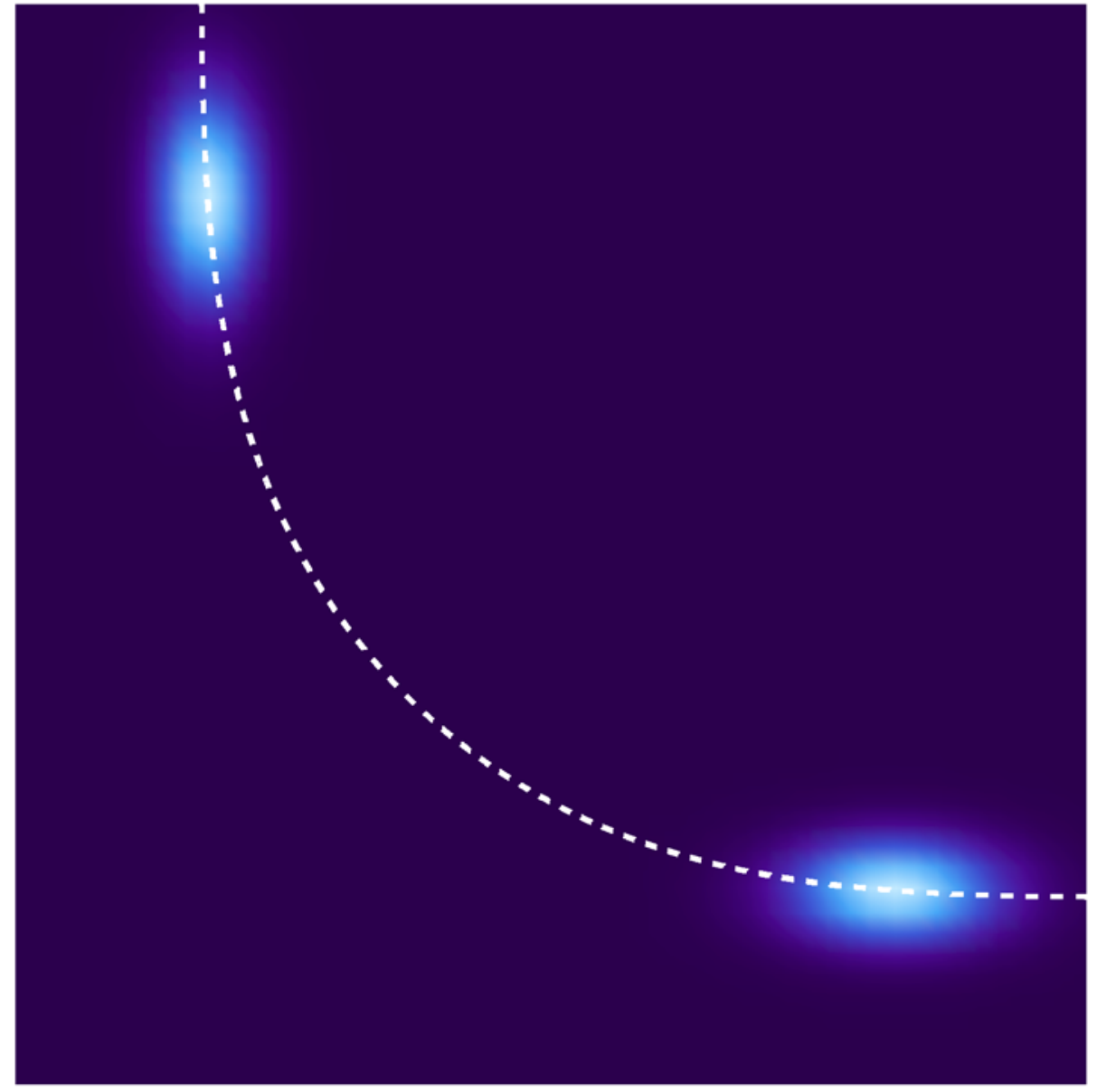} \vspace{1ex}
\end{minipage}%
\begin{minipage}[c]{3.25cm}%
\includegraphics[width=3cm]{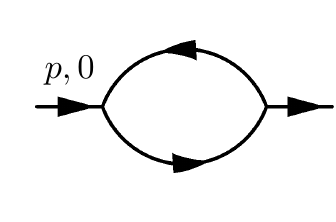} \vspace{1ex}
\end{minipage}

\caption{\label{fig:bare} (Color online) \textcolor{black}{Real part of the
bare polarization bubble in the static limit. The electron dispersion
corresponds to the usual one of }Bi2212 Ref.\ \cite{Norman07}\textcolor{black}{.
In panel a) we have depicted the bare polarization bubble, showing
a maximum at the diagonal wave vector $\left(Q_{\mathbf{0}},Q_{\mathbf{0}}\right)$,
as well as some features around $\left(\pi,\pi\right)$. In panel
b) we show the same polarization, but restricted to the domain of
validity of the SU(2) fluctuations, shown in the panel c). Remarkably,
we observe that the intensity gets displaced, with the emergence of
a peak at $\mathbf{Q}=\left(0,0\right)$, and also the predominance
of the axial wave vectors $\left(0,Q_{\mathbf{0}}\right)$, and $\left(Q_{\mathbf{0}},0\right)$,
compared to the diagonal one.} }
\end{figure}

We start with a simple study of the bare polarization plotted in Fig.\ref{fig:bare}a)
with the band structure of Bi2212.

\begin{align}
\Pi_{bare}^{a}\left(p,0\right) & =-T\sum_{k}G_{k}G_{k+p},
\end{align}

with $k=\left(\mathbf{k,\varepsilon}\right)$, and $G_{k}^{-1}=i\varepsilon_{n}-\xi_{\mathbf{k}}$.
Here we notice the well-known features corresponding to a maximum
along the diagonal, at the wave vector $\left(Q_{0},Q_{0}\right)$,
as well as some structure lying close to the $\left(\pi,\pi\right)$
region. Nothing particular is visible on the axes, apart from the
line corresponding to the $\pf$ wave vectors, but overall, the value
of the polarization on the axes is less important than on the diagonal.
In Fig.\ref{fig:bare}b), we give the same study of the bare polarization,
but with a width of integration in \textbf{k}-space restricted to
the SU(2) hot regions (insert in Fig.\ref{fig:bare}c)) 
\begin{align}
\Pi_{bare}^{b}\left(p,0\right) & =-T\sum_{k}M_{0,k}M_{0,k+p}G_{k}G_{k+p}.
\end{align}
The result is drastically different from the bare polarization, with
the emergence of a structure at $\mathbf{Q}=\left(0,0\right)$, accompanied
by a drastic increase of the pics along the axes, which precisely
correspond to the wave vectors observed experimentally $\left(0,Q_{0}\right)$
and $\left(Q_{0},0\right)$. This effect of the SU(2) regions on the
pairing fluctuations is the generic feature that we describe in this
section. It shows that, even if we start with a diagonal modulation
vector, at rather high energy, upon the effect of the SU(2) pairing
fluctuations the wave vector is tilting along the axes.

\subsection{Vertex corrections}

\begin{figure}
a)%
\begin{minipage}[c]{6.25cm}%
\includegraphics[width=7cm]{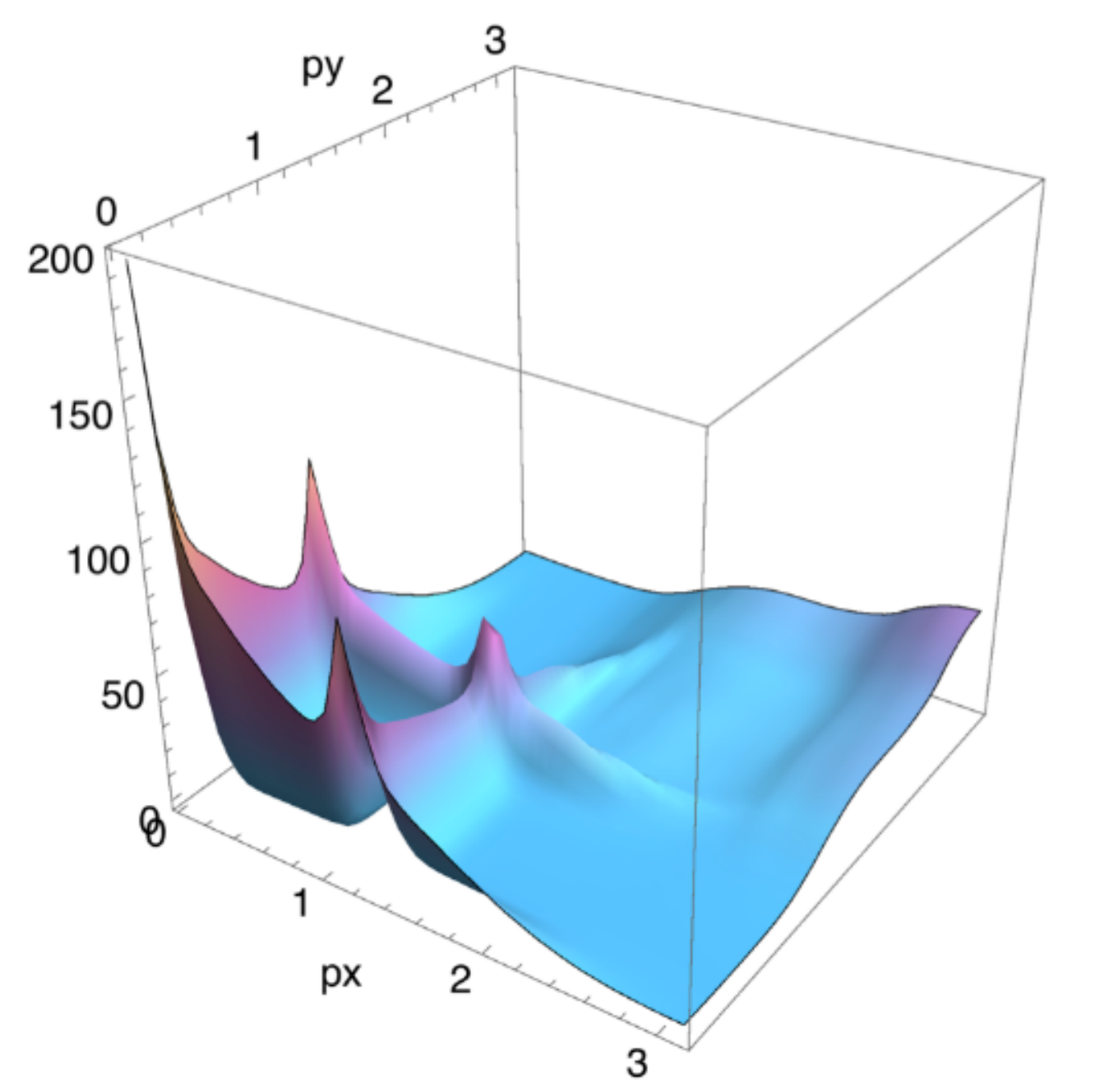} \vspace{1ex}
\end{minipage}

b) %
\begin{minipage}[c]{3.25cm}%
\includegraphics[width=2.5cm]{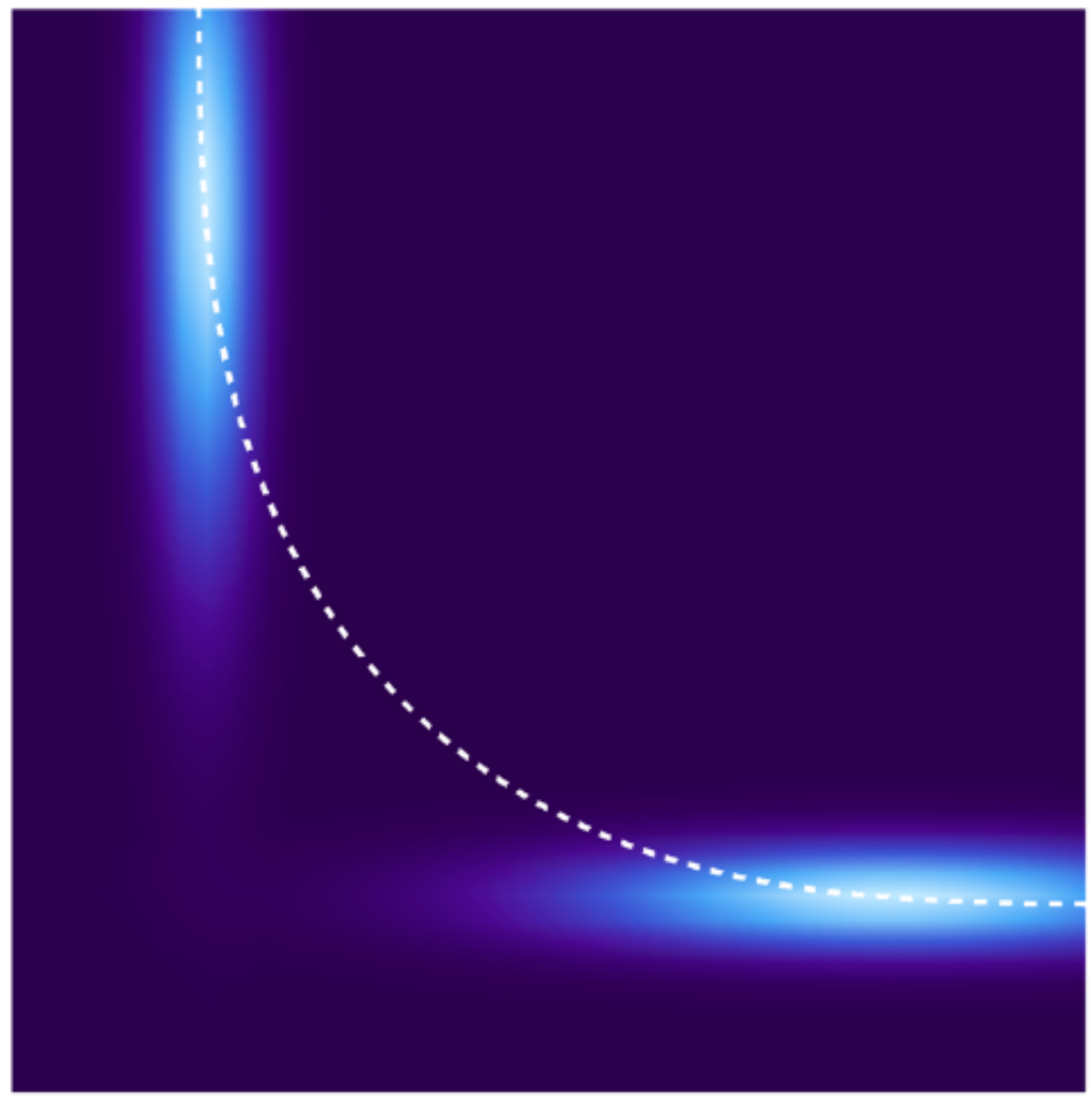} \vspace{1ex}
\end{minipage}%
\begin{minipage}[c]{4.25cm}%
\includegraphics[width=4cm]{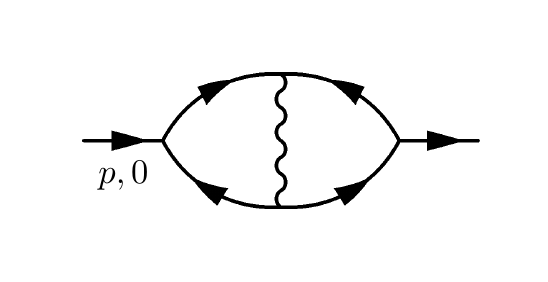} \vspace{1ex}
\end{minipage}

\textcolor{black}{\caption{\label{fig:vertex} (Color online) \textcolor{black}{a) Real part
of the polarization bubble, with one vertex correction, in the static
limit. The electron dispersion corresponds to the usual one of }Bi2212 Ref.\ \cite{Norman07}\textcolor{black}{. The charge response
is depicted in the case where the fluctuations are centered at the
hot-spots as shown in b). In this case, the peak along the diagonal
remains but with a lower intensity than the response along $\left(0,Q_{0}\right)$
and $\left(Q_{0},0\right)$.}}
} 
\end{figure}

\begin{figure}
a)%
\begin{minipage}[c]{6.25cm}%
\includegraphics[width=7cm]{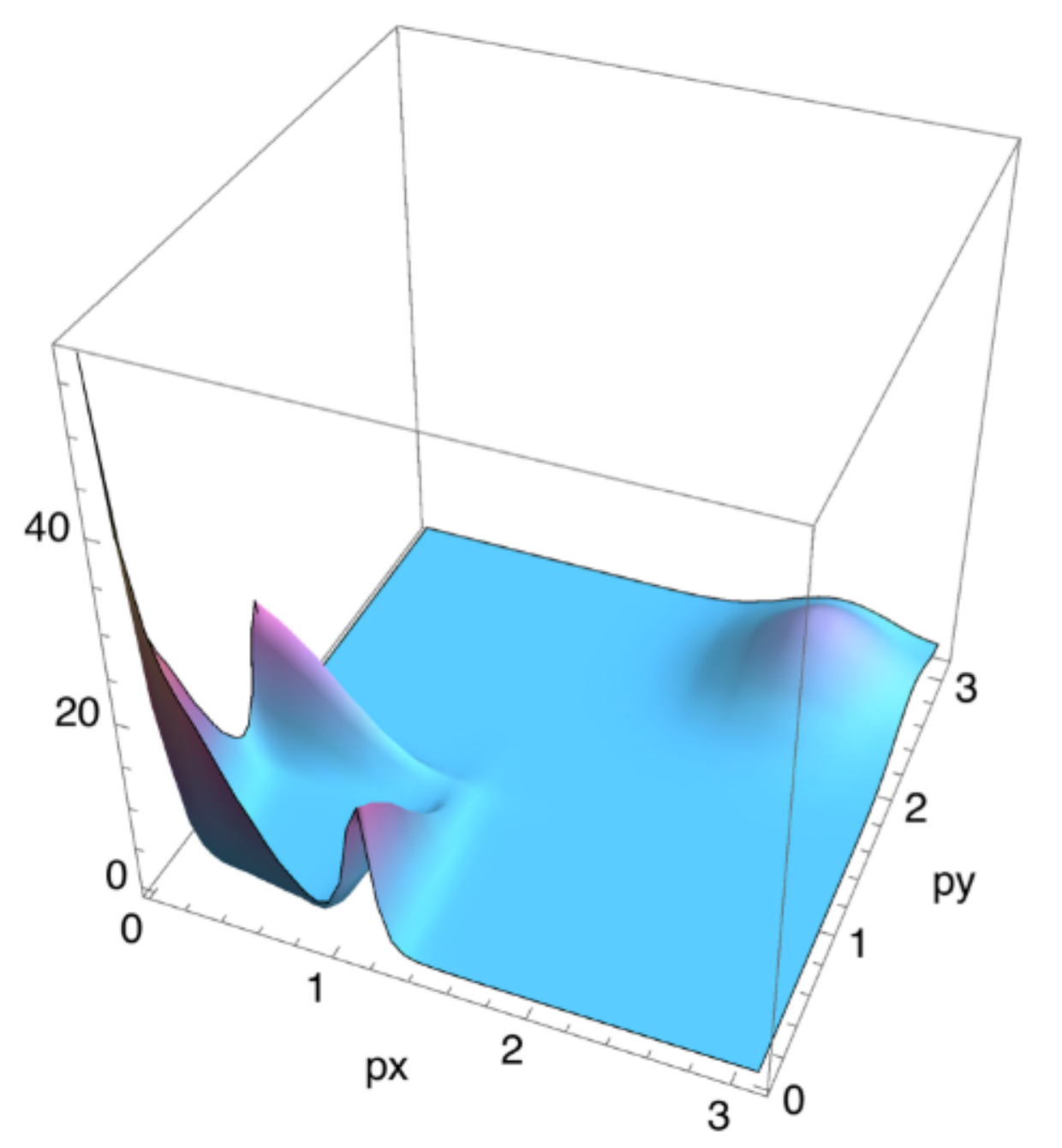} \vspace{1ex}
\end{minipage}

b) %
\begin{minipage}[c]{3.25cm}%
\includegraphics[width=2.5cm]{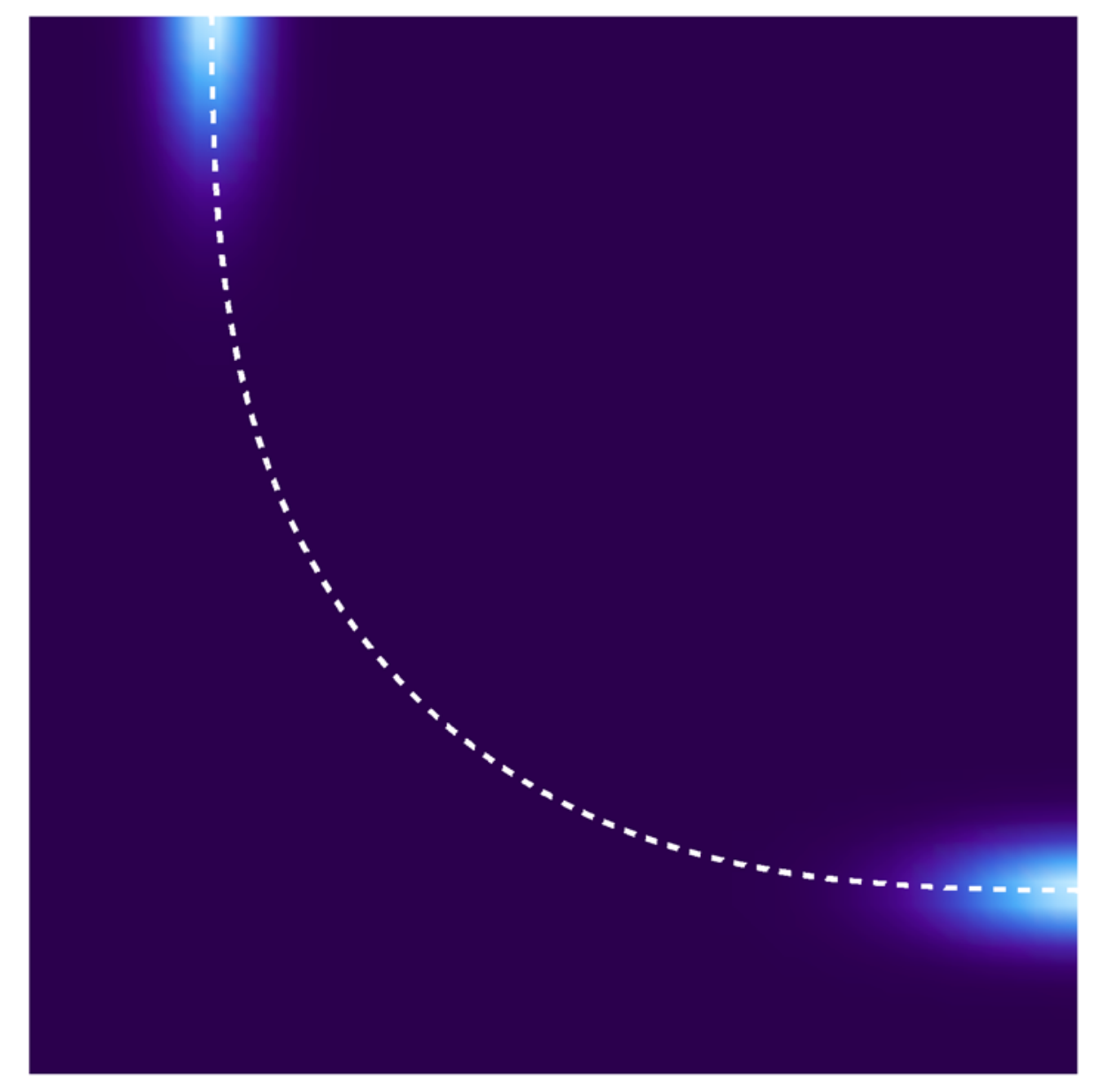} \vspace{1ex}
\end{minipage}%
\begin{minipage}[c]{4.25cm}%
\includegraphics[width=4cm]{Fig11c} \vspace{1ex}
\end{minipage}

\textcolor{black}{\caption{\label{fig:su2fluct} (Color online) \textcolor{black}{Real part of
the polarization bubble, with one vertex correction, in the static
limit. The electron dispersion corresponds to the usual one of }Bi2212 Ref.\ \cite{Norman07}\textcolor{black}{. The charge response
is depicted in the case where the fluctuations are centered at the
Zone Edge, with a rather small extension. In this case, the peak
along the diagonal is completely lost, while only the response along
$\left(0,Q_{0}\right)$ and $\left(Q_{0},0\right)$ is retained.}}
} 
\end{figure}

We provide the study of the SU(2) pairing fluctuations in Figs.\ref{fig:vertex}
and\ref{fig:su2fluct}.

\begin{align}
\Pi_{v}\left(p,0\right) & =-T\sum_{k,q}\pi_{k,k+p,q}^{s}G_{k}G_{k+p}G_{-k-q}G_{-k-p-q},
\end{align}
with the four variables $k=\left(\mathbf{k,\varepsilon}\right)$,
$q=\left(\mathbf{q},\omega\right)$, $G_{k}^{-1}=i\varepsilon_{n}-\xi_{\mathbf{k}}$
and $G_{-k}^{-1}=-i\varepsilon_{n}-\xi_{\mathbf{-k}}$. The form of
the SU(2) pairing propagator $\langle\Delta_{k,q}^{\dagger}\Delta_{k',q'}\rangle=\pi_{k,k',q}^{s}\delta_{\mathbf{q},\mathbf{q'}}$
has been defined in section \ref{sec:SU(2)-fluctuations-coupled},
Eqns.(\ref{eq:24},\ref{eq:nlsm2-1-1-1},\ref{eq:31}):

\begin{align}
\pi_{k,k',q}^{s} & =M_{0,k}M_{0,k'}\frac{\pi_{0}}{J_{0}\omega_{n}^{2}+J_{1}({\bf {v}}\cdot{\bf {q}})^{2}-a_{0}}.\label{eq:82}
\end{align}

The presence of the vertex factors $M_{0,k}$ and $M_{0,k'}$ in Eqn.(\ref{eq:82})
restrict the summation over $\mathbf{k}$ to the anti-nodal region
of the Brillouin Zone. We observe that the same physical effects as
the ones present in Fig.\ref{fig:bare}b) are present in Fig.\ref{fig:vertex}a).
The shape of the SU(2) hot regions doesn't really affect the two main
observable effects, as we show in Appendix \ref{sec:Test-of-various}
as long as the hot regions are centered around the host spots. We
observe the emergence of a pic at zero wave vector as well as the
predominance of the response along the axes over the response on the
diagonal. As we show in Fig. \ref{fig:AL}, the inclusion of Aslamazov-Larkin
diagrams doesn't change the conclusion.

In Fig.\ref{fig:su2fluct}a), the same study is performed for hot
regions centered at the Zone Edge. In this case the effect is even
more pronounced, with nothing left on the diagonal, but a range of
wave vectors which dominate around \textcolor{black}{$\left(0,Q_{\mathbf{0}}\right)$,
and $\left(Q_{\mathbf{0}},0\right)$, with a line still visible at
the wave vectors $\pf$. }

\subsection{Implications for the phase diagram of the cuprates}

When comparing the bare polarization Fig.\ref{fig:bare}a) with the
effects of the SU(2) paring in Figs.\ref{fig:vertex} and \ref{fig:su2fluct},
two main effects are noticeable. First, we observe a shift of the
spectral weight from the diagonal to the axes, with the formation
of CDW instabilities around the experimentally observed wave vectors
$\left(0,Q_{0}\right)$ and $\left(Q_{0},0\right)$ and second, we
see the emergence of a peak at $\mathbf{Q}=\left(0,0\right)$. The
effect of the SU(2) pairing fluctuations is thus twofold, with the
peak at $\mathbf{Q}=\left(0,0\right)$ hinting at the presence of a
nematic precursor around the temperature $T^{*}$ \cite{Lawler10,Mesaros:2011jj},
while the peaks at finite wave vectors along the axes correspond to
the CDW modulations observed experimentally.

A comment with respect to the center of mass symmetry
is in order. The SU(2) fluctuations considered here do not favor any
specific symmetry, whether it is s (or s'), or d-wave. Since those
fluctuations connect gap equations around $\mathbf{q}=0$, the gaps
are stabilized around one $\mathbf{k}$-point in the Brillouin zone
instead of coupling two anti-nodal regions as it is the case for the
AF coupling. The d-wave symmetry of the $\mathbf{Q}=\left(0,0\right)$
and $\left(0,Q_{0}\right)$ and $\left(Q_{0},0\right)$ orders comes
from the original AF correlations describes in section \ref{sec:The-SU(2)-dome}.
The SU(2) fluctuations are lifting the degeneracy whereas the d-wave
character of the charge instabilities remain. This has the important
consequence that axial orders are systematically accompanied with
a Pomeranchuck, or nematic instability within our model.

In Appendix \ref{sec:Nematicity-and-charge}, we argue that there
is no vertex/ self-energy cancellation in the case of a wavy fluctuations
line in the Cooper channel, which leads us to consider the $\mathbf{Q}=\mathbf{0}$
peak as a real effect which accompanies the CDW modulations at finite
wave vectors. In Fig.\ref{fig:su2fluct}, when the SU(2) original
wave vector has rotated around the zone edge, we observe a dispersion
line around $\pf$, with a finite range of quasi-degenerate wave vectors,
and no well-defined peak at a preferential one. We argue in the next
section \ref{sec:Effect-of-the}, that this corresponds to the formation
of excitonic pairs, which can take many wave vectors around $\pf$
in the anti-nodal region around the zone edge. These excitonic pairs
are instrumental in the formation of the pseudo-gap as they will proliferate
with temperature, leading to a gapping out of the anti-nodal region
at $T^{*}$.

An important point, is that the non-linear $\sigma$-model provides
a strong constraint between the charge and SC channels. This in turn
gives strong mode-mode coupling. The thermodynamics of such a model
typically produces some phase separation \cite{Gorkov:1987vw,Gorkov:1989ga,Schmalian:2000fn,Terletska:2011dd},
which leads to the creation of patches of charge modulations. Whether
the particle hole pairs have many wave vectors $\pf$ or condense
only to one or two wave vectors $\left(0,Q_{0}\right)$ and $\left(Q_{0},0\right)$,
depends on the degree of fluctuations in the system. A lot of SU(2)
fluctuations produce the emergence of excitonic patches with multiple
wave vectors, whereas when the fluctuations are frozen, the bosons
condense to one wave vector. The detailed statistical study of the
phase separation is left for for later work, while we give in the
next section \ref{sec:Effect-of-the} a derivation of the formation
of patches of excitonic pairs with multiple wave vectors $\pf$.

\section{\label{sec:Effect-of-the}Preformed Particle-Hole Pairs }

Non-linear $\sigma$-models are the theoretical
tools for describing fluctuations from a non-abelian group. As we
saw in the previous section \ref{sec:The-SU(2)-line}, on the case
of the SU(2) pseudo-spin symmetry, the corresponding non-linear $\sigma$-model
is O(4). Three phases describe the fluctuations : the SC phase, the
CDW phase and the phase rotating between the two modes. The goal of
this section is to focus on the emergence of local modes coming from
the non-linear coupling between the two modes. These local singularities
have been described as ``skyrmions'' or static topological defects
of the theory in the past. We defer a thorough study of these for
future work and focus here on a more pedestrian approach leading to
similar local droplets- or patches of particle-hole pairs.

\subsection{Particle-hole pairing formation\label{sub:Particle-hole-pairing-formation}}

In this section we study the possibility of the formation of particle
hole pairs,-or excitons, with strong binding energy. We will show
that the excitons are bosons with a quasi-degenerate line of finite
momenta around $\pf$ in the anti-nodal region. We proceed as for
the study of Cooper pairing and show that a logarithm is present in
the solution of the Schr\"{o}dinger equations, which is cut-off only by
the curvature of the Fermi surface. Hence, for flat enough regions
of the electronic dispersion, the excitonic pairing can occur.

\subsubsection{Wave function}

The wave function for the particle-hole pair writes 
\begin{align}
\psi_{\mathbf{r,}\mathbf{r}'}^{Ch} & =\sum_{\tilde{\mathbf{k}},\mathbf{P}}e^{-i\mathbf{P}\cdot\left(\frac{\mathbf{r+}\mathbf{r}'}{2}\right)}e^{i\tilde{\mathbf{k}}\cdot\left(\mathbf{r-}r'\right)}\chi_{\tilde{\mathbf{k}},\mathbf{P}},\label{eq:A1-2}
\end{align}

with $\tilde{\mathbf{k}}=-\mathbf{k}+\mathbf{P}$, $\mathbf{P}=\left\{ 2\mathbf{k}_{F}\right\} $
scans the wave vectors represented in Fig.\ref{fig:spec}a). In order
to discuss the most generic solution, we first set $\chi_{\tilde{\mathbf{k}},\mathbf{P}}=cte$.
The summation over $\mathbf{P}$ in Eqn. (\ref{eq:A1-2}) has an important
consequence to localize the center of mass of the excitonic pair at
the point zero (with this representation). The structure in \textbf{$\mathbf{k}$
}then gives the finite extension for the local patch as well as some
possible intrinsic pattern. Note that the wave function for a Cooper
pair $\Delta\sim\sigma\left\langle \psi_{\mathbf{k},\sigma}\psi_{-\mathbf{k},\sigma}\right\rangle $with
modulation vector $\mathbf{P}$ takes a very similar form

\begin{align}
\psi_{\mathbf{r,}r'}^{SC} & =\sum_{\mathbf{k}}e^{-i\mathbf{P}\cdot\left(\frac{\mathbf{r+}\mathbf{r}'}{2}\right)}e^{i\mathbf{k}\cdot\left(\mathbf{r-}r'\right)}\Delta_{0}.\label{eq:A1-1-1}
\end{align}

\subsubsection{The Schr\"{o}dinger equation}

We focus now on the particle-hole instability in manner of Cooper
pairing. We start from the Fermi liquid and look whether a particle-hole
pair of the form given by Eqn.(\ref{eq:A1-2}) can destabilize the
ground state. The equation of motion for $\psi_{\mathbf{r},\mathbf{r'}}$
writes 
\begin{align}
\left[-\frac{\hbar^{2}}{2m}\left(\partial_{\mathbf{r}}^{2}-\partial_{\mathbf{r'}}^{2}\right)+V\left(\mathbf{r},\mathbf{r'}\right)\right]\psi_{\mathbf{r},\mathbf{r'}} & =E\psi_{\mathbf{r},\mathbf{r'}}.\label{eq:85-1}
\end{align}

We study the potential term by taking the average over space $\left\langle \right\rangle _{\mathbf{r}.\mathbf{r'}}$
of Eqn.(\ref{eq:85-1}), which is equivalent to taking the $\left(\mathbf{k},\mathbf{k'}\right)=0$
component in momentum space. 
\begin{align}
\left\langle V_{\mathbf{r},\mathbf{r'}}\psi_{\mathbf{r},\mathbf{r'}}\right\rangle  & =\sum_{\mathbf{k},\mathbf{k'},\mathbf{k_{1}},\mathbf{k'_{1}}}V_{\mathbf{k},\mathbf{k'}}\psi_{\mathbf{k_{1}},\mathbf{k'_{1}}}\sum_{\mathbf{r},\mathbf{r'}}e^{i\left(\mathbf{k}+\mathbf{k_{1}}\right)\cdot\mathbf{r}}e^{-i\left(\mathbf{k'}+\mathbf{k'_{1}}\right)\cdot\mathbf{r'}}\nonumber \\
 & =\sum_{\mathbf{k},\mathbf{k'}}V_{\mathbf{k},\mathbf{k'}}\psi_{\mathbf{k},\mathbf{k'}}.
\end{align}

Within the change of variables $\mathbf{k}\rightarrow\mathbf{\tilde{k}}-\mathbf{P}/2;\mathbf{k'}\rightarrow\tilde{\mathbf{k}}'+\mathbf{P}/2,$
and $\mathbf{P}=2\mathbf{k}_{F}$, we get the following equations
\begin{align}
\left(E-\frac{\hbar^{2}}{m}\mathbf{k}_{F}\cdot\tilde{\mathbf{k}}\right)\chi_{\tilde{\mathbf{k}},\mathbf{p}} & =C,\nonumber \\
C=\sum_{\tilde{\mathbf{k}}}V_{\mathbf{\tilde{k}},\tilde{\mathbf{k}}'}\chi_{\tilde{\mathbf{k}},\mathbf{P}},
\end{align}
where $V_{\mathbf{\tilde{k}},\tilde{\mathbf{k}}'}=\int_{\mathbf{r}}V\left(r\right)e^{i\left(\tilde{\mathbf{k}}-\tilde{\mathbf{k}}'\right)\cdot\mathbf{r}}$
is an attractive potential coming from the pairing fluctuations. With
$\omega_{F}$ the width of the fluctuation spectrum, we model

\begin{align}
\bar{V} & =-\frac{V}{L^{2}}, & \mbox{ if }0<\tilde{\mathbf{k}},\tilde{\mathbf{k}'}<\frac{\omega_{F}}{\rho_{0}};\nonumber \\
 & =0 & \mbox{ elsewhere}.
\end{align}
Herein, $\rho_{0}$ is the electronic density of states at the Fermi
level. Eqn.(\ref{eq:8}) can then easily be solved, leading to the
bonding energy 
\begin{align}
E & =-2\hbar\omega_{F}e^{-2/\left(\rho_{0}V\right)}.\label{eq:9-2-1}
\end{align}
The formation of particle-hole pairs at multiple $2\mathbf{k}_{F}$
wave vectors is a logarithmic instability of the Fermi liquid in the
presence of an attractive potential. In the standard BCS theory, the
coupling between density and phase fluctuations is weak. In some specific
cases, however, like the attractive Hubbard model, density and phase
couple strongly and our model likewise predicts the emergence of s-wave
excitonic patches. Within the SU(2) scenario for Cuprates, the typical
scale associated with the pairing fluctuations is strong, of order
of the formation of the SU(2) dome, and can naturally be associated
with the PG scale $T^{*}$.

\subsection{Integrating the SU(2) fluctuations}

In order to derive the effective action for the sub-leading orders,
we integrate out the SU(2) fluctuations, averaging now over the effective
modes $\hat{Q}$ : 
\begin{align}
Z_{fin} & =e^{-S_{fin}}, & \mbox{with }S_{fin}=-\frac{1}{2}\left\langle \left(S_{int}\right)^{2}\right\rangle _{Q}.
\end{align}

From Eqn.(\ref{eq:B2}) we have 
\begin{align}
S_{fin} & =-\frac{1}{8}Tr\int_{x,x',x_{1},x_{1}'}\left\langle \overline{\Psi}_{x}\hat{\overline{Q}}_{x,x'}\Psi_{x'}\overline{\Psi}_{x_{1}}\hat{Q}_{x_{1},x_{1}'}\Psi_{x'_{1}}\right\rangle _{Q},
\end{align}
where $\hat{\overline{Q}}$,$\hat{Q}$ are defined in Eqns.(\ref{eq:9},\ref{eq:10}).
Using the more practical Wigner- transform defined in Eqn.(\ref{eq:18}),
we get

\begin{align}
S_{fin} & =-\frac{1}{8}Tr\sum_{k,q,k',q'}\left\langle \overline{\Psi}_{k+q}\hat{\overline{M}}_{k,q}\Psi_{k}\overline{\Psi}_{k'+q'}\hat{M}_{k',q'}\Psi_{k'}\right\rangle _{Q}.\label{eq:34}
\end{align}

Disentangling Eqn.(\ref{eq:34}) is quite lengthy but the result
produces a sum of an effective component in the SC channel (\ref{eq:35})
and another one in the charge channel (\ref{eq:36}) $S_{fin}=S_{fin}^{a}+S_{fin}^{b}$,
with

\begin{align}
S_{fin}^{a} & =-\frac{1}{2}Tr\sum_{k,q,k',q',\sigma,\sigma'}\sigma\sigma'\left\langle \Delta_{k,q}^{\dagger}\Delta_{k',q'}\right\rangle _{Q}\nonumber \\
 & \times\psi_{k+q/2,\sigma}^{\dagger}\psi_{-k+q/2,-\sigma}^{\dagger}\psi_{-k'+q'/2,-\sigma'}\psi_{k'+q'/2,\sigma'},\label{eq:35}\\
S_{fin}^{b} & =-\frac{1}{2}Tr\sum_{k,q,k',q',\sigma,\sigma'}\left\langle \chi_{k,q}^{\dagger}\chi_{k',q'}\right\rangle _{Q}\nonumber \\
 & \times\psi_{k+q/2,\sigma}^{\dagger}\psi_{k+Q_{0}-q/2,\sigma}\psi_{k'+Q_{0}-q'/2,\sigma'}^{\dagger}\psi_{k'+q'/2,\sigma'}.\label{eq:36}
\end{align}

Note that since the bosonic propagator Eqn.(\ref{eq:nlsm2-1}) conserves
the number of particles, there is no mixed term in the above Eqns.
The forms of $\left\langle \Delta_{k,q}^{\dagger}\Delta_{k',q'}\right\rangle _{Q}$
and $\left\langle \chi_{k,q}^{\dagger}\chi_{k',q'}\right\rangle _{Q}$
are identical, up to a mass term, and are given by the non-linear
$\sigma$-model Eqns.(\ref{eq:24},\ref{eq:nlsm2-1-1-1}) 
\begin{align}
\langle\Delta_{k,q}^{\dagger}\Delta_{k',q'}\rangle_{Q} & =\pi_{k,k',q}^{s}\delta_{\mathbf{q},\mathbf{q'}}\label{eq:deltaprop}\\
\left\langle \chi_{k,q}^{\dagger}\chi_{k',q'}\right\rangle _{Q} & =\pi_{k,k',q}^{c}\delta_{\mathbf{q},\mathbf{q'}},
\end{align}
where the form of the SU(2) propagator has been defined in section
\ref{sec:SU(2)-fluctuations-coupled}, Eqns.(\ref{eq:24},\ref{eq:nlsm2-1-1-1},\ref{eq:31}):
\begin{align}
\pi_{k,k',q}^{c} & =M_{0,k}M_{0,k'}\frac{\pi_{0}}{J_{0}\omega_{n}^{2}+J_{1}({\bf {v}}\cdot{\bf {q}})^{2}+a_{0}},\label{eq:pidef}\\
\pi_{k,k',q}^{s} & =M_{0,k}M_{0,k'}\frac{\pi_{0}}{J_{0}\omega_{n}^{2}+J_{1}({\bf {v}}\cdot{\bf {q}})^{2}-a_{0}},
\end{align}

where $a_{0}$ is the mass term from Eqn.(\ref{eq:31}),

\subsection{\label{sub:Effect-of-the}Excitonic patches}

One can now perform a second Hubbard-Stratonovich transformation in
order to decouple $S_{\text{fin}}^{a}[\psi]$ in Eq.(\ref{eq:35})
in the charge channel, to get

$S_{\text{fin}}^{a}[\psi]=S_{0}^{a}[\chi]+S_{1}^{a}[\psi,\chi]$,
with 
\begin{align}
S_{0}^{a}[\chi] & =-\sum_{kk'q,\sigma}\pi_{k,k',q}^{^{S}-1}\chi_{-\sigma,q-k,q-k'}\chi_{\sigma,k,k'},\label{eq:S1}\\
S_{1}^{a}[\psi,\chi] & =\sum_{kk',\sigma}[\chi_{-\sigma,-k+q,-k'+q}\psi_{\mathbf{k,}\sigma}^{\dagger}\psi_{\mathbf{k',\sigma}}\nonumber \\
 & +\chi_{\sigma,k,k'}\psi_{\mathbf{-k+q,}-\sigma}^{\dagger}\psi_{\mathbf{-k'+q},-\sigma}\Bigr].
\end{align}
Stationarity of the free energy leads to 
\begin{equation}
\chi_{\sigma,k,k'}=\sum_{q}\pi_{k,k',q}^{s}\langle\psi_{\mathbf{-k}+\mathbf{q},\sigma}^{\dagger}\psi_{\mathbf{-k'}+\mathbf{q},\sigma}\rangle.\label{eq:chidef}
\end{equation}
We will drop in the following the spin label of the $\chi_{k,k'}$
field, since both spin configurations are degenerate. Together with
the bare fermionic action Eq.\ (\ref{eq:bare}) one obtains the effective
action 
\begin{equation}
S_{0}^{a}[\psi]+S_{1}^{a}[\psi,\chi]=-\sum_{kk'\sigma}\overline{\tilde{\psi}}_{\mathbf{k}}\hat{G}_{k,k'}^{-1}\tilde{\psi}_{\mathbf{k'}}
\end{equation}
with the two component fermionic field 
\begin{equation}
\tilde{\psi}_{\mathbf{k}}=(\psi_{\mathbf{k,-\sigma}},\psi_{\mathbf{k',\sigma}})^{T}.
\end{equation}
and the inverse propagator 
\begin{align}
\hat{G}_{k,k'}^{-1} & =\left(\!\!\begin{array}{cc}
(i\epsilon_{n}-\xi_{\kv}) & -\chi_{-\sigma,k,k'}\\
-\chi_{\sigma,k,k'} & (i\epsilon_{n}+\xi_{k'})
\end{array}\!\!\right).\label{eq:invpropg}
\end{align}

\subsection{Gap equation for the charge order}

\begin{figure}[h]
\begin{minipage}[c]{4.25cm}%
\includegraphics[width=4cm]{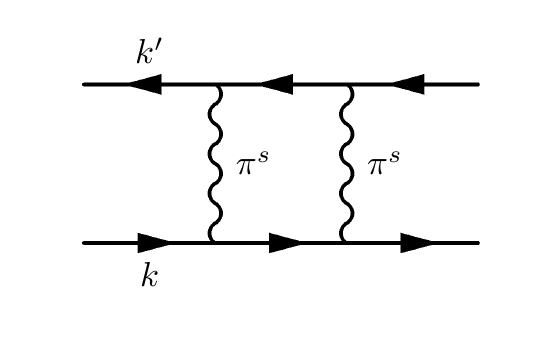} \vspace{1ex}
\end{minipage}

\textcolor{black}{\caption{\label{fig:diags4} (Color online) \textcolor{black}{Infinite ladder
series corresponding to the gap equations $\left(\ref{eq:chieq3}\right)$.}}
} 
\end{figure}

The gap equation to study the charge ordering our system, stems directly
from the Dyson equation for the fermionic propagator \begin{subequations}
\begin{align}
\hat{G}_{k,k'}^{-1} & =\hat{G}_{0}^{-1}-\hat{\Sigma}_{k,k'},\\
\mbox{ with } & \hat{G}_{0}^{-1}=\left(\begin{array}{cc}
i\epsilon_{n}-\xi_{k}\\
 & i\epsilon_{n}-\xi_{k'}
\end{array}\right),\\
\mbox{and } & \hat{\Sigma}_{k,k'}=\left(\begin{array}{cc}
 & \chi_{k,k'}\\
\chi_{k,k'}
\end{array}\right).
\end{align}
\end{subequations} 
\begin{equation}
\hat{G}_{k,k'}=-\langle{\cal T}\tilde{\psi}_{k}\overline{\tilde{\psi}}_{k'}\rangle,
\end{equation}
is obtained by inverting Eq.\ (\ref{eq:invpropg}) and one finds
\begin{align}
[\hat{G}_{k,k'}]_{12} & =-\langle\psi_{\mathbf{k,}-\sigma}^{\dagger}\psi_{\mathbf{k',-\sigma}}\rangle\nonumber \\
 & =-\frac{\chi_{k,k'}}{(i\epsilon_{n}-\xi_{{\bf {k}}})(i\epsilon_{n}'-\xi_{k'})-\chi_{k,k'}^{2}}.\label{eq:chieq2}
\end{align}

which finally yields (see Fig. \ref{fig:diags4}) 
\begin{equation}
\chi_{k,k'}=\sum_{q}\pi_{k,k',q}^{s}[\hat{G}_{q-k,q-k'}]_{12}.\label{eq:chieq3}
\end{equation}

The maximum solution comes from the denominator in Eqn.(\ref{eq:chieq2}),
and especially from the \textbf{(k},\textbf{k')}- point close to the
Fermi surface. In order to keep the solution tractable without loosing
some physical effects, we neglect the frequency dependence of $\chi$.
We can then easily compute the Matsubara sum at $T=0$. We find 
\begin{equation}
\chi_{{\bf {k}},{\bf {k}}'}=\bar{\pi}_{0}\sum_{{\bf {q}}}I({\bf {k}},{\bf {k}}',{\bf {q}})\label{eq:chieq4}
\end{equation}
with 
\begin{equation}
I({\bf {k}},{\bf {k}}',{\bf {q}})=\frac{\chi_{q-k,q-k'}}{2}\Biggl[\frac{(|\omega_{1}|+|\omega_{2}|+r)\sgn(\omega_{1}\omega_{2})-r}{r(|\omega_{1}|+|\omega_{2}|)(|\omega_{1}|+r)(|\omega_{2}|+r)}\Biggr]
\end{equation}
and \begin{subequations} 
\begin{align}
r & =\sqrt{\bar{J}_{1}\xi_{{\bf {q}}}^{2}+\bar{a}_{0}}\\
\omega_{1/2} & =(\xi_{{\bf {q-k}}}+\xi_{{\bf {q-k'}}})/2\pm\!\sqrt{\!(\xi_{{\bf {q-k}}}-\xi_{{\bf {q-k'}}})^{2}\!/\!4+\!\chi_{q-k,q-k'}^{2}}\qquad
\end{align}
\end{subequations} where $\bar{\pi}_{0}=\pi_{0}/J_{0}$, $\bar{J}_{1}=J_{1}/J_{0}$
and $\bar{a}_{0}=a_{0}/J_{0}$.

\subsubsection{Dependence of the solution on the exciton wave vectors}

\begin{figure}
\begin{minipage}[c]{4.25cm}%
\includegraphics[width=4cm]{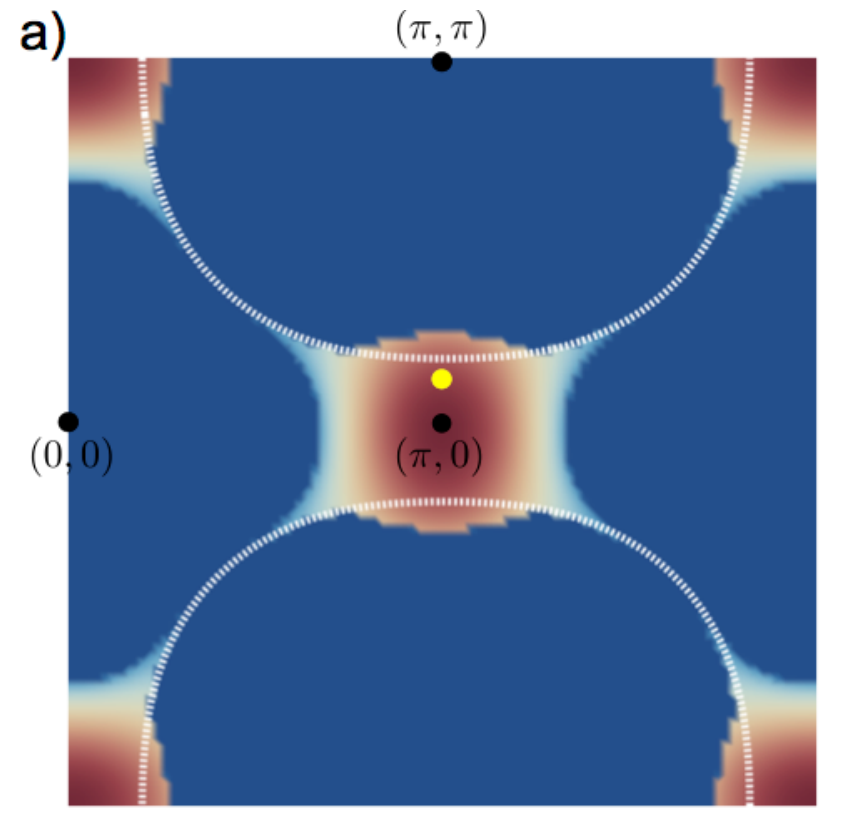} \vspace{1ex}
\end{minipage}%
\begin{minipage}[c]{4.25cm}%
\includegraphics[width=4cm]{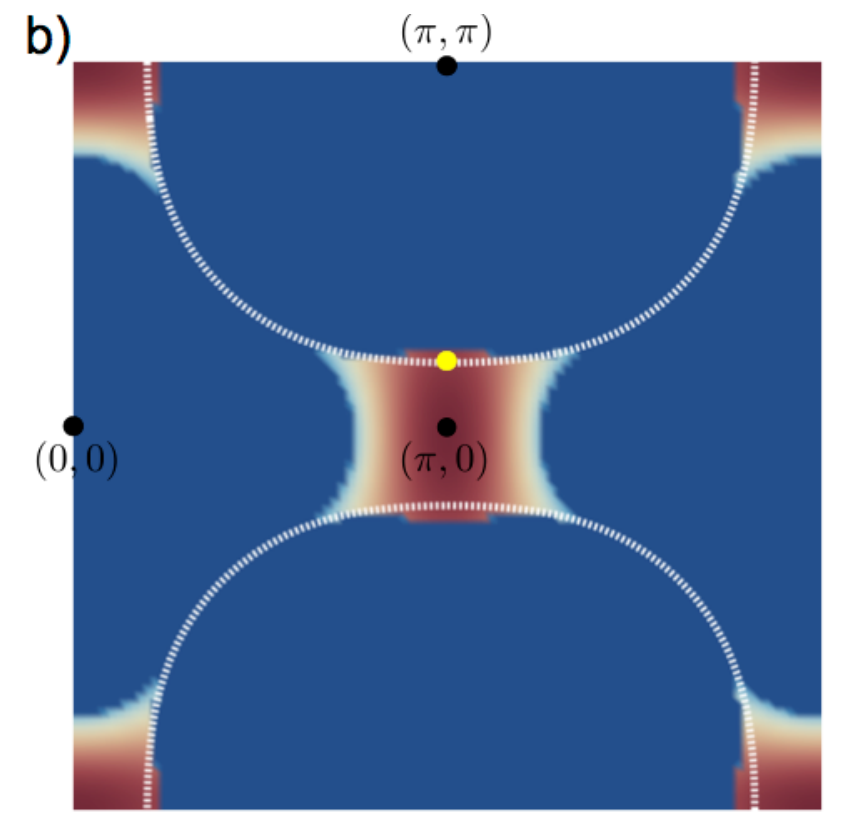} \vspace{1ex}
\end{minipage}

\begin{minipage}[c]{4.25cm}%
\includegraphics[width=4cm]{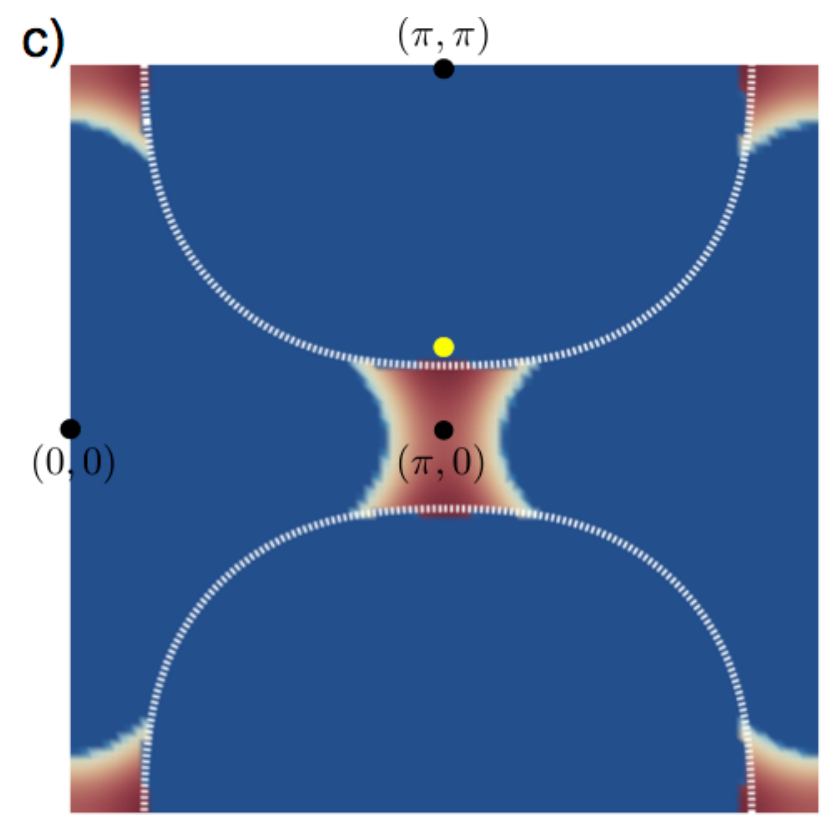} \vspace{1ex}
\end{minipage}

\begin{centering}
\includegraphics[scale=0.7]{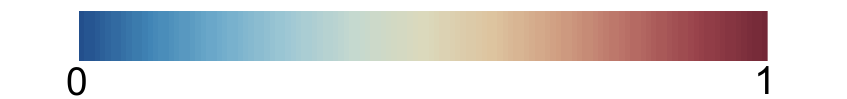} 
\par\end{centering}

\caption{\label{fig:kpk} (Color online) \textcolor{black}{Density plots of
the charge order parameter $|\chi_{k,k'}|$ obtained from a numerical
solution of Eqn.(\ref{eq:chieq4}) as a function of the coupling vector
$\mathbf{P=k'-k}$. Red, white and blue represents respectively high,
intermediate and low values of $\chi$. The white dotted line represents
the Fermi surface. We have fixed the $k$ point in a) the point $(\pi,0)$,
b) the point on the Fermi surface at the zone edge and c) a point
shifted a little from the Fermi surface (see the yellow point). The
interaction is $\bar{\pi_{0}}=0.3$ and the mass $\bar{a_{0}}=5$.
The form of the solution only depends on the $k$ point.}}
\end{figure}

In a first study, we want to find all wave vectors $(\mathbf{k},\mathbf{k'})$
which give the maximum response. For this task, we solve Eq.(\ref{eq:chieq4})
for the charge ordering parameter $\chi$ numerically for arbitrary
coupling vector ${\bf {P}={\bf {k}'-{\bf {k}}}}$. We further make
the approximation $\chi_{q-k,q-k'}\approx\chi_{-k,-k'}$ on the right
hand side of Eq.(\ref{eq:chieq4}). Fixing then a reference point
${\bf {k}}$ in the first BZ, we solve the mean-field equation upon
varying the coupling vector ${\bf {P}={\bf {k}'-{\bf {k}}}}$ and
look for the points ${\bf {k}'}$ where the solution is maximal. As
already suspected from Eq.(\ref{eq:chieq4}), nonzero solutions are
obtained only when both points ${\bf {k}}$ and ${\bf {k}'}$ are
situated close to the FS. The numerical solution in Fig.\ref{fig:kpk}
shows that for all couplings $\{{\bf {P}}\}$ connecting two points
of the Fermi surface, the height of non-zero $\chi$ is very similar.

\subsubsection{Solution for $\mathbf{k'}-\mathbf{k}=\pf$}

\begin{figure}[h]
\begin{minipage}[c]{4.25cm}%
\includegraphics[width=4cm]{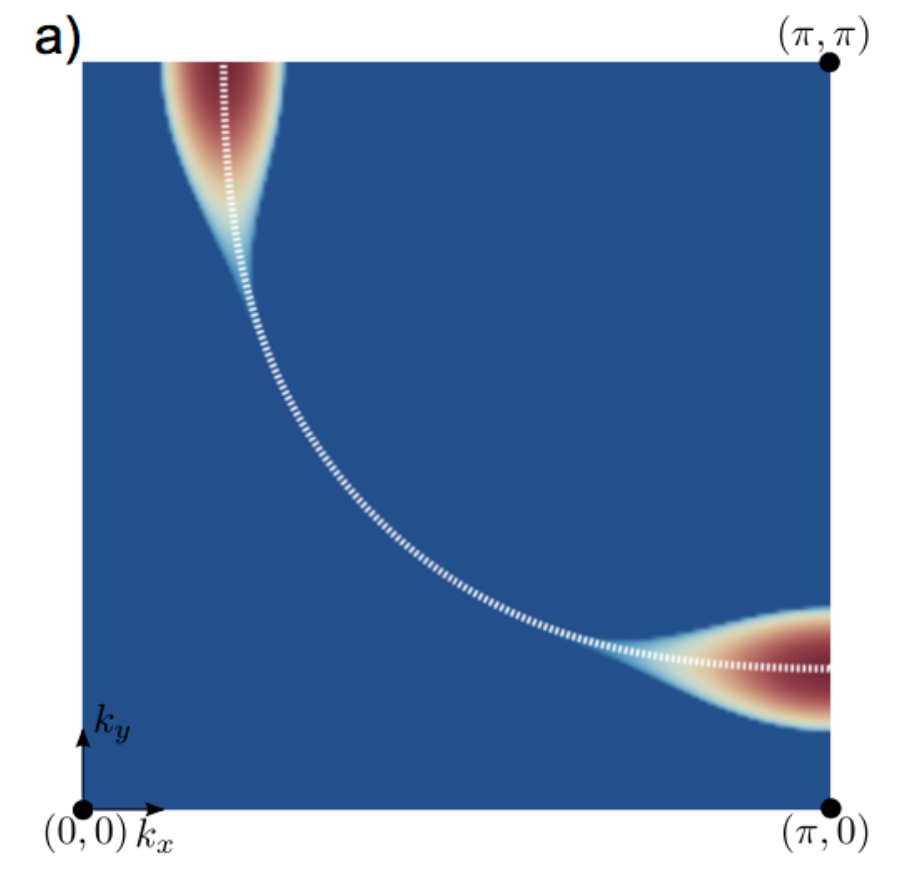} \vspace{1ex}
\end{minipage}%
\begin{minipage}[c]{4.25cm}%
\includegraphics[width=4cm]{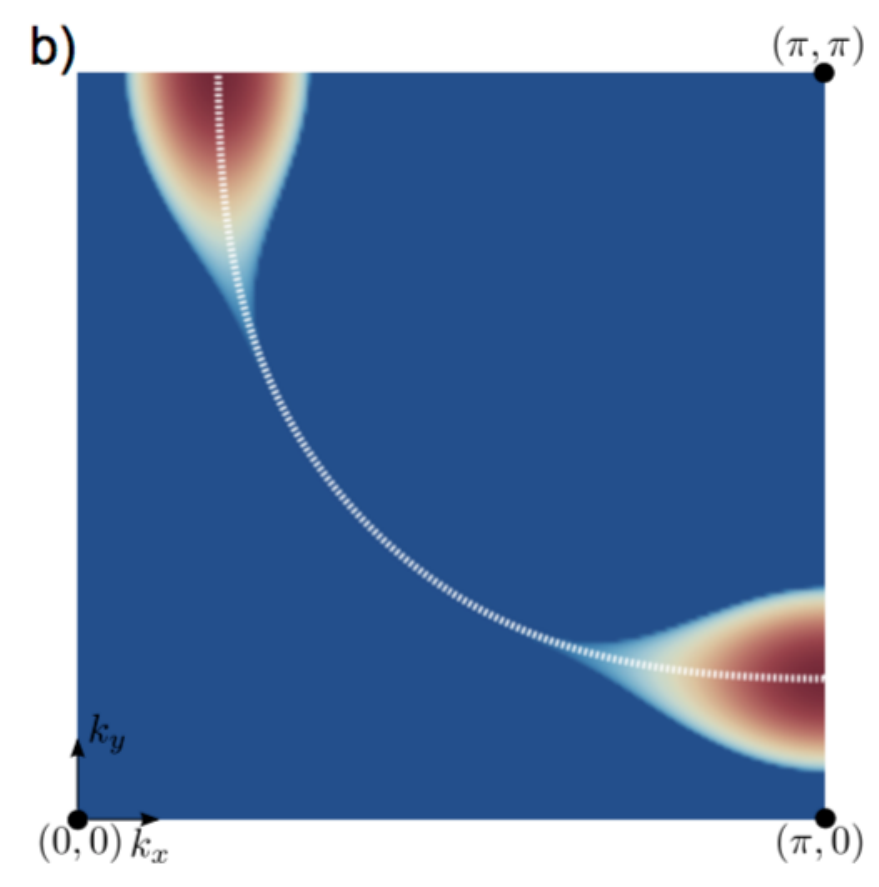} \vspace{1ex}
\end{minipage}

\begin{minipage}[c]{4.25cm}%
\includegraphics[width=4cm]{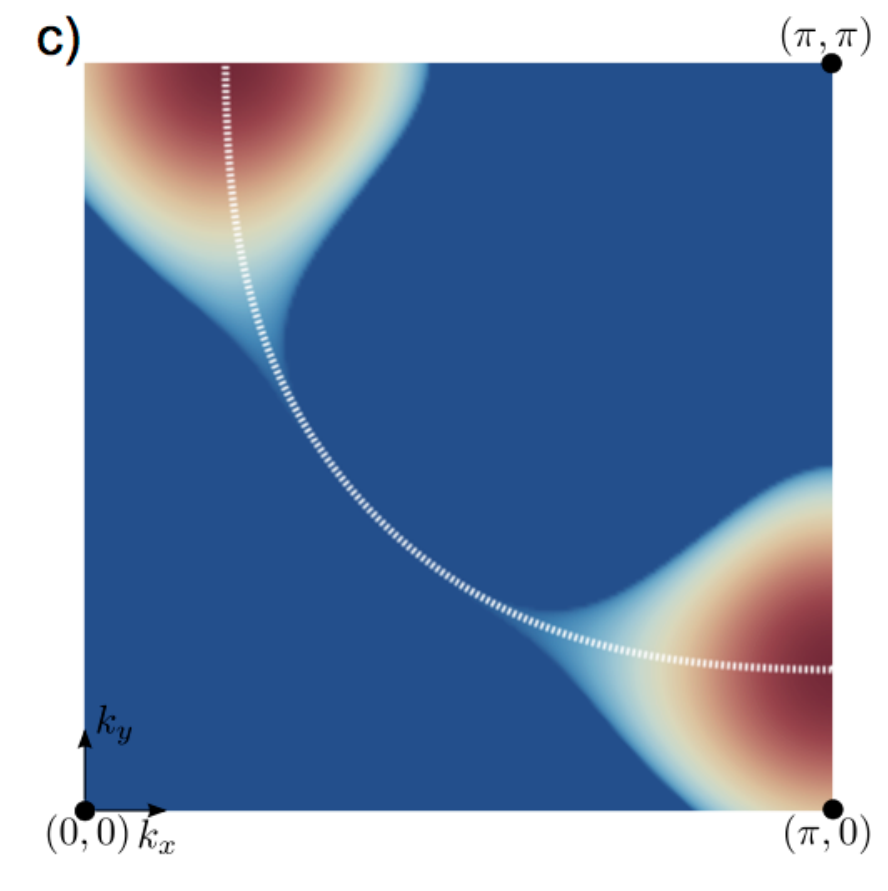} \vspace{1ex}
\end{minipage}

\begin{centering}
\includegraphics[scale=0.7]{Fig14d} 
\par\end{centering}

\caption{\label{fig:2pf} (Color online) \textcolor{black}{Density plots of
the charge order parameter $|\chi_{k,k'}|$ obtained from a numerical
solution of Eqn.(\ref{eq:chieq4}) in the first BZ for the coupling
vector $\mathbf{P=-2k}$ for different interaction strength a) $\bar{\pi_{0}}=0.01$
b) $\bar{\pi_{0}}=0.03$ c) $\bar{\pi_{0}}=0.3$ and the mass $\bar{a_{0}}=5$.
Red, white and blue represents respectively high, intermediate and
low values of $|\chi_{k,k'}|$. The white dotted line represents the
Fermi surface. The magnitude as well as the size of the gap increases
with the strength of the interaction.}}
\end{figure}

In Fig.\ref{fig:2pf}, we present the solution of the gap equations
for a range of wave vectors connecting points around the Fermi surface
in the anti-nodal region. Hence we called these wave vectors $\pf$.
We observe that, as the coupling strength is increased, the $\pf$
wave vectors are able to gap out the entire anti-nodal region the
BZ. The main idea here, is that the SU(2) pairing fluctuations not
only rotate the original charge modulation wave vector from the diagonal
to the axes, but gives space for a range of wave vectors to participate
to the electron-hole pairing. As shown in section \ref{sub:Particle-hole-pairing-formation},
the set of $\pf$ leas to a logarithm in the direction perpendicular
to the Fermi surface, and hence leads to the formation of preformed
pairs. Each $\pf$ wave vector shares only a small portion of the
phase space in momentum \textbf{k}, around the Fermi surface, therefore
the logarithmic divergence is finally cut by curvature in the direction
transverse to the corresponding Fermi wave vector. 

A remark about the symmetry of the charge modulations
in the patch are in order here. The $\pf$ order comes from $\pi_{q}^{s}$
in Eq.(\ref{eq:chieq3}) and thus connect the ordering parameters
$\chi_{k,k'}$ around the same points in the Brillouin zone. The solution
of the gap equation (\ref{eq:chieq3}) alone cannot distinguish between
d-wave and s or s' order. But as before we have seen in section \ref{sec:The-SU(2)-dome},
that d-wave $\pf$ order already emerges from short range AF correlations.
The SU(2) fluctuations have thus to be seen as an additional force
action on top of AF correlations, which altogether leads to the stabilization
of the d-wave symmetry for the $\pf$ droplets. Hence, the effect
of the SU(2) fluctuations will be to select the involution A) Eqn.(\ref{eq:8})
as the preferential SU(2) partner of d-wave SC.

The spreading of the wave vectors is typical of the formation of patches
in real space, that we have described in a previous paper \cite{Montiel16}.
The patches have an internal modulation structure very close to the
checkerboard observed experimentally. They can be frozen, or fluctuate
at a temperature closer to the PG. The detailed study of this intricate
dynamics goes beyond the scope of this paper. It relies on the existence
of the constraint in the non-linear $\sigma$-model, which creates
strong effective correlations between the charge and SC modes, and
finally a type of mode-mode coupling in the charge sector. Tis in
turn typically leads to phase separation, and entropic effects. In
section \ref{sub:The-physical-lines} we give a heuristic picture
of the phase diagram in this approach.

\subsubsection{Solution for $\mathbf{k}'=-\mathbf{k}$}

For this type of involution, we do not find any formation of a gap.
It is not surprising, since it doesn't imply any band crossing of
the electronic dispersion, as already noticed in Fig.\ref{fig:spec}c').
Indeed, the involution which is sending $\mathbf{k}\rightarrow-\mathbf{k}$
leaves the electronic dispersion $\xi_{\mathbf{k}}$invariant.

\begin{figure}[h]
\begin{minipage}[c]{4.25cm}%
\includegraphics[width=4cm]{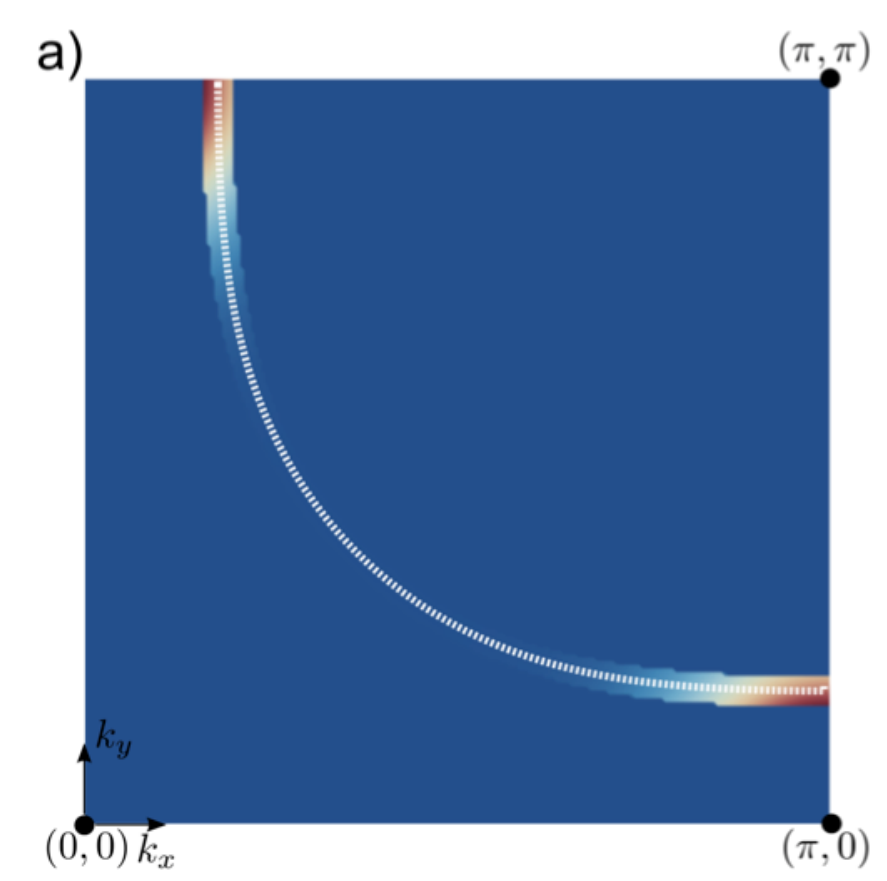} \vspace{1ex}
\end{minipage}%
\begin{minipage}[c]{4.25cm}%
\includegraphics[width=4cm]{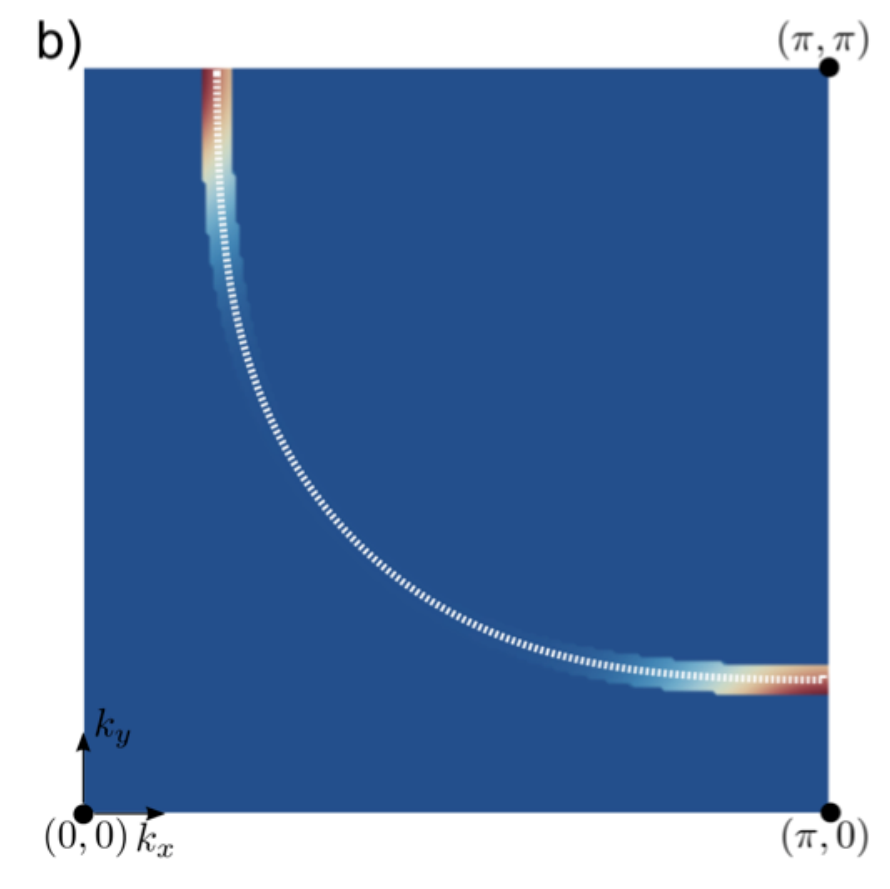} \vspace{1ex}
\end{minipage}

\begin{minipage}[c]{4.25cm}%
\includegraphics[width=4cm]{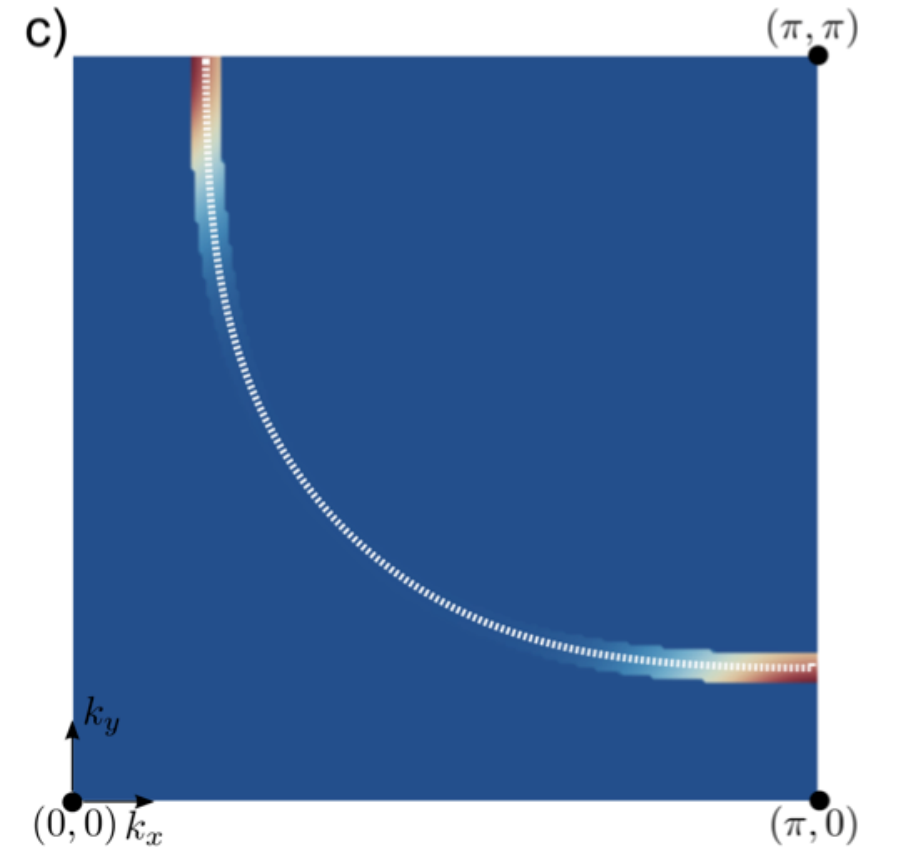} \vspace{1ex}
\end{minipage}

\begin{centering}
\includegraphics[scale=0.7]{Fig14d} 
\par\end{centering}

\caption{\label{fig:kmk-1} (Color online) \textcolor{black}{Density plots
of the charge order parameter $|\chi_{k,k'}|$ obtained from a numerical
solution of Eqn.(\ref{eq:chieq4}) in the first BZ for the coupling
vector $\mathbf{P=-2k}$ for different interaction strength a) $\bar{\pi_{0}}=0.03$
b) $\bar{\pi_{0}}=0.3$ c) $\bar{\pi_{0}}=3$ and the mass $\bar{a_{0}}=5$.
Red, white and blue represents respectively high, intermediate and
low values of $|\chi_{k,k'}|$. The white dotted line represents the
Fermi surface. The magnitude of the peak increases with the interaction
contrary to the size of the gap.}}
\end{figure}

The parameters in Figs.\ \ref{fig:kmk-1} are: $\bar{\pi}_{0}=0.3$,
$\bar{J}_{1}=10^{-6}$, $\bar{a}_{0}=5$ in units of the band gap
and we take a constant mass $\bar{a}_{0}$ over the BZ. The dispersion
is approximated by a tight-binding dispersion $\xi_{{\bf {k}}}$ with
parameter set ``tb2'' from Ref.\cite{Norman07} and the chemical
potential adjusted to account for 10\% hole doping. In fact, the shape
of the numerical solution of $\chi$ turns out to be not very sensitive
to the model parameters $\bar{J}_{1}$ and $\bar{a}_{0}$.

\section{Pair Density Wave (PDW)}

In this section we look at the potential generation
of other types of order, and in particular, of the PDW order from
the SU(2) fluctuations.

\subsection{Gap equation}

The decoupling of $S_{int}^{b}$ in Eqn. (\ref{eq:36}) follows the
same steps as for Eqn.(\ref{eq:35}) Performing the Hubbard-Stratonovich
transformation, we get $S_{\text{fin}}^{b}[\psi]=S_{0}^{b}[\chi]+S_{1}^{b}[\psi,\chi]$,
with

\begin{align}
S_{0}^{b}[\chi] & =-\sum_{kk'q,\sigma}\pi_{k,k',q}^{^{c}-1}\Delta_{-\sigma,k,k'+Q_{0}}^{a\dagger}\Delta_{\sigma,k+Q_{0}+q,k'+q}^{b},\\
S_{1}^{b}[\psi,\chi] & =\sum_{kk',\sigma}\Bigl[\Delta_{-\sigma,k,k'+Q_{0}}^{a\dagger}\sigma\psi_{\mathbf{k}+\mathbf{Q_{0}}+\mathbf{q},-\sigma}\psi_{\mathbf{k'}+\mathbf{q},\sigma}\nonumber \\
 & +\sigma\psi_{\mathbf{k},-\sigma}^{\dagger}\psi_{\mathbf{k'}+\mathbf{Q_{0}},-\sigma}^{\dagger}\Delta_{\sigma,k+Q_{0}+q,k'+q}^{b}\Bigr].
\end{align}

Herein $\Delta_{-\sigma,k,k'+Q_{0}}^{a}$ is the modulated superconducting
field whose condensation leads to the PDW state. We follow closely
the last section (\ref{sub:Effect-of-the}) to derive the corresponding
equations for the PDW channel:

Again, the stationarity of the free energy leads to 
\begin{equation}
\Delta_{\sigma,k,k'+Q_{0}}^{a}=\sum_{q}\pi_{k,k',q}^{c}\langle\sigma\psi_{\mathbf{k}+\mathbf{q},-\sigma}\psi_{\mathbf{k'}+\mathbf{Q_{0}}+\mathbf{q},\sigma}\rangle.\label{eq:chidef-1}
\end{equation}
As developed in the previous section, we obtain the effective action
\begin{equation}
S_{0}^{b}[\psi]+S_{1}^{b}[\psi,\chi]=-\sum_{kk'\sigma}\overline{\tilde{\psi}}_{\mathbf{k}\sigma}\hat{G}_{k,k'}^{-1}\tilde{\psi}_{\mathbf{k',\sigma}}
\end{equation}
with the four-component fermionic field 
\begin{equation}
\tilde{\Psi}_{\mathbf{k},\sigma}=(\psi_{\mathbf{k,-\sigma}},\psi_{\mathbf{k'}+\mathbf{Q_{0}},\sigma}^{\dagger},\psi_{\mathbf{k'},-\sigma},\psi_{\mathbf{k}+\mathbf{Q_{0}},\sigma}^{\dagger})^{T},
\end{equation}
and the conjugation 
\begin{align}
\overline{\tilde{\Psi}} & =(\psi_{\mathbf{k,-\sigma}}^{\dagger},-\psi_{\mathbf{k'}+\mathbf{Q_{0}},\sigma},\psi_{\mathbf{k'},-\sigma}^{\dagger},-\psi_{\mathbf{k}+\mathbf{Q_{0}},\sigma}),
\end{align}
and the inverse propagator now writes 
\begin{align*}
\hat{G}_{k,k'}^{-1} & =\left(\begin{array}{cc}
\hat{G}^{b,-1}\\
 & \hat{G}^{a,-1}
\end{array}\right)
\end{align*}

\begin{align}
\hat{G}^{a,-1} & =\left(\!\!\begin{array}{cc}
(i\epsilon_{n}-\xi_{\kv}) & -\sigma\Delta_{\sigma,k,k'+Q_{0}}^{a}\\
\sigma\Delta_{\sigma,k,k'+Q_{0}}^{a\dagger} & (-i\epsilon_{n}-\xi_{\mathbf{k'}+\mathbf{Q_{0}}})
\end{array}\!\!\right),\label{eq:invpropg-2}
\end{align}

and 
\begin{align}
\hat{G}^{b,-1} & =\left(\!\!\begin{array}{cc}
(i\epsilon_{n}-\xi_{\mathbf{k'}}) & \sigma\Delta_{\sigma,k+Q_{0},k'}^{b}\\
-\sigma\Delta_{\sigma,k+Q_{0},k'}^{b\dagger} & (-i\epsilon_{n}+\xi_{\mathbf{k}+\mathbf{Q_{0}}})
\end{array}\!\!\right).\label{eq:invpropg-2-1}
\end{align}

\begin{figure}[h]
\begin{minipage}[c]{4.25cm}%
\includegraphics[width=4cm]{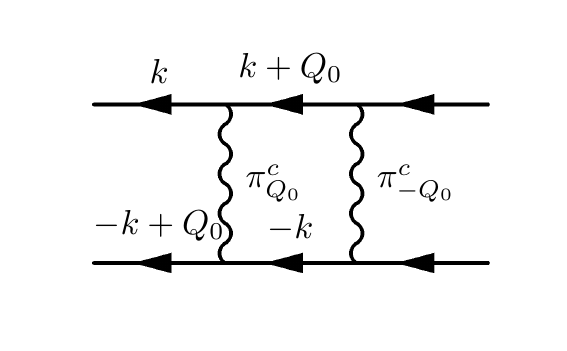} \vspace{1ex}
\end{minipage}

\textcolor{black}{\caption{\label{fig:diags5} (Color online) \textcolor{black}{Infinite ladder
series corresponding to the gap equations $\left(\ref{eq:chieq3-1},\ref{eq:chieq3-1-1}\right)$.}}
} 
\end{figure}

The gap equation to study the charge ordering in our system stems
directly from the Dyson equation for the fermionic propagators \begin{subequations}
\begin{align}
\hat{G}^{a,-1} & =\hat{G}_{0,a}^{-1}-\hat{\Sigma}_{a},\\
\mbox{ with } & \hat{G}_{0,a}^{-1}=\left(\begin{array}{cc}
i\epsilon_{n}-\xi_{\mathbf{k}}\\
 & -i\epsilon_{n}-\xi_{\mathbf{k'}+\mathbf{Q_{0}}}
\end{array}\right),\\
\mbox{and } & \hat{\Sigma}_{a}=\left(\begin{array}{cc}
 & \sigma\Delta_{-\sigma,k,k'+Q_{0}}^{a}\\
-\sigma\Delta_{\sigma,k,k'+Q_{0}}^{a\dagger}
\end{array}\right),
\end{align}

and 
\begin{align}
\hat{G}^{b,-1} & =\hat{G}_{0,b}^{-1}-\hat{\Sigma}_{b},\\
\mbox{ with } & \hat{G}_{0,b}^{-1}=\left(\begin{array}{cc}
i\epsilon_{n}-\xi_{\mathbf{k}+\mathbf{Q_{0}}}\\
 & -i\epsilon_{n}-\xi_{\mathbf{k'}}
\end{array}\right),\\
\mbox{and } & \hat{\Sigma}_{b}=\left(\begin{array}{cc}
 & -\sigma\Delta_{-\sigma,k+Q_{0},k'}^{b}\\
\sigma\Delta_{\sigma,k+Q_{0},k'}^{b\dagger}
\end{array}\right).
\end{align}

\end{subequations} 
\begin{equation}
\hat{G}_{k,k'}=-\langle{\cal T}\tilde{\Psi}_{k}\overline{\tilde{\Psi}}_{k'}\rangle,
\end{equation}
is obtained by inverting Eq.\ (\ref{eq:invpropg-2}) and one finds
\begin{align}
[\hat{G}_{k,k'}^{b}]_{12} & =-\sigma\langle\psi_{\mathbf{k}+\mathbf{Q_{o}},-\sigma}\psi_{\mathbf{k',-\sigma}}\rangle\nonumber \\
 & =-\frac{\Delta_{-\sigma,k+Q_{0},k'}^{b}}{(i\epsilon_{n}-\xi_{{\bf {k}}+\mathbf{Q_{0}}})(-i\epsilon_{n}'-\xi_{\mathbf{k}'})+\left|\Delta_{-\sigma,k+Q_{0},k'}^{b}\right|^{2}}.\label{eq:chieq2-1}
\end{align}
modulated superconducting field $\Delta_{-\sigma,k,k'+Q_{0}}^{a}$
finally yields (see Fig. \ref{fig:diags5}) 
\begin{equation}
\Delta_{-\sigma,k,k'+Q_{0}}^{a}=-\sum_{q}\pi_{k,k',q}^{c}[\hat{G}_{k+q,k'+q}^{b}]_{12}.\label{eq:chieq3-1}
\end{equation}

The same derivation goes for the field $\Delta_{\sigma,k,k'+Q_{0}}^{b}$,
leading to

\begin{equation}
\Delta_{\sigma,k+Q_{0},k'}^{b}=-\sum_{q}\pi_{k,k',q}^{c}[\hat{G}_{k+q,k'+q}^{a}]_{12},\label{eq:chieq3-1-1}
\end{equation}
with 
\begin{align}
[\hat{G}_{k,k'}^{a}]_{12} & =-\sigma\langle\psi_{\mathbf{k,}\sigma}\psi_{\mathbf{k'}+\mathbf{Q_{0}},\mathbf{\sigma}}\rangle\nonumber \\
 & =-\frac{\Delta_{\sigma,k,k'+Q_{0}}^{a}}{(i\epsilon_{n}-\xi_{{\bf {k}}})(-i\epsilon_{n}'-\xi_{\mathbf{k}'+\mathbf{Q_{0}}})+\left|\Delta_{\sigma,k,k'+Q_{0}}^{a}\right|^{2}}.\label{eq:chieq2-1-1}
\end{align}

\subsection{Numerical solution}

\begin{figure}[h!]
\begin{minipage}[c]{4.25cm}%
a)\includegraphics[width=3.95cm]{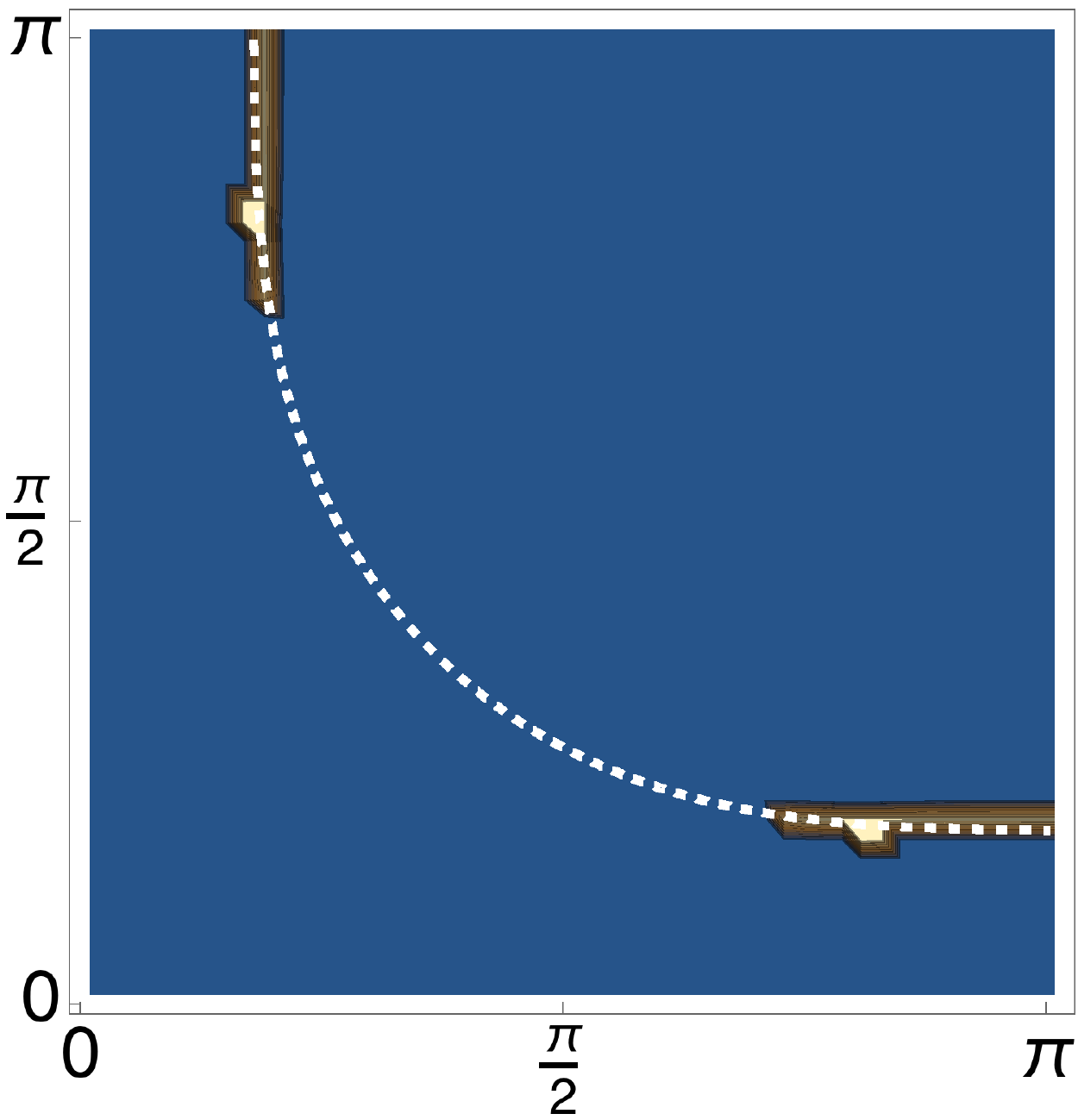} \vspace{1ex}
\end{minipage}%
\begin{minipage}[c]{4.25cm}%
b)\includegraphics[width=3.95cm]{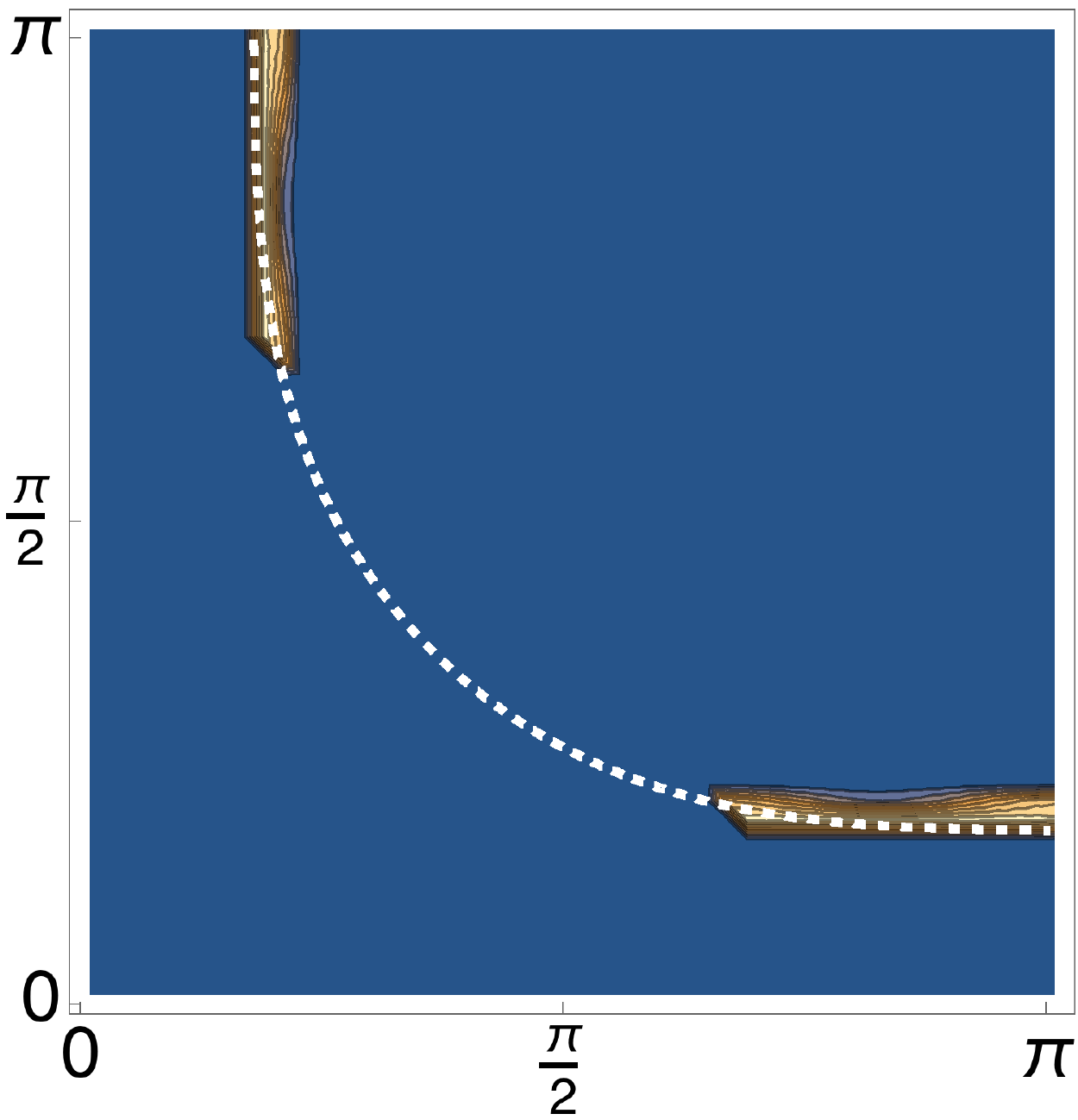} \vspace{1ex}
\end{minipage}

\begin{minipage}[c]{4.25cm}%
c)\includegraphics[width=3.95cm]{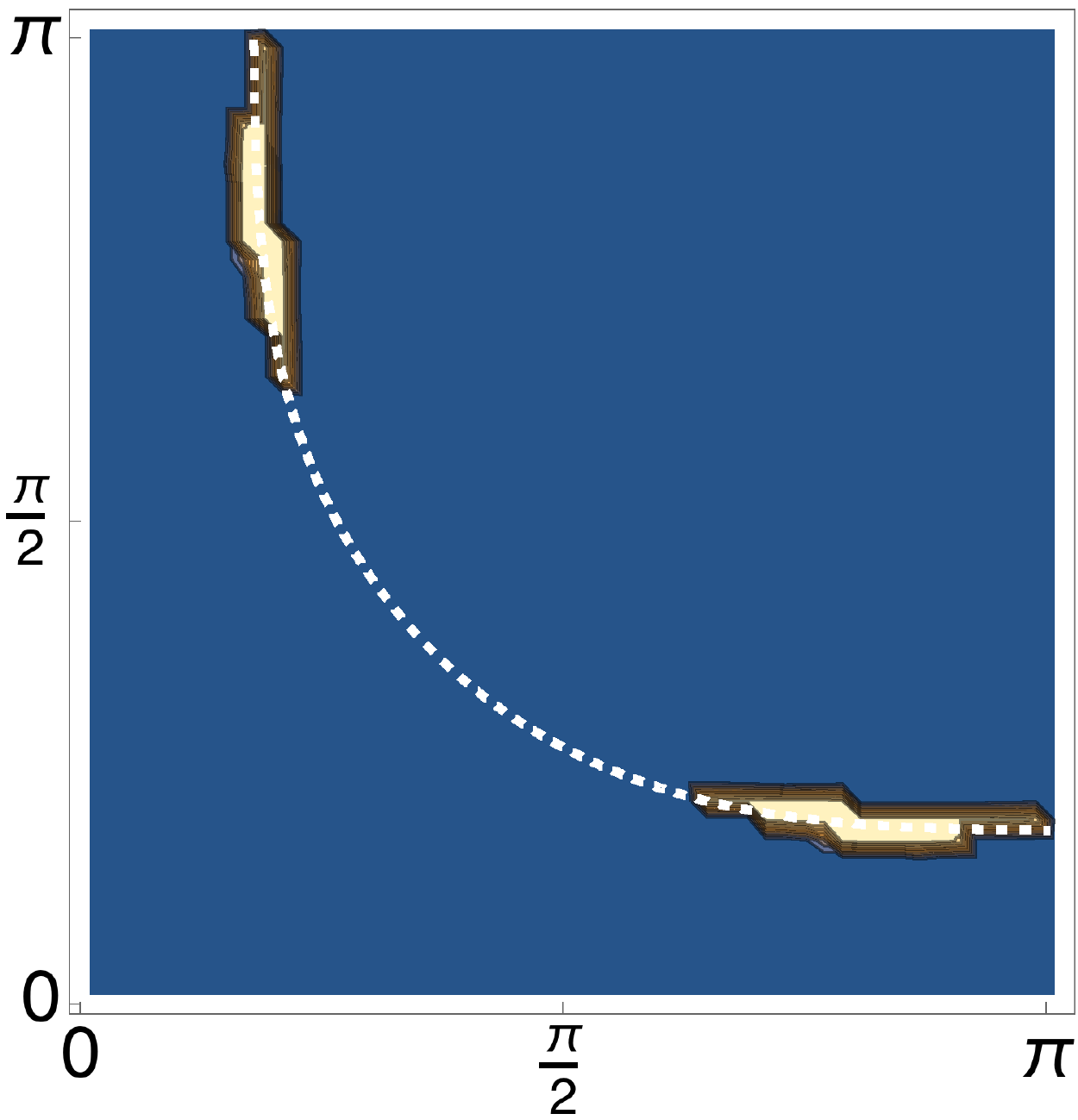} \vspace{1ex}
\end{minipage}

\caption{\label{fig:PDW} (Color online) \textcolor{black}{Density plot of
the solution of Eqns. ( \ref{eq:chieq3-1}) and (\ref{eq:chieq3-1-1}),
searching for a finite center of mass pairing around the resulting
wave vectors a) $\left(Q_{0},0\right)$, b) $\left(0,Q_{0}\right)$
and c) $\left(Q_{0},Q_{0}\right)$. Vanishing solutions are color-coded
in blue while non-vanishing points are depicted in yellow. We obtain
a finite response in the anti-nodal region. We are not able to distinguish
here between wave vectors along the axes and on the diagonal, but
on the other hand our result give wave vectors starting with the first
frequency $\mathbf{Q_{0}}$ and not from the second harmonics $2\mathbf{Q_{0}}$
\cite{Wang15b}.}}
\end{figure}

The solution of Eqns.(\ref{eq:chieq2-1},\ref{eq:chieq3-1}) is given
in Fig.\ref{fig:PDW}\textcolor{black}{. We observe the formation
of a PDW order, or superconducting order with finite resulting momentum.
The wave vector obtained from the SU(2) fluctuations directly depends
on our starting point wave vector for the charge channel. When we
start with a diagonal wave vector $\mathbf{Q}_{0}$, which led to
the effective actions Eqns.(\ref{eq:nlsm2-1-1-1},\ref{eq:24},\ref{eq:31}),
this in turn led to a diagonal wave vector for the PDW instability
as see in Fig.\ref{fig:PDW}c). However, starting with the two axial
wave vectors $\mathbf{Q}_{x}$ and $\mathbf{Q}_{y}$ (with $\mathbf{Q}_{x}\sim0.3\pi/a$
and $\mathbf{Q}_{y}\sim0.3\pi/a$) forming the checkerboard structure
observed in experiments, we obtain the formation of a similar PDW
instability, but with axial wave vectors, respectively $\mathbf{Q}_{y}$
and $\mathbf{Q}_{x}$, as depicted in Figs.\ref{fig:PDW}a) and b).}

An important point concerns the symmetry of the
PDW order generated this way. As was commented previously, since the
SU(2) propagator $\pi_{q}^{c}$ is centered around $\mathbf{q=0}$
in Eq.(\ref{eq:chieq3-1-1}) the SU(2) fluctuations alone do not select
any specific symmetry, whether it is s' or d-wave. Contrarily to the
previous modulations, which were already generated by the AF correlations
described in section \ref{sec:The-SU(2)-dome}, here the PDW order
parameter is directly emerging from the SU(2) fluctuations. Hence
the symmetry could be either s' or d wave at this stage of the theory.
It is possible that the experimental context, like the presence of
strong disorder finally selects the s' symmetry, as recently seen
in experiments \cite{Hamidian16}. Another important
point in our study, is modulation wave vector associated with the
PDW order is precisely the wave vector of the CDW observed experimentally,
and not its second, or higher harmonics. This is in agreement with
the findings of the key STM experiment, showing the presence of a
very small PDW order with the same wave vector as the charge order
\cite{Hamidian16}. The generation of PDW starting from a CDW instability
has been described in detailed in a recent work, using a Ginzburg-Landau
formalism\cite{Wang15b}. Here we give an alternative way to generate
the PDW instability, starting from a formalism which involves the
SU(2) fluctuations. Note that the form of the SU(2) fluctuations obtained
in Eqns.(\ref{eq:nlsm2-1-1-1},\ref{eq:24},\ref{eq:31}) remains
unchanged, when the starting wave vector is varied.

\section{Global phase diagram for Cuprate superconductors\label{sec:Global-phase-diagram}}

\subsection{The physical lines\label{sub:The-physical-lines}}

\subsubsection{Generic trends}

\begin{figure}[t]
\includegraphics[width=7.6cm]{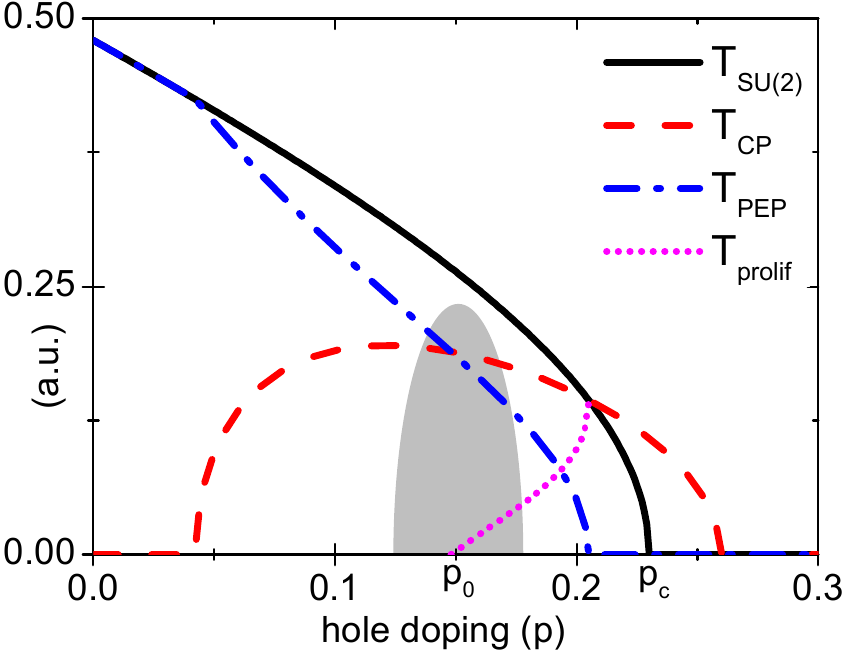} \vspace{1ex}
 \caption{\label{fig:sup} (Color online) \textcolor{black}{Temperature-hole
(T,p) phase diagram calculated from the simplified Ginzburg-Landau
model. The SU(2) order parameter critical temperature (solid line)
follows the PG critical temperature $T^{*}$. The temperature scale
corresponding to Cooper pairing $T_{CP}$ has the form of a dome like
in real compounds. The PEP critical temperature (dash-dotted line)
decreases with doping. The \textquotedbl{}excitonic\textquotedbl{}
patches proliferation temperature vanishes in the underdoped regime
and increases at doping close to $0.12$.}}
\end{figure}

\textcolor{black}{In order to obtain the Temperature-hole doping phase
diagram, we developed a minimal model based on Ginzburg-Landau functional.
The free energy depends on the SU(2) order parameter $\Delta_{SU(2)}$
and writes : 
\begin{align}
F_{SU(2)}=a_{SU(2)}\Delta_{SU(2)}^{2}+\frac{g_{SU(2)}}{2}\Delta_{SU(2)}^{4}.\label{free_su2}
\end{align}
where we assume that $\Delta_{SU(2)}$ is homogeneous and $a_{SU(2)}$
and $g_{SU(2)}$ are energy parameters. Minimizing the free energy
(\ref{free_su2}) as regards to $\Delta_{SU(2)}$ gives the relation
$\Delta_{SU(2)}=\sqrt{\frac{a_{SU(2)}}{g_{SU(2)}}}$. We assume that
the magnitude of the order parameter at zero temperature is proportional
to the critical temperature. The doping dependence of the critical
temperature of the SU(2) can be reproduce by parametrizing the energy
parameters as : $a_{SU(2)}=\bar{a}_{SU(2)}(p_{SU(2)}^{c}-p)$ and
$g_{SU(2)}=1+m_{SU(2)}(p-p_{SU(2)}^{c})$ where $p_{SU(2)}^{c}$ is
the hole doping where the SU(2) is expected to disappear. In our model,
the temperature where the SU(2) symmetry disappear is associated with
$T^{*}$.}

\textcolor{black}{In this simple model, the Copper Pairing (CP) and
Preformed Excitonic Pairing (PEP) energy scales are related by SU(2)
symmetry. From a theoretical point of view, it means that both SC
and PhP order parameter are constrained as exposed in the relation
(\ref{constrain}) that writes $\Delta_{SU(2)}=\sqrt{\Delta_{CP}^{2}+\Delta_{PEP}^{2}}$
where $\Delta_{CP}$ and $\Delta_{PEP}$ are the CP and PEP gap scales,
respectively. Considering, in a mean field heuristic picture, a coexisting
SC and PhP phase, we can write the Free energy as: 
\begin{align}
F_{1}=a_{CP}\Delta_{CP}^{2}+\frac{g_{CP}}{2}\Delta_{CP}^{4}+a_{PEP}\Delta_{PEP}^{2}+\frac{g_{PEP}}{2}\Delta_{PEP}^{4}.\label{free_coex}
\end{align}
where $a_{CP}$, $g_{CP}$, $a_{PEP}$ and $g_{PEP}$ are energy parameters.
Taking into account the $SU(2)$ constraint of Eq.(\ref{constrain}),
we can replace the order parameter in Eq.(\ref{free_coex}) by the
relation $\Delta_{PEP}=\sqrt{\Delta_{SU(2)}^{2}-\Delta_{CP}^{2}}$
where $\Delta_{SU(2)}$ has been determined from equation (\ref{free_su2}).
Minimizing the Free energy as regards to the CP energy scale, one
can find the expression of the SC and PhP scales as 
\begin{align}
\Delta_{CP}=\sqrt{\frac{a_{CP}-a_{PEP}+g_{PEP}\Delta_{SU(2)}^{2}}{g_{PEP}+g_{CP}}}\nonumber \\
\Delta_{PEP}=\sqrt{\Delta_{SU(2)}^{2}-\Delta_{CP}^{2}}
\end{align}
where $a_{CP}=l_{CP}(p_{CP}^{c}-p)$, $a_{PEP}=l_{PEP}(p_{PEP}^{c}-p)$,
$g_{PEP}=M+p$ and $g_{CP}=M+Mp$ with $M,L_{PEP}$ and $l_{SC}$
are free parameters.}

The phase diagram of Fig.\ref{fig:sup}, can be understood as follows.
The black line $T_{SU(2)}$ corresponds to the SU(2)-dome, depicted
in section \ref{sec:The-SU(2)-dome}, Fig. \ref{fig:dome}. It defines
the upper energy scale, above which we lose the SU(2) fluctuations.
The temperature $T_{CP}$ ( dashed line, red) corresponds to the typical
energy for Cooper pairing. The temperature $T_{PEP}$ (dashed-dotted
line, blue) is the typical energy associated with the formation of
particle-hole pairs. It is proportional to the SU(2) fluctuations,
as shown in sections \ref{sec:SU(2)-fluctuations-coupled} and \ref{sec:Effect-of-the}.
The particle hole pairing strength is driven by the SU(2) propagator
for the non-linear $\sigma$- model, as visible in Eqn.(\ref{eq:chieq3}).
Note that at the point where $T_{CP}=T_{SU(2)}$, $T_{PEP}=0$ since
there is space for fluctuations. As doping decreases, $T_{PEP}$ increases,
and crosses $T_{CP}$ in the middle, for a doping $0.010<p_{c}<0.013$.
We note that there is a threshold in temperature, above which the
Patches of Excitons start to proliferate. For $p<p_{c}$, since $T_{PEP}>T_{CP}$
the excitonic patches have a higher binding energy, and entropic effects
due to the finite size of the patches, leads to a proliferation temperature
very close to zero. Alternatively, for $p>p_{c}$, since $T_{PEP}<T_{CP}$,
there is a competition between Cooper pairing with the tendency to
form a global SC state, and formation of excitonic patches. In this
case the competition holds between the two states, and the proliferation
temperature $T_{prolif}$ ( dotted magenta line) gradually increases
up to $T_{SU(2)}$ around optimal doping. A simple evaluation of the
proliferation temperature is given in the next paragraph.

Interestingly, the phase diagram of Fig.\ref{fig:sup} singles out
an intermediate critical doping $p_{c}\simeq0.12$ which differentiates
two regions in the underdoped regime. For $p<p_{c}$, the picture
is of a complete fractionalization of the Fermi surface, with at $T=0$
a SC order around the nodes and the anti-nodal region fully gapped
out by excitonic patches. It is in line with a ``two-gaps'' picture.
On the other hand, for $p>p_{c}$, at $T=0$ the system in in the
SC state, with excitonic patches starting to proliferate as temperature
is raised. It is a one gap picture, which becomes fully valid at $p_{0}\simeq0.21$,
where the SC state is gapping out the Fermi surface up to the energy
$T_{SU(2)}$.

The doping $p_{c}\simeq0.12$ is thus the doping at which the two
scales of formation of the excitonic patches and Cooper pairs are
equal $T_{CP}=T_{PEP}$. It is conceivable that around this doping,
the SU(2) symmetry is strong enough to produce phase separation, but
the SU(2) fluctuations are frozen enough so that we observe experimentally
one (or two) resulting modulation wave vectors around $\left(Q_{0},0\right)$
and $\left(0,Q_{0}\right)$extending up to ten lattice sites, and
thus experimentally detectable. A real space picture of this scenario
is given in Ref.\cite{Montiel16}.

\paragraph{2. Proliferation temperature}

We give here a simple derivation of the proliferation temperature.
We have a competition between Cooper Pairing, leading to the formation
of a global SC state, and a local state of particle hole excitonic
patches, each carrying a specific entropy. Suppose there is $n_{p}$
excitonic patches and thus $1-n_{p}$ Cooper pairs (to simplify the
discussion we took a ``two-fluids'' only, model). The global Free
energy writes 
\begin{align}
F & =-\left(1-n_{p}\right)\frac{a_{CP}^{2}}{4g}+n_{p}\left(-\frac{a_{PEP}^{2}}{4g}+T\log n_{p}\right),\label{eq:132}
\end{align}

where $g$ is a high energy coupling constant that we take equal for
the two fluids, $-a_{CP}^{2}/(4g)$ and $-a_{PEP}^{2}/(4g)$ come
from the mean-field minimization for each order parameter, coming
for example from Eqn.(\ref{free_coex}). We have $a_{CP}=T-T_{CP}$
and $a_{p}=T-T_{PEP}$. Minimizing Eqn.(\ref{eq:132}) with respect
to $n_{p}$ leads to 
\begin{align}
\log n_{p} & =-\frac{1}{T}\frac{a_{CP}^{2}-a_{PEP}^{2}}{4g},\label{eq:133}
\end{align}
which leads to a temperature above which $n_{p}\simeq1$, also called
proliferation temperature 
\begin{align*}
T_{prolif} & =\begin{cases}
\frac{a_{CP}^{2}-a_{PEP}^{2}}{4g}, & \mbox{if }\left|a_{CP}\right|>\left|a_{PEP}\right|\\
0, & \mbox{elsewhere }.
\end{cases}
\end{align*}

\subsection{Strange Metal phase}

\begin{figure}[b]
\begin{minipage}[c]{8.25cm}%
\includegraphics[width=8cm]{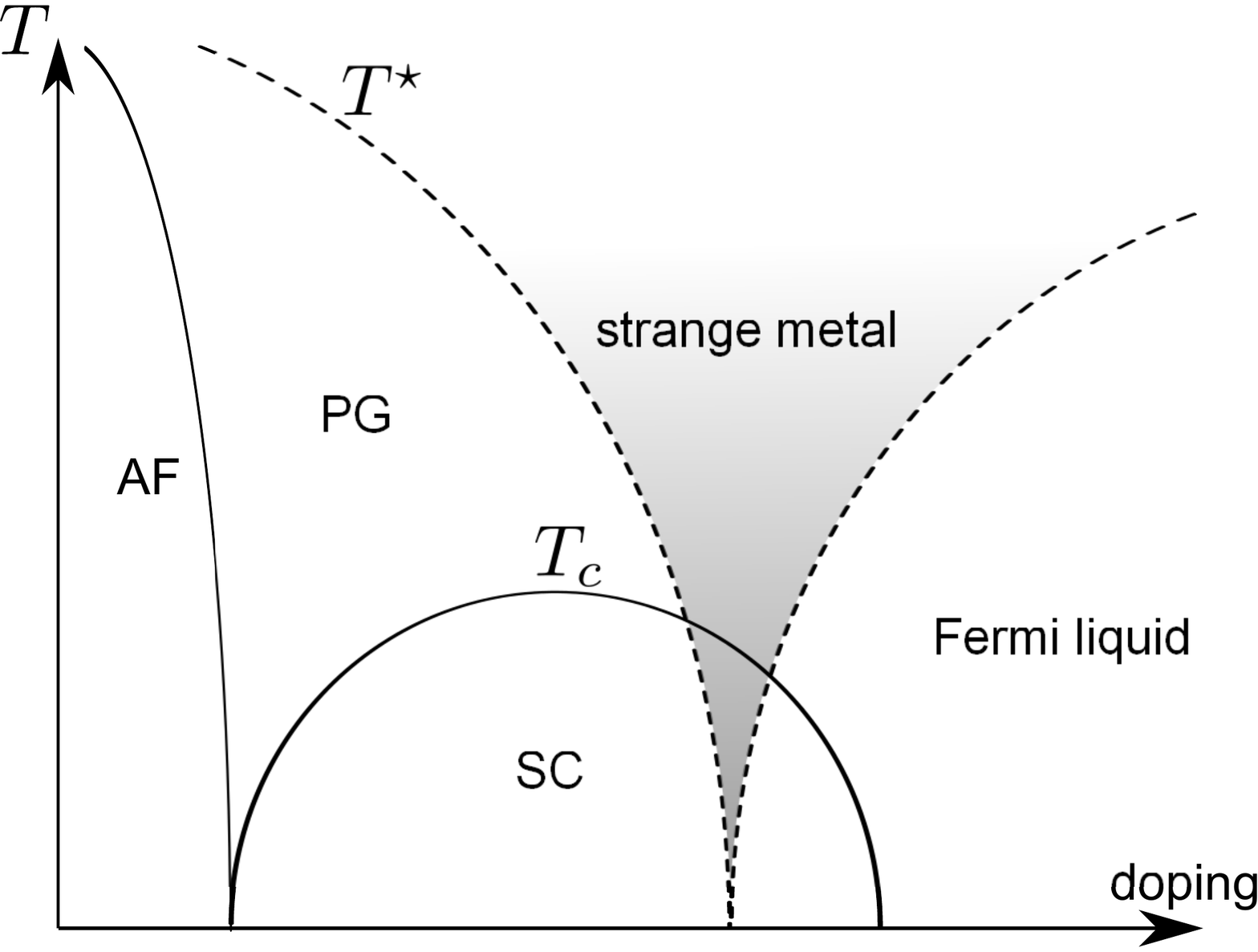} \vspace{1ex}
\end{minipage}\caption{\label{fig:diag_phas} (Color online) Schematic phase diagram of cuprate
superconductors as a function of hole doping and temperature T where
PG is the pseudogap phase, AF the antiferromagnetic phase and SC the
superconducting phase. In the grey shaded strange-metal phase, the
electrical resistivity scales linearly with T.}
\end{figure}

\begin{figure}[tb]
\centering \includegraphics[width=9cm]{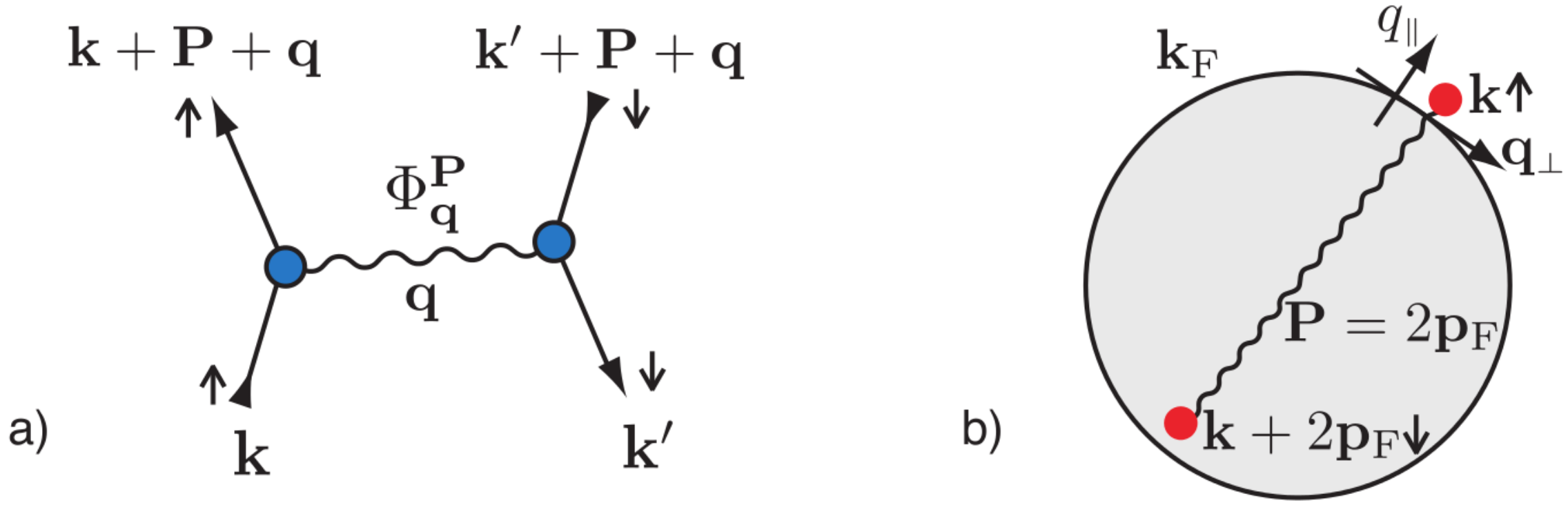} \vspace{0.5ex}
 c) \centering \includegraphics[width=9cm]{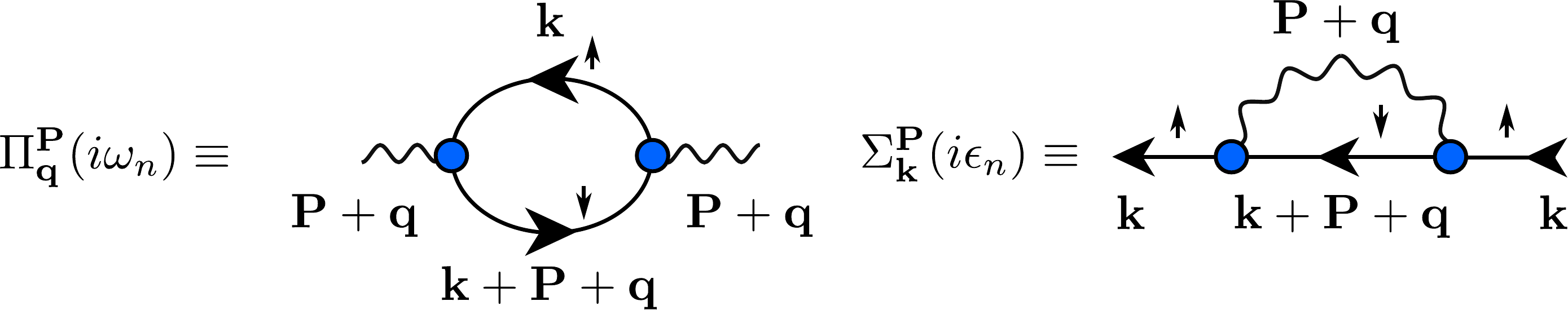} \vspace{-2.5ex}
 \caption{\label{fig:scattering} (Color online) a) Graphical representation
of the interaction in Eq.\ (\ref{eq:sbfull}). The wavy line represents
the bosonic propagator $\Phi_{{\bf {q}}}^{\pv}$ at criticality for
$|{\bf {q}}|\ll|{\bf P}|$. b) RES scattering between two electrons
close to the FS at ${\bf {k}}$ and ${\bf {k}}+\pf$ according to
Eq.\ (\ref{eq:sbfull}). c) Diagrammatic representation of the one-loop
bosonic polarization $\Pi_{{\bf {q}}}$ and the fermionic self-energy
$\Sigma_{{\bf {q}}}$ for the RPE mode.}
\end{figure}

This part explores the consequences of the RES for the phase diagram
of cuprates (see Fig.\ \ref{fig:diag_phas}) when this mode becomes
critical. There are experimental indications that the electric transport
in this system has both 2D and 3D character. We will thus calculate
the resistivity $\rho$ in $d=3$ and $d=2$, in the absence of a
gap and show that it differs from the usual Fermi liquid $T^{2}$
scaling with a typical $T/\log T$ behavior. Therefore, we evaluate
the bosonic polarization induced by critical $\chi$-modes. Details
can be found in Appendix \ref{sec:Bosonic-polarization-bubble}. At
quadratic order in the excitonic fluctuations, we obtain the following
effective interaction 
\begin{equation}
\!\!S_{\text{crit}}[\psi]\!\!=\!\!\!\sum_{kk'qP,\sigma}\!\!\!\Phi_{q}^{\pv}\ \psi_{\sigma,\mathbf{k}}^{\dagger}\ \psi_{\sigma,\mathbf{k}+\mathbf{P}+\mathbf{q}}\ \psi_{-\sigma,\mathbf{k'}}^{\dagger}\ \psi_{-\sigma,\mathbf{k'}-\mathbf{P}-\mathbf{q}},\label{eq:sbfull}
\end{equation}
see Fig.\ \ref{fig:scattering}a,b), with $\Phi_{q}^{\pv}=\langle\chi_{-P-q}\chi_{P+q}\rangle$.
The form of above interaction corresponds to a coupling with a collection
of bosons (see also Eqn.(\ref{eq:deff})). The renormalized bosonic
propagator follows from Dyson's equation $[\Phi_{\mathbf{q}}^{\pv}(\Omega)]^{-1}=q^{2}+m-\Pi_{\mathbf{q}}^{\pv}(\Omega)$.
Therein, the bare propagator is assumed to have Ornstein-Zernike form.
The retarded bosonic polarization, $\Pi_{{\bf {q}}}^{\mathbf{P}}=\Pi'_{{\bf {q}}}+i\Pi''_{{\bf {q}}}$
in Fig.\ \ref{fig:scattering}c), evaluated for ${\bf {P}}=\pf$,
yields $\Pi'_{{\bf {q}}}(\Omega)=c\biggl[(\Omega+q_{\xa})\ln|\Omega+q_{\xa}|-(\Omega-q_{\xa})\ln|\Omega-q_{\xa}|\biggr]$
and $\Pi''_{{\bf {q}}}(\Omega)=\pi c\Bigl[(\Omega+q_{\xa})\theta(-\Omega-q_{\xa})+(\Omega-q_{\xa})\theta(\Omega-q_{\xa})\Bigr]$.
With $\Omega$ we denote real frequencies and $c$ is a non-universal
factor depending on the details of the dispersion.

Next, we calculate the electronic self-energy depicted in Fig. \ref{fig:scattering}c).
Here again we refer the reader to the Appendix \ref{sec:Fermionic-self-energy}
for details. Note that the self-energy requires a summation over all
ordering vectors ${\bf P}$. Up to logarithms, each $\mathbf{P}$-wave
gives the same contribution. In the quantum critical regime, we have,
the scaling behavior $\Pi'_{\mathbf{q}}\left(\Omega\right)\sim2cq_{\parallel}\ln\left|\Omega\right|$,
and $\Pi''_{\mathbf{q}}\left(\Omega\right)\sim\pi c\Omega$. We use
this scaling law to evaluate the self-energy of an electron scattering
through a single bosonic mode written in Matsubara form as $\Phi_{\mathbf{P}}^{-1}\left(i\omega_{n}\right)=\gamma\left|\omega_{n}\right|-v_{\parallel}q_{\parallel}\ln\left|\omega_{n}\right|+v_{\perp}q_{\perp}^{2}/2$.
The evaluation is performed in $d=3$ and at the first order in the
leading singularity we obtain $\Sigma\left(i\epsilon_{n}\right)=i\epsilon_{n}/\left(4\pi v_{\parallel}v_{\perp}\ln\left|\epsilon_{n}\right|\right)$.
We note that -with logarithmic corrections, this form is typical of
a marginal Fermi liquid \cite{Varma89} and can account for the properties
of the strange metal phase depicted in Fig.\ \ref{fig:diag_phas}.
In $d=2$ the self-energy scales like $\Sigma\left(i\epsilon_{n}\right)\sim i\sqrt{\left|\epsilon_{n}\right|}Sgn\left(\epsilon_{n}\right)$.

We turn now to the discussion of the relaxation time for electron-electron
scattering process from a semiclassical Boltzmann treatment. Details
are given in Appendix \ref{sec:Boltzmann-treatment}). The Boltzmann
equation for the non-equilibrium electron distribution $f_{\mathbf{k}}$
writes \cite{Abrikosov88,Paul13} 
\begin{equation}
\left(\frac{\partial f_{\mathbf{k}}}{\partial t}\right)_{\text{collisions}}\hspace{-2ex}=-e\mathbf{E}\cdot\nabla_{\mathbf{k}}f_{\mathbf{k}}=-I_{ei}\left[f_{\mathbf{k}}\right]-I_{ee}\left[f_{\mathbf{k}}\right],\label{eq:boltzmann-1}
\end{equation}
where $e$ is the elementary charge, $\mathbf{E}$ a static electric
field and $I_{ei}$ respectively $I_{ee}$ are the electron-impurity
respectively electron-electron collision integrals. The electron-electron
collision integral is obtained from Fermi's golden rule 
\begin{align}
 & I_{ee}\left[f_{\mathbf{k}}\right]=\frac{1}{V}\sum_{\mathbf{q}}\int_{-\infty}^{\infty}\!\!d\Omega\,\text{Im}\Phi_{\mathbf{q}}\left(\Omega\right)\delta\left(\epsilon_{\mathbf{k}}-\epsilon_{{{\bf {k}}+\pv-{\bf {q}}}}-\Omega\right)\times\nonumber \\
 & \!\!\!\Bigl[f_{\mathbf{k}}\left(1-f_{{{\bf {k}}+{\pv}-{\bf {q}}}}\right)(1+\nb\left(\Omega\right))-\left(1-f_{\mathbf{k}}\right)f_{{{\bf {k}}+\pv-{\bf {q}}}}\nb\left(\Omega\right)\Bigr],\label{eq:iee}
\end{align}
with $\nb(x)=(\exp{(x/T)}-1)^{-1}$ and we drop the contribution from
$I_{ei}$. Relaxation-time approximation amounts to set $f_{\mathbf{k}}\simeq f_{0,{\bf {k}}}-g_{{\bf {k}}}f_{0,{\bf {k}}}(1-f_{0,{\bf {k}}})$
where $f_{0}$ is the equilibrium distribution and $g_{{\bf {k}}}=\tau e\mathbf{E}\cdot\mathbf{v}_{{\bf {k}}}/T$.
In this approximation Eq.\ (\ref{eq:iee}) becomes 
\begin{align}
 & I_{ee}\left[f_{\mathbf{k}}\right]=\frac{1}{V}\sum_{\mathbf{q}}\!\!\int_{-\infty}^{\infty}\!\!\!\!\!\!d\Omega\,\text{Im}\Phi_{\mathbf{q}}\left(\Omega\right)\nb\left(\Omega\right)f_{0,{\bf {k}}+\pv-\mathbf{q}}(1-f_{0,{\bf {k}}})\nonumber \\
 & \qquad\times\left(g_{{\bf {k}}+\pv-\mathbf{q}}-g_{{\bf {k}}}\right)\delta\left(\epsilon_{\mathbf{k}}-\epsilon_{{\bf {k}}+\pv-\mathbf{q}}-\Omega\right).\label{eq:iee2-1}
\end{align}
We see from Eq.\ (\ref{eq:iee2-1}) that this theory has a non-vanishing
imbalance velocity factor, since for ${\bf {q}}=0$, $\left(g_{{\bf {k}}+\pv}-g_{{\bf {k}}}\right)\neq0$.
This implies that no additional $T$ dependence arises from the angular
part of the integral. To make the connection of the scattering time
$\tau$ and resistivity $\rho$, we write the electrical current density
as ${\bf {J}}=-2e\langle{\bf {v}}\rangle$ and note its connection
to $\rho$ via ${\bf {J}}=\rho^{-1}{\bf {E}}$. For small electric
fields, $\rho\sim\tau^{-1}$ and solving above Boltzmann equation
for $\tau$ yields $\tau^{-1}\sim T/\ln(T)$, such that $\rho\sim T/\ln(T)$.
In $d=2$, the resistivity scales like $\rho\sim\sqrt{T}$. The scaling
forms given here are valid in the anti-nodal region of the BZ, while
the nodal region will provide a Fermi liquid-like $T^{2}$ law both
in the PG phase and in the strange metal phase. The study of the strong
anisotropy of the scattering rates along the Fermi surface, and interplay
between the $T^{2}$ and anomalous behaviors is the subject of active
experimental investigations \cite{Hussey:2011kp,husseycupratescriticality,Barisic:2015tg}
and we will devote a further detailed development of our theory to
address the issue.

\section{Conclusion}

This paper has been devoted to the study of the implications of the
non-linear $\sigma$-model which describes the fluctuations of the
SU(2) rotation matrix between the d-wave SC state and the d-wave charge
order. One important result of this paper is that, when interacting
with the conduction electrons, hot regions are created in the Brillouin
Zone, and in particular a line where the SU(2) fluctuations are massless-
which we called an SU(2) line, was found, crossing the Fermi surface
at the AF hot spots. The main effect of the SU(2) pairing fluctuations
on the charge sector, is to tilt the modulation wave vector from the
diagonal to the axial wave vectors $\left(0,Q_{0}\right)$ and $\left(Q_{0},0\right)$.
Secondarily, SU(2) fluctuations affect the SC sector by creating a
small PDW instability on top of already existing d-wave superconducting
phase. Lastly, the SU(2) pairing fluctuations lead to the formation
of preformed particle-hole pairs, that we have called ``excitons'',
for which a range of $\pf$ wave vectors are allowed. The intrinsic
constraint of the non-linear $\sigma$-model creates some strong mode
coupling within the charge sector and thus the creation of excitonic
patches, which then proliferate up to the PG temperature $T^{*}$.
We give a preliminary description of the phase diagram of the cuprates,
including a preliminary theory for the anomalous transport properties
in the strange metal part of the phase diagram. Implications of the
theory for various experimental probes are left for future publications.

The ultimate goal of this theory is to address all possible experimental
results, but this goes beyond the scope of this paper. Note that,
within the SU(2) scenario, a few experimental issues have already
been addressed. The phase diagram as a function of magnetic field
and temperature was considered in Ref.\cite{Meier13}, the structure
of the modulations inside a vortex core in Ref.\cite{Einenkel14}.
A study of the Raman $A_{1g}$ mode was given in Ref.\cite{Montiel15a},
and the gapping out of the Fermi surface in the anti-nodal region
seen in ARPES was described in Ref.\cite{Montiel:2016it}. The findings
of modulations up to $T^{*}$seen, by STM in Bi2212, as well as the
resonance of neutron scattering in Hg1201 will be addressed in forthcoming
publications \cite{Montiel16}.

The presence of electron pockets in the AN zone of
the first BZ is a very ongoing debate in the cuprate community \cite{Taillefer:2016iijj}. The
Hall resistivity measurements in YBCO have shown a change in the carrier density $n$ (between the $n=p$ regime at low doping to $n=1-p$ regime at high doping) around $p=0.19$ \cite{Taillefer:2016ih,Storey:2016}. This change of regime has been associated with the opening of the PG around the hot spot, in the AN zone of the first BZ that should be present at any doping \cite{Taillefer:2016iijj}. Our scenario can provide an explanation of the change in the carrier density that will be  addressed in a shortcoming work \cite{Morice17}. Therefore, the absence of such electron pockets in the AN zone of the first BZ at any doping could be one check of the validity of our scenario.

Moreover, one specific signature of the preformed particle-hole pairs
could be the photoluminescence signal. It is well-known in semiconductor
physics that excitons exhibit a photoluminescence signal \cite{Klingshirnbook}. In
our model, the stabilization of such exciton particle-hole pair patches could
also lead to such photoluminescence signal. The exploration and the calculation
of such signature is left to later work.

The SU(2) scenario presented here should be applicable for materials where the SC and the CDW  states are close in energy. Another condition should be the presence of an interaction that could stabilize such SU(2) fluctuations (like short-range AF correlations). The underdoped cuprate compounds are currently the best candidates wherein these two conditions are present.

\begin{acknowledgments}
We thank Y.\ Sidis, I. Paul, H. Alloul and Ph. Bourges, for stimulating
discussions. This work has received financial support from LabEx PALM
(ANR-10-LABX-0039-PALM), ANR project UNESCOS ANR-14-CE05-0007, as
well as the grant Ph743-12 of the COFECUB. The authors also like to
thank the IIP (Natal, Brazil), the Aspen Center for Physics and the
Perimeter Institute (Ontario, Canada), where parts of this work were
done, for hospitality. This work is supported by the ERC, under grant
agreement AdG-694651-CHAMPAGNE. 
\end{acknowledgments}

\appendix

\section{Mean-Field gap equations: strong coupling case\label{sec:Mean-Field-gap-equations:}}

\begin{figure}[H]
\begin{minipage}[c]{3.8cm}%
a)\includegraphics[width=3.6cm]{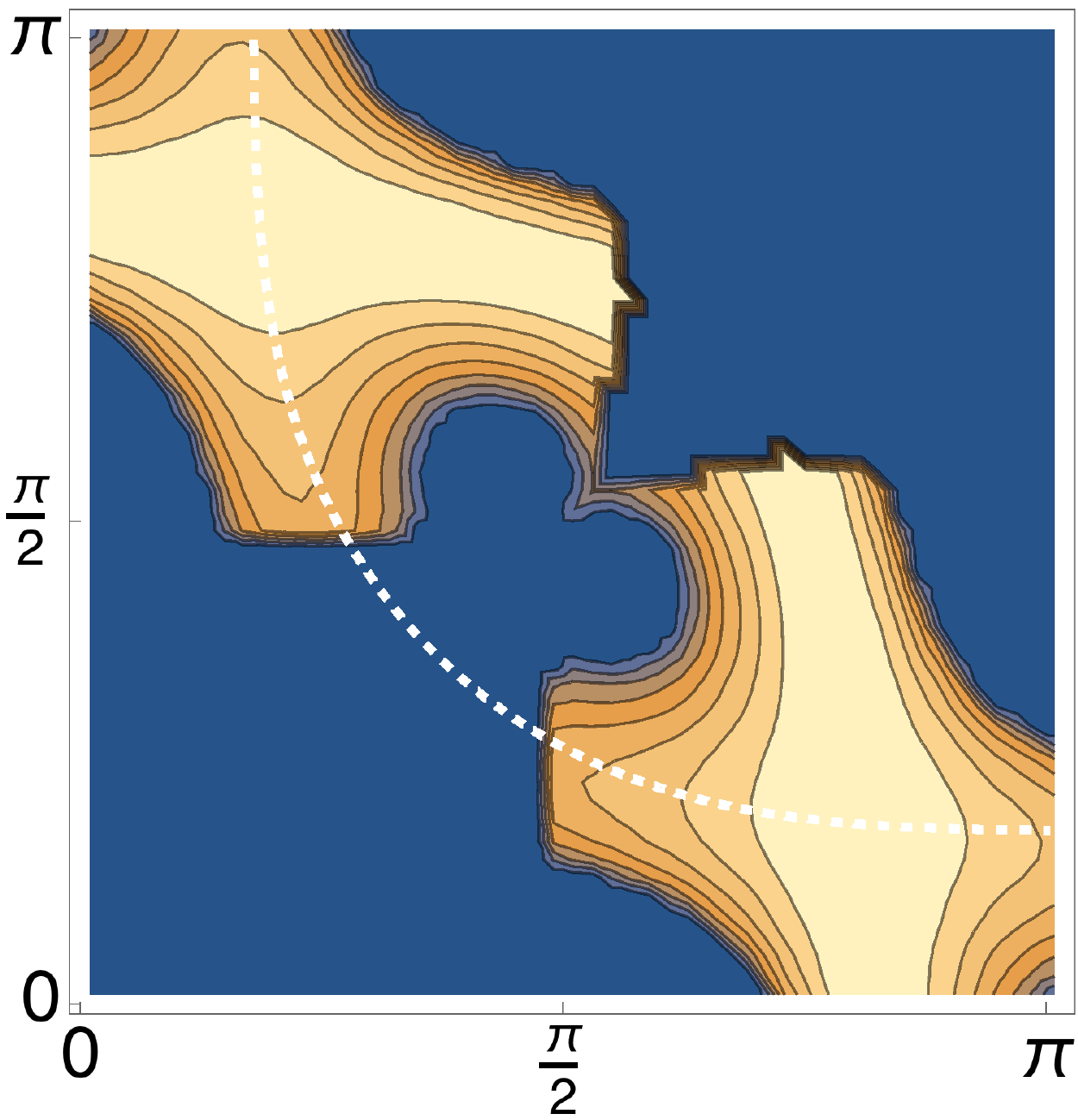} \vspace{1ex}
\end{minipage}%
\begin{minipage}[c]{3.8cm}%
b)\includegraphics[width=3.6cm]{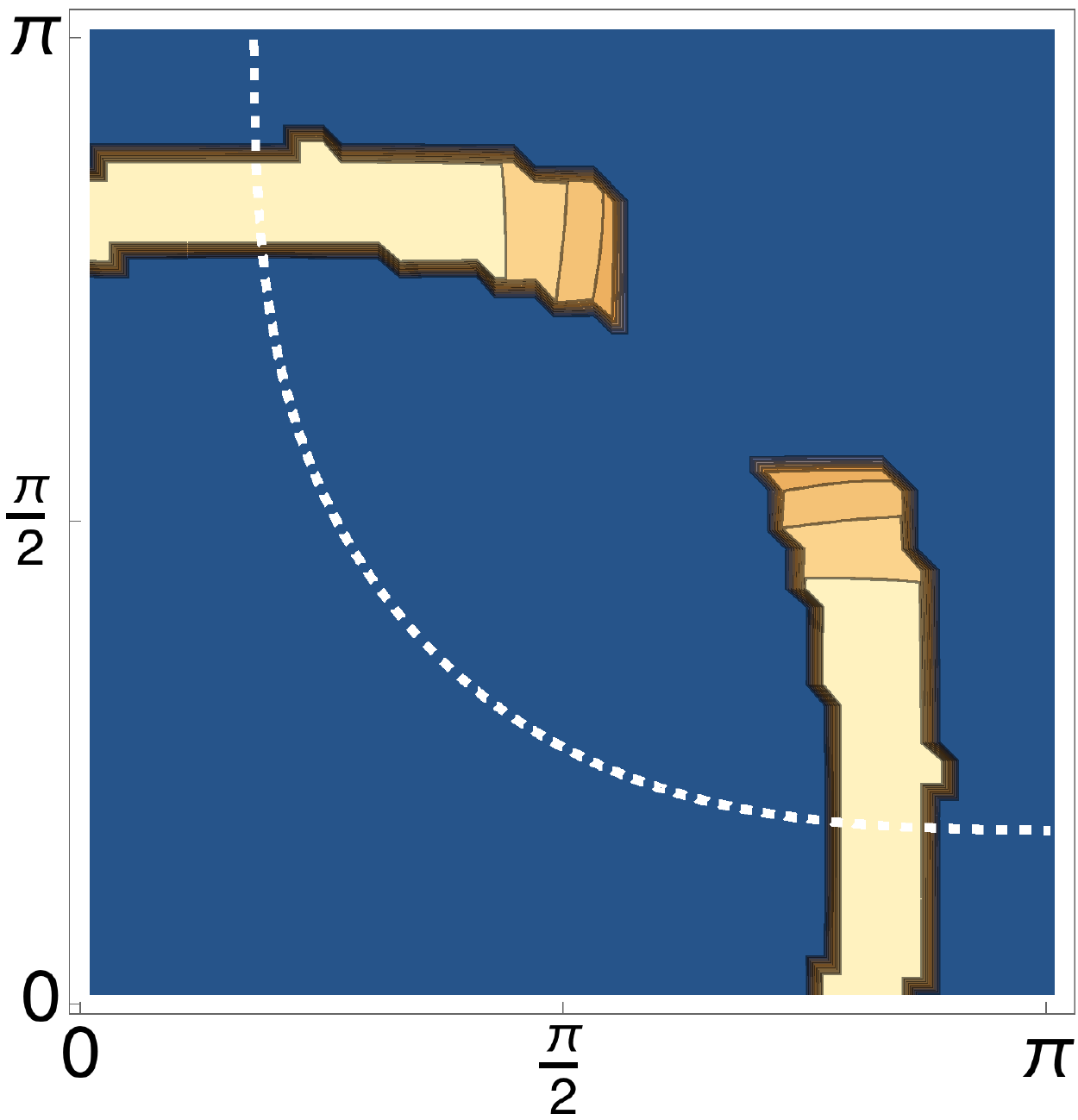} \vspace{1ex}
\end{minipage}

\begin{minipage}[c]{3.8cm}%
c)\includegraphics[width=3.6cm]{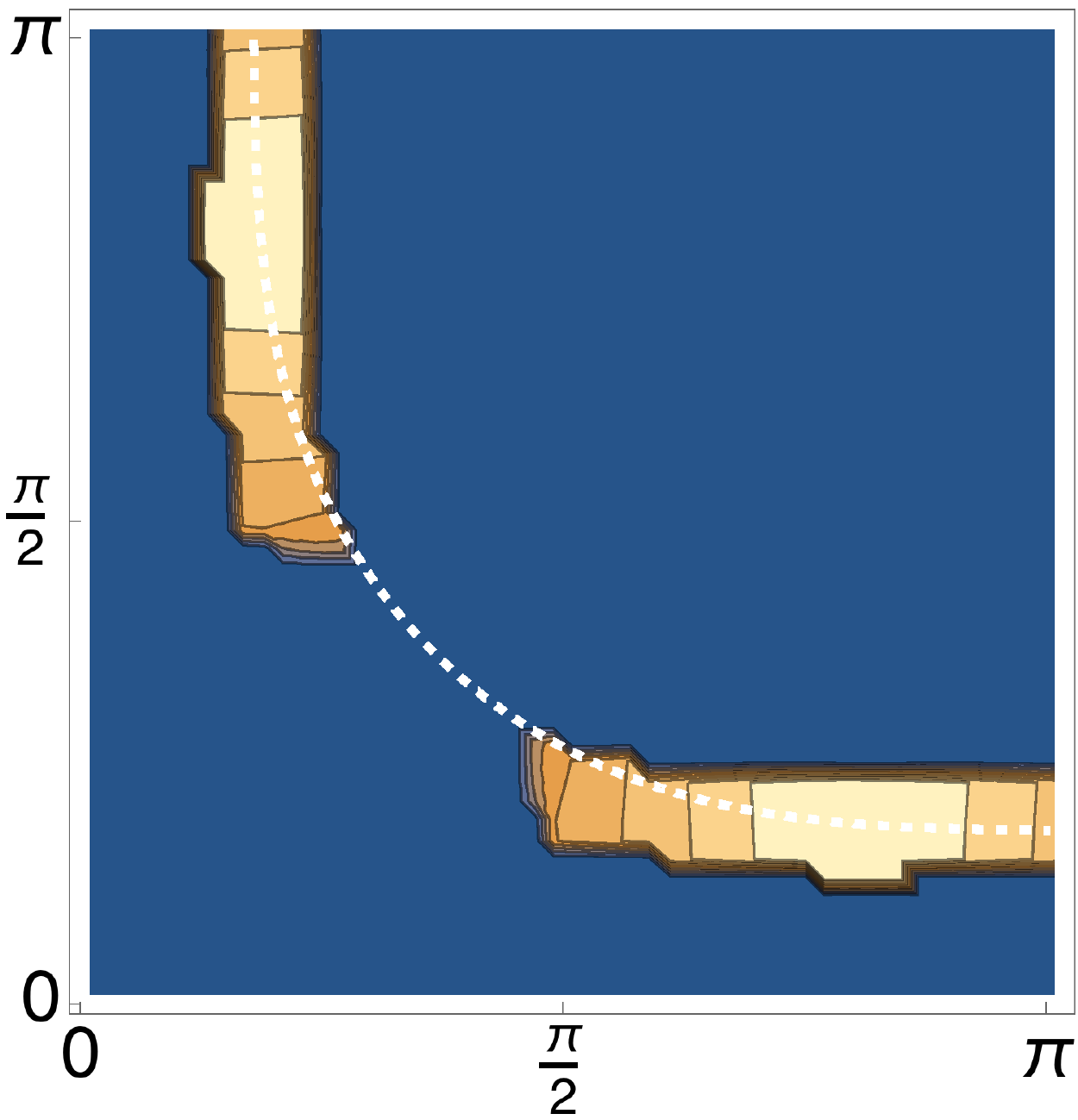} \vspace{1ex}
\end{minipage}%
\begin{minipage}[c]{3.8cm}%
d)\includegraphics[width=3.6cm]{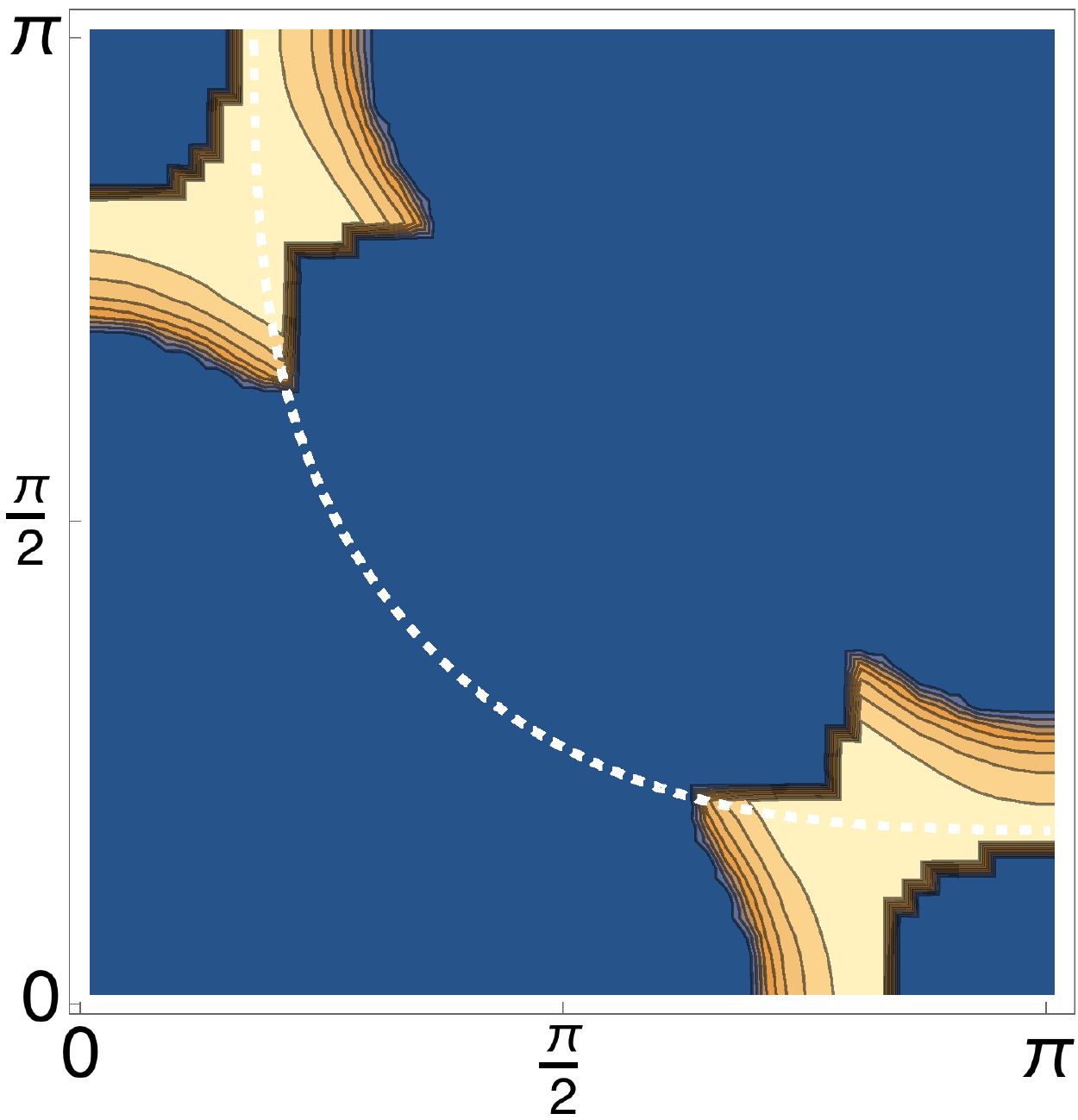} \vspace{1ex}
\end{minipage}

\begin{minipage}[c]{3.8cm}%
e)\includegraphics[width=3.6cm]{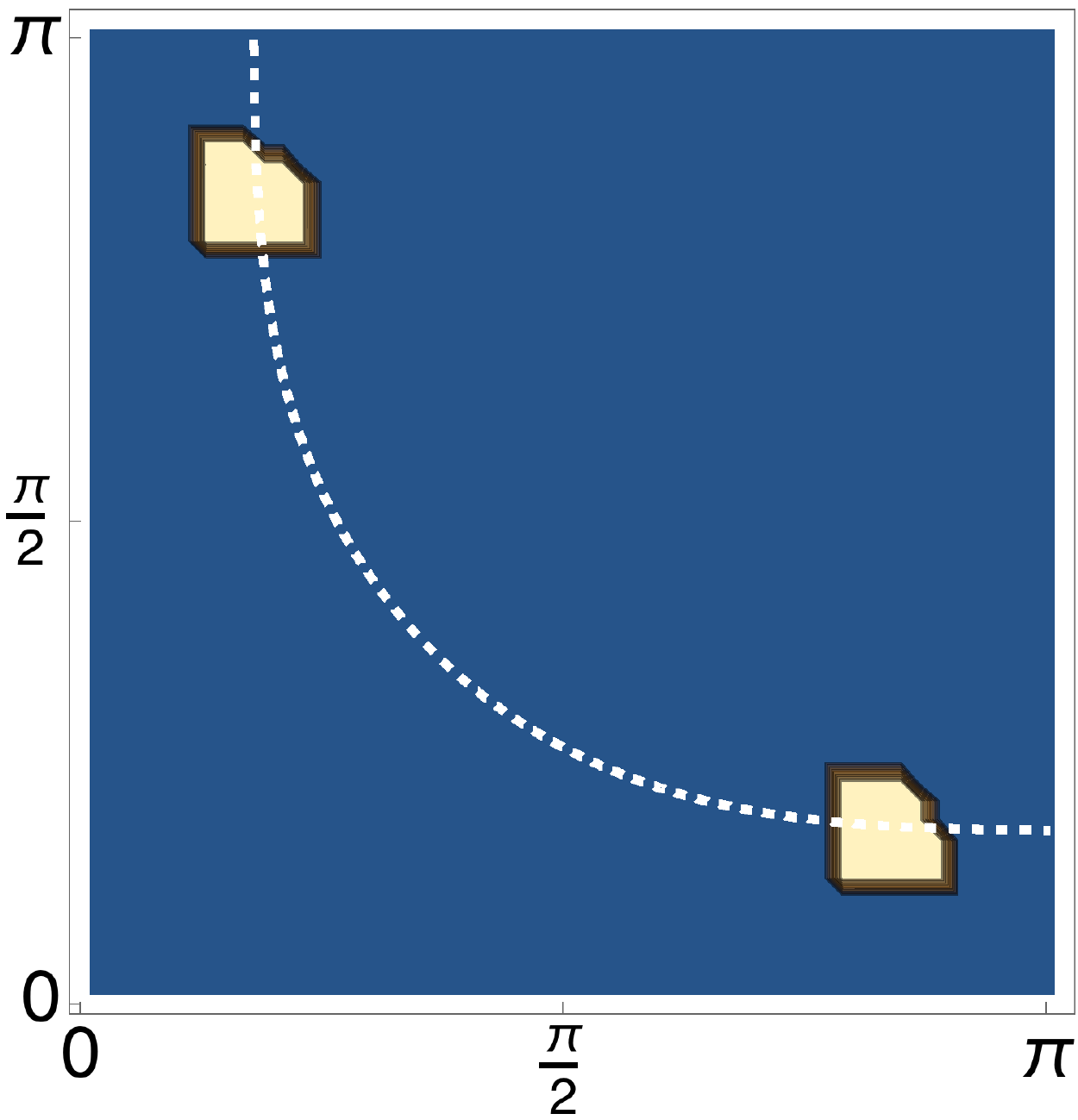} \vspace{1ex}
\end{minipage}%
\begin{minipage}[c]{3.8cm}%
f)\includegraphics[width=3.6cm]{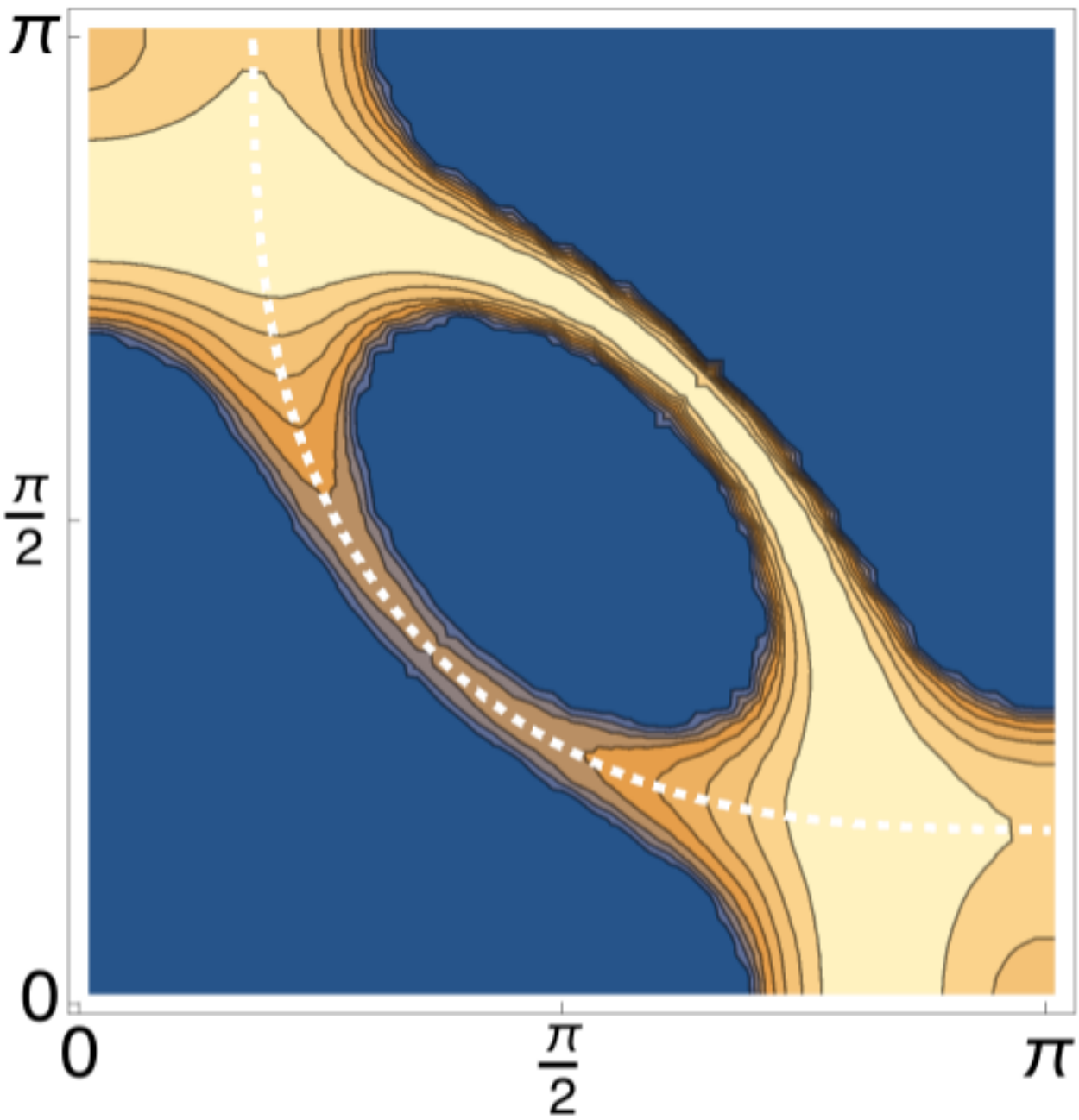} \vspace{1ex}
\end{minipage}

\caption{\label{fig:gaps2} (Color online) \textcolor{black}{Solution of the
gap equations, in the case of a strong AF coupling, from Eqns. (\ref{eq:chiQ0})
and (\ref{eq:SCgap}) for various modulation wave vectors with a)
the diagonal wave vector $\left(Q_{\mathbf{0}},Q_{\mathbf{0}}\right)$
linking two hot spots, b) the axial wave vector$\left(Q_{\mathbf{0}},0\right)$
and c) $\left(0,Q_{\mathbf{0}}\right)$ which are observed experimentally,
d) the AF wave vector $\left(\pi,\pi\right)$ and e) the null wavec
vector. The solution of the SC gap equation is given in f). The calculations
are made on the band structure of Bi2212 form Ref.\cite{Norman07}
(see details in the text for the band parameters). Vanishing solutions
are color-coded in blue while non-vanishing points are depicted in
yellow. The calculations are made within the approximation $J_{\overline{\mathbf{q}}}=J\delta\left(\overline{\mathbf{q}}\right)$,
with $J=0.9$, which restricts the q-integration at the vector $\left(\pi,\pi\right)$.
The energy units, if not stated otherwise, are in eV.}}
\end{figure}

We give in Fig.\ref{fig:gaps2} the solution of the gap equations
(\ref{eq:chiQ0},\ref{eq:SCgap}) in the strong coupling case, for
which we have taken $J=0.9$ eV. WE observe, as the coupling is increased,
a pronounced difference between the SC solution and the CDW solutions,
in that the SC solution gaps out the entire Fermi surface whereas
the CDW solution remain confined in the anti-nodal regions of the
Brillouin Zone.

\section{Zone Edge hot lines\label{sec:Zone-Edge-hot}}

\begin{figure*}[t]
\begin{minipage}[c]{4.25cm}%
\includegraphics[width=4cm]{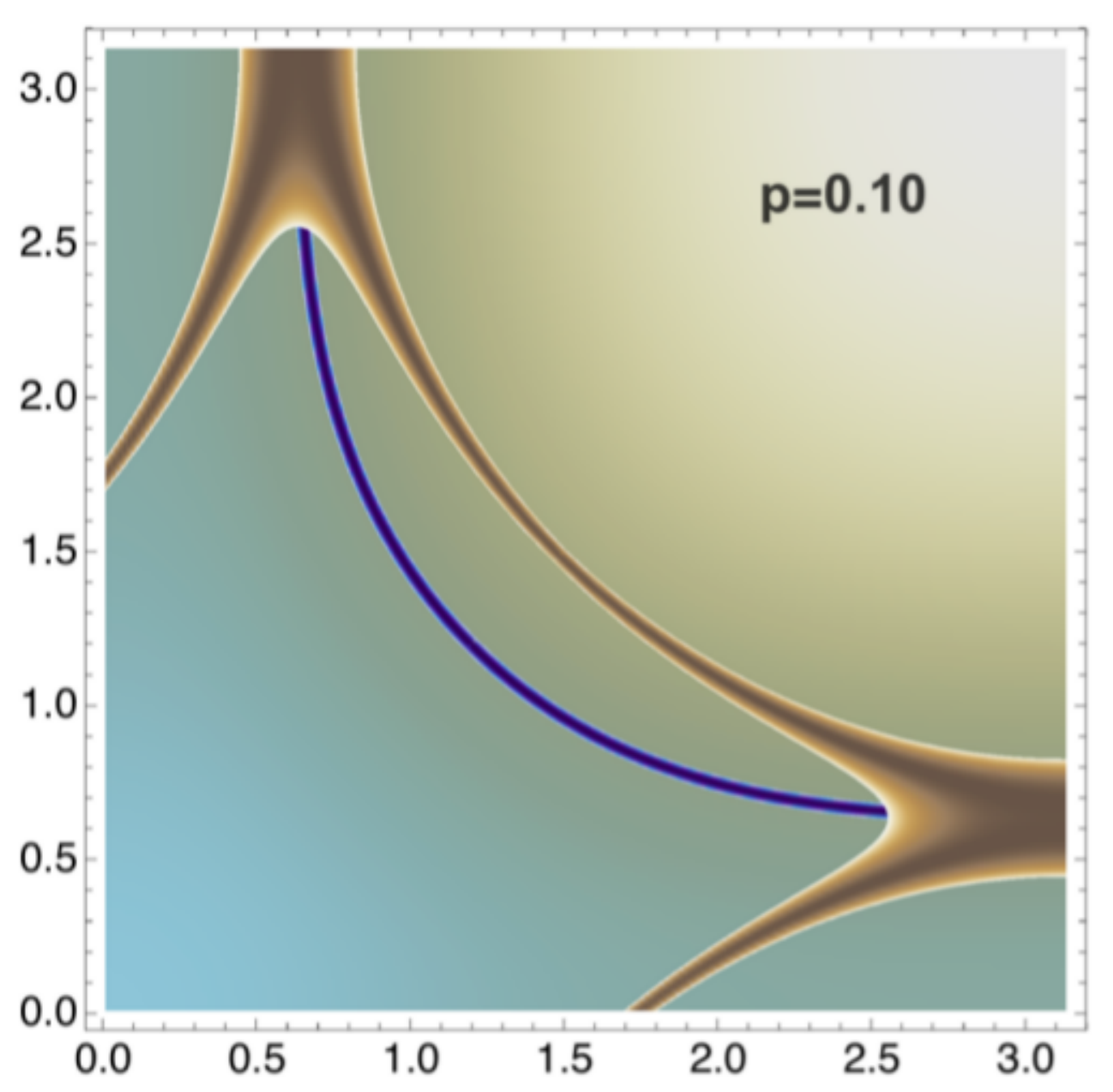} \vspace{1ex}
\end{minipage}%
\begin{minipage}[c]{4.25cm}%
\includegraphics[width=4cm]{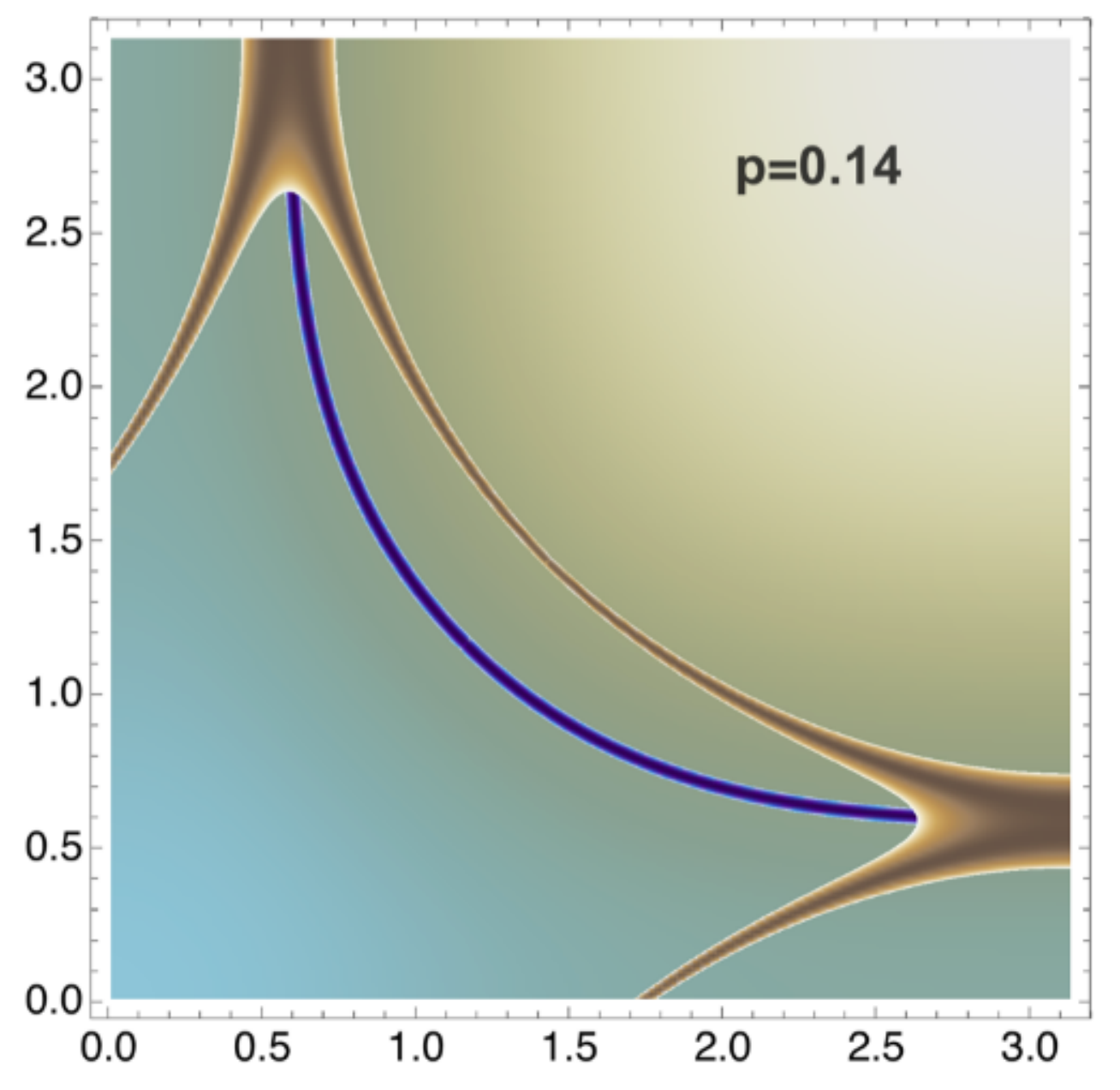} \vspace{1ex}
\end{minipage}%
\begin{minipage}[c]{4.25cm}%
\includegraphics[width=4cm]{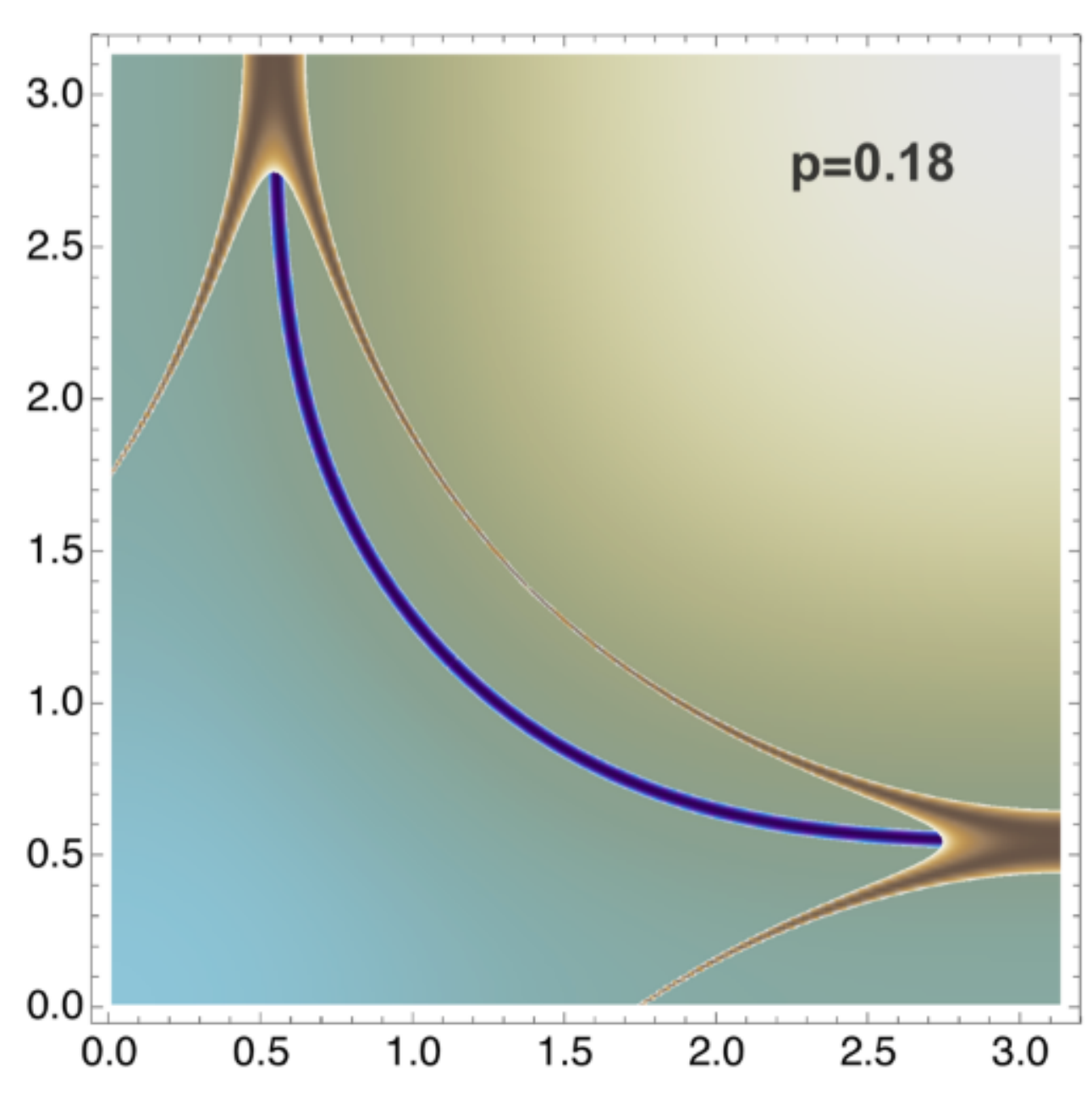} \vspace{1ex}
\end{minipage}%
\begin{minipage}[c]{4.25cm}%
\includegraphics[width=4cm]{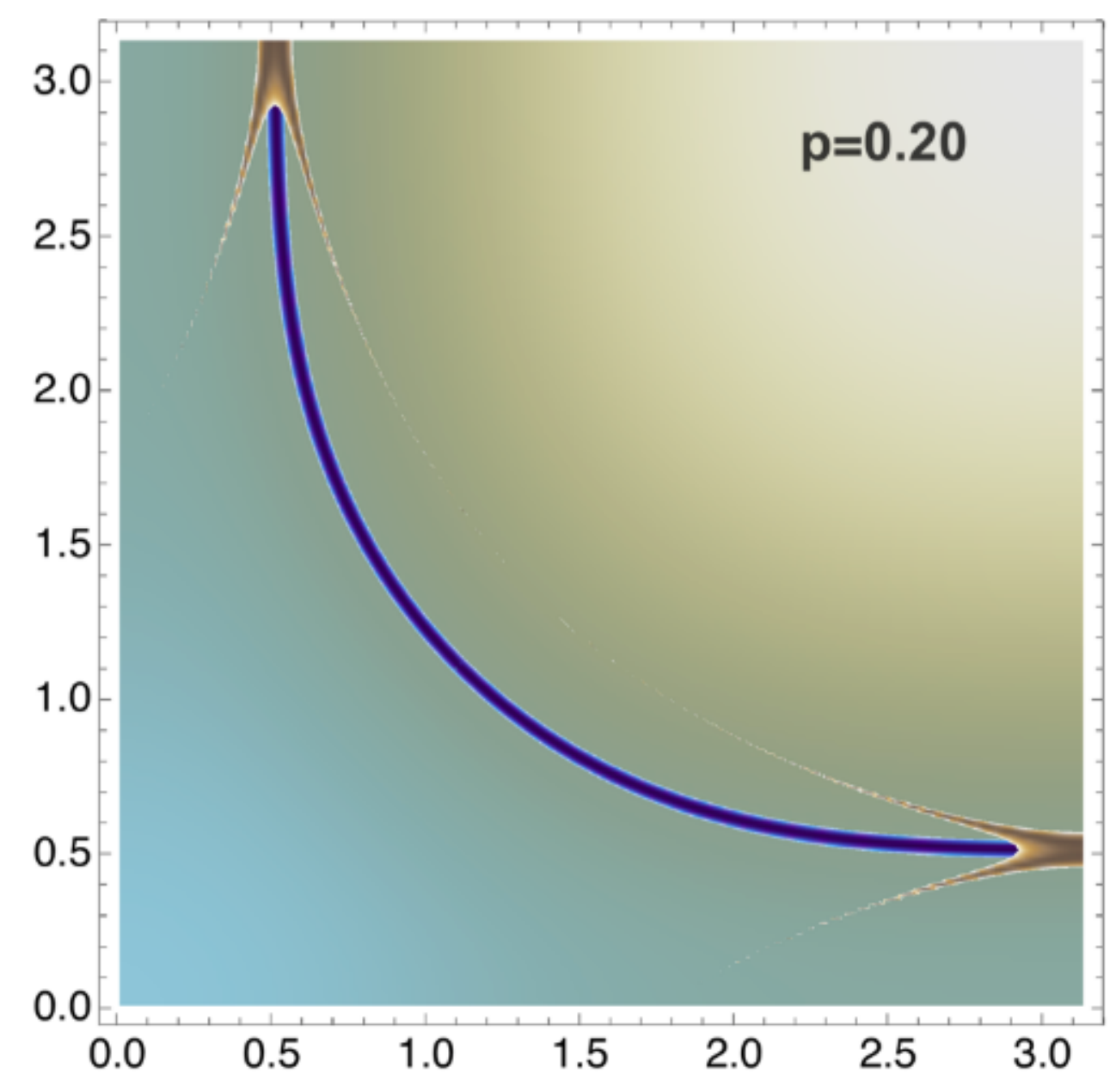} \vspace{1ex}
\end{minipage}

\caption{\label{fig:dopZE} (Color online) Evolution of the SU(2) fluctuations
as the function of the hole-doping. We present two set of curves for
the Zone Edge wave vector\textcolor{black}{{} $\pf$}. Note that
in this case, the mass has a minimum lies at the ZE. The electron
dispersion is modelized in tight-binding approximation for Bi2212 Ref.\ \cite{Norman07}
(parameter set tb2). }
\end{figure*}

We present in Fig.\ref{fig:dopZE} the evolution of the hot regions
for a charge wave vector located at the Zone Edge. Note that the fractionalization
of the Fermi surface is also efficient in that case, with a minimum
of the mass ( or infinite fluctuations) located at the Zone Edge.

\section{Test of various anisotropies\label{sec:Test-of-various}}

\begin{figure}[h]
a)%
\begin{minipage}[c]{6.25cm}%
\includegraphics[width=7cm]{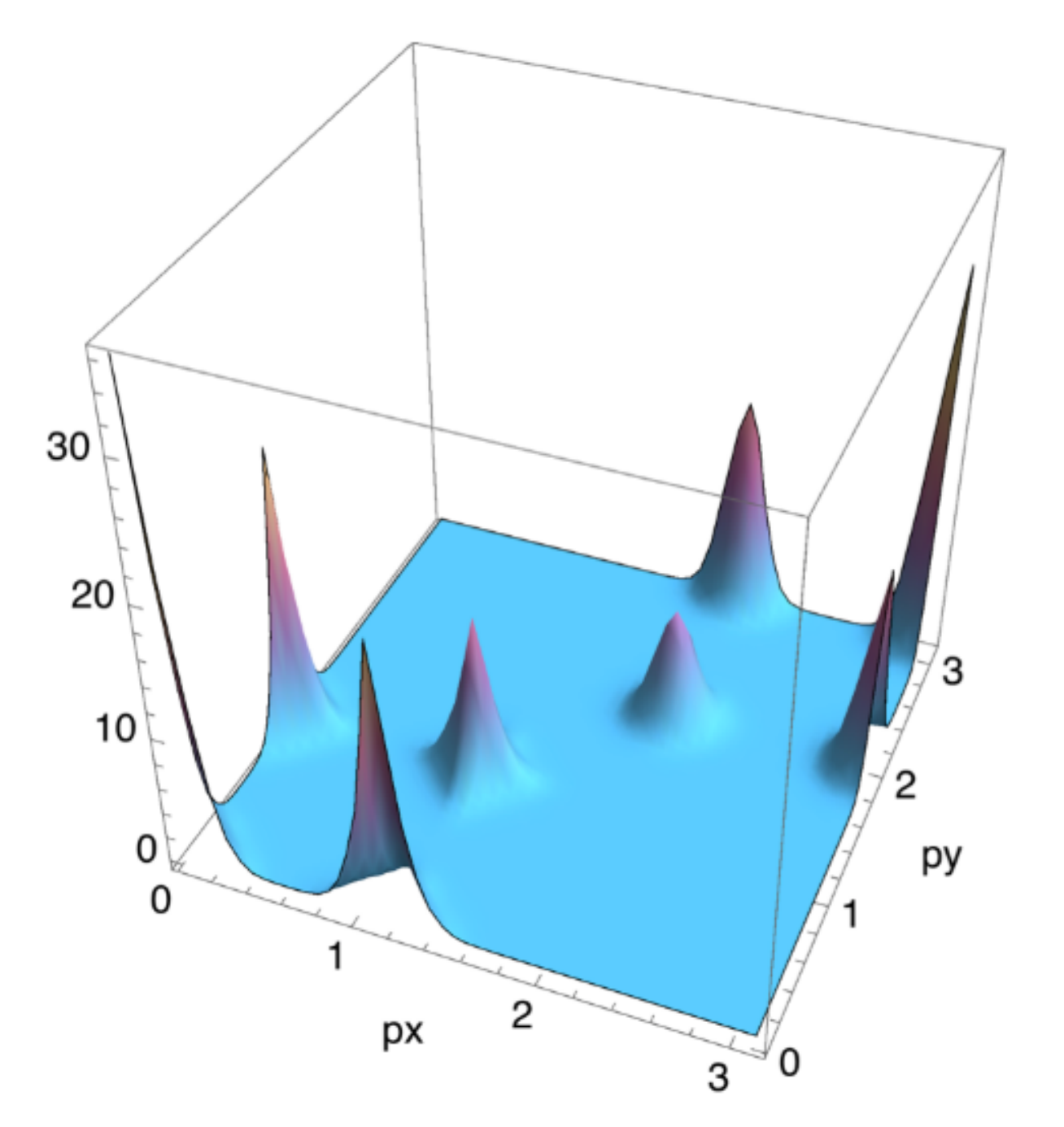} \vspace{1ex}
\end{minipage}

c) %
\begin{minipage}[c]{3.25cm}%
\includegraphics[width=2.5cm]{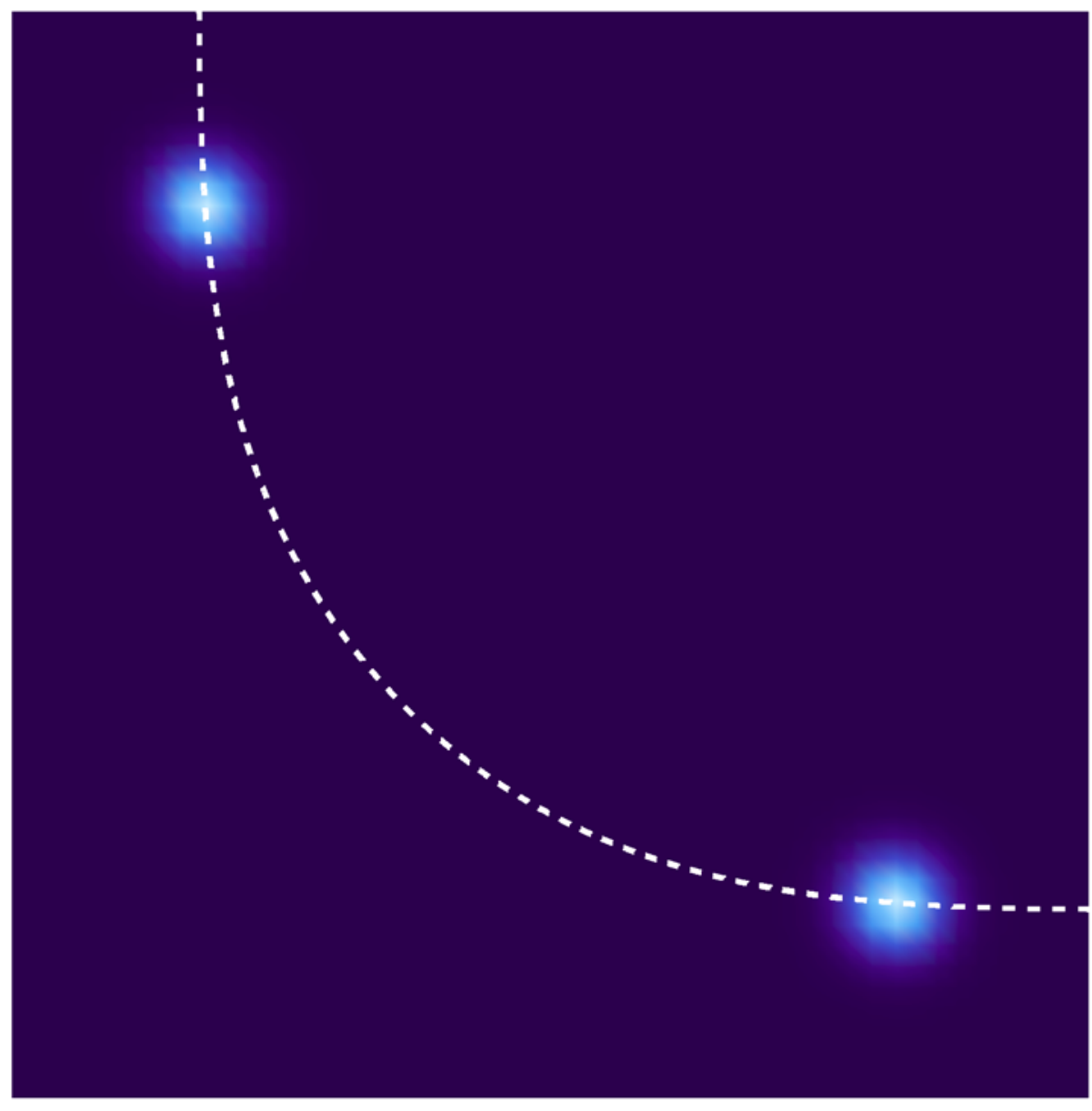} \vspace{1ex}
\end{minipage}%
\begin{minipage}[c]{4.25cm}%
\includegraphics[width=4cm]{Fig11c} \vspace{1ex}
\end{minipage}

\textcolor{black}{\caption{\label{fig:anis1} (Color online) \textcolor{black}{Real part of the
polarization bubble, with one vertex correction, in the static limit.
The electron dispersion corresponds to the usual one of } Bi2212
Ref.\ \cite{Norman07}\textcolor{black}{. We show here that for a
dispersion centered exclusively at the hot spots (panel c), the response
on the axes $\left(0,Q_{0}\right)$ and $\left(Q_{0,}0\right)$ has
still a stronger amplitude than the response on the diagonal $\left(Q_{0},Q_{0}\right)$.}}
} 
\end{figure}

\begin{figure}[h]
a)%
\begin{minipage}[c]{6.25cm}%
\includegraphics[width=7cm]{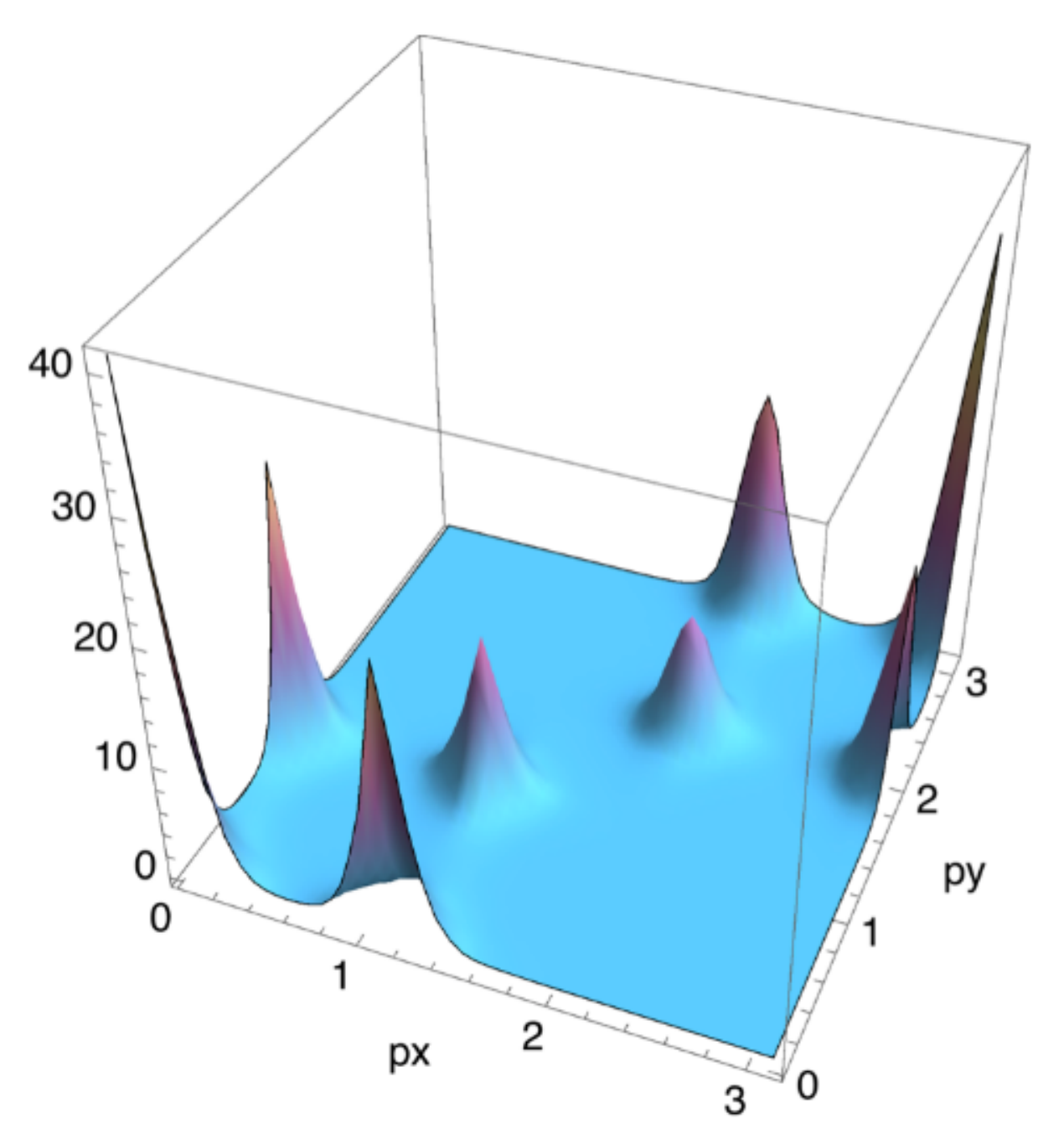} \vspace{1ex}
\end{minipage}

c) %
\begin{minipage}[c]{3.25cm}%
\includegraphics[width=2.5cm]{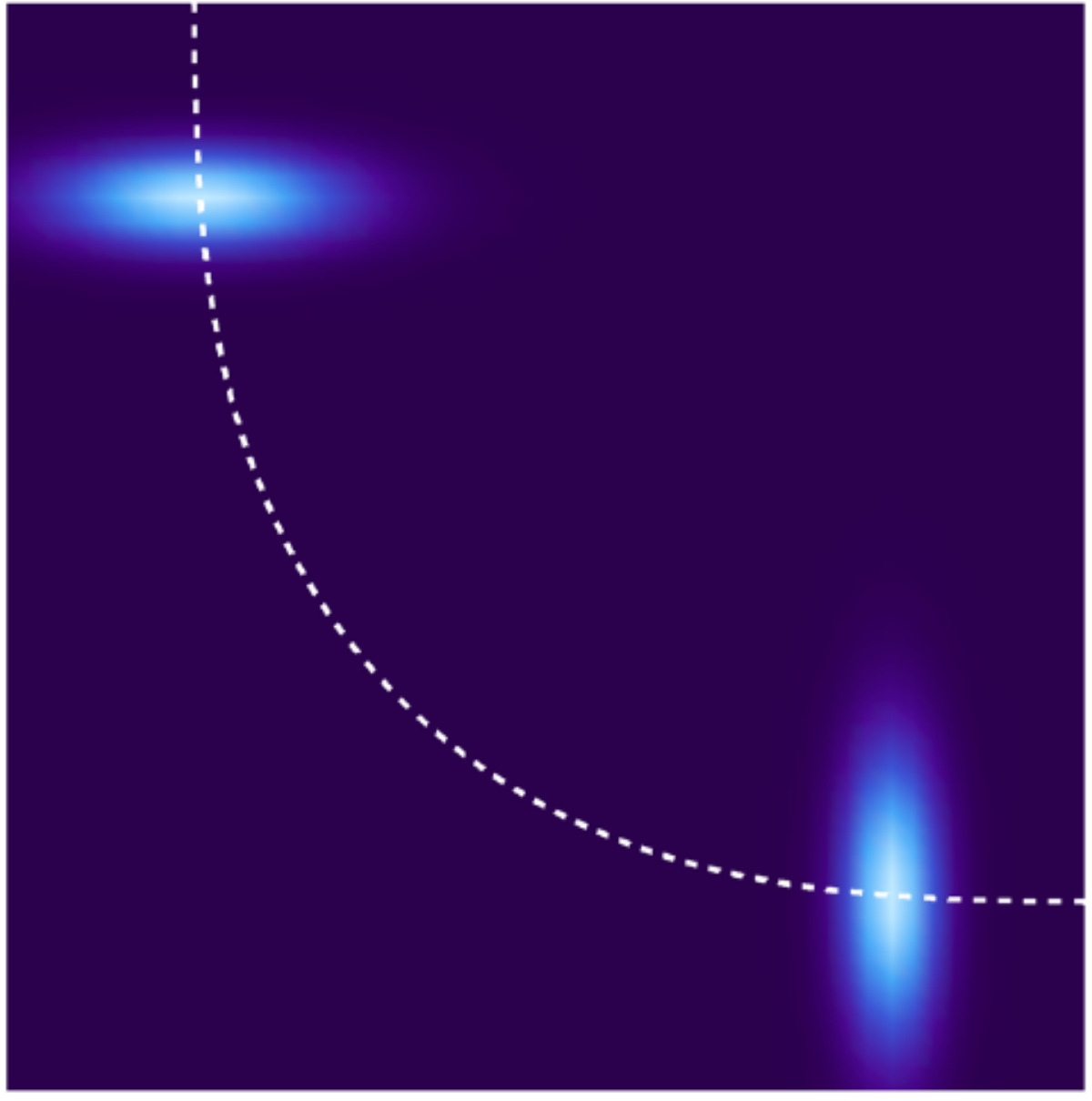} \vspace{1ex}
\end{minipage}%
\begin{minipage}[c]{4.25cm}%
\includegraphics[width=4cm]{Fig11c} \vspace{1ex}
\end{minipage}

\textcolor{black}{\caption{\label{fig:anis2} (Color online) \textcolor{black}{Real part of the
polarization bubble, with one vertex correction, in the static limit.
The electron dispersion corresponds to the usual one of }Bi2212
Ref.\ \cite{Norman07}\textcolor{black}{. Same study as in Fig.\ref{fig:anis1},
but for a ``flat pancake'' shape of the SU(2) fluctuations. }}
} 
\end{figure}

This Appendix gives a thorough study of the effect of the SU(2) hot
regions in the response of the charge susceptibility. We look at various
types of anisotropy (Figs.\ref{fig:anis1} and \ref{fig:anis2}) in
the size and shape of the hot regions. The main conclusion is that
in all cases, the axial response is favored compared to the diagonal
one.

\begin{figure}[h]
a)%
\begin{minipage}[c]{6.25cm}%
\includegraphics[width=7cm]{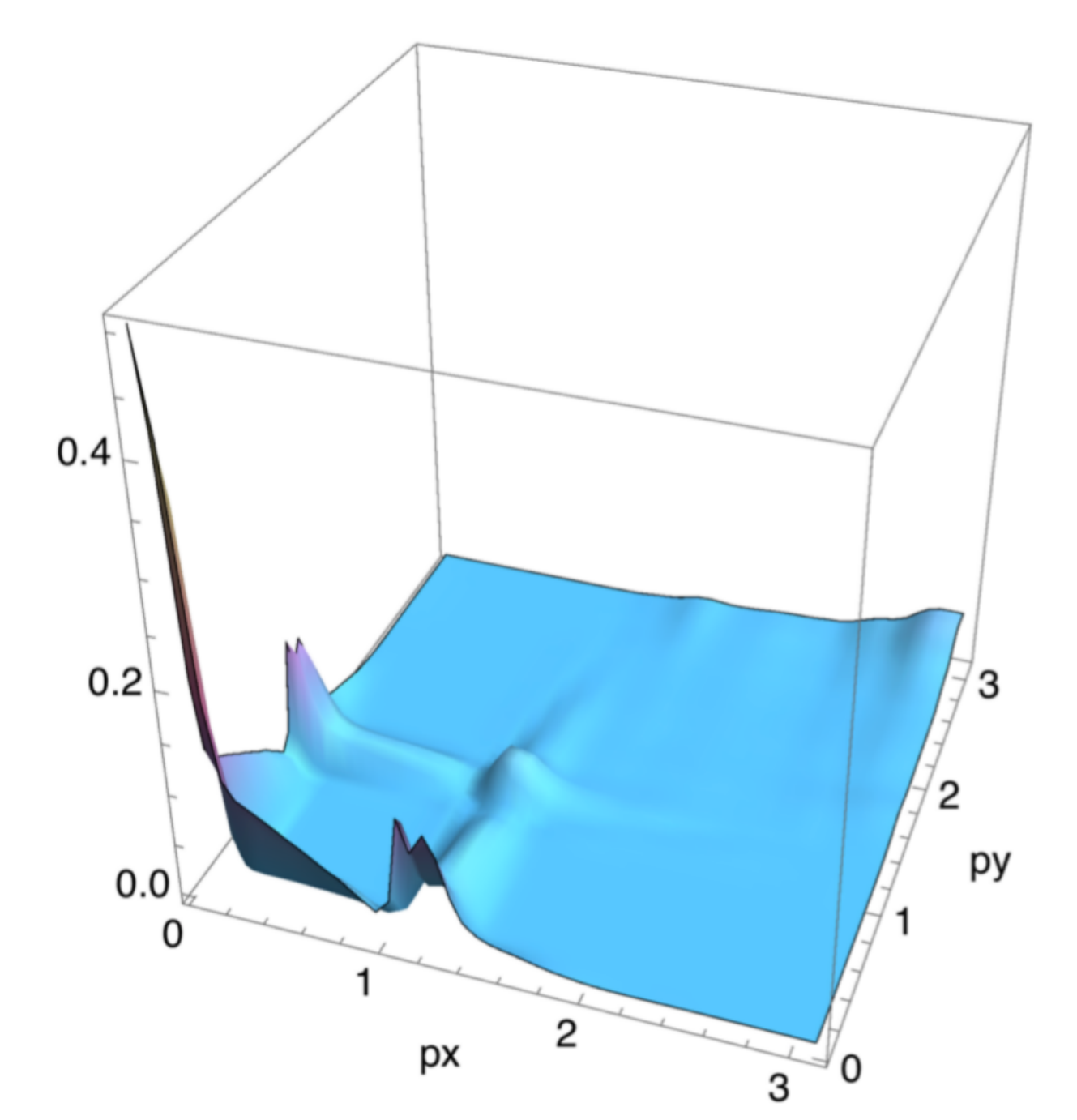} \vspace{1ex}
\end{minipage}

c) %
\begin{minipage}[c]{3.25cm}%
\includegraphics[width=2.5cm]{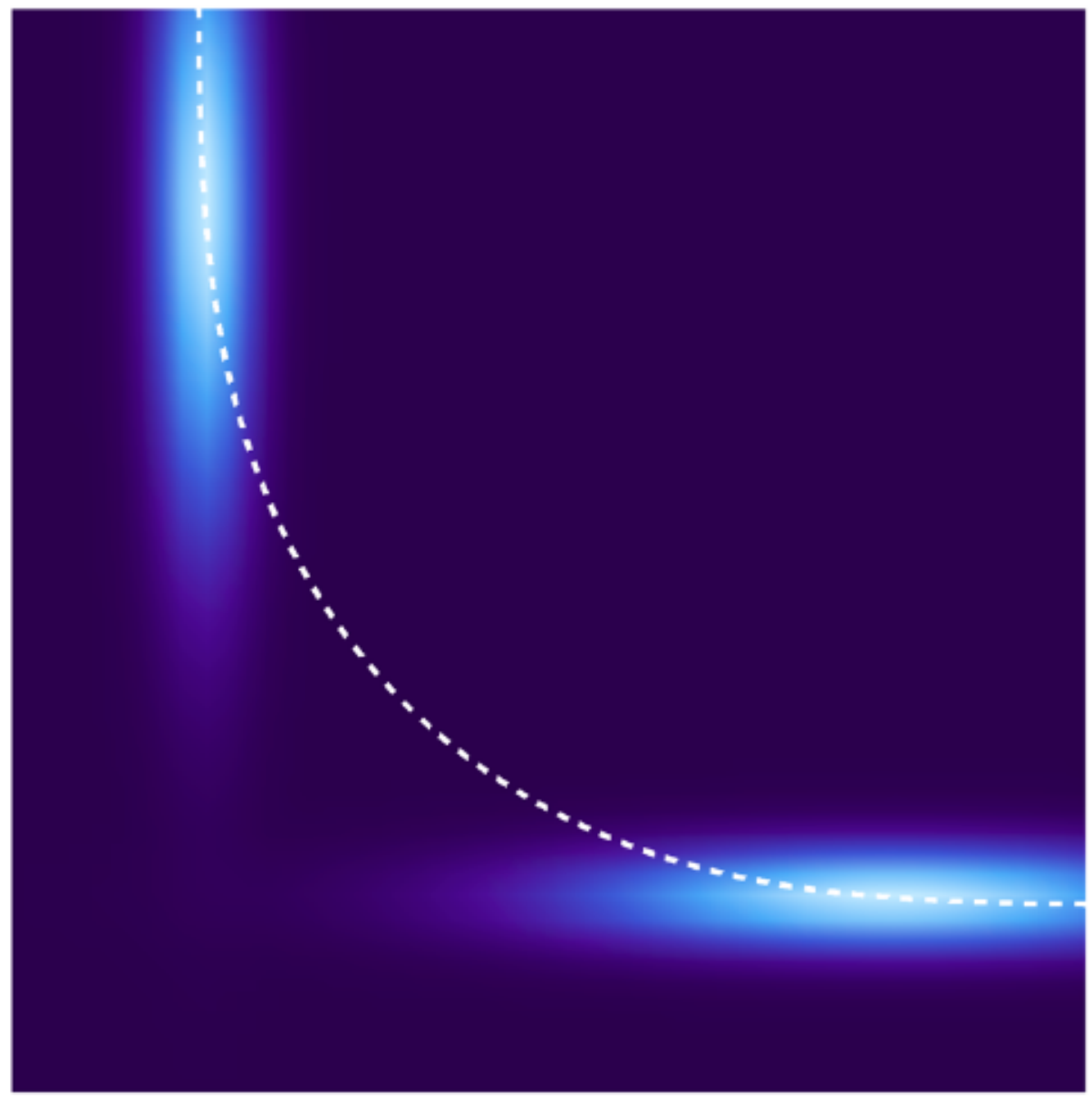} \vspace{1ex}
\end{minipage}%
\begin{minipage}[c]{4.25cm}%
\includegraphics[width=4cm]{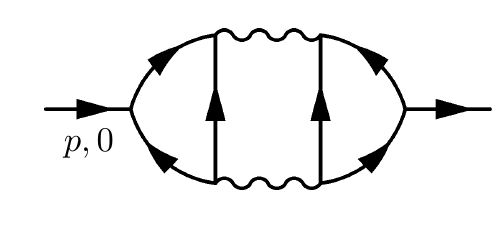} \vspace{1ex}
\end{minipage}

\textcolor{black}{\caption{\label{fig:AL} (Color online) \textcolor{black}{Real part of the
polarization bubble, corresponding to the Aslamozov-Larkin diagrams,
in the static limit. The electron dispersion corresponds to the usual
one of }Bi2212 Ref.\ \cite{Norman07}\textcolor{black}{. This
contribution typically behaves in the same way as the polarization
wiht vertex corrections. }}
} 
\end{figure}

The typical form of the Aslamazov-Larkin polarization is shown in
Fig.\ref{fig:AL}.

\begin{align}
\Pi_{AL}\left(p,0\right) & =-T\sum_{q}\pi_{q}^{s}\pi_{p+q}^{s}\left(B_{q}\right)^{2},\\
\mbox{with }B_{q} & =\sum_{k}G_{k}G_{k+p}G_{-k-q},
\end{align}
with the four variables $k=\left(\mathbf{k,\varepsilon}\right)$,
$q=\left(\mathbf{q},\omega\right)$, $G_{k}^{-1}=i\varepsilon_{n}-\xi_{\mathbf{k}}$
and $G_{-k}^{-1}=-i\varepsilon_{n}-\xi_{\mathbf{-k}}$. This contribution
typically behaves in the same manner as the vertex corrections. We
show it here for completeness for one typical form of the SU(2) hot
region.

\section{Nematicity and charge conservation\label{sec:Nematicity-and-charge}}

\subsection{Cancellation in the Fermi liquid case}

\begin{figure}[H]
\begin{minipage}[c]{4.25cm}%
a)

\includegraphics[width=4cm]{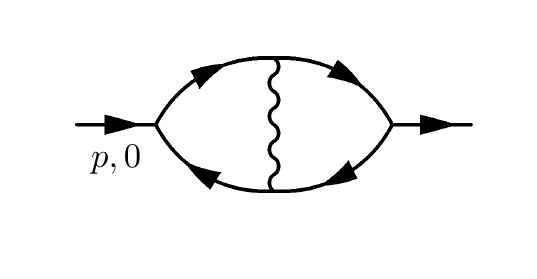} \vspace{1ex}
\end{minipage}%
\begin{minipage}[c]{4.25cm}%
b)

\includegraphics[width=4cm]{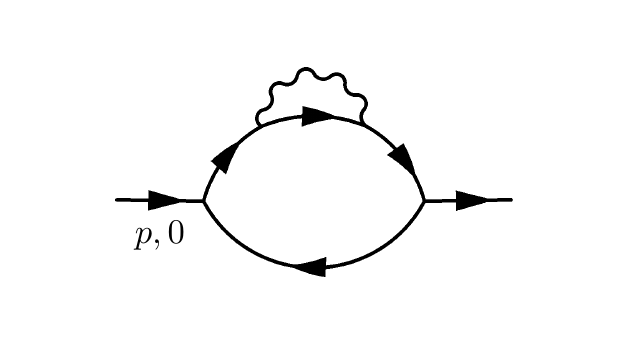} \vspace{1ex}
\end{minipage}

\textcolor{black}{\caption{\label{fig:diags2} (Color online) \textcolor{black}{The standard
vertex self-energy cancellation due to charge conservation. The sum
of the diagrams $a)+2b)=0$.}}
} 
\end{figure}

We first consider the Fermi liquid case depicted in Fig.\ref{fig:diags2}a)
and b). There is a well-known cancellation between the two diagrams,
\begin{align}
I_{a}+2I_{b}=0,\label{eq:136}
\end{align}
that we reproduce here for completeness. We have 
\begin{align}
I_{a} & =\sum_{k,q}F_{q}G_{k}G_{k+p}G_{k+q}G_{k+p+q},\\
I_{b} & =\sum_{k,q}F_{q}G_{k}^{2}G_{k+p}G_{k+q},
\end{align}

where $k,q$ stand for the 4-vector $\mathbf{k},\mathbf{q}$, $F_{q}$
is the boson line, $G_{k}$ is the Fermionic Green's function, with
$G_{k}^{-1}=i\varepsilon_{n}+\Sigma\left(\varepsilon_{n}\right)-\xi_{\mathbf{k}}$.
We use the decoupling trick 
\begin{align}
G_{k}G_{k'} & =\frac{G_{k}-G_{k'}}{G_{k'}^{-1}-G_{k}^{-1}}.\label{eq:138}
\end{align}

Using Eqn.(\ref{eq:138}) we have 
\begin{align}
I_{a} & =\sum_{k,q}F_{q}[\frac{G_{k}G_{k+p}-G_{k}G_{k+p+q}}{H_{q}^{2}}\nonumber \\
 & \frac{-G_{k+p}G_{k+q}+G_{k+q}G_{k+p+q}}{H_{q}^{2}}]\\
I_{b} & =\sum_{k,q}F_{q}\left[\frac{G_{k}^{2}G_{k+q}}{H_{q}}-G_{k+p}\left(\frac{G_{k}-G_{k+q}}{H_{q}^{2}}\right)\right],
\end{align}

where $H_{q}=G_{k+p}^{-1}-G_{k+p+q}^{-1}\simeq-i\omega+\Sigma(\varepsilon_{n})-\Sigma\left(\varepsilon_{n}+\omega_{n}\right)+\mathbf{v}_{F}\cdot\mathbf{q}$.
We observe that terms of the kind $\int_{\mathbf{k}}G_{k}^{n}G_{k+p}=0$
for all integers $n>1$ because, since the external momentum $p=\left(\mathbf{p},0\right)$
carries no frequency, the poles are in the same half-plane, which
leads to ( in the limit where $p\rightarrow0$) 
\begin{align}
I_{a} & =-2\sum_{q}F_{q}\frac{G_{k}G_{k+q}}{H_{q}^{2}},\\
I_{b} & =\sum_{q}F_{q}\frac{G_{k}G_{k+q}}{H_{q}^{2}}.
\end{align}
This in turn gives the result Eqn.(\ref{eq:136}).

\subsection{Absence of cancellation in the case of SC lines}

\begin{figure}
a)%
\begin{minipage}[c]{4.25cm}%
\includegraphics[width=4cm]{Fig11c} \vspace{1ex}
\end{minipage}b)%
\begin{minipage}[c]{4.25cm}%
\includegraphics[width=4cm]{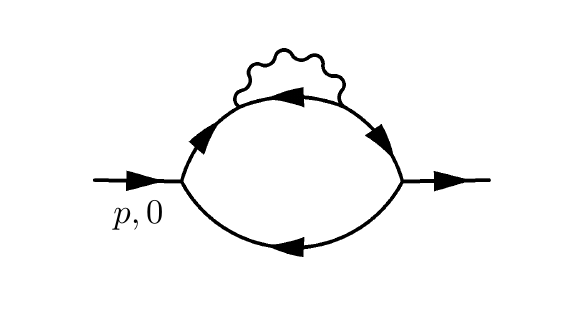} \vspace{1ex}
\end{minipage}

\textcolor{black}{\caption{\label{fig:diags3} (Color online) \textcolor{black}{Vertex corrections
and self-energy corrections at the one loop level. We check in the
text that there is no cancellation in the case of pairing lines $a)+2b)\protect\neq0$.}}
} 
\end{figure}

We now apply the same recipes to the diagrams of Fig.\ref{fig:diags3},
and see that the cancellation doesn't hold in this case. We have 
\begin{align}
I_{a} & =\sum_{k,q}F_{q}G_{k}G_{k+p}G_{-k-q}G_{-k-p-q},\\
I_{b} & =\sum_{k,q}F_{q}G_{k}^{2}G_{k+p}G_{-k-p},
\end{align}

with $G_{-k}^{-1}=-i\varepsilon_{n}+\Sigma\left(-\varepsilon_{n}\right)-\xi_{-\mathbf{k}}$.
Using Eqn.(\ref{eq:138}), this can be cast into 
\begin{align}
I_{a} & =\sum_{k,q}F_{q}\left[\frac{-G_{k}G_{k+p}+G_{-k-q}G_{k+p}}{\overline{H}_{k,q}}+\frac{G_{k}^{2}G_{k+p}}{\overline{H}_{k,q}}\right],\\
I_{b} & =\sum_{k,q}F_{q}\left[\frac{G_{k}^{2}+G_{-k-q}^{^{2}}-2G_{k}^{3}}{\overline{H}_{k,q}}\right],
\end{align}

with $\overline{H}_{k,q}=G_{-k-q}^{-1}-G_{k}^{-1}\simeq-2i\varepsilon_{n}-i\omega+\Sigma\left(-\varepsilon_{n}-\omega_{n}\right)-\Sigma\left(\varepsilon_{n}\right)+\mathbf{v}_{F}\cdot q+2\xi_{\mathbf{k}}$.
To check the non-cancellation of diagrams, we can take the difference
$dI=I_{a}+2I_{b}$, which gives, in the limit where $p\rightarrow0$,
\begin{align}
dI & =\sum_{k,q}F_{q}\left[\frac{G_{-k-q}^{2}-G_{k}^{2}}{\overline{H}_{k,q}}+2\frac{G_{k}^{3}}{\overline{H}_{k,q}}\right].\label{eq:148}
\end{align}

The first term in Eqn.(\ref{eq:148}) vanish after a change of variables,
and we end up with a non-zero contribution

\begin{align}
dI & =2\sum_{k,q}F_{q}\frac{G_{k}^{2}G_{k+p}}{\overline{H}_{k,q}}.\label{eq:149}
\end{align}

\begin{figure}[h]
\begin{minipage}[c]{6.25cm}%
\includegraphics[width=6cm]{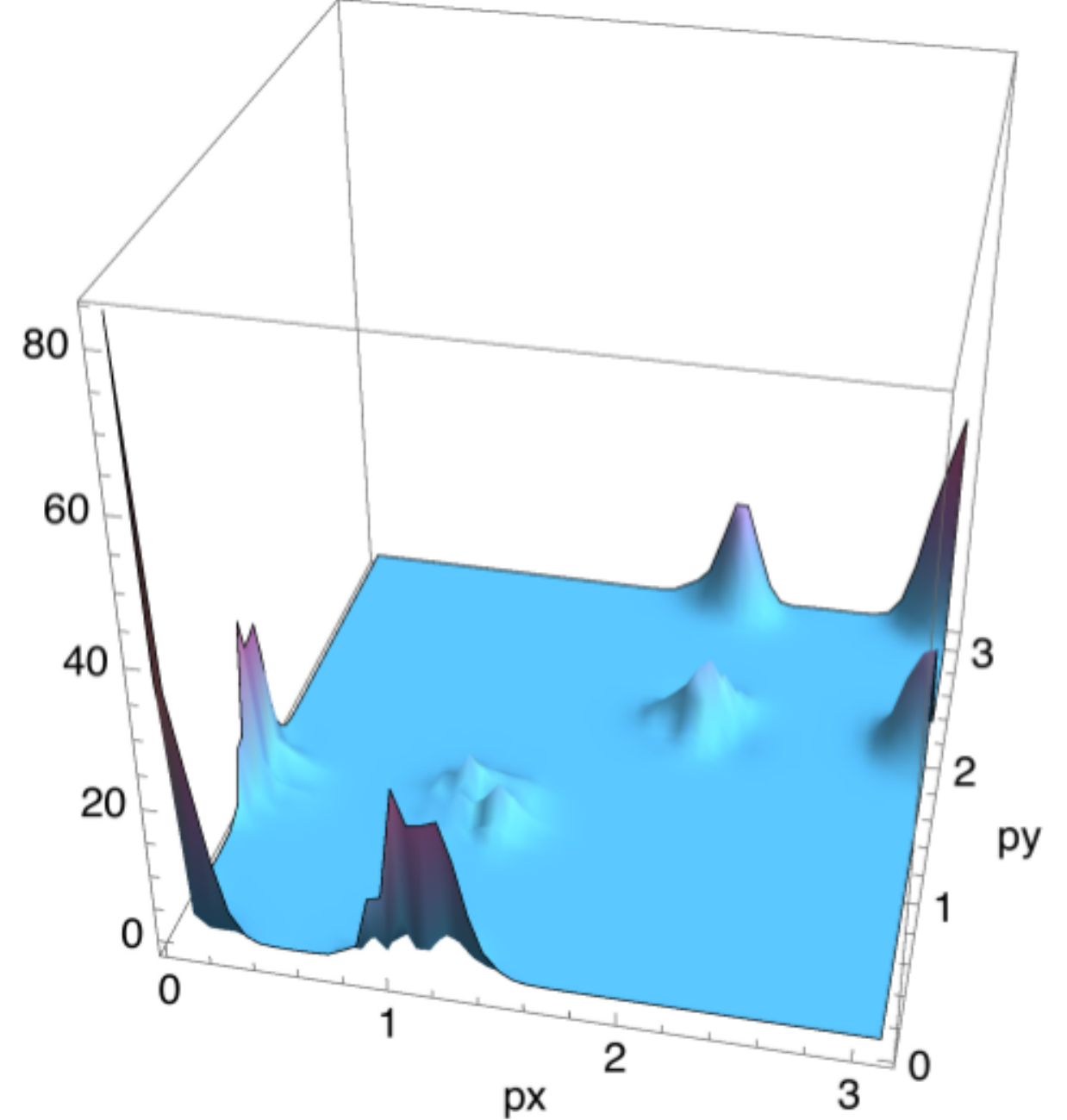} \vspace{1ex}
\end{minipage}

\begin{minipage}[c]{3.25cm}%
\includegraphics[width=2.5cm]{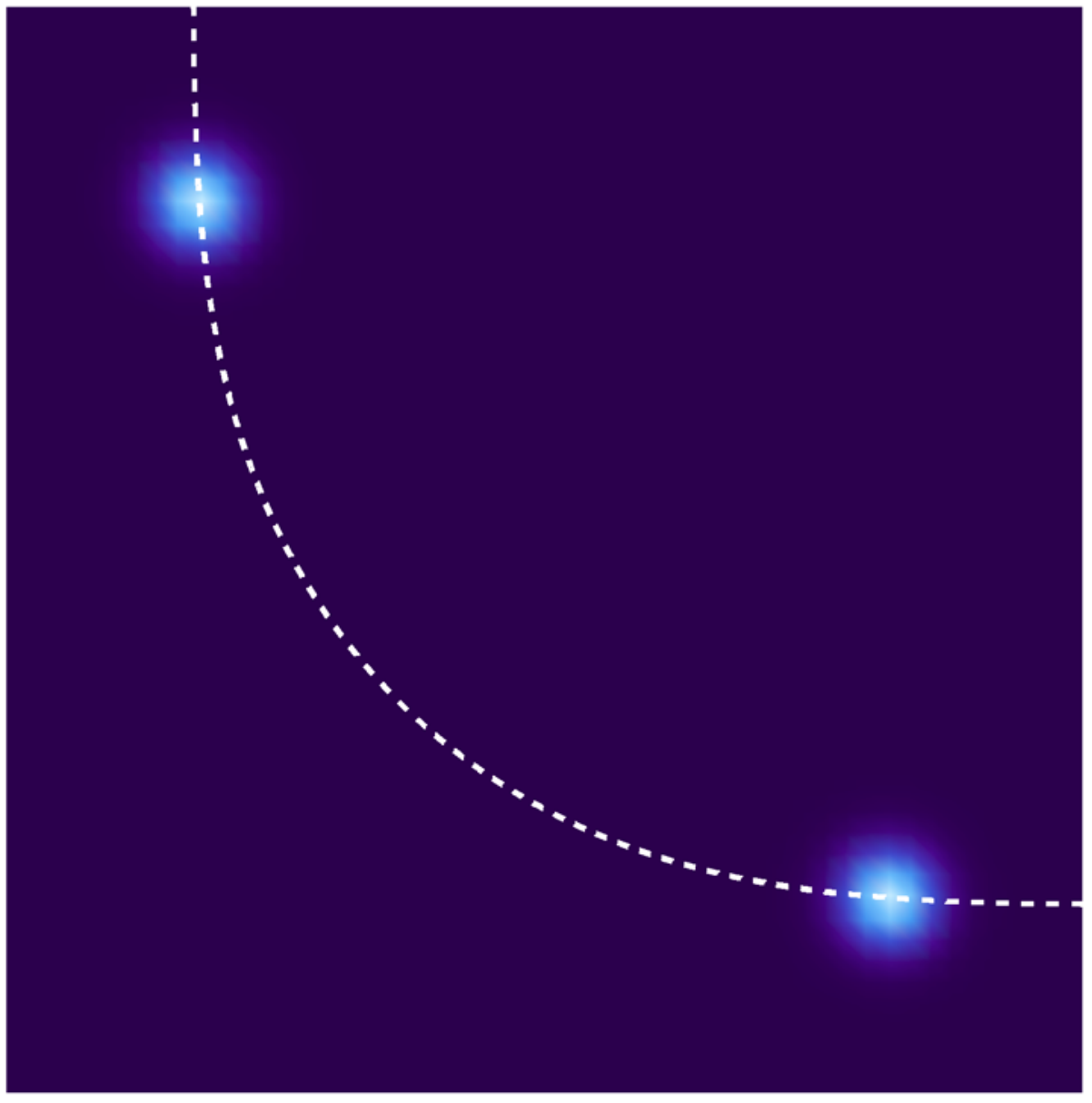} \vspace{1ex}
\end{minipage}

\textcolor{black}{\caption{\label{fig:cancellation} (Color online) \textcolor{black}{Plot of
$dI$ in Eqn.(\ref{eq:149}). The electron dispersion corresponds
to the usual one of }Bi2212 Ref.\ \cite{Norman07}\textcolor{black}{. }}
} 
\end{figure}

To fix the ideas, we plot $dI$ in Eqn.(\ref{eq:149}) in Fig.\ref{fig:cancellation}.

\section{Gaussian fluctuations}

We study now the case where the charge order becomes critical for
some region of oxygen doping, under the SC dome. Fluctuations of the
charge mode are treated at the Gaussian level. The effective action
$S_{1}$ Eqn. (\ref{eq:S1}), which describes the coupling of the
fermion to the excitonic patches, can be written 
\begin{align}
S_{1} & =\frac{1}{2}\sum_{\sigma,k,p}\chi_{k,k+p}\psi_{k,\sigma}^{\dagger}\psi_{k+p,\sigma},
\end{align}
where $p=P+q$ is the fluctuation around the ordering wave vector
$\mathbf{P}=\pf$ and $\chi_{k,k+p}$ is the bosonic charge mode with
wave vector $\mathbf{P}$. This charge order is peculiar in the sense
that $\chi$ depends not only on the slow fluctuations around the
wave vector $\mathbf{P}$ but as well on the fast momentum $\mathbf{k}$.
We account for this dependence through a form factor via the definition
\begin{align}
\chi_{k,k+p} & \equiv\chi_{p}F_{k},
\end{align}
where $F_{k}$ is a form factor having a finite extension around the
nesting point $\mathbf{k}_{n}$ associated with $\mathbf{P}$. The
form of the boson propagator $\Phi_{q}=\left\langle \mathcal{T}\chi_{-p}\chi_{p}\right\rangle $is
given by 
\begin{align}
\exp\left[-S_{\Phi}\right] & =\left\langle \exp\left[-S_{1}\right]\right\rangle _{\psi} & =\exp\left[\frac{1}{2}\left\langle S_{1}^{2}\right\rangle _{\psi}\right].
\end{align}
Expanding then to the second order in $\chi$ we obtain 
\begin{align}
S_{\Phi} & =-\frac{1}{2}\!\sum_{k,p,\sigma,k',p',\sigma'}\!\!\!\!\!\!\chi_{p}\chi_{p'}F_{k}F_{k'}\left\langle \psi_{k,\sigma}^{\dagger}\psi_{k+p,\sigma}\psi_{k',\sigma'}^{\dagger}\psi_{k'+p',\sigma'}\right\rangle _{\psi}.
\end{align}
The contraction of indices leads to the conditions $\sigma'=\sigma$,
$p'=-p$ and $k'=k+p$. From which we get 
\begin{align}
S_{\Phi} & =\frac{1}{2}\sum_{p}\chi_{-p}\Phi_{p}^{-1}\chi_{p}\\
\mbox{ with } & \Phi_{p}^{-1}=\Phi_{0,p}^{-1}-\Pi_{p},\nonumber \\
 & \Pi_{p}=-T\sum_{\varepsilon,k,\sigma}F_{k}F_{k+p}G_{k}G_{k+p}.
\end{align}
For the evaluation of the bosonic bubble $\Phi_{p}$ the form factors
have been neglected in the evaluation of $\Pi_{p}$. $\Phi_{0,p}^{-1}=\omega^{2}+q^{2}+m$
is a high energy contribution of the Ornstein-Zernike type. The resulting
scattering of the fermions around the bosonic charge mode writes 
\begin{align}
\exp\left[-S_{crit}\right] & =\left\langle \exp\left[-S_{1}\right]\right\rangle _{\chi} & =\exp\left[\frac{1}{2}\left\langle S_{1}^{2}\right\rangle _{\chi}\right],
\end{align}
with 
\begin{align}
S_{crit} & =-\frac{1}{2}\!\sum_{k,k',\sigma,\sigma',p,p'}\!\!\!\!\!\!\!\!\!\!\left\langle \chi_{p}\chi_{p'}\right\rangle _{\chi}F_{k}F_{k'}\psi_{k,\sigma}^{\dagger}\psi_{k+p,\sigma}\psi_{k',\sigma'}^{\dagger}\psi_{k'+p',\sigma'},\\
 & =-\frac{1}{2}\sum_{k,k',\sigma,\sigma',p}\Phi_{p}F_{k}F_{k'}\psi_{k,\sigma}^{\dagger}\psi_{k+p,\sigma}\psi_{k',\sigma'}^{\dagger}\psi_{k'-p,\sigma'}.\nonumber 
\end{align}

\section{Bosonic polarization bubble\label{sec:Bosonic-polarization-bubble}}

The renormalized bosonic propagator follows from Dyson's equation
\begin{equation}
[\Phi_{\mathbf{q}}(\Omega)]^{-1}=q^{2}+m-\Pi_{\mathbf{q}}(\Omega)\label{eq:deff}
\end{equation}
Therefore, we evaluate the bosonic polarization $\Pi_{\qv}$ with
${\bf P}=\pf$ as depicted in Fig.\ 4c) of the main text. The diagram
writes 
\begin{align}
\Pi_{\qv}(i\omega_{m}) & =-\frac{1}{\beta V}\sum_{\epsilon_{n},\kv}G(i\epsilon_{n}+i\omega_{m}/2,\kv+\qv/2)\nonumber \\
 & \qquad\times G(i\epsilon_{n}-i\omega_{m}/2,\kv+\pv-\qv/2),\label{eq:pidiagr}
\end{align}
with temperature $T=\beta^{-1}$ and $V$ being the volume of the
system. We will evaluate the diagram using bare Green functions of
the form 
\begin{equation}
G(i\epsilon_{n},\kv)^{-1}=i\epsilon_{n}-\xi_{\kv},
\end{equation}
with $\xi_{{\bf {k}}}=\epsilon_{{\bf {k}}}-\mu$, where $\epsilon_{{\bf {k}}}$
is the fermion dispersion and $\mu$ their chemical potential. It
is convenient to measure the three dimensional momentum vector $\kv\equiv(k_{\xa},\kv_{\xe})$
(where $k_{\xa}$ is a scalar and $\kv_{\xe}$ is a two dimensional
vector) relative to a point on the Fermi surface. The coordinate system
is then chosen such, that the two components of $\kv_{\xe}$ are oriented
perpendicular to the surface normal vector and the $k_{\xa}$ component
parallel to it, as depicted in Fig.\ 4b) of the main text. The dispersion
is then approximated by 
\begin{equation}
\xi_{\kv}=v_{\xa}k_{\xa}+v_{\xe}k_{\xe}^{2}/2.\label{eq:xidef}
\end{equation}
and from the construction follows $\xi_{\kv+\pv}=\xi_{-\kv}$. The
momentum transfer $\qv$ between the two coupled electrons is quasi
one-dimensional in parallel direction only, so that we calculate $\Pi_{\qv}$
for $\qv\equiv(q_{\xa},0)$. Equation (\ref{eq:pidiagr}) then yields
\begin{align}
 & \Pi_{\qv}(i\omega_{m})=\nonumber \\
 & -\frac{1}{\beta V}\sum_{\epsilon_{n},\kv}\frac{1}{i(\epsilon_{n}+\omega_{m}/2)-v_{\xa}(k_{\xa}+q_{\xa}/2)-v_{\xe}k_{\xe}^{2}/2}\nonumber \\
 & \qquad\times\frac{1}{i(\epsilon_{n}-\omega_{m}/2)+v_{\xa}(k_{\xa}-q_{\xa}/2)-v_{\xe}k_{\xe}^{2}/2},\label{eq:piexpl}
\end{align}
Evaluating the Matsubara sum at $T=0$ gives 
\begin{align}
 & \Pi_{\qv}(i\omega_{m})=-\frac{1}{v_{\xa}v_{\xe}V}\nonumber \\
 & \qquad\times\sum_{\kv}\frac{\theta(2k_{\xa}+q_{\xa}+k_{\xe}^{2})-\theta(-2k_{\xa}+q_{\xa}+k_{\xe}^{2})}{i\omega_{m}-2k_{\xa}},\label{eq:wsum}
\end{align}
where $\theta(x)$ denotes the Heaviside step function and we have
also rescaled the parallel momenta to measure it in units of $v_{\xa}$
as well as the perpendicular momenta by $\sqrt{v_{\xe}}$. We can
now perform the analytic continuation to real frequencies $\Omega$
by setting $i\omega_{m}=\Omega+i\delta$ and we write the retarded
self-energy as $\Pi_{\qv}(\Omega)\equiv\Pi_{\qv}(\Omega+i\delta)$
for convenience. In the infinite volume limit and for $\Omega\neq|q_{\xa}|$
the polarization yields 
\begin{align}
 & \Pi_{\qv}(\Omega)=\nonumber \\
 & -\frac{c}{\pi}\int\!d\kv_{\xe}\!\biggl[P\!\int_{-k_{\xe}^{2}}^{\Lambda}\!\frac{dy}{\Omega+q_{\xa}-y}-P\!\int_{-\Lambda}^{k_{\xe}^{2}}\!\frac{dy}{\Omega-q_{\xa}-y}\nonumber \\
 & \qquad-i\pi\bigl(\theta(k_{\xe}^{2}+\Omega+q_{\xa})-\theta(k_{\xe}^{2}-\Omega+q_{\xa})\bigr)\biggr],
\end{align}
with $c^{-1}=(4\pi)^{2}v_{\xa}v_{\xe}$, $P$ stands for the Cauchy
principal value and $\Lambda$ is an UV cut-off. The special cases
for $\Omega=|q_{\xa}|$ is not explicated here but can be obtained
in the same fashion. After integrating over the parallel component
we get 
\begin{align}
 & \Pi_{\qv}(\Omega)=-\frac{c}{\pi}\int d\kv_{\xe}\biggl[\ln|k_{\xe}^{2}+\Omega+q_{\xa}|+\ln|k_{\xe}^{2}-\Omega+q_{\xa}|\nonumber \\
 & \qquad-i\pi\bigl(\theta(k_{\xe}^{2}+\Omega+q_{\xa})-\theta(k_{\xe}^{2}-\Omega+q_{\xa})\bigr)\biggr],
\end{align}
where some non-universal contributions that depend solely on $\Lambda$
were dropped. The remaining two-dimensional integral over $\kv_{\xe}$
is now easily computed in polar coordinates. Regularizing again the
UV sector by a cut-off $\Lambda$ we find (up to some pure cut-off contributions)
\begin{subequations} 
\begin{align}
\Pi_{{\bf {q}}}(\Omega) & =\Pi'_{{\bf {q}}}(\Omega)+i\Pi''_{{\bf {q}}}(\Omega),\\
\Pi'_{{\bf {q}}}(\Omega) & =c\biggl[(\Omega+q_{\xa})\ln|\Omega+q_{\xa}|-(\Omega-q_{\xa})\ln|\Omega-q_{\xa}|\biggr],\\
\Pi''_{{\bf {q}}}(\Omega) & =\pi c\Bigl[(\Omega+q_{\xa})\theta(-\Omega-q_{\xa})+(\Omega-q_{\xa})\theta(\Omega-q_{\xa})\Bigr],\qquad
\end{align}
\label{eq:pianalyt} \end{subequations} where $\Pi'$ and $\Pi''$
denote the real, respectively imaginary part of the polarization.
In a last step we invert Dysons equation (\ref{eq:deff}) to obtain
the imaginary part of $\Phi_{\mathbf{q}}$, which is responsible for
damping. Because the leading scattering contribution to the resistivity
comes from the low momentum and frequency transfer of the critical
bosonic mode, one can drop the entire $\Phi_{0,\mathbf{q}}^{-1}(\Omega)$
term compared to $\Pi_{\mathbf{q}}$. One obtains 
\begin{equation}
\text{Im}\Phi_{{\bf {q}}}(\Omega)\simeq\frac{\Pi''_{{\bf {q}}}(\Omega)}{(\Pi'_{{\bf {q}}}(\Omega))^{2}+(\Pi''_{{\bf {q}}}(\Omega))^{2}}\label{eq:imdeff0}
\end{equation}
and we can write $\text{Im}\Phi_{{\bf {q}}}$ in compact form as 
\begin{equation}
\text{Im}\Phi_{{\bf {q}}}(\Omega)=\frac{1}{\pi c}b_{{\bf {q}}}(\Omega)\label{eq:imdeffcomp}
\end{equation}
with 
\begin{align}
b_{{\bf {q}}}(\Omega)=\frac{s\theta(-s)+t\theta(t)}{\pi^{-2}[s\ln|s|-t\ln|t|]^{2}+[s\theta(-s)+t\theta(t)]^{2}}\label{eq:imdefffull}
\end{align}
where $s=\Omega+q_{\xa}$ and $t=\Omega-q_{\xa}$.

The study of this paper corresponds to the regime $q_{\parallel\leq}\Omega$.
Reporting this approximation in Eqns.\ (\ref{eq:pianalyt}) we get
\begin{subequations} 
\begin{align}
\Pi'_{{\bf {q}}}(\Omega) & \simeq2cq_{\xa}\ln|\Omega|,\label{eq:pifa1}\\
\Pi''_{{\bf {q}}}(\Omega) & =\pi c\Omega.\qquad\label{eq:pifb1}
\end{align}
\end{subequations}

\section{Fermionic self-energy\label{sec:Fermionic-self-energy}}

The fermionic self-energy is 
\begin{align}
\Sigma_{\kv}(i\epsilon_{n}) & =-\frac{1}{\beta V}\sum_{\omega_{m},\qv}\!\!\Phi_{\qv}(i\omega_{m})G(i\epsilon_{n}+i\omega_{m},\kv+\qv),\label{eq:sigdiagr}
\end{align}
with the fermionic and bosonic Green functions \begin{subequations}
\begin{align}
(G(i\epsilon_{n},\kv))^{-1}=i\epsilon_{n}-v_{\xa}k_{\xa}-v_{\xe}k_{\xe}^{2}/2,\\
\Phi_{\qv}^{-1}(i\omega_{m})=\gamma\left|\omega\right|_{m}-v_{\xa}q_{\xa}\log\left|\overline{\omega}_{m}\right|+v_{\xe}q_{\xe}^{2}/2,
\end{align}
\end{subequations} where the form of the bosonic Green's function
is taken from Eqns. (\ref{eq:pifa1}-\ref{eq:pifb1}).

The notation $\overline{\omega}_{n}=\omega_{n}/\Lambda$ where $\Lambda$
is an UV cut-off. To simplify the calculation, we will not calculate
the full above self-energy, but the truncated one 
\begin{equation}
\Delta\Sigma(i\epsilon_{n})=\Sigma_{\kv=0}(i\epsilon_{n})-\Sigma_{\kv=0}(0).
\end{equation}
With above formulas, the truncated self-energy writes 
\begin{align}
 & \Delta\Sigma(i\epsilon_{n})=-\frac{1}{\beta V}\sum_{\omega_{m},\qv}\frac{1}{\gamma\omega_{m}^{2}-v_{\xa}q_{\xa}+v_{\xe}q_{\xe}^{2}/2}\nonumber \\
 & \,\,\times\left(\frac{1}{i\epsilon_{n}+i\omega_{m}-v_{\xa}q_{\xa}-v_{\xe}q_{\xe}^{2}/2}-\frac{1}{i\omega_{m}-v_{\xa}q_{\xa}-v_{\xe}q_{\xe}^{2}/2}\right),\label{eq:sigdiagr2}
\end{align}
and the sums are evaluated in infinite volume and vanishing temperature
limit. Rescaling the variables according to $v_{\xa}q_{\xa}=x$ and
$v_{\xe}q_{\xe}^{2}/2=y$, the self-energy follows as 
\begin{align}
 & \Delta\Sigma(i\epsilon_{n})=c_{2}T\sum_{\omega_{n}}\,I_{\omega},\label{eq:sigdiagr3}
\end{align}
with 
\begin{align}
I_{\omega} & =\int_{-\infty}^{\infty}dx\int_{0}^{\infty}dy\frac{1}{\gamma\left|\omega_{n}\right|-x\log\overline{\omega}_{n}+y}\nonumber \\
 & \times\frac{i\epsilon_{n}}{(i\epsilon_{n}+i\omega_{n}-x-y)(i\omega_{n}-x-y)}
\end{align}
and $c_{2}^{-1}=4\pi^{2}v_{\xa}v_{\xe}$. We first perform the integration
in x, performed in the complex plane. Two types of poles are present,
the cones coming from the fermionic Green's functions on the right,
and the ones coming from the boson propagator.

\subsection{Poles from the Fermionic Green's function}

The contribution of the poles from the fermionic Green's function
are taken at $x_{0}=i\omega_{n}-y$ and give a contribution 
\begin{align}
I_{1} & =\int_{0}^{\infty}dy\frac{2i\pi Sgn\left(\epsilon_{n}\right)\theta\left[\left|\epsilon_{n}\right|-\left|\omega_{n}\right|\right]}{\gamma\left|\omega_{n}\right|+y-x_{0}\log\overline{\omega}_{n}}.
\end{align}
We keep the term proportional to $\log\overline{\omega}_{n}$ in the
denominator, and get 
\begin{align}
I_{1} & =\int_{0}^{\infty}dy\frac{2i\pi Sgn\left(\epsilon_{n}\right)\theta\left[\left|\epsilon_{n}\right|-\left|\omega_{n}\right|\right]}{\left(-i\omega_{n}+y\right)\log\overline{\omega}_{n}},\nonumber \\
 & =\frac{2i\pi Sgn\left(\epsilon_{n}\right)\theta\left[\left|\epsilon_{n}\right|-\left|\omega_{n}\right|\right]}{\log\overline{\omega}_{n}}\log\left(\frac{\Lambda}{-i\omega_{n}}\right).
\end{align}
Reporting in (\ref{eq:sigdiagr3}) and symmetrizing with respect to
$\omega_{n}$ leads to 
\begin{align}
\Delta\Sigma(i\epsilon_{n}) & =c_{2}T\sum_{\omega_{n}>0}i\pi Sgn\left(\epsilon_{n}\right)\theta\left[\left|\epsilon_{n}\right|-\left|\omega_{n}\right|\right],\nonumber \\
 & =c_{2}i\pi\epsilon_{n}.
\end{align}
Note that the logarithmic singularity has been lost here, up to the
UV cut-off. This self-energy will this not produce a significant shortening
of the electron lifetime.

\subsection{Poles from the boson propagator }

The contribution from the boson propagator is taken at $x_{0}=\left(\gamma\left|\omega_{n}\right|+y\right)/\log\left|\omega_{n}\right|$
and writes 
\begin{align}
I_{2} & =\int_{0}^{\Lambda}dy\frac{-i\pi}{\log\left|\omega_{n}\right|}\frac{i\epsilon_{n}}{(i\epsilon_{n}+i\omega_{n}-x_{0}-y)(i\omega_{n}-x_{0}-y)}.
\end{align}
Taking only the term not proportional to $1/\log\left|\omega_{n}\right|$,
we get 
\begin{align}
I_{2} & =\int_{0}^{\Lambda}dy\frac{-i\pi}{\log\left|\omega_{n}\right|}\frac{i\epsilon_{n}}{(i\epsilon_{n}+i\omega_{n}-y)(i\omega_{n}-y)}.
\end{align}
The integration over y is done exactly and leads to 
\begin{align}
I_{2} & =\frac{-i\pi}{\log\left|\omega_{n}\right|}\log\left|\frac{\overline{\omega}_{n}}{\overline{\epsilon}_{n}+\overline{\omega}_{n}}\right|.\label{eq:i2a}
\end{align}
The form (\ref{eq:i2a}) is non-vanishing only in the limit $\left|\omega_{n}\right|\leq\left|\epsilon_{n}\right|$.
Expanding in this limit we get 
\begin{align}
I_{2} & =\frac{i\pi}{\log\left|\omega_{n}\right|}\log\left|\overline{\epsilon}_{n}\right|,\label{eq:i2a-1}
\end{align}
and reporting in Eqn. (\ref{eq:sigdiagr3}) we obtain 
\begin{align}
\Delta\Sigma(i\epsilon_{n}) & =c_{2}T\sum_{\left|\omega_{n}\right|\leq\left|\epsilon_{n}\right|}i\pi\frac{\log\left|\overline{\epsilon}_{n}\right|}{\log\left|\omega_{n}\right|},\\
 & =c_{2}i\pi li(\overline{\epsilon_{n}})\log\left|\overline{\epsilon_{n}}\right|,\nonumber 
\end{align}
where the function logarithmic integral $li(x)$ is defined by 
\[
li(x)=\int_{0}^{x}\frac{dt}{\ln t}.
\]
Expanding by part and considering the regime where $x\ll1$, we get
\begin{equation}
li(x)=\frac{x}{\log x}+\frac{x}{(\log x)^{2}}+\mathcal{O}\left[\frac{x}{\left(\log x\right)^{3}}\right].\label{eq:logint}
\end{equation}
The second term in Eqn. (\ref{eq:logint}) provides the desired singularity,
and we obtain 
\begin{align}
\Delta\Sigma(i\epsilon_{n}) & \simeq c_{2}\frac{i\pi\epsilon_{n}}{\log\left|\epsilon_{n}\right|},
\end{align}
which, up to a logarithm, is the form for the electron self-energy
in the strange metal phase.

\section{Boltzmann treatment\label{sec:Boltzmann-treatment}}

\subsection{Relaxation time from the Boltzmann equation}

The Boltzmann equation for the non-equilibrium electron distribution
$f_{\mathbf{k}}$ writes \cite{Abrikosov88,MahanBook,Paul13} 
\begin{equation}
\left(\frac{\partial f_{\mathbf{k}}}{\partial t}\right)_{\text{collisions}}\hspace{0ex}=-e\mathbf{E}\cdot\nabla_{\mathbf{k}}f_{\mathbf{k}}=-I_{ei}\left[f_{\mathbf{k}}\right]-I_{ee}\left[f_{\mathbf{k}}\right],\label{eq:boltzmann}
\end{equation}
where $e$ is the elementary charge, $\mathbf{E}$ a static electric
field which is supposed to be small and $I_{ei}$ respectively $I_{ee}$
are the electron-impurity respectively electron-electron collision
integrals. We make the approximation to consider only electron-electron
scattering such that $I_{ei}=0$. The equilibrium distribution of
non-interacting fermions at temperature $T$ is $f_{0,{\bf {k}}}=(\exp(\beta\xi_{{\bf {k}}})+1)^{-1}$.
The distributions are normalized such, that 
\begin{equation}
\int\frac{d{\bf {k}}}{(2\pi)^{3}}f_{{\bf {k}}}=\int\frac{d{\bf {k}}}{(2\pi)^{3}}f_{0,{\bf {k}}}=2n_{0},\label{eq:f0norm}
\end{equation}
for $T\rightarrow0$ and with $n_{0}=k_{\text{F}}^{3}/(3\pi^{2})$.
For small electric field ${\bf E}$ one can develop $f_{{\bf {k}}}$
around the equilibrium distribution $f_{0,{\bf {k}}}$ and the so-called
relaxation-time approximation amounts to express the collision integral
by $I_{ee}\left[f_{\mathbf{k}}\right]=(f_{\mathbf{k}}-f_{0,{\bf {k}}})/\tau$,
where $\tau$ is the relaxation time. Plugging this Ansatz in the
kinetic equation (\ref{eq:boltzmann}) we find 
\begin{align}
f_{\mathbf{k}} & \simeq f_{0,{\bf {k}}}+\tau e\mathbf{E}\cdot\nabla_{\mathbf{k}}f_{\mathbf{k}}\simeq f_{0,{\bf {k}}}+\tau e\mathbf{E}\cdot\mathbf{v}_{{\bf {k}}}\frac{\partial f_{0,{\bf {k}}}}{\partial\epsilon_{{\bf {k}}}}\nonumber \\
 & =f_{0,{\bf {k}}}-g_{{\bf {k}}}f_{0,{\bf {k}}}(1-f_{0,{\bf {k}}}),\label{eq:ff0dev}
\end{align}
with $g_{{\bf {k}}}=\beta\tau e\mathbf{E}\cdot\mathbf{v}_{{\bf {k}}}$.
Note, that the standard approximations, to replaced $f_{{\bf {k}}}$
on the right-hand-side again by $f_{0,{\bf {k}}}$, neglecting the
momentum dependence of $\tau$ and the dispersion relation for free
fermions $\epsilon_{{\bf {k}}}=k^{2}/(2m)$ 
now solve the Boltzmann equation (\ref{eq:boltzmann}) within relaxation-time
approximation and for small electric field.

The electron-electron collision integral is evaluated using Fermi's
golden rule which yields 
\begin{align}
 & I_{ee}\left[f_{\mathbf{k}}\right]=\frac{1}{V}\sum_{\mathbf{q}}\int_{-\infty}^{\infty}\!\!d\Omega\,\text{Im}\Phi_{\mathbf{q}}\left(\Omega\right)\delta\left(\epsilon_{\mathbf{k}}-\epsilon_{{\bf {k}+\pv-{\bf {q}}}}-\Omega\right)\nonumber \\
 & \times\Bigl[f_{\mathbf{k}}\left(1-f_{{\bf {k}+\pv-{\bf {q}}}}\right)(1+\nb\left(\Omega\right))-\left(1-f_{\mathbf{k}}\right)f_{{\bf {k}+\pv-{\bf {q}}}}\nb\left(\Omega\right)\Bigr].
\end{align}
Here $\nb(\Omega)=(\exp(\beta\Omega)-1)^{-1}$ is the Bose function.
Using the fact that $\text{Im}\Phi_{\mathbf{q}}\left(-\Omega\right)=-\text{Im}\Phi_{\mathbf{q}}\left(\Omega\right)$
we can rewrite above integral as 
\begin{align}
I_{ee}\left[f_{\mathbf{k}}\right] & =\frac{1}{V}\sum_{\mathbf{q}}\!\!\int_{-\infty}^{\infty}\!\!\!\!\!\!d\Omega\,\text{Im}\Phi_{\mathbf{q}}\left(\Omega\right)\nb\left(\Omega\right)\nonumber \\
 & \times\Bigl[f_{\mathbf{k}}\left(1-f_{{\bf {k}+\pv}-\mathbf{q}}\right)\delta\left(\epsilon_{{\bf {k}+\pv}-\mathbf{q}}-\epsilon_{\mathbf{k}}-\Omega\right)\nonumber \\
 & \quad-f_{{\bf {k}}+\pv-\mathbf{q}}\left(1-f_{\mathbf{k}}\right)\delta\left(\epsilon_{\mathbf{k}}-\epsilon_{{\bf {k}+\pv}-\mathbf{q}}-\Omega\right)\Bigr].
\end{align}
From Eq.\ (\ref{eq:ff0dev}) and since by definition $I_{ee}\left[f_{0,\mathbf{k}}\right]=0$
for the equilibrium distribution $f_{0,\mathbf{k}}$, we can rewrite
the collision integral as 
\begin{align}
 & I_{ee}\left[f_{\mathbf{k}}\right]=\frac{1}{V}\sum_{\mathbf{q}}\!\!\int_{-\infty}^{\infty}\!\!\!\!\!\!d\Omega\,\text{Im}\Phi_{\mathbf{q}}\left(\Omega\right)\nb\left(\Omega\right)\nonumber \\
 & \quad\times f_{0,{\bf {k}+\pv-\mathbf{q}}}(1-f_{0,{\bf {k}}})\left(g_{{\bf {k}+\pv-\mathbf{q}}}-g_{{\bf {k}}}\right)\delta\left(\epsilon_{\mathbf{k}}-\epsilon_{{\bf {k}+\pv}-\mathbf{q}}-\Omega\right),\label{eq:iee2}
\end{align}
where contributions $\sim E^{2}$ have been dropped. We see from Eq.\ (\ref{eq:iee2})
that this theory has a non-vanishing imbalance velocity factor, since
for ${\bf {q}}=0$, $\left(g_{{\bf {k}}+\pv}-g_{{\bf {k}}}\right)\neq0$.
The non-vanishing velocity imbalance factor provides that no additional
$T$ dependence arises from the angular part of the integral.

Setting $\vk=-{\bf {v}}_{{\bf {k}}+\pv}\simeq\kf/m\equiv\vf$ such
that the relaxation time approximation simplifies to \begin{subequations}
\begin{align}
g_{{\bf {k}}} & \simeq\beta\tau e{\bf {E}}\cdot\vf,\\
g_{{\bf {k}'}-\mathbf{q}} & \simeq-\beta\tau e{\bf {E}}\cdot\vf,
\end{align}
\end{subequations} we further multiply the collision integral with
$\vk\cdot\mathbf{e}$ where $\mathbf{e}$ is a unit vector in the
direction of $\mathbf{E}$ and sum over $\kv$. We find 
\begin{align}
\frac{1}{V}\sum_{\mathbf{k}}(\vk\cdot\mathbf{e})I_{ee}\left[f_{\mathbf{k}}\right] & =-2\beta\tau e\vf\cdot\mathbf{E}\,\tilde{I},\label{eq:ieekv}
\end{align}
and 
\begin{align}
 & \tilde{I}=\frac{1}{V^{2}}\sum_{\mathbf{k},\mathbf{q}}(\vk\cdot\mathbf{e})\int_{-\infty}^{\infty}\!\!\!\!\!\!d\Omega\,\text{Im}\Phi_{\mathbf{q}}\left(\Omega\right)\nb\left(\Omega\right)\nonumber \\
 & \quad\times\Bigl[f_{0,{\bf {k}}+\pv-\mathbf{q/2}}(1-f_{0,{\bf {k+q/2}}})\delta\left(\epsilon_{\mathbf{k+q/2}}-\epsilon_{{\bf {k}+\pv-\mathbf{q/2}}}-\Omega\right)\Bigr].\label{eq:tauint}
\end{align}
if we approximate $\vk\approx v_{F}$, such that $\tilde{I}=v_{\textrm{F}}I$
with 
\begin{align}
 & I=\frac{1}{V^{2}}\sum_{\mathbf{k},\mathbf{q}}\int_{-\infty}^{\infty}\!\!\!\!\!\!d\Omega\,\text{Im}\Phi_{\mathbf{q}}\left(\Omega\right)\nb\left(\Omega\right)\nonumber \\
 & \quad\times\Bigl[f_{0,{\bf {k}}+\pv-\mathbf{q/2}}(1-f_{0,{\bf {k+q/2}}})\delta\left(\epsilon_{\mathbf{k+q/2}}-\epsilon_{{\bf {k}}+\pv-\mathbf{q/2}}-\Omega\right)\Bigr].\label{eq:tauint}
\end{align}
For convenience, the electric field was oriented in the direction
of $\kf$ and we have symmetrize the last expression in ${\bf {q}}$.

The left-hand side of the Boltzmann Eq.\ (\ref{eq:boltzmann}) yields
\begin{equation}
-e\mathbf{E}\cdot\nabla_{\mathbf{k}}f_{\mathbf{k}}=-e\mathbf{E}\cdot\mathbf{v}_{\mathbf{k}}\frac{\partial f_{0,{\bf {k}}}}{\partial\epsilon_{{\bf {k}}}}.\label{eq:boltzlhs}
\end{equation}
where again the relaxation-time approximation Eq.\ (\ref{eq:ff0dev})
was used and we developed for small electric field ${\bf {E}}$. From
the definition of the equilibrium distribution $f_{0}$ we find for
zero temperature 
\begin{equation}
\frac{\partial f_{0,{\bf {k}}}}{\partial\epsilon_{{\bf {k}}}}=-\delta(\xi_{{\bf {k}}}).\label{eq:t0deriv}
\end{equation}
Moreover, let us define the density of states for free fermions, which
yields in $d=3$ and for vanishing temperature 
\begin{equation}
\nu(\epsilon)=\frac{1}{V}\sum_{{\bf {k}}}\delta(\epsilon-\epsilon_{{\bf {k}}})=\frac{m}{2\pi^{2}}\left(2m\epsilon\right)^{1/2}.\label{eq:dos}
\end{equation}
We then have 
\begin{align}
 & \frac{1}{V}\sum_{{\bf {k}}}\frac{\partial f_{0,{\bf {k}}}}{\partial\xi_{{\bf {k}}}}=-\frac{1}{V}\sum_{{\bf {k}}}\delta(\epsilon_{{\bf {k}}}-\epsilon_{\text{F}})\nonumber \\
 & =-\frac{1}{V}\sum_{{\bf {k}}}\delta(\epsilon_{{\bf {k}}}-\epsilon_{\text{F}})\int_{-\infty}^{\infty}d\epsilon\,\delta(\epsilon-\epsilon_{{\bf {k}}})\nonumber \\
 & =-\int_{-\infty}^{\infty}d\epsilon\,\delta(\epsilon-\epsilon_{\text{F}})\nu(\epsilon)=-\rho_{0},\label{eq:dos1}
\end{align}
with $\rho_{0}=\nu(\epsilon_{\text{F}})=v_{\text{F}}/(2\pi^{2})$
and with $\epsilon_{\text{F}}=k_{\text{F}}^{2}/(2m)$.

It is again advantageous to multiply Eq.\ (\ref{eq:boltzlhs}) by
${\bf {v}_{{\bf {k}}}}$ and to sum over ${\bf {k}}$. With the identity
${\sum_{{\bf {k}}}({\bf {E}\cdot{\bf {v}_{k}){\bf {v}_{{\bf {k}}}}}}}={\bf {E}}/3\sum_{{\bf {k}}}v_{{\bf {k}}}^{2}$
and the help of Eqns.\ (\ref{eq:t0deriv},\ref{eq:dos1}) we find
\begin{equation}
\frac{1}{V}\sum_{{\bf {k}}}\mathbf{v}_{{\bf {k}}}(\mathbf{E}\cdot\mathbf{v}_{{\bf {k}}})\frac{\partial f_{0,{\bf {k}}}}{\partial\epsilon_{{\bf {k}}}}=-\frac{\rho_{0}v_{\text{F}}^{2}{\bf {E}}}{3}.\label{eq:boltzlhs2}
\end{equation}
Multiplying above equation by ${\bf {E}}$ and together with Eq.\ (\ref{eq:ieekv}),
we can solve the Boltzmann equation for the relaxation time 
\begin{equation}
\tau^{-1}=6\beta I/\rho_{0}.\label{eq:bztau}
\end{equation}

\subsection{Connection between resistivity and relaxation time}

The electrical current density is \cite{MahanBook} 
\begin{equation}
{\bf {J}}=-2e\langle{\bf {v}}\rangle=-2e\frac{1}{V}\sum_{{\bf {k}}}f_{{\bf {k}}}{\bf {v}_{\bd{k}},\label{eq:currentj}}
\end{equation}
where the factor 2 accounts for the spin. Inserting Eqns.\ (\ref{eq:ff0dev})
in Eq.\ (\ref{eq:currentj}), the average with the equilibrium
distribution $f_{0,{\bf {k}}}$ vanishes and with Eqns.\ (\ref{eq:t0deriv},\ref{eq:boltzlhs2})
one finds 
\begin{equation}
{\bf {J}}=-\frac{2\tau e^{2}}{V}\sum_{{\bf {k}}}\vk({\bf {E}}\cdot\vk)\frac{\partial f_{0,{\bf {k}}}}{\partial\epsilon_{{\bf {k}}}}=\frac{2\tau\rho_{0}e^{2}v_{\text{F}}^{2}{\bf {E}}}{3}.\label{eq:currentj2}
\end{equation}
The conductivity $\sigma$ is defined as ${\bf {J}}=\sigma{\bf {E}}$,
such that the resistivity $\rho=\sigma^{-1}$ follows from the above
equation as 
\begin{equation}
\rho=3/(2\rho_{0}e^{2}v_{\text{F}}^{2}\tau).\label{eq:resistivity}
\end{equation}
Combining Eqns.\ (\ref{eq:bztau}) and (\ref{eq:resistivity}) the
resistivity is 
\begin{equation}
\rho=\left(\frac{3}{e\rho_{0}v_{\text{F}}}\right)^{2}\beta I.\label{eq:resbm0}
\end{equation}
The last step is to calculate the integral in Eq.\ (\ref{eq:tauint})
with $\text{Im}F_{{\bf {q}}}$ that we derived previously in section
\ref{sec:Bosonic-polarization-bubble}.

\subsection{Evaluation of the collision integral}

To evaluate the integral $I$ in Eq.\ (\ref{eq:tauint}) we obtain
from the dispersion relation Eq.\ (\ref{eq:xidef}) \begin{subequations}
\begin{align}
\xi_{\kv+\qv/2} & =v_{\xa}(k_{\xa}+q_{\xa}/2)+v_{\xe}k_{\xe}^{2}/2,\\
\xi_{{\bf {k}}+\pv-\qv/2} & =-v_{\xa}(k_{\xa}-q_{\xa}/2)+v_{\xe}k_{\xe}^{2}/2,
\end{align}
\end{subequations} such that 
\begin{align}
 & I=\frac{1}{V^{2}}\sum_{\mathbf{k},\mathbf{q}}\int_{-\infty}^{\infty}\!\!\!\!\!\!d\Omega\,\text{Im}\Phi_{\mathbf{q}}\left(\Omega\right)\nb\left(\Omega\right)\nonumber \\
 & \quad\times\Bigl[f_{0,{\bf {k}}+\pv-\mathbf{q/2}}(1-f_{0,{\bf {k+q/2}}})\delta\left(2v_{\xa}k_{\xa}-\Omega\right)\Bigr].\label{eq:tauint3}
\end{align}
Now the integration over $\Omega$ is trivially performed and we take
the infinite volume limit $V\rightarrow\infty$. This amounts to replace
\begin{subequations} 
\begin{align}
\frac{1}{V}\sum_{\mathbf{k}} & \rightarrow\int\frac{d\mathbf{k}}{(2\pi)^{3}}\equiv\frac{1}{(2\pi)^{3}}\int_{-\infty}^{\infty}dk_{\xa}\int d\mathbf{k}_{\xe},\\
\frac{1}{V}\sum_{\mathbf{q}} & \rightarrow\frac{1}{2\pi}\int_{-\infty}^{\infty}dq_{\xa},
\end{align}
\end{subequations} which yields 
\begin{align}
 & I=\frac{1}{(2\pi)^{4}}\int_{-\infty}^{\infty}dk_{\xa}\int d\mathbf{k}_{\xe}\int_{-\infty}^{\infty}dq_{\xa}\,\text{Im}\Phi_{\mathbf{q}}\left(2v_{\xa}k_{\xa}\right)\nonumber \\
 & \quad\times\nb\left(2v_{\xa}k_{\xa}\right)\Bigl[f_{0,{\bf {k}}+\pv-\mathbf{q/2}}(1-f_{0,{\bf {k+q/2}}})\Bigr].\label{eq:tauint4}
\end{align}
For small temperatures we further approximate 
\begin{equation}
f_{0,{\bf {k}}}\rightarrow\theta(-\xi_{{\bf {k}}})=1-\theta(\xi_{{\bf {k}}}),
\end{equation}
and we rescale the variables as 
\begin{equation}
2v_{\xa}k_{\xa}=x,\qquad v_{\xe}k_{\xe}^{2}=y,\qquad v_{\xa}q_{\xa}=q,
\end{equation}
in order to obtain 
\begin{align}
I & =c_{1}\int_{-\infty}^{\infty}dx\int_{0}^{\infty}dy\int_{-\infty}^{\infty}dq\,\text{Im}\Phi_{q}\left(x\right)\nb\left(x\right)\nonumber \\
 & \qquad\quad\times\theta(x+y+q)\theta(x-y-q),\label{eq:tauint5}
\end{align}
with $c_{1}=(4(2\pi)^{3}v_{\xa}^{2}v_{\xe})^{-1}$. The integral over
$y$ yields 
\begin{align}
 & \int_{0}^{\infty}dy\theta(x+y+q)\theta(x-y-q)\nonumber \\
 & \quad=(x-q)\theta(x+q)\theta(x-q)+2x\theta(x)\theta(x-q)\theta(-x-q).
\end{align}
Substituting this integral in the above equation (\ref{eq:tauint5})
gives $I=I_{1}+I_{2}$, with 
\begin{equation}
I_{1}=c_{1}\int_{0}^{\infty}dx\int_{-x}^{x}dq\,(x-q)\,\text{Im}\Phi_{q}\left(x\right)\nb\left(x\right),\label{eq:tauint6}
\end{equation}
and 
\begin{equation}
I_{2}=c_{1}\int_{0}^{\infty}dx\int_{-\infty}^{-x}dq\,2x\,\text{Im}\Phi_{q}\left(x\right)\nb\left(x\right).\label{eq:tauint6a}
\end{equation}
To extract the temperature scaling from $I$ we first rescaling the
variables as $\beta x=\tilde{x}$ and $\beta q=\tilde{q}$. With Eq.\ (\ref{eq:imdeffcomp})
the integrals in Eqns.\ (\ref{eq:tauint6},\ref{eq:tauint6a}) can
be written as $I_{1/2}=(2\pi^{2}v_{\xa}\beta^{2})^{-1}\tilde{I}_{1/2}$,
where 
\begin{align}
\tilde{I}_{1} & =\int_{0}^{\infty}d\tilde{x}\int_{-\tilde{x}}^{\tilde{x}}d\tilde{q}\,\frac{1}{e^{\tilde{x}}-1}\nonumber \\
 & \qquad\quad\times\frac{\tilde{t}^{2}}{\pi^{-2}[\tilde{s}\ln\tilde{s}-\tilde{t}\ln\tilde{t}+2\tilde{q}\ln T]^{2}+\tilde{t}^{2}},\label{eq:dimint}
\end{align}
and 
\begin{align}
\tilde{I}_{2} & =\int_{0}^{\infty}d\tilde{x}\int_{\tilde{x}}^{\infty}d\tilde{q}\,\frac{1}{e^{\tilde{x}}-1}\nonumber \\
 & \qquad\quad\times\frac{4\tilde{x}^{2}}{\pi^{-2}[\tilde{s}\ln\tilde{s}-\tilde{t}\ln(-\tilde{t})+2\tilde{q}\ln T]^{2}+4\tilde{x}^{2}}.\label{eq:diminta}
\end{align}
We have also used $c_{1}/c=1/(2\pi v_{\xa})$ which follows from above
definitions and defined $\tilde{s}=\tilde{x}+\tilde{q}$ and $\tilde{t}=\tilde{x}-\tilde{q}$.
Both integrands in Eqns.\ (\ref{eq:dimint},\ref{eq:diminta}) are
strictly positive within the integral bounds such that the resistivity,
which follows from Eq.\ (\ref{eq:resbm0}) as 
\begin{equation}
\rho=\frac{T}{2v_{\xa}}\left(\frac{3}{\pi e\rho_{0}v_{\text{F}}}\right)^{2}\tilde{I},\label{eq:resbm1}
\end{equation}
is positive and physically meaningful. In the last step we will evaluate
$\tilde{I}$ in the absence and then in the presence of logarithmic
corrections.

First, we will evaluate $\tilde{I}_{1/2}$ in Eqns.\ (\ref{eq:dimint},\ref{eq:diminta})
without logarithmic corrections, that is we set all logarithms to
one and neglect the explicit $\ln T$ term. One finds 
\begin{equation}
\tilde{I}_{1}\simeq\int_{0}^{\infty}d\tilde{x}\int_{-\tilde{x}}^{\tilde{x}}d\tilde{q}\,\frac{1}{e^{\tilde{x}}-1}=2\int_{0}^{\infty}d\tilde{x}\,\frac{\tilde{x}}{e^{\tilde{x}}-1}=\frac{\pi^{2}}{3},\label{eq:dimintlog1}
\end{equation}
and 
\begin{equation}
\tilde{I}_{2}\simeq\int_{0}^{\infty}d\tilde{x}\,\frac{\tilde{x}^{2}}{e^{\tilde{x}}-1}\int_{\tilde{x}}^{\infty}d\tilde{q}\frac{1}{(\tilde{q}/\pi)^{2}+\tilde{x}^{2}}=\frac{\pi^{3}\arctan\pi}{6},\label{eq:dimintlog2}
\end{equation}
and we have also neglected the $(2\tilde{q}/\pi)^{2}$ term in the
denominator of $\tilde{I}_{1}$ of Eq.\ (\ref{eq:dimint}) to perform
the integration. From Eq.\ (\ref{eq:resbm1}), the resistivity follows
as 
\begin{equation}
\rho=\frac{3(1+(\pi\arctan\pi)/2)T}{2v_{\xa}(e\rho_{0}v_{\text{F}})^{2}},\label{eq:resbm2}
\end{equation}
such that $\rho\sim T$.

Second, we evaluate $\tilde{I}_{1/2}$ in Eqns.\ (\ref{eq:dimint},\ref{eq:diminta})
including logarithmic corrections. Since $\tilde{q}<\tilde{x}$ in
$\tilde{I}_{1}$, we set $\tilde{x}-\tilde{q}\simeq\tilde{x}$. One
finds from Eq.\ (\ref{eq:dimint}) 
\begin{align}
\tilde{I}_{1}\simeq\int_{0}^{\infty}d\tilde{x}\int_{-\tilde{x}}^{\tilde{x}}d\tilde{q}\,\frac{1}{e^{\tilde{x}}-1}\frac{\tilde{x}^{2}}{(2\tilde{q}/\pi)^{2}[\ln\tilde{x}+\ln T]^{2}+\tilde{x}^{2}}.\label{eq:dimint1}
\end{align}
Above $\tilde{q}$ integral has the typical scale 
\begin{equation}
\tilde{q}_{\text{typ}}=-\frac{\pi\tilde{x}}{2(\ln\tilde{x}+\ln T)},
\end{equation}
such that the main contribution to the integral comes from the $|\tilde{q}|\leq\tilde{q}_{\text{typ}}$
sector while the tail $\tilde{x}\geq|\tilde{q}|>\tilde{q}_{\text{typ}}$
is only sub-leading. We have chosen the sign such that $\tilde{q}_{\text{typ}}$
is positive for $\tilde{x}<1$ and $T<1$. Neglecting the contribution
from the tails we obtain 
\begin{align}
\tilde{I}_{1} & \simeq\int_{0}^{\infty}d\tilde{x}\frac{1}{e^{\tilde{x}}-1}\int_{-\infty}^{\infty}d\tilde{q}\,\frac{1}{1+(\tilde{q}/\tilde{q}_{\text{typ}})^{2}}\nonumber \\
 & =-\frac{\pi^{2}}{2}\int_{0}^{\infty}d\tilde{x}\frac{\tilde{x}}{e^{\tilde{x}}-1}\,\frac{1}{\ln\tilde{x}+\ln T}.\label{eq:dimint2}
\end{align}
We split the remaining integral in two contributions $\tilde{I}_{1}\simeq\tilde{I}_{\gg T}+\tilde{I}_{\gg\tilde{x}}$,
such that the first one $\tilde{I}_{\gg T}$ captures the limit $1>\tilde{x}\gg T$
while the second one $\tilde{I}_{\gg\tilde{x}}$ the $1>T\gg\tilde{x}$
limit. For the $1>\tilde{x}\gg T$ limit 
\begin{align}
\tilde{I}_{\gg T} & \simeq-\frac{\pi^{2}}{2\ln T}\int_{T}^{\infty}d\tilde{x}\frac{\tilde{x}}{e^{\tilde{x}}-1}\nonumber \\
 & \simeq-\frac{\pi^{2}}{2\ln T}\int_{T}^{1}d\tilde{x}=-\frac{\pi^{2}}{2}\frac{1}{\ln T}+\mathcal{O}(T/\ln T).\label{eq:dimint3}
\end{align}
The $1>T\gg\tilde{x}$ limit gives 
\begin{align}
\tilde{I}_{\gg\tilde{x}} & \simeq-\frac{\pi^{2}}{2}\int_{0}^{T}d\tilde{x}\frac{\tilde{x}}{e^{\tilde{x}}-1}\frac{1}{\ln\tilde{x}}\nonumber \\
 & \simeq-\frac{\pi^{2}}{2}\int_{0}^{T}\frac{d\tilde{x}}{\ln\tilde{x}}\propto-\frac{T}{\ln T}.\label{eq:dimint4}
\end{align}
Next, we evaluate $\tilde{I}_{2}$ in Eq.\ (\ref{eq:diminta}) including
logarithmic corrections. Since $\tilde{q}>\tilde{x}$, we set $\tilde{q}\pm\tilde{x}\simeq\tilde{q}$
in the logarithms. One finds from Eq.\ (\ref{eq:diminta}) 
\begin{align}
\tilde{I}_{2}\simeq\int_{0}^{\infty}d\tilde{x}\int_{\tilde{x}}^{\infty}d\tilde{q}\,\frac{1}{e^{\tilde{x}}-1}\frac{\tilde{x}^{2}}{(\tilde{q}/\pi)^{2}[\ln\tilde{q}+\ln T]^{2}+\tilde{x}^{2}}.\label{eq:dimint2a}
\end{align}
Again, we split the integral in two contributions $\tilde{I}_{2}\simeq\tilde{I}_{\gg T}+\tilde{I}_{\gg\tilde{x}}$,
such that the first one $\tilde{I}_{\gg T}$ captures the limit $\tilde{x}\gg T$
while the second one $\tilde{I}_{\gg\tilde{x}}$ the $T\gg\tilde{x}$
limit and in both cases $T\ll1$. For the $\tilde{x}\gg T$ limit
\begin{align}
\tilde{I}_{\gg T} & \simeq\int_{T}^{\infty}d\tilde{x}\int_{\tilde{x}}^{\infty}d\tilde{q}\,\frac{\tilde{x}^{2}}{e^{\tilde{x}}-1}\frac{1}{(\tilde{q}\ln T/\pi)^{2}+\tilde{x}^{2}}\nonumber \\
 & \simeq-\frac{\pi^{2}}{2\ln T}\int_{T}^{\infty}d\tilde{x}\,\frac{\tilde{x}}{e^{\tilde{x}}-1}\simeq-\frac{\pi^{2}}{2}\frac{1}{\ln T}+\mathcal{O}(T/\ln T),\label{eq:dimint2b}
\end{align}
similar to Eq.\ (\ref{eq:dimint3}) and for $T\gg\tilde{x}$ one
obtains 
\begin{align}
\tilde{I}_{\gg\tilde{x}} & \simeq\int_{0}^{T}d\tilde{x}\int_{\tilde{x}}^{\infty}d\tilde{q}\,\frac{\tilde{x}^{2}}{e^{\tilde{x}}-1}\frac{1}{(\tilde{q}\ln T/\pi)^{2}+\tilde{x}^{2}}\nonumber \\
 & \simeq-\frac{\pi^{2}}{2\ln T}\int_{0}^{T}d\tilde{x}=-\frac{\pi^{2}}{2}\frac{T}{\ln T}.\label{eq:dimint3b}
\end{align}
The total integral is therefore 
\begin{equation}
\tilde{I}_{1}=-\frac{\pi^{2}}{\ln T}+\mathcal{O}(T/\ln T),
\end{equation}
such that the resistivity with Eq.\ (\ref{eq:resbm1}) yields 
\begin{equation}
\rho=-\frac{1}{2v_{\xa}}\left(\frac{3}{e\rho_{0}v_{\text{F}}}\right)^{2}\frac{T}{\ln T}+\mathcal{O}(T^{2}/\ln T).
\end{equation}
Because $T$ is small, the second contribution is subdominant and
the resistivity scales like $\rho\sim T/|\ln T|$, with a logarithmic
correction compared to Eq.\ (\ref{eq:resbm2}).

\bibliographystyle{apsrev4-1}
\bibliography{Cuprates}

\end{document}